\documentclass{pasj00}
\draft
\usepackage{rotating}
\usepackage{supertabular}
\usepackage{lscape}
\usepackage{pdflscape}

\begin{document}
\SetRunningHead{Author(s) in page-head}{Running Head}

\title{Using the Markov Chain Monte Carlo method to study the physical
properties GeV-TeV BL Lac objects}



%
 \author{%
   Longhua \textsc{Qin}\altaffilmark{1,2,3,5}
   Jiancheng \textsc{Wang}\altaffilmark{1,3,5}
   Chuyuan \textsc{Yang}\altaffilmark{1,3,5}
   Zunli \textsc{Yuan}\altaffilmark{1,3,5}
   Shiju \textsc{Kang}\altaffilmark{4}
   AND
   Jirong \textsc{Mao}\altaffilmark{1,3,5}
}
 \altaffiltext{1}{Yunnan Observatory, Chinese Academy of Sciences, Kunming,
Yunnan Province 650011, PR China}
 \email{qlh@ynao.ac.cn}
  \altaffiltext{2}{University of Chinese Academy of Sciences,Beijing, PR China }
 \altaffiltext{3}{Key Laboratory for the Structure and Evolution of Celestial Objects,
Chinese Academy of Sciences,  Kunming, PR China}
 \altaffiltext{4}{School of Electrical Engineering, Liupanshui Normal University, Liupanshui, Guizhou, 553004, China}
 \altaffiltext{5}{Center for Astronomical Mega-Science, Chinese Academy of Sciences, 20A Datun Road, Chaoyang District, Beijing, 100012, China}

\maketitle

\begin{abstract}

We fit the spectral energy distributions (SEDs) of 46 GeV - TeV
BL Lac objects in the frame of leptonic one-zone synchrotron self-Compton
(SSC) model and investigate the physical properties of these
objects. We use the Markov Chain Monte Carlo (MCMC) method to obtain the basic parameters, such as magnetic
field (B), the break energy of the relativistic electron
distribution ($\gamma'_{\rm{b}}$) and the electron energy spectral
index. Based on the modeling results, we support the following
scenarios on GeV-TeV BL Lac objects: (1) Some sources have large Doppler
factors, implying other radiation mechanism should be considered. (2)
Comparing with FSRQs, GeV-TeV BL Lac objects have weaker magnetic field and
larger Doppler factor, which cause the ineffective cooling and shift the
SEDs to higher bands. Their jet powers are around $4.0\times
10^{45}~\rm{ erg\cdot s}^{-1}$, comparing with radiation power,
$5.0\times 10^{42}~\rm{ erg\cdot s}^{-1}$, indicating that only a small
fraction of jet power is transformed into the emission power. (3)
For some BL Lacs with large Doppler factors, their jet components could have
two substructures, e.g., the fast core and the slow sheath. For
most GeV-TeV BL Lacs, Kelvin-Helmholtz instabilities are suppressed
by their higher magnetic fields, leading few micro-variability or
intro-day variability in the optical bands. (4) Combined with a sample of
FSRQs, an anti-correlation between the peak luminosity $L_{\rm {pk}}$
and the peak frequency $\nu_{\rm {pk}}$ is obtained, favoring the blazar
sequence scenario. In addition, an anti-correlation between the
jet power $P_{\rm {jet}}$ and the break Lorentz factor $\gamma_{\rm
{b}}$ also supports the blazar sequence.

\end{abstract}

\section{INTRODUCTION}

Blazars are the subclasses of radio-lond Active Galactic Nuclei
(AGNs), subdivided based on their emission lines: the flat spectrum
quasars (FSRQs) have strong broad emission lines while BL Lac objects
(BL Lacs) have weak or absent optical emission lines (EW$\leq 5
\mbox{\AA}$) \citep{urr99}. Their broadband emission is
mainly dominated by non-thermal components originated from a
relativistic jet aligned with our line of sight \citep{urr99}, and
shows two humps. The low hump, falling into IR and X-rays, is
explained with the relativistic electron synchrotron radiation; the
high peak, located at MeV and TeV bands, is explained by
the lepton or the hadron models \citep{bott2010,bott2013,cao2013b,zheng2016}. Blazars often exhibit
strong and fast variability across all electromagnetic spectrum.
The location of synchrotron peak ($\nu_{\rm {sy}}$) is used
to classify blazars as the low-synchrotron-peaked (LSP; $\nu_{\rm
{sy}}<$ 10$^{14}$Hz), the intermediate-synchrotron-peaked (ISP; 10$^{14}$Hz
$\leq\nu_{\rm {sy}} \leq$ 10$^{15}$Hz, and the high-synchrotron-peaked blazars (HSP; $\nu_{\rm {sy}} >$
10$^{15}$Hz) by \cite{abdo2010a}.

BL Lac objects are thought to be ``blue'' quasars with weak or no
external seed photons plus an inefficient accretion disk
\citep{nara1997,bland1999}, their SEDs suffer less
contamination by external photons and give us an opportunity to
explore the intrinsic physical properties of emitting region as well
as the jet. Comparing with the FSRQs, BL Lacs have lower jet power and
inefficient accretion ratio. The SEDs of BL Lacs modeled by
a certain radiation mechanism allow us to investigate
the physical properties. With a large number of blazars,
some authors suggested that the jet comprises a dominant
proton component and a small fraction of jet power is radiated if
there is one proton per electron \citep{cel2008,yan2014} , and this assumption was also
analyzed by \cite{tan2015}.

For the blazar sequence \citep{fos1998,kubo1998}, it is explained as that the
radiative cooling is stronger in more powerful blazar. The blazar
sequence is formally expressed as the anti-correlation between the
peak luminosity ($L_{\rm pk}$) and the peak frequency of the
synchrotron component ($\nu_{\rm pk}$) or the anti-correlation
 between the jet power $P_{\rm {jet}}$ and the break Lorentz factor $\gamma_{\rm
{b}}$. Some authors suggest that the
sequence is a result of the selection effect \citep{pad2003,nie2006,chen2011,gio2005,gio2012}. However, other authors
propose that the blazar sequence still holds theoretically \citep{Ghi1998,bott2002,fin2013}.

TeV BL Lacs usually show a less amount
of optical variability than the LBLs do. A few sources of them
have large Doppler factor $\delta_{\rm D}$, which are not consistent
with ones by VLBI observations \citep{piner2014}. A new physical scenario
has been proposed to fit their SEDs \citep{tav2008,chen2017}, in which
a lower $\delta_{\rm D}$ is needed and supported by
observations \citep{gian2011}.

The statistical study of physical properties on BL Lacs is needed.
Some authors \citep{zhang2012,yan2014,ino2016, ding2017}) use the
samples of TeV-GeV BL Lacs to obtain the physical properties by
fitting the SEDs. However, the number of objects in samples is small
(\cite{zhang2012}), and the method to fit the SEDs needs the better
error evaluation \citep{yan2014,man2011,man2012}. In addition,
\cite{ding2017} also use a sample of TeV BL Lacs to investigate the
physical properties based on a log-parabolic spectrum of electron
energy distributions (EEDs). This type of EEDs could reflect
stochastic acceleration in the jet. Considering the impact of the
EEDs on the SEDs \citep{yan2013,qin2017}, we use the broken
power-law spectrum of EEDs to fit the SEDs of a sample of BL Lacs
that contains HBL, IBL and LBL. It is noted that this type of EEDs
could be produced in the emitting region and is commonly used to fit
the SEDs of blazars.

For BL Lac objects, the simplest model is the homogeneous one-zone
SSC model. This model has been considerable successes in reproducing
the broadband SEDs of all classes of blazars
\citep{Ghi2010b,zhang2012,xiong2014}, in which the SEDs with two
bumps are assumed to be produced by the synchrotron and the inverse
Compton (IC) emissions of ultra-relativistic particles
\citep{fin2008,Ghi2010b}. In addition, the high-energy gamma-ray
photons are attenuated due to the extragalactic background light
(EBL) absorption \citep{per2008}.  The observed VHE flux in the
energy $E_{\rm{\gamma}}$ is given by
$f_{\rm{obs}}(E_{\rm{\gamma}})=f_{\rm{int}}(E_{\rm{\gamma}})\times
e^{-\rm{\tau}{(E_{\rm{\gamma}},z)}}$, where $f_{\rm{obs}}$ and
$f_{\rm{int}}$ are the observed and intrinsic flux respectively, and
$\rm{\tau}{(E_{\rm{\gamma}},z)}$ is the optical depth of $E_\gamma$
photon which depends on the choice of the EBL template. In the
paper, we use the EBL model proposed by \citep{raz2009} and
\citep{fin2010} to rebuild the SEDs. We explore the high-dimensional
model parameters using the MCMC method in fitting (quasi-)
simultaneous multi-band spectra based on one-zone SSC scenario. The
MCMC method used here is adapted from the public code
``CosmoMC''\footnote{http://cosmologist.info/cosmomc/} offered by
\cite{lew2002,mac2003}. For details, please refer the papers
\citep{mac2003,yuan2011,yan2013,yuan2016} and a review in Sec. 2.

Throughout this work, we take Hubble constant $H_0$ = 70 km$\cdot$s$^{-1} \cdot$Mpc$^{-1}$, $\Omega_{\rm M}$
=0.3, and $\Omega_{\Lambda}$=0.7 to calculate the luminosity distance.

\section{MODEL AND STRATEGY }

In the SSC scenario, we apply an one-zone spherical blob of the jet
filled with the uniform magnetic field $B$, moving
with velocity $\beta = \nu/c$ Lorentz factor $\Gamma = (1-\beta^2)^{-1/2}$ at a small angle ($\theta$) to the line of
sight, where $c$ is the speed of light. The SED is
produced by both the synchrotron radiation and the SSC process, while the observed SED is strongly enhanced
by a relativistic Doppler factor given by $\delta_{\rm D} = [\Gamma(1 -
 {\beta}\rm {cos}\theta)]^{-1}$, where $\delta_{\rm D} \approx \Gamma$ if $\theta \approx 1/\Gamma$.

We assume the size $R'_{\rm b}$ of a blob to be calculated by $R'_{\rm b} \approx \delta_{\rm D} t_{\rm {v,min}} c(1+z)^{-1}$,
where $ t_{\rm {v,min}}$ is the minimum variability time-scale. Note that the primes are used for the quantities in
the rest frame of the black hole, while the unprimed quantities are defined in the observer frame or the blob's frame.
The electron spectrum is described by a broken power-law distribution
with the form

\begin{equation}
N(\gamma')=\left\{
    \begin{array}{ll}
    K'_e\gamma'^{-p_{1}}    &\mbox{$\gamma'_{\rm {min}}\leq\gamma'\leq\gamma'_{\rm{b}}$}\\
    K'_e{\gamma'_{b}}^{p_{2}-p_{1}}\gamma'^{-p_{2}}   &\mbox{$\gamma'_{\rm{b}}<\gamma'\leq\gamma'_{\rm {max}}$}\\
    \end{array},
\right.
\end{equation}
where $\gamma'_{b}$ is the break Lorentz factor, $p_{1,2}$ is the spectral
index below and above $\gamma'_{\rm {b}}$, $K'_e$ is the normalization factor. Note that the magnetic field $B$ is defined
in the blob's frame.

The synchrotron flux ($\nu F_\nu$) is given by \citep{sau2004,fin2008}
\begin{equation}
f_{\rm {\epsilon}}^{\rm {syn}}=\frac{\sqrt{3}\delta_{\rm D}^{4}\epsilon'e^{3}B}{4\pi
d_{\rm L}^2}\int_{1}^{\infty}d\gamma'N(\gamma')R(x),
\end{equation}
where $e$ is the fundamental charge, $h$ is the Planck constant, $d_{\rm{L}}$ is the luminosity distance with the redshift $z$,
$\epsilon'=[h\nu](1+z)m_ec^2]/\delta_{\rm D}$ is the dimensionless energy of synchrotron photons,
$m_e$ and $c$ are the mass of electron and the speed of light. Other quantities in equation (2) are
$x=4\pi\varepsilon' m_e^2c^4/3eBh\gamma'^2$,
$R(x)=2x^2\{K_{4/3}(x)K_{1/3}(x)-0.6x[K_{4/3}^2(x)-K_{1/3}^2(x)]\}$,
and $K_{\alpha}(x)$ is the modified $\alpha$-order Bessel function, and its numerical integration can be found in \cite{fin2008}.

We use the SSC model described by \citep{fin2008}
\begin{equation}
f_{\epsilon_{s}}^{\rm {ssc}}=\frac{9\sigma_{\rm T}\epsilon'_{s}}{16\pi
R'_{b}}\int_{0}^{\infty}d\epsilon_s'\frac{f_\epsilon^{\rm syn}}{\epsilon'^{3}}\int_{\gamma'_{\rm {min}}}^{\gamma'_{\rm {max}}}
d\gamma'\frac{N(\gamma')}{\gamma'^{2}}F(q,\Gamma_e),
\end{equation}
where $\sigma_{\rm T}$ is the Thomson cross-section, $\epsilon_s'=[h\nu_s](1+z)m_ec^2]/\delta_{\rm D}$ is the dimensionless energy
of IC scattered photons and the function $F(q,\Gamma_e)$ is given by
\begin{eqnarray}
F(q,\Gamma_e)=[2q\rm{ln}q+(1+2q)(1-q)+\frac{1}{2}\frac{(\Gamma_e q)^2}{(1+\Gamma_e q)}(1-q)]\quad
  \nonumber \\
\quad(\frac{1}{4\gamma'}<q<1),
\end{eqnarray}
where $q=\frac{\varepsilon'/\gamma'}{\Gamma_e(1-\varepsilon'/\gamma')}$ and $\Gamma_e=4\varepsilon'\gamma'$.
In the GeV-TeV regime, the above function has already considered the KN effect, which makes the IC inefficiency.

As shown above, there are nine parameters in the SSC model, including
the size of blob $R'_{\rm{b}}$, the magnetic field $B$, the Doppler
factor $\delta_{\rm {D}}$, and the electron spectrum
($p_1, p_2, \gamma'_{\rm {min}},\gamma'_{\rm {b}}, \gamma'_{\rm {max}},
K'_{\rm{e}}$). $\gamma'_{\rm {min}}$ is always poorly constrained by
the SED modelling. The $\gamma'_{\rm {min}}$ of some sources are
getting from \cite{zhang2012}, and are
set via the method offered by \cite{tav2000}. For the sources
not included in \cite{zhang2012}, we also use the method offered by
\cite{tav2000} to obtain the $\gamma'_{\rm {min}}$. In this
method, the $\gamma'_{\rm {min}}$ is obtained by modeling the radio
to X-ray data based on the particle distribution with a power law. If no observational data is
available to constrain $\gamma'_{\rm {min}}$, we set $\gamma'_{\rm
{min}}$ as 5.0 based on the pretreatment. From $t_{\rm {v,min}}$, we
can get the blob's size. For the sources without minimum
variability, we simply set $t_{\rm v,min}$  as one day
\citep{Ghi1998,fos2008,cao2013}. Because the model is not sensitive to
$\gamma'_{\rm {max}}$, $\gamma'_{\rm {max}} =100\gamma'_{\rm
{b}}$ is adopted.

The MCMC technique is well suitable to search multi-dimensional
parameter space and obtains the uncertainties of the model
parameters based on the observational data
\citep{yuan2011,yan2014,yan2015}. According to the Bayes' Theorem,
the posterior probability of a model with a set of parameters
(hereafter $\vec{\theta}$) upon the data (hereafter $\cal D$) is
given by
\begin{equation}
\cal P (\vec{\theta}\mid D)~\propto {\cal L}(D\mid \vec{\theta})\cal P(\vec{\theta}),
\end{equation}
where $ {-\rm {ln}\cal L}(D\mid \vec{\theta})~\propto
{\sum\limits_{i=1}^{N}(\frac{f_{\rm i}-f_{\rm {obs}}}{\sigma_{\rm
{obs}}})^2}$ is the likelihood function, and $f_{\rm i}$ is the
model flux in the different band and $N$ is the number of data
associated with the band, $f_{\rm {obs}}$ and $\sigma_{\rm {obs}}$  are
the flux of observational data and its variance respectively. The
MCMC ensures that the probability functions of model parameters can
be asymptotically obtained by the number density of samples.
Comparing with the least-square fitting method, the MCMC can give
the better error evaluation and the confident levels (C.L.) of
parameters. Furthermore, for a complex model, the MCMC can
obtain the fitting results much faster than the chi-square minimization does.

After calculations, two probability distributions can be obtained.
The maximum probability is exactly the same as the best-fit one
obtained by minimizing the likelihood. The marginalized
probability distributions is the probability distribution of the
parameters contained in the subset. It gives the probabilities of
various values of the parameters in the subset and reflects the confident levels of
parameters. To get the same result as in the best-fit method, the
marginalized probability distributions require the large number
density of samples to run in the calculation procedure. In the paper, we use the
best-fit parameters to rebuild the SEDs and give the confident levels of the
parameters in the 68\%. It is noted that if the parameters are
constrained well, then two types of distributions will have the
similar shape and interval.

\section{APPLICATIONS}
Our sample contains 46 $Fermi$ BL Lacs objects, in which the
board-band SEDs cover from radio, optical, X-ray to $\gamma$-ray
bands. The different types of blazars are from Roma-BZCAT catalog
\citep{mas2010} and TeVCat \footnote{http://tevcat.uchicago.edu},
which contain 32 HBLs, 10 IBLs and 4 LBLs. In the paper, for some BL
Lacs with bad $Fermi$ data (such as only flux upper limits), their
(quasi-) simultaneous $Fermi$ data are not used to reproduce the
SEDs. Instead, for these objects, we add the GeV gamma-ray data
which are from the $Fermi$-LAT 4-year Point Source(3FGL) catalog
\citep{ace2015}. Although these data are not simultaneous with the
optical, X-ray and TeV bands, they will give a rough constrain on
the SEDs and do not affect our results. It is noted that we do not
include the 4-year Point data to fit the objects at the flare stage
in our sample. The rest data in many bands are from other
instruments such as KAV, $Suzaku$, $Beppo$SAX, $Swift$, H.E.S.S and
MAGIC. The simultaneous or quasi-simultaneous SEDs data of BZB
J1058+5628, OT 081, PKS 0048-09, PKS 0851+202, 1H 1013+498,  BZB
J0033-1921 and PG 1246+586 are taken from \cite{gio2012}. The
optical-UV and other band data of PKS 0426-380, 4C 01.28 are gotten
from \cite{gio2012} and \cite{abdo2010a} respectively. The data of
B3 2247+381, 1ES 1215+303,  RBS 0413, 1H 0414+009  PKS 0447-439  are
obtained from \cite{ale2012a}, \cite{ale2012b}, \cite{ali2012a},
\cite{ali2012b}, and \cite{pra2012}. The SEDs data of remaining
objects are taken from the literatures compiled by \cite{zhang2012}.

\begin{table*}
\caption{Model parameters derived in the one-zone SSC model. The mean values and the marginalized 68\%
confidence intervals (CI) are reported \label{Table:1}.}
\centering
\tiny
\tabcolsep 1.2mm
\begin{tabular}{l|ccc ccccc ccc}
\hline \hline
Source name$^a$ & $z$ & $ B(0.1 G)$ & $\gamma'_{\rm {min}}$  &Log[${\gamma'_{\rm {b}}}$] & $\delta_{\rm D}(10)$& Log[K${'_e}$] & $p_1$ & $p_2$ & $t_{\rm {var}}^b$& Log$\frac{M_{\rm B}}{M_{\hbox{$\odot$}}}$&Ref$^c$ \\
~[1] & [2] & [3] & [4] & [5] & [6] & [7] &  [8] & [9] & [10] & [11] & [12] \\
\hline
    1ES0229+200&0.140& 0.010&  5.00&  6.16&  6.09& 53.36&  2.02&  2.62&24.0&9.2&~1~\\
    (68\% CI)&~-~& 0.010~-~ 0.024&~-~& 5.99~-~ 6.25& 4.79~-~ 6.02&53.19~-~53.55& 2.00~-~ 2.09& 2.29~-~ 3.10&~-~&~-~&~-~\\
    1ES0347-121&0.185& 0.016&100.00&  5.30&  7.94& 51.43&  1.68&  2.92&12.0$^1$&8.7&~1~\\
    (68\% CI)&~-~& 0.016~-~ 0.027&~-~& 5.15~-~ 5.35& 6.79~-~ 8.00&50.58~-~52.14& 1.51~-~ 1.84& 2.84~-~ 2.99&~-~&~-~&~-~\\
    1ES0806+524&0.138& 0.377&  5.00&  4.91&  2.44& 54.87&  2.45&  4.07&24.0&8.9&~2~\\
    (68\% CI)&~-~& 0.335~-~ 0.766&~-~& 4.70~-~ 4.90& 1.83~-~ 2.58&53.38~-~55.00& 2.12~-~ 2.44& 3.83~-~ 4.21&~-~&~-~&~-~\\
    1ES1011+496&0.212& 0.559&  5.00&  5.11&  2.83& 52.30&  1.81&  4.22&24.0&8.3&~-~\\
    (68\% CI)&~-~& 0.267~-~ 0.865&~-~& 5.05~-~ 5.24& 2.45~-~ 3.56&52.07~-~52.68& 1.77~-~ 1.88& 4.14~-~ 4.43&~-~&~-~&~-~\\
    1ES1101-232&0.186& 0.030&  5.00&  5.58&  7.99& 52.07&  1.87&  3.56&12.0$^1$&0.0&~-~\\
    (68\% CI)&~-~& 0.030~-~ 0.039&~-~& 5.47~-~ 5.61& 7.35~-~ 8.00&51.47~-~52.34& 1.75~-~ 1.93& 3.47~-~ 3.62&~-~&~-~&~-~\\
  1ES1101-232 f&0.186&17.420&  5.00&  4.98&  0.51& 54.30&  2.41&  4.56&12.0$^1$&0.0&~-~\\
  (68\% CI)&~-~&12.888~-~70.174&~-~& 4.68~-~ 4.97& 0.23~-~ 0.62&50.83~-~54.13& 1.61~-~ 2.38& 3.57~-~ 4.45&~-~&~-~&~-~\\
    1ES1215+303&0.130& 0.100&100.00&  4.31&  3.54& 53.18&  1.85&  3.57&24.0&8.5&~3~\\
    (68\% CI)&~-~& 0.100~-~ 0.140&~-~& 4.22~-~ 4.37& 3.12~-~ 3.51&53.03~-~53.37& 1.80~-~ 1.91& 3.53~-~ 3.62&~-~&~-~&~-~\\
   1ES1218+30.4&0.184& 0.174&100.00&  4.74&  4.33& 48.41&  1.00&  3.60&12.0$^1$&8.6&~1~\\
   (68\% CI)&~-~& 0.136~-~ 0.309&~-~& 4.68~-~ 4.80& 3.59~-~ 4.64&48.47~-~49.14& 1.00~-~ 1.17& 3.53~-~ 3.72&~-~&~-~&~-~\\
    1ES1959+650&0.047& 0.689&  5.00&  5.51&  2.84& 53.85&  2.46&  4.33&10.0$^1$&8.1&~1~\\
    (68\% CI)&~-~& 0.461~-~ 0.878&~-~& 5.46~-~ 5.59& 2.60~-~ 3.30&53.75~-~53.99& 2.44~-~ 2.48& 4.19~-~ 4.50&~-~&~-~&~-~\\
    1ES2344+514&0.044& 0.650&100.00&  4.84&  1.34& 51.74&  1.86&  3.64&12.0$^2$&8.8&~1~\\
    (68\% CI)&~-~& 0.638~-~ 1.991&~-~& 4.64~-~ 4.93& 0.79~-~ 1.36&48.81~-~53.68& 1.17~-~ 2.30& 3.56~-~ 3.73&~-~&~-~&~-~\\
  1ES2344+514 f&0.044& 0.022&100.00&  6.30&  3.47& 54.86&  2.37&  4.07&12.0$^2$&8.8&~1~\\
  (68\% CI)&~-~& 0.019~-~ 0.286&~-~& 5.71~-~ 6.23& 1.47~-~ 2.54&54.23~-~54.75& 2.32~-~ 2.40& 2.94~-~ 5.00&~-~&~-~&~-~\\
     1H0414+009&0.287& 0.101&100.00&  5.01&  5.42& 53.68&  2.10&  3.79&24.0&0.0&~-~\\
     (68\% CI)&~-~& 0.100~-~ 0.117&~-~& 4.96~-~ 5.03& 5.13~-~ 5.44&53.31~-~53.98& 2.02~-~ 2.17& 3.75~-~ 3.84&~-~&~-~&~-~\\
     1H1013+498&0.212& 3.261&100.00&  4.59&  1.44& 54.01&  2.24&  3.82&24.0&0.0&~-~\\
     (68\% CI)&~-~& 2.682~-~ 4.492&~-~& 4.54~-~ 4.62& 1.30~-~ 1.52&53.94~-~54.13& 2.22~-~ 2.29& 3.71~-~ 3.91&~-~&~-~&~-~\\
          3C66A&0.444& 0.385&170.00&  4.60&  4.48& 52.91&  1.79&  5.07&12.0$^1$&8.6&~3~\\
          (68\% CI)&~-~& 0.328~-~ 0.475&~-~& 4.56~-~ 4.62& 4.11~-~ 4.80&52.36~-~53.20& 1.66~-~ 1.87& 5.01~-~ 5.13&~-~&~-~&~-~\\
     B32247+381&0.119& 0.310&100.00&  5.07&  2.34& 52.98&  2.05&  4.01&24.0&0.0&~-~\\
     (68\% CI)&~-~& 0.208~-~ 0.491&~-~& 5.02~-~ 5.17& 1.99~-~ 2.65&52.86~-~53.20& 2.01~-~ 2.12& 3.86~-~ 4.31&~-~&~-~&~-~\\
  BZBJ0033-1921&0.610& 0.101&100.00&  4.27&  5.42& 50.02&  1.02&  3.34&24.0&0.0&~-~\\
  (68\% CI)&~-~& 0.100~-~ 0.342&~-~& 4.08~-~ 4.25& 3.77~-~ 5.29&50.16~-~52.04& 1.00~-~ 1.53& 3.31~-~ 3.35&~-~&~-~&~-~\\
  BZBJ1058+5628&0.143& 1.642&100.00&  4.08&  1.43& 54.18&  2.25&  3.68&24.0&8.7&~3~\\
  (68\% CI)&~-~& 2.755~-~12.155&~-~& 3.72~-~ 3.99& 0.71~-~ 1.20&51.67~-~54.21& 1.62~-~ 2.30& 3.61~-~ 3.71&~-~&~-~&~-~\\
      H1426+428&0.129& 0.101&200.00&  5.47&  1.26& 48.56&  1.00&  2.86&24.0&9.1&~1~\\
      (68\% CI)&~-~& 0.100~-~ 0.174&~-~& 5.51~-~ 5.82& 0.79~-~ 1.13&48.63~-~49.37& 1.00~-~ 1.17& 2.78~-~ 2.95&~-~&~-~&~-~\\
      H2356-309&0.165& 0.014& 5.00&  5.79&  5.65& 54.94&  2.33&  3.39&24.0&8.6&~1~\\
      (68\% CI)&~-~& 0.017~-~ 0.065&~-~& 5.53~-~ 5.75& 3.44~-~ 5.30&54.43~-~54.89& 2.27~-~ 2.35& 3.31~-~ 3.46&~-~&~-~&~-~\\
         MRK421&0.030& 0.546& 20.00&  4.84&  3.83& 50.24&  1.67&  4.05& 3.0$^4$&8.3&~1~\\
         (68\% CI)&~-~& 0.507~-~ 0.641&~-~& 4.80~-~ 4.85& 3.62~-~ 3.94&50.18~-~50.29& 1.66~-~ 1.68& 4.01~-~ 4.08&~-~&~-~&~-~\\
       MRK421 f&0.030& 0.116& 20.00&  5.07&  8.00& 49.95&  1.57&  3.50& 3.0$^4$&8.3&~1~\\
       (68\% CI)&~-~& 0.114~-~ 0.134&~-~& 5.04~-~ 5.11& 7.54~-~ 8.00&49.90~-~50.05& 1.56~-~ 1.59& 3.44~-~ 3.55&~-~&~-~&~-~\\
         MRK501&0.034& 0.495&200.00&  5.40&  1.83& 55.00&  2.61&  3.99&12.0$^5$&9.2&~1~\\
         (68\% CI)&~-~& 0.331~-~ 0.724&~-~& 5.31~-~ 5.48& 1.57~-~ 2.13&54.70~-~55.26& 2.54~-~ 2.66& 3.89~-~ 4.12&~-~&~-~&~-~\\
       MRK501 f&0.034& 1.238&  5.00&  5.64&  1.94& 48.57&  1.48&  2.71& 1.0$^6$&9.2&~1~\\
       (68\% CI)&~-~& 1.085~-~ 1.428&~-~& 5.61~-~ 5.67& 1.85~-~ 2.03&48.36~-~48.78& 1.44~-~ 1.53& 2.67~-~ 2.75&~-~&~-~&~-~\\
         Mkn180&0.045& 0.432&100.00&  4.87&  1.35& 51.69&  1.80&  3.62&24.0$^7$&8.2&~2~\\
         (68\% CI)&~-~& 0.218~-~ 5.271&~-~& 4.57~-~ 4.84& 0.42~-~ 1.26&50.25~-~52.75& 1.44~-~ 2.02& 3.50~-~ 3.68&~-~&~-~&~-~\\
     PG1553+113&0.360& 0.101&100.00&  4.73&  7.76& 53.02&  1.83&  3.97&24.0&8.6&~3~\\
     (68\% CI)&~-~& 0.100~-~ 0.128&~-~& 4.67~-~ 4.76& 7.07~-~ 7.70&52.36~-~53.42& 1.68~-~ 1.93& 3.94~-~ 4.00&~-~&~-~&~-~\\
    PKS0447-439&0.107& 0.762&100.00&  4.70&  2.06& 55.23&  2.49&  4.36&24.0&8.8&~3~\\
    (68\% CI)&~-~& 0.616~-~ 0.906&~-~& 4.63~-~ 4.76& 1.90~-~ 2.24&54.77~-~55.53& 2.38~-~ 2.56& 4.23~-~ 4.49&~-~&~-~&~-~\\
    PKS1424+240&0.160& 0.811&100.00&  4.52&  3.72& 54.99&  2.33&  5.56&24.0$^8$&0.0&~-~\\
    (68\% CI)&~-~& 0.677~-~ 0.938&~-~& 4.48~-~ 4.54& 3.52~-~ 3.95&54.62~-~55.21& 2.24~-~ 2.38& 5.34~-~ 5.73&~-~&~-~&~-~\\
    PKS2005-489&0.071& 0.115&200.00&  4.39&  8.00& 54.19&  2.37&  3.20&10.0$^1$&8.1&~5~\\
    (68\% CI)&~-~& 0.116~-~ 0.139&~-~& 4.34~-~ 4.40& 7.46~-~ 8.00&53.88~-~54.48& 2.30~-~ 2.45& 3.17~-~ 3.23&~-~&~-~&~-~\\
    PKS2155-304&0.116& 0.407&400.00&  4.43&  8.00& 51.31&  1.81&  4.06& 2.0$^{10}$&8.7&~6~\\
    (68\% CI)&~-~& 0.397~-~ 0.430&~-~& 4.39~-~ 4.47& 7.90~-~ 8.00&50.69~-~51.72& 1.66~-~ 1.92& 4.02~-~ 4.10&~-~&~-~&~-~\\
  PKS2155-304 f&0.116& 0.183&200.00&  4.95&  7.65& 54.37&  2.37&  4.02& 2.0$^{10}$&8.7&~6~\\
  (68\% CI)&~-~& 0.183~-~ 0.423&~-~& 4.63~-~ 5.05& 5.39~-~ 8.00&51.97~-~54.97& 1.80~-~ 2.50& 3.90~-~ 4.25&~-~&~-~&~-~\\
        RBS0413&0.190& 0.104&100.00&  5.18&  3.19& 53.75&  2.14&  3.42&24.0&8.0&~1~\\
        (68\% CI)&~-~& 0.100~-~ 0.225&~-~& 4.98~-~ 5.24& 2.44~-~ 3.09&52.54~-~54.26& 1.89~-~ 2.27& 3.29~-~ 3.53&~-~&~-~&~-~\\
\hline
\end{tabular}
\end{table*}

\begin{table*}

\begin{flushleft}
\hspace{0.5cm}Table 1. Continue--\\
\end{flushleft}
\centering
\tiny
\tabcolsep 1.2mm
\begin{tabular}{l|ccc ccccc ccc}
\hline \hline
Source name$^a$ & $z$ & $ B(0.1 G)$ & $\gamma'_{\rm {min}}$  &Log[${\gamma'_{\rm {b}}}$] & $\delta_{\rm D}(10)$& Log[K${'_e}$] & $p_1$ & $p_2$ & $t_{\rm {var}}^b$& Log$\frac{M_{\rm B}}{M_{\hbox{$\odot$}}}$&Ref$^c$ \\
~[1] & [2] & [3] & [4] & [5] & [6] & [7] &  [8] & [9] & [10] & [11] & [12] \\
\hline
   RGBJ0152+017&0.080& 0.180& 5.00&  5.04&  1.77& 56.16&  2.70&  3.27&24.0&0.0&~-~\\
   (68\% CI)&~-~& 0.101~-~ 2.434&~-~& 4.54~-~ 5.18& 0.52~-~ 1.50&51.53~-~55.35& 1.70~-~ 2.56& 3.07~-~ 3.27&~-~&~-~&~-~\\
   RGBJ0710+591&0.125& 0.118&800.00&  4.99&  3.32& 54.83&  2.46&  2.37&24.0&8.3&~1~\\
   (68\% CI)&~-~& 0.100~-~ 0.700&~-~& 4.82~-~ 5.00& 1.85~-~ 3.14&54.03~-~54.72& 2.31~-~ 2.47& 2.36~-~ 2.45&~-~&~-~&~-~\\
\hline
     S50716+714&0.300& 0.393&100.00&  3.90&  9.97& 52.03&  1.78&  3.91& 3.0$^9$&8.6&~7~\\
     (68\% CI)&~-~& 0.368~-~ 0.655&~-~& 3.80~-~ 3.89& 8.31~-~10.00&50.04~-~52.44& 1.24~-~ 1.90& 3.88~-~ 3.93&~-~&~-~&~-~\\
   S50716+714 f&0.300& 0.115&100.00&  4.24&  9.50& 52.14&  1.68&  3.91& 3.0$^9$&8.6&~7~\\
   (68\% CI)&~-~& 0.100~-~ 0.678&~-~& 3.96~-~ 4.22& 4.58~-~ 9.07&50.03~-~52.32& 1.00~-~ 1.73& 3.83~-~ 3.94&~-~&~-~&~-~\\
    PKS0851+202&0.306& 1.481&100.00&  3.49&  2.41& 54.10&  1.87&  5.16&24.0&8.7&~3~\\
    (68\% CI)&~-~& 1.467~-~ 1.505&~-~& 3.49~-~ 3.50& 2.38~-~ 2.43&54.08~-~54.12& 1.86~-~ 1.88& 5.13~-~ 5.18&~-~&~-~&~-~\\
 PKS0048-09&0.634& 0.207&100.00&  3.96&  5.16& 54.44&  1.98&  3.74&24.0&0.0&~-~\\
     (68\% CI)&~-~& 0.165~-~ 0.269&~-~& 3.85~-~ 4.03& 4.63~-~ 5.72&54.06~-~54.71& 1.84~-~ 2.08& 3.70~-~ 3.78&~-~&~-~&~-~\\
     PG1246+586&0.847& 0.080&100.00&  4.36&  7.62& 53.14&  1.72&  4.00&24.0&0.0&~-~\\
     (68\% CI)&~-~& 0.079~-~ 0.162&~-~& 4.26~-~ 4.37& 6.09~-~ 8.00&53.12~-~53.24& 1.72~-~ 1.77& 3.97~-~ 4.03&~-~&~-~&~-~\\
    BLLacertae&0.069& 1.556& 50.00&  3.43&  2.82& 52.09&  1.78&  4.00& 3.0$^3$&8.2&~1~\\
     (68\% CI)&~-~& 1.395~-~ 1.734&~-~& 3.41~-~ 3.45& 2.71~-~ 2.94&51.99~-~52.16& 1.74~-~ 1.82& 3.96~-~ 4.03&~-~&~-~&~-~\\
   BLLacertae f&0.069& 2.265& 50.00&  3.68&  2.02& 53.59&  2.33&  3.51& 3.0$^3$&8.2&~1~\\
   (68\% CI)&~-~& 0.821~-~25.789&~-~& 3.51~-~ 4.54& 1.05~-~ 2.13&53.57~-~54.59& 2.34~-~ 3.00& 3.28~-~ 3.99&~-~&~-~&~-~\\
\hline
           Wcom&0.102& 0.128&200.00&  4.43&  3.73& 53.42&  2.01&  3.66&12.0$^{11}$&7.4&~4~\\
           (68\% CI)&~-~& 0.115~-~ 0.198&~-~& 4.38~-~ 4.51& 3.26~-~ 3.86&53.34~-~53.67& 1.98~-~ 2.10& 3.62~-~ 3.79&~-~&~-~&~-~\\
         Wcom f&0.102& 0.440&200.00&  4.18&  2.06& 53.61&  2.00&  3.50&12.0$^{11}$&7.4&~4~\\
         (68\% CI)&~-~& 0.101~-~ 0.709&~-~& 4.14~-~ 4.29& 1.68~-~ 2.94&53.18~-~53.89& 1.88~-~ 2.08& 3.46~-~ 3.56&~-~&~-~&~-~\\
        PKS0426-380&1.110& 0.004&100.00&  3.25& 12.00& 53.25&  1.00&  2.90&24.0&8.6&~3~\\
    (68\% CI)&~-~& 0.004~-~ 0.005&~-~& 3.22~-~ 3.30&11.95~-~12.00&53.22~-~53.33& 1.00~-~ 1.03& 2.87~-~ 2.96&~-~&~-~&~-~\\
       4C01.28&0.890& 0.116&100.00&  3.96&  6.21& 55.90&  2.27&  4.82&24.0&0.0&~-~\\
        (68\% CI)&~-~& 0.103~-~ 0.132&~-~& 3.91~-~ 4.03& 5.72~-~ 6.56&55.70~-~56.21& 2.20~-~ 2.39& 4.52~-~ 5.05&~-~&~-~&~-~\\
          OT081&0.322& 0.073&100.00&  4.12&  4.61& 55.07&  2.05&  5.37&24.0&8.7&~3~\\
          (68\% CI)&~-~& 0.071~-~ 0.077&~-~& 4.10~-~ 4.14& 4.53~-~ 4.62&55.05~-~55.14& 2.05~-~ 2.08& 5.11~-~ 5.54&~-~&~-~&~-~\\
   PKS1717+177&0.137& 0.101&100.00&  4.68&  1.53& 55.73&  2.37&  4.72&24.0&8.5&~3~\\
    (68\% CI)&~-~& 0.100~-~ 0.110&~-~& 4.65~-~ 4.75& 1.48~-~ 1.53&55.70~-~55.79& 2.36~-~ 2.39& 4.60~-~ 5.57&~-~&~-~&~-~\\

\hline
\end{tabular}
\vskip 0.4 true cm
\small
{\it a: }{The $``\rm f"$ represents the high stage of object.}
{\it b: }{The minimum variability timescales refer to the following references: (1)\cite{zhang2012}; (2)\cite{acc2011}; (3)\cite{rav2002}; (4)\cite{bla2005}; (5)\cite{and2009a}; (6)\cite{tav2001}; (7)\cite{alb2006}; (8)\cite{acc2010}; (9)\cite{fos2006}; (10)\cite{aha2009}; (11)\cite{acc2009}}
{\it c: }{The black hole mass refer to the following references: (1)\cite{woo2002};  (2)\cite{wu2002};  (3)\cite{Ghi2010a};  (4)\cite{lia2003}; (5)\cite{wag2008};  (6)\cite{aha2009};  (7)\cite{and2009b}}
\end{table*}

\section{RESULTS AND DISCUSSION}

We have studied the sample that contains 39 GeV-TeV $Fermi$ BL Lacs
, in which 7 sources have the flare states. As demonstrated above,
we use a simple SSC model to reproduce the SEDs, in which the
parameters are listed in Table 1 and the SEDs are shown in the
figures of the appendix. From the figures, for 5 HBLs, such as 1ES
1011+496, 1H 0414+009, BZBJ 0033-1921, BZBJ 1058+5628, MRK 421, 1
IBL S5 0716-714 and 1 LBL OT 081, the simple SSC model does not well
fit their SEDs. For some objects, such as PG 1246+586 and  PKS
0851+202, their VHE ($\geq$ 100 GeV) have a upward trend, and their
origin is still debated. In addition, for these objects with 4-year
$Fermi$-LAT data, such as H 2356-309 and PKS 2005-489, their SEDs in
GeV band are not well fitted.

\subsection{Distributions of Model Parameters}
From the Table 1 and Fig. 1, the redshift distribution of GeV-TeV BL Lacs extends from 0.03 to 1.11, the derived black hole masses are around $10^8 M_{\hbox{$\odot$}}$. It is found that the values of $B$ are more extreme than that given by \cite{Ghi2010a} and \cite{zhang2012}, and most of them are around 0.01, except for 1ES 1101-232 in the high stage, because its emission mechanism is still on debate. It is implied that lower magnetic field could lead to ineffective cooling and $\gamma'_b$ will shift to higher value causing hard Gamma rays. The Doppler factor $\delta_{\rm D}$ for most sources is clustered at 11.6, however, a few are larger than 50 and even reach to 100. In addition, the Doppler distribution seems to have a double-hump, which could be caused by the limited numbers of the sample or the unsuitable SSC model, and we will further discuss this problem later. It is noted that the objects in our sample are all low redshift objects.

\begin{figure}
\begin{center}
\includegraphics[width=50mm]{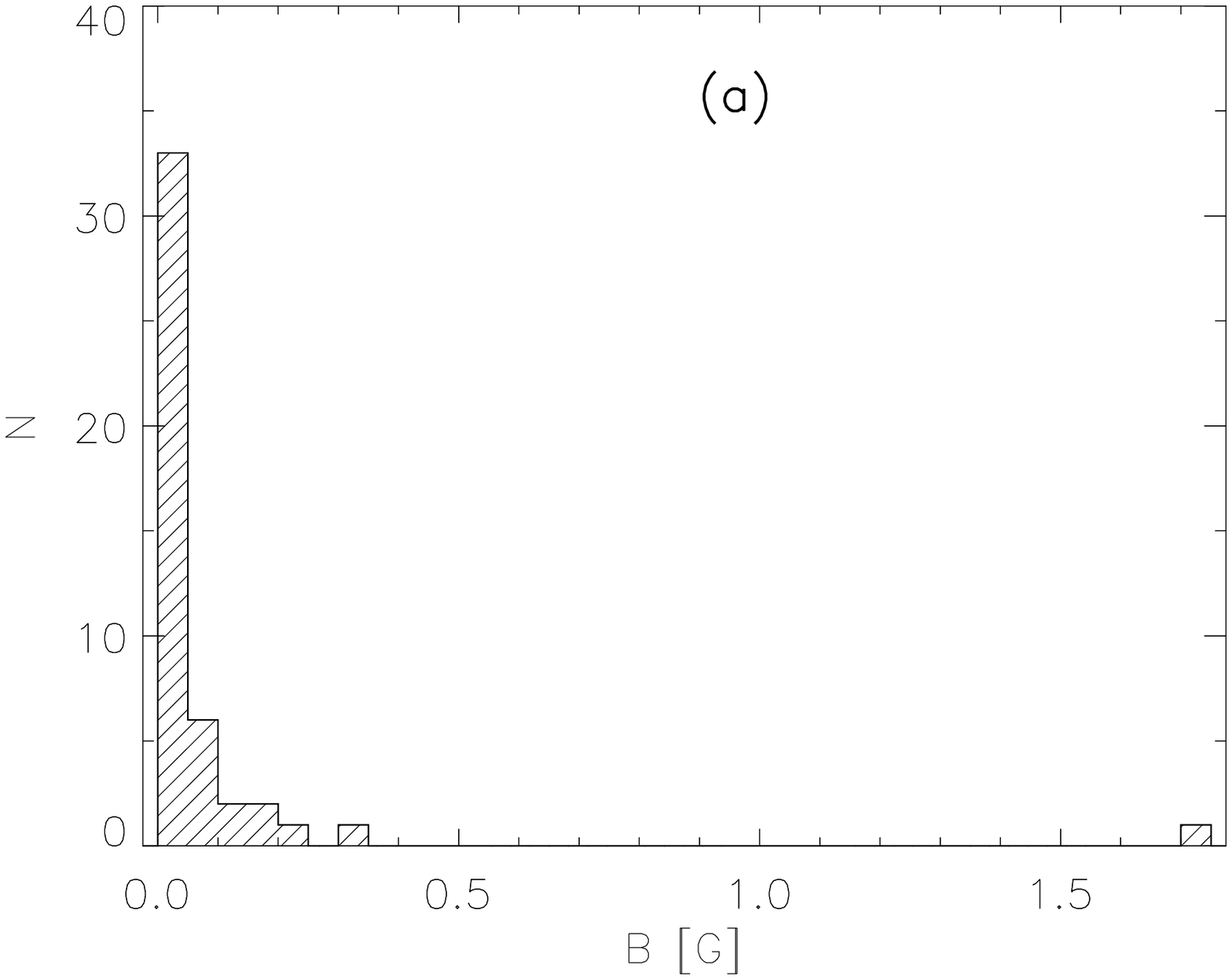}
\includegraphics[width=50mm]{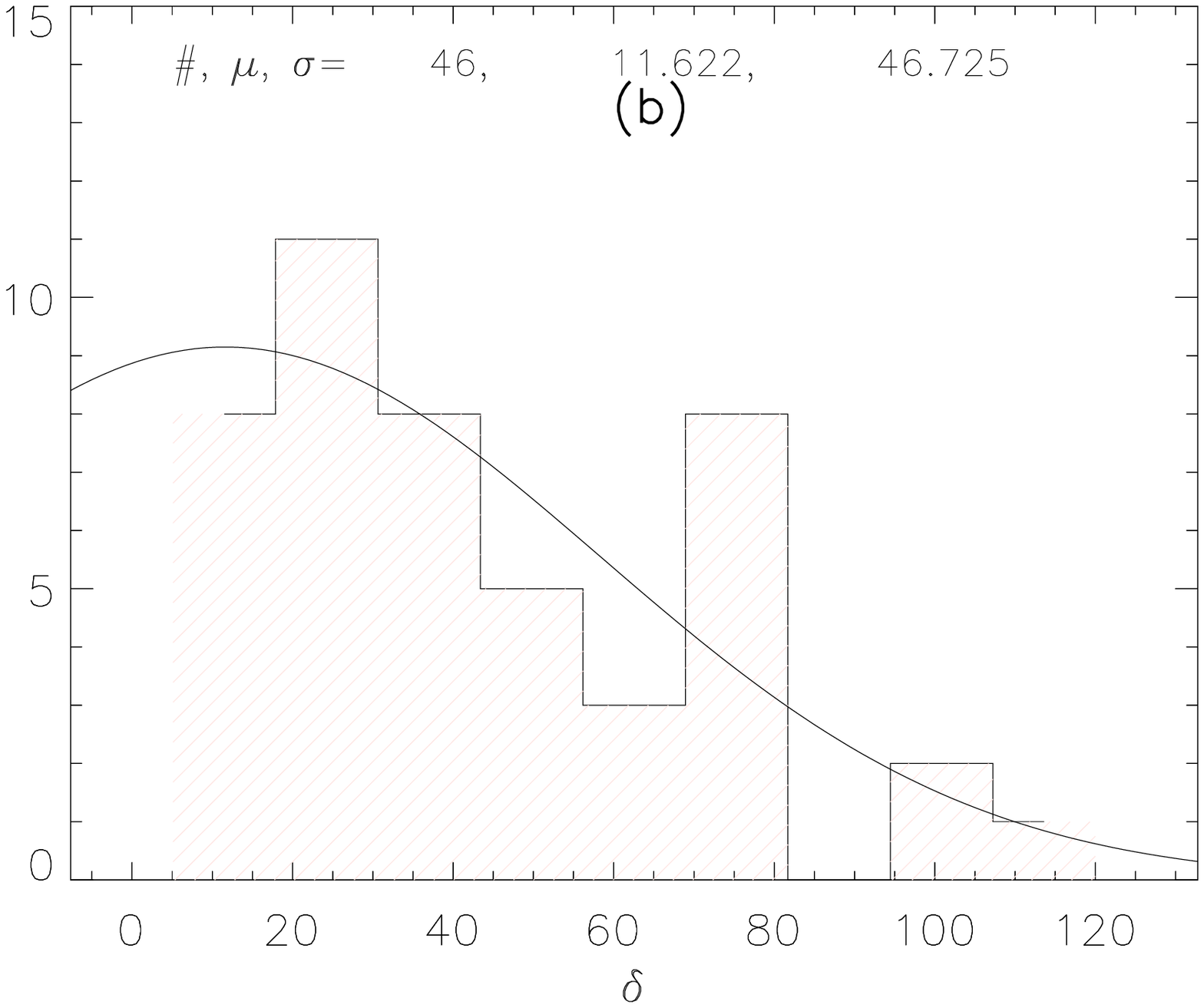}

\includegraphics[width=50mm]{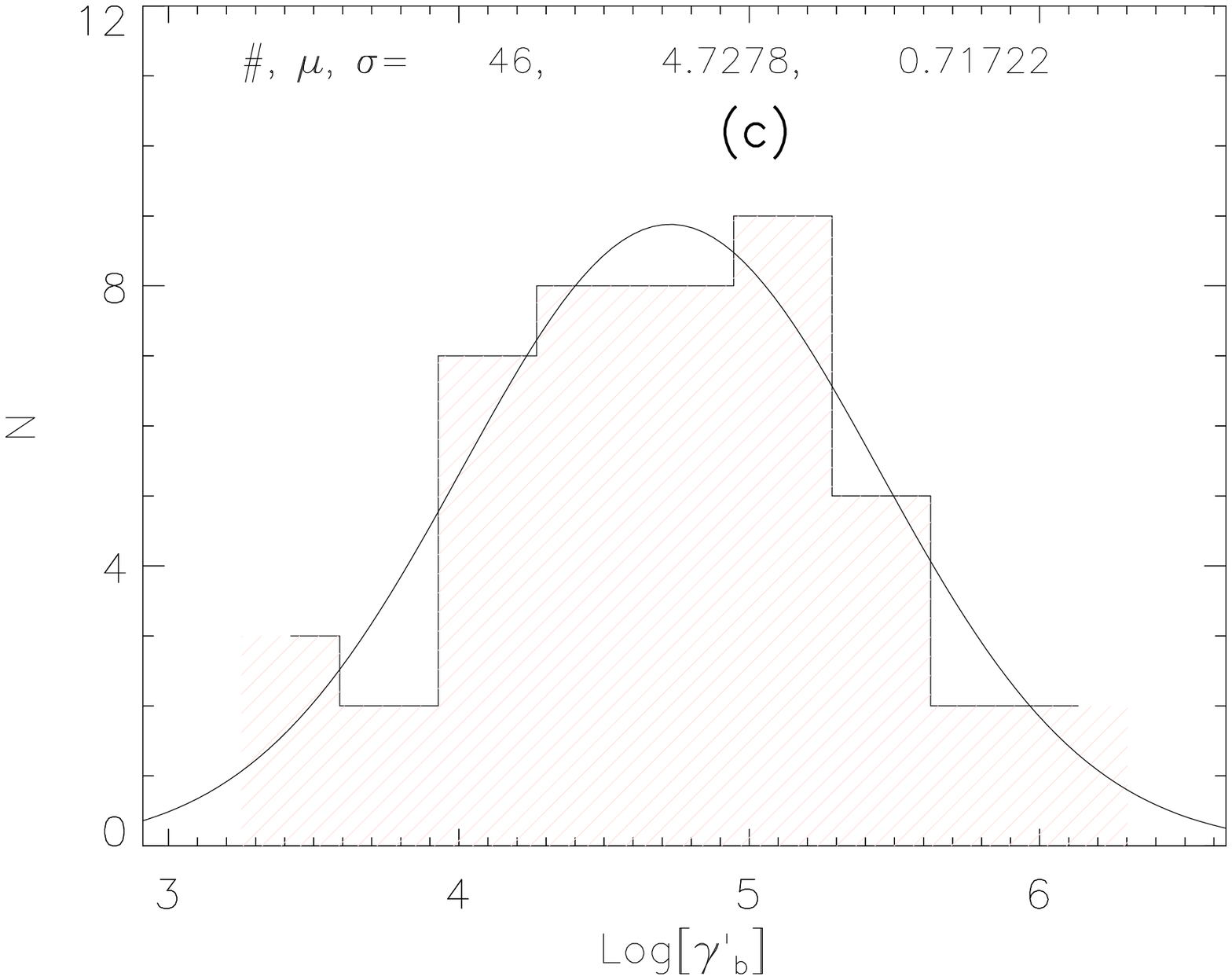}
\includegraphics[width=50mm]{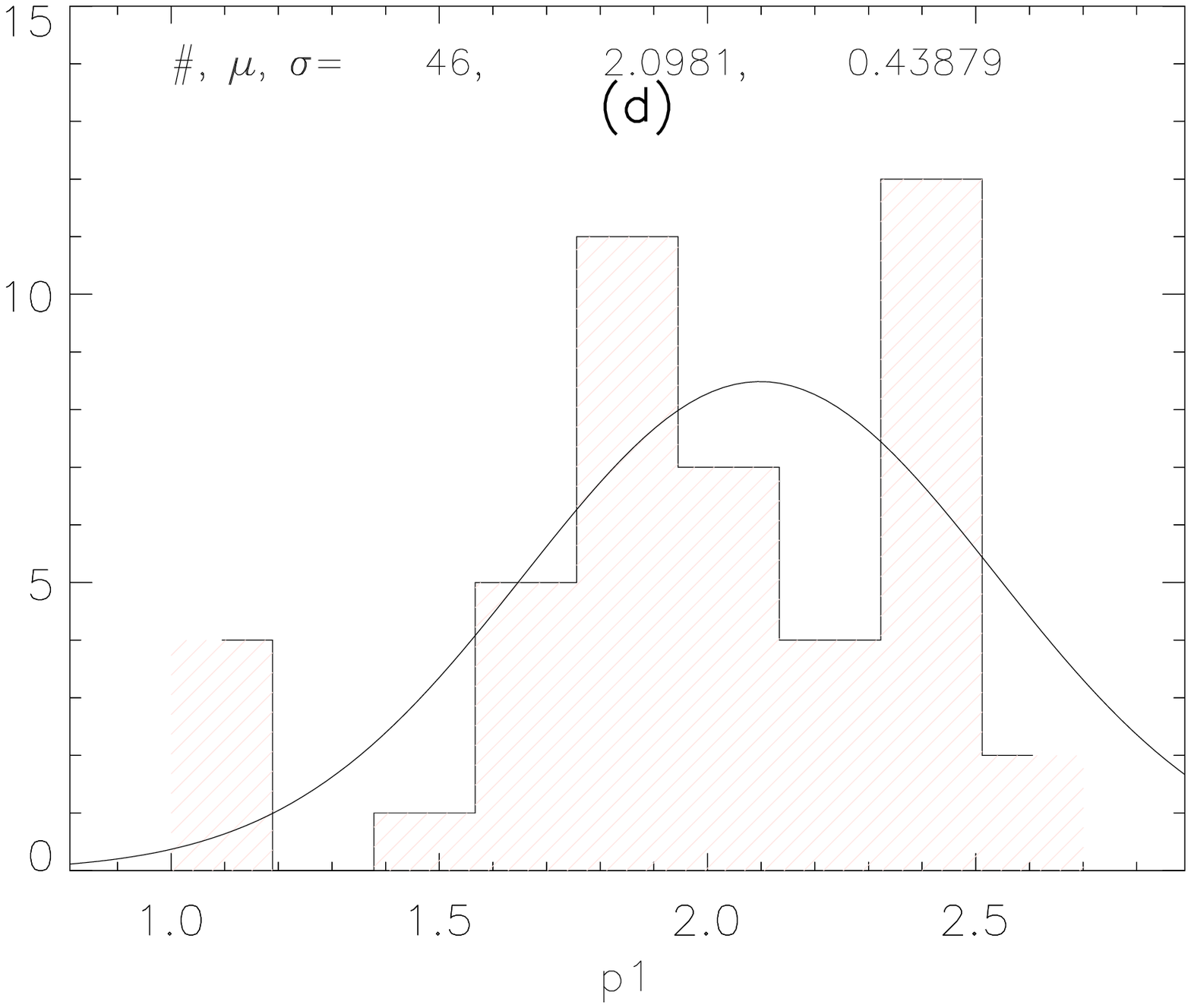}
\includegraphics[width=50mm]{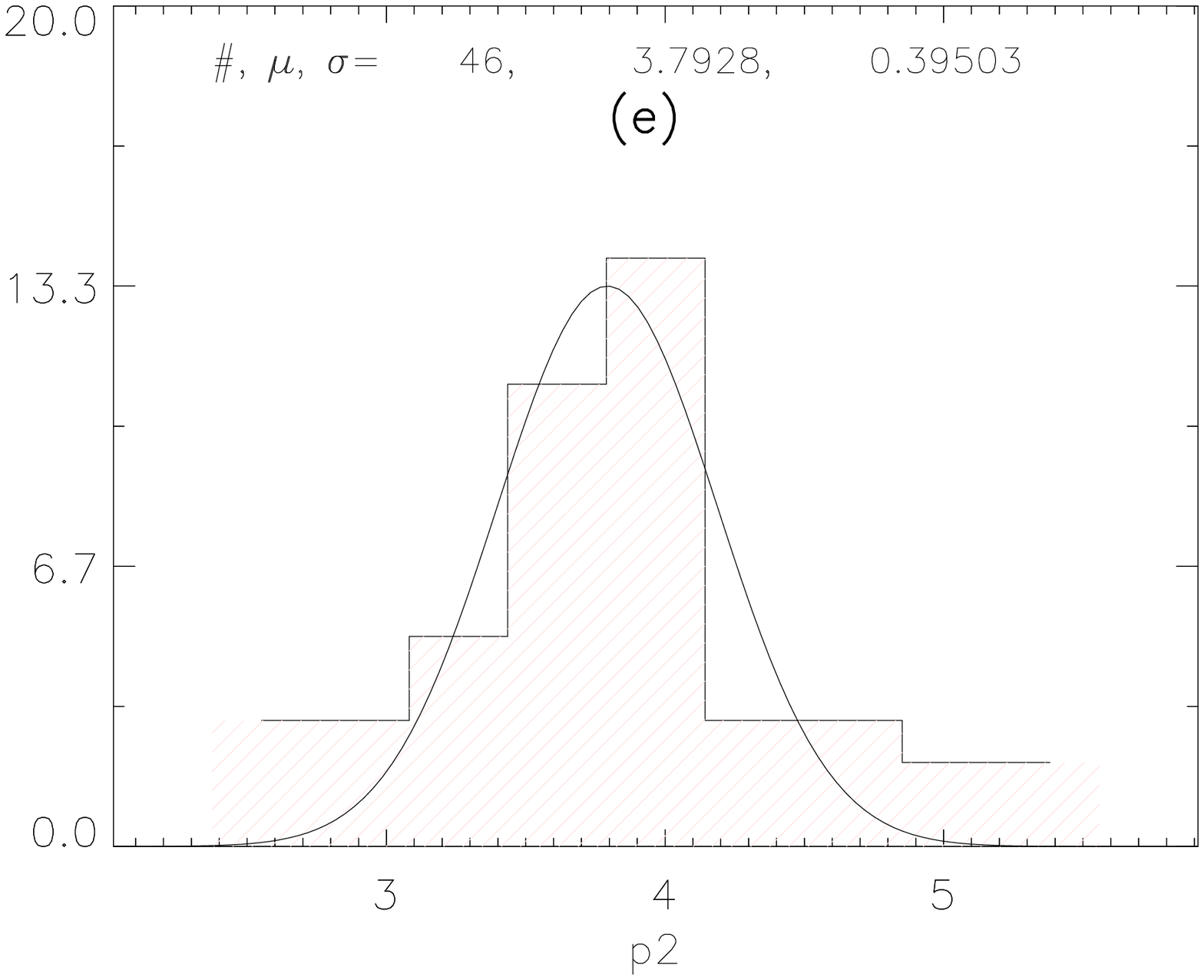}
\end{center}
\caption{(a) and (b) The distributions of the magnetic field B and the
beaming factor $\delta_{\rm D}$. (c) The distribution of the break
Lorentz factor $\gamma_b$. (d) and (e) The spectral index $p_1$ and
$p_2$. The Gauss function is used to fit the modeling parameter.
``\#'', ``$\mu$'' and ``$\sigma$'' represent the number of the BL Lac object,
the mean and the standard deviation of the modeling parameter respectively.}

\end{figure}

\subsection{Properties of Jet}
We calculate the jet power and the radiative power using the parameters obtained from our model. The jet power can be estimated by
$P_{\rm {jet}}=\pi R'^2_{\rm {b}}\Gamma^2c(U'_{\rm{e}}+U'_{\rm {p}}+U'_{\rm {B}})$, where the emitting electron $U'_{\rm {e}}$, the cold proton $U'_{\rm {p}}$ and the magnetic field $U'_{\rm {B}}$ are given by
\begin{equation}
U'_{\rm {e}}=m_{\rm {e}} c^2 \int{n(\gamma)\gamma}{d\gamma},
\end{equation}
\begin{equation}
U'_{\rm {p}}=m_{\rm {p}} c^2 \int{n(\gamma)}{d\gamma},
\end{equation}
\begin{equation}
U'_{\rm {B}}=\frac{B^2}{8\pi},
\end{equation}
where $n(\gamma$)=${ N(\gamma)}/{\frac{4\pi R'^3_{\rm b}}{3}}$ obtained by the assumption of
one proton per emitting electron. For
the radiative power $P_{\rm {r}}$, it can be estimated by the
total non-thermal luminosity as
\begin{equation}
P_{\rm {r}}=L'\Gamma^2 \approx L \frac{\Gamma^2}{\delta^4},
\end{equation}
where $L$ is obtained by the SED fits. The results are reported in Table 2 and Fig. 2. It is found that the values of $Log P_{\rm {r}}$,$Log P_{\rm B}$, $Log P_{\rm e}$, $Log P_{\rm p}$ and $Log P_{\rm {jet}}$ are clustered at 42.7, 42.6, 44.8, 45.5 and 45.6 respectively. The distribution of radiative power shows a bimodal and could be caused by the limited numbers of the sample. In our sample, we can see from the table: (1)there is $P_e > P_B$, showing that the jets are particle-dominated; (2) there is $P_r \sim P_B$, indicating that the Poynting flux accounts for the observed radiation; (3) there are $P_r < P_e $ and $P_p \sim P_{jet}$, suggesting that an additional energy reservoir of protons is needed to accelerate electrons (\cite{tan2015}). Most sources have the equipartition parameter $U'_{\rm e}/U'_{\rm B}$ to be $1-10^3$, but some sources have $U'_{\rm e}/U'_{\rm B}$ to be larger than several thousands, and even to $10^6$, implying that one zone SSC model could be unsuitable.

\begin{figure}
\begin{center}
\includegraphics[width=16cm,height=14cm]{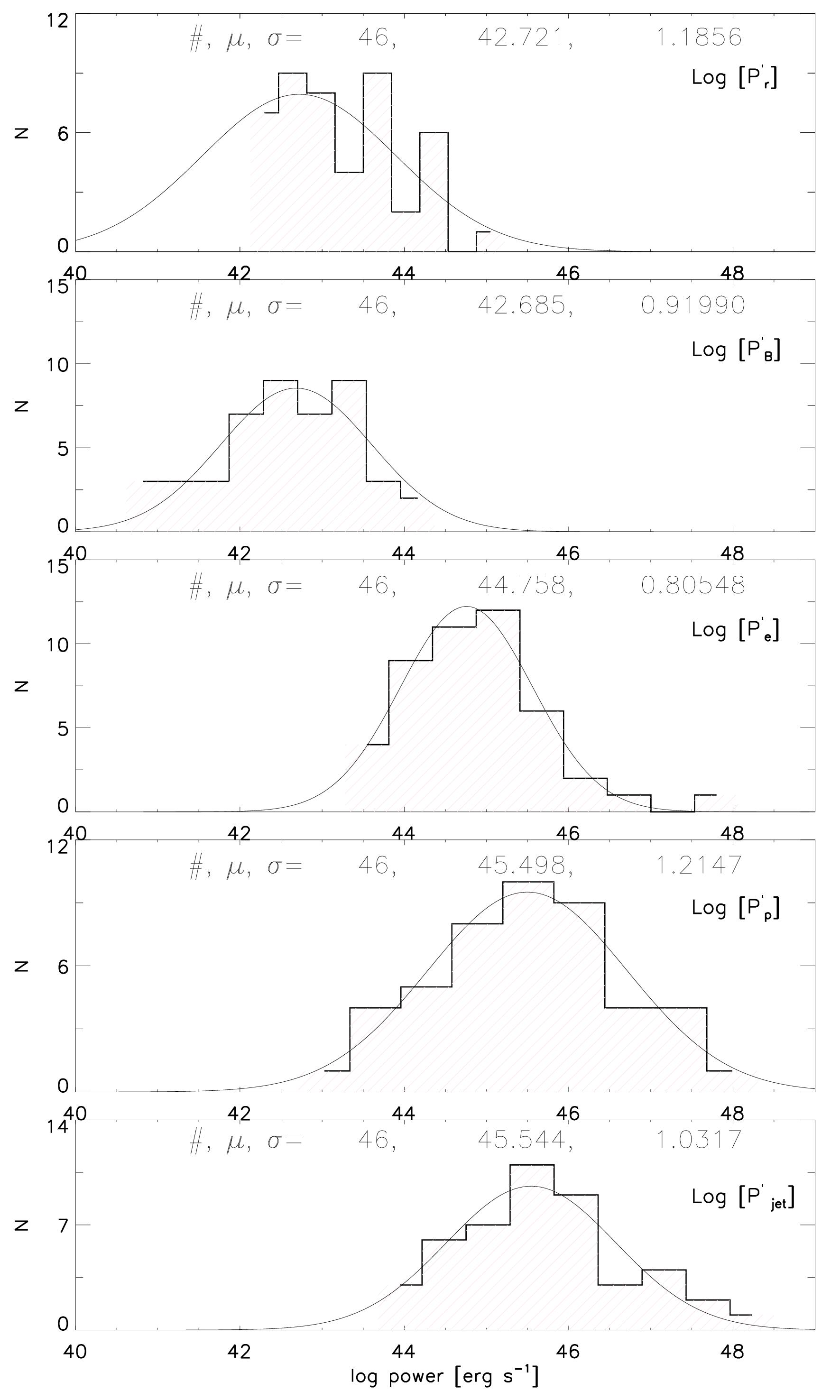}
\end{center}
\caption{The distributions of radiative power ($P'_{\rm {r}}$), Poynting flux ($P'_{\rm {B}}$), electron ($P'_{\rm {e}}$), proton ($P'_{\rm {p}}$) and total jet power ($P'_{\rm {jet}}$). Symbols are the same as in Fig.1.}

\end{figure}
\begin{table*}

\caption{The ratios of the energy densities of relativistic electrons to magnetic field in the emitting regions, the jet powers in the forms of bulk motion of electrons, protons and Poynting flux and the radiative powers \label{Table:2}.}
\centering
\scriptsize
\tabcolsep 1.2mm
\begin{tabular}{l|lll llll}
\hline \hline
Source name$^a$ & $U'_{\rm e}/U'_{\rm B}$ & Log $P_{\rm r}$ & Log $P_{\rm B}$ & Log $P_{\rm e}$ & Log $P_{\rm p}$  &  Log $P_{\rm {jet}}$ \\
~[1] & [2] & [3] & [4] & [5] & [6] & [7] \\
\hline
      1ES0229+200&  4991.4&42.9&41.4&45.1&47.0&47.0\\
    1ES0347-121&  1555.6&42.3&41.7&44.8&44.6&45.0\\
    1ES0806+524&   144.6&42.9&43.0&45.2&47.1&47.1\\
    1ES1011+496&     6.3&43.9&43.5&44.3&45.3&45.4\\
    1ES1101-232&   337.4&42.4&42.2&44.7&45.9&45.9\\
  1ES1101-232 f&    23.2&45.2&43.0&44.3&46.4&46.4\\
    1ES1215+303&   267.0&42.7&42.5&44.9&45.3&45.4\\
   1ES1218+30.4&    34.2&42.9&42.7&44.2&43.4&44.3\\
    1ES1959+650&    30.5&42.7&43.1&44.6&46.6&46.6\\
    1ES2344+514&    33.5&42.6&41.9&43.4&43.7&43.9\\
  1ES2344+514 f& 44187.8&42.2&40.6&45.3&46.0&46.0\\
     1H0414+009&    61.7&43.0&43.1&44.9&45.4&45.5\\
     1H1013+498&     1.7&44.3&43.9&44.1&44.8&44.9\\
          3C66A&   130.6&44.5&43.3&45.4&45.5&45.7\\
     B32247+381&    15.4&42.8&42.8&44.0&44.4&44.5\\
     BLLacertae&    70.2&42.9&42.7&44.6&45.3&45.4\\
   BLLacertae f&   114.0&43.2&42.5&44.5&45.6&45.6\\
  BZBJ0033-1921&   193.4&43.5&42.9&45.2&44.7&45.3\\
  BZBJ1058+5628&     7.9&43.5&43.3&44.2&44.9&45.0\\
      H1426+428&  4485.9&43.7&40.7&44.4&42.7&44.4\\
      H2356-309& 14317.8&42.1&41.6&45.7&47.4&47.4\\
         MRK421&    21.2&42.5&42.4&43.7&44.2&44.3\\
       MRK421 f&    93.5&42.4&42.3&44.3&44.4&44.6\\
         MRK501&   107.1&42.6&42.2&44.2&44.8&44.9\\
       MRK501 f&   394.6&43.8&41.0&43.5&43.5&43.8\\
         Mkn180&    13.4&42.4&42.2&43.3&43.4&43.7\\
     PG1553+113&    42.8&43.6&43.7&45.3&45.6&45.8\\
    PKS0447-439&    26.5&43.5&43.3&44.8&45.6&45.6\\
    PKS1424+240&     7.2&44.2&44.4&45.2&46.0&46.0\\
    PKS2005-489&    36.8&42.4&43.3&44.9&45.3&45.5\\
    PKS2155-304&    37.6&42.8&43.0&44.6&44.4&44.8\\
  PKS2155-304 f&  3724.8&43.7&42.2&45.8&46.2&46.3\\
        RBS0413&   188.8&42.7&42.3&44.6&45.1&45.2\\
   RGBJ0152+017&  8161.4&42.6&41.8&45.7&47.6&47.6\\
   RGBJ0710+591&    56.1&43.0&42.5&44.3&44.1&44.5\\
\hline
     S50716+714&    53.5&43.2&43.5&45.3&45.7&45.8\\
   S50716+714 f&  2678.4&43.5&42.4&45.8&46.0&46.2\\
   PKS0851+202&    22.4&44.5&44.0&45.4&46.0&46.1\\
     PKS0048-09&   331.8&44.2&43.5&46.0&46.5&46.6\\
     PG1246+586&   468.7&43.8&43.2&45.9&46.1&46.3\\
\hline
           Wcom&   466.7&42.7&42.2&44.9&45.1&45.3\\
         Wcom f&   375.4&43.7&42.3&44.8&45.1&45.3\\
    PKS0426-380&5394623&44.3&41.3&48.1&48.3&48.5\\
        4C01.28&  3603.1&44.2&43.1&46.7&47.4&47.5\\
          OT081&  5157.9&43.7&42.5&46.3&46.8&46.9\\
    PKS1717+177& 27975.9&43.4&41.0&45.5&46.2&46.3\\

\hline
\end{tabular}
\vskip 0.4 true cm

\end{table*}

As mentioned in Section 4.1, a few sources have large $\delta_{\rm
D}$, which are not consistent with ones by VLBI observations (e.g.,
\cite{piner2014}). Several authors proposed the jet model with
spine-sheath \citep{tav2008} or the decelerating-jet model
\citep{geo2003} to reduce the extreme $\delta_{\rm D}$ (and bulk
Lorentz factor). For Swift J1644+57 and Swift J2058+05
\citep{mim2015}, the rapid X-ray variability could be originated
from the internal jet, similar to the blazar geometry of normal AGNs
\citep{bloom2011}. However, the radio flux increased gradually over
several months could be caused by the shock interaction between the
fast core and the dense external gas surrounding the SMBH
\citep{gian2011}, which is similar to GRB afterglows, indicating
that the relativistic jet may contain a fast core and slow sheath.
In addition, HBLs usually show weak micro-variability or intro-day
variability in the optical bands, supporting the model offered by
\cite{gaur2012} that the Kelvin-Helmholtz instability
\citep{rom1999} is suppressed when $B > B_{\rm c}$, where $B_{\rm
c}$ is the critical value of the axial magnetic fields in sub-parsec
to parsec scale jets which is given by \citep{rom1995}
\begin{equation}
B_{\rm c}=\frac{\sqrt{4\pi nm_{\rm e}c^2(\Gamma^2-1)}}{\Gamma},
\end{equation}
where $n$ and $m_{\rm e}$ are the electron density and rest mass
respectively, and $\Gamma$ is the bulk Lorentz factor. If $\Gamma
\gg 1$, then $B_{\rm c} \sim \sqrt{4\pi nm_{\rm e}c^2}$. Although we
use simple SSC model to reproduce the SEDs, as demonstrated above,
but one-zone jet model could be unreasonable for sources with
extreme bulk Lorentz, and the spine-sheath jet model is approved. In
our sample, the larger $\delta_{\rm D}$ corresponds the lower $B$.
This also well known for extreme blazar such as 1ES 0229+200 and 1ES
0347-121(e.g., \cite{tan2014}), so there should be $B < B_{\rm
{real}}$ for two component model. We roughly take $B/B_{\rm c}$ to
check the instability shown in Fig. 3. It is found that the
Kelvin-Helmholtz instability is suppressed for most GeV-TeV BL Lac
objects, which show few micro-variability in the optical bands. BL
Lacs with $B/B_{\rm c}<$ 1 contain all LBLs and some HBLs, they may
involve the IC radiation process from weak extra field.

\begin{figure}
\begin{center}
\includegraphics[width=14cm,height=8cm]{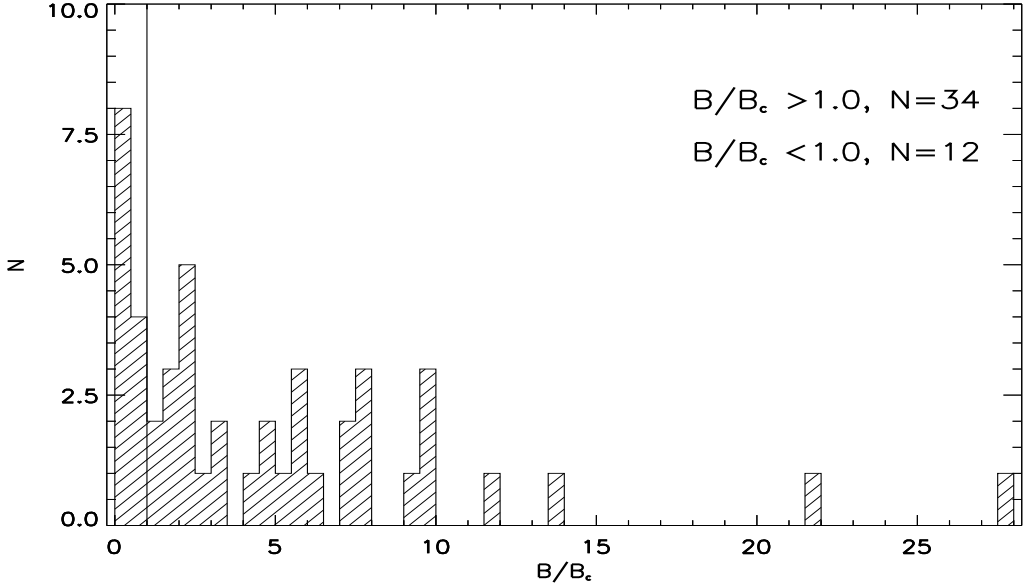}
\end{center}
\caption{Kelvin-Helmholtz instabilities in sub-parsec to parsec
scale jets will be suppressed if the magnetic field $B$ exceeds the
critical value $B_c$. The vertical line represents the value of
$B/B_c$ =1.0. }

\end{figure}

\subsection{Relativistic electron distributions and Accretion rates}
For electron distributions,  $p_{\rm 1}$ and $p_{\rm 2}$ are
clustered at 2.1 and 3.7 respectively. This result favors that
$N'(\gamma') \propto \gamma'^{-2}$ or $p_{\rm 1} \sim 2$ for
$\gamma' < \gamma'_{\rm b}$, which
 is expected in the slow-cooling regime \citep{fin2013}. In addition, one can find that for BZB J0033-1921 and PKS 0426-380,
 their $p_{\rm 1}$ are smaller than 1.3 and deviate from the standard picture that $N'(\gamma') \sim \gamma'^{-q}$ below $\gamma'_{\rm b}$ in
 slow-cooling regime if electrons are injected with $\sim \gamma^{-q}$. However, standard shock acceleration theories predict that q $\sim$ 2
 or $N'(\gamma') \sim \gamma'^{-2}$ in the fast-cooling regime.  Some authors introduced new acceleration mechanisms, such as
 magnetic reconnection  \citep{guo2014,guo2016} and collisionless shocks \citep{asa2015}, or the KN effect on the SSC \citep{bo2014} and
 IC \citep{yan2016} emissions to obtain a hard electron distribution, but for these objects with the low magnetization
 ($\sigma \leq 1$, $\sigma=\frac{B}{\sqrt{4\pi nm_{\rm e}c^2}}$), like 1ES 2344-514 and PKS 0426-380, the magnetic reconnection is inefficient in a matter dominated flow.

\begin{figure}
\begin{center}
\includegraphics[width=14cm,height=8cm]{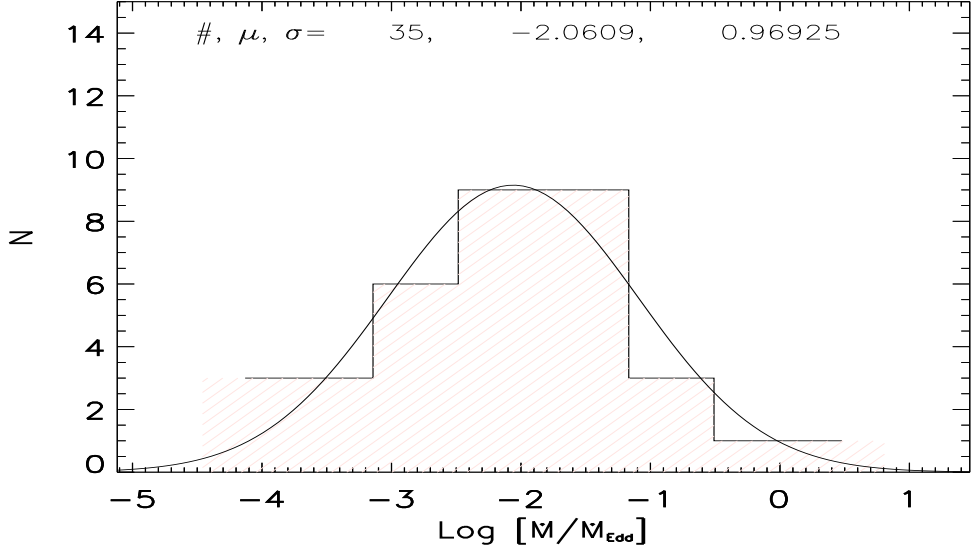}
\end{center}
\caption{Distributions of accretion rates in Eddington units. The Gauss fit result is plotted on the picture. Symbols are the same as in Fig.1. }

\end{figure}

Fig. 4 shows the accretion rates $\dot{M}$ in Eddington unit, in
which we set $\dot{M}= \frac{P_{\rm {jet}}}{c^2}$ \citep{ghi08}. It
is found that $Log [\dot{M}/\dot{M_{\rm Edd}}]$ is clustered at
-2.07, and only 5 sources have accretion rates larger than -1 which
is a $''$divide$''$ value between FSRQs and BL Lacs
\citep{Ghi2010a}. It is interesting to note that PKS0426-380 and
OT081 (LBL) with high accretion rates could show LBL to be different
from IBL/HBL and need complex emission mechanism such as
IC\citep{tan2013}.

\subsection{The blazar sequence}
In the paper, we constructed a large sample of GeV-TeV BL Lacs plus FSRQs obtained from \cite{kang2014} to test the blazar sequence.  $\nu_{\rm {pk}}$ is obtained by
\begin{equation}
    \nu_{\rm {pk}}=\frac{4}{3}\gamma'_b\nu_{\rm {L}}\frac{\delta_{\rm D}}{1+z},
\end{equation}
where $\nu_{\rm L}$ is the Larmor frequency.

The relation of $\nu_{\rm {pk}}-L_{\rm {pk}}$ for the sample is plotted in
Fig. 5. It is found that FSRQs are located at the top-left region, and BL Lacs cover at the down-right region.
Our sample also contains some objects in the flare stage. Using the correlation
analysis, we find that there are a strong anti-correlations between
$\nu_{\rm {pk}}$ and $L_{\rm {pk}}$ in our sample. The sample
sources with low stage could have stronger anti-correlations than
that for the sources with flare stage. In addition, we also plot
the relation of $\gamma'_{\rm b}-P'_{\rm {jet}}$ based on BL Lacs,
shown on the right panel of Fig. 5.

\begin{figure}
\begin{center}
\includegraphics[width=16cm,height=9.5cm]{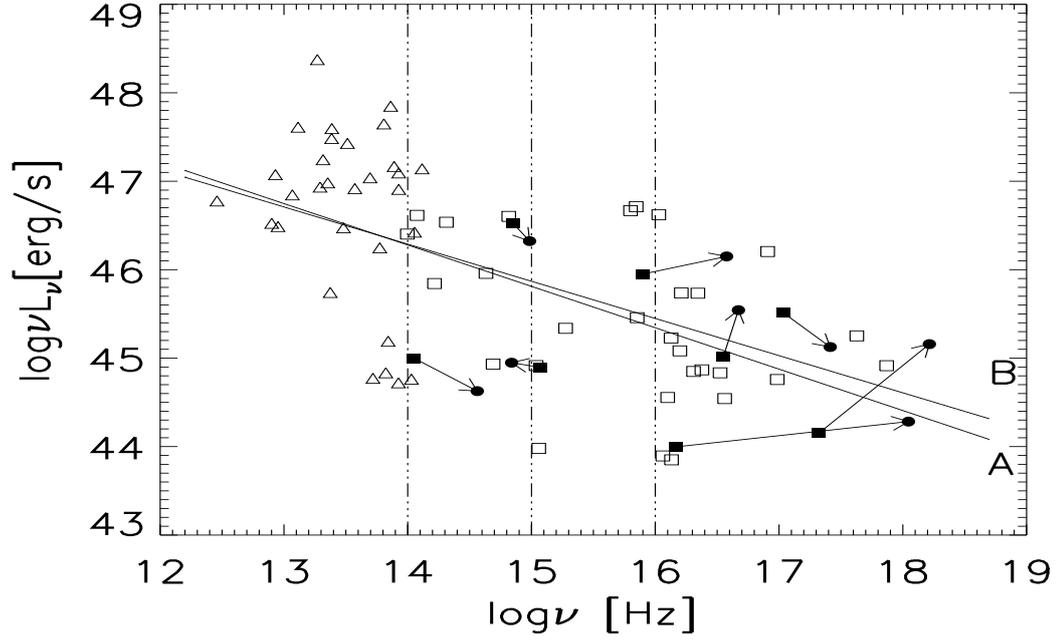}
\includegraphics[width=16cm,height=9.5cm]{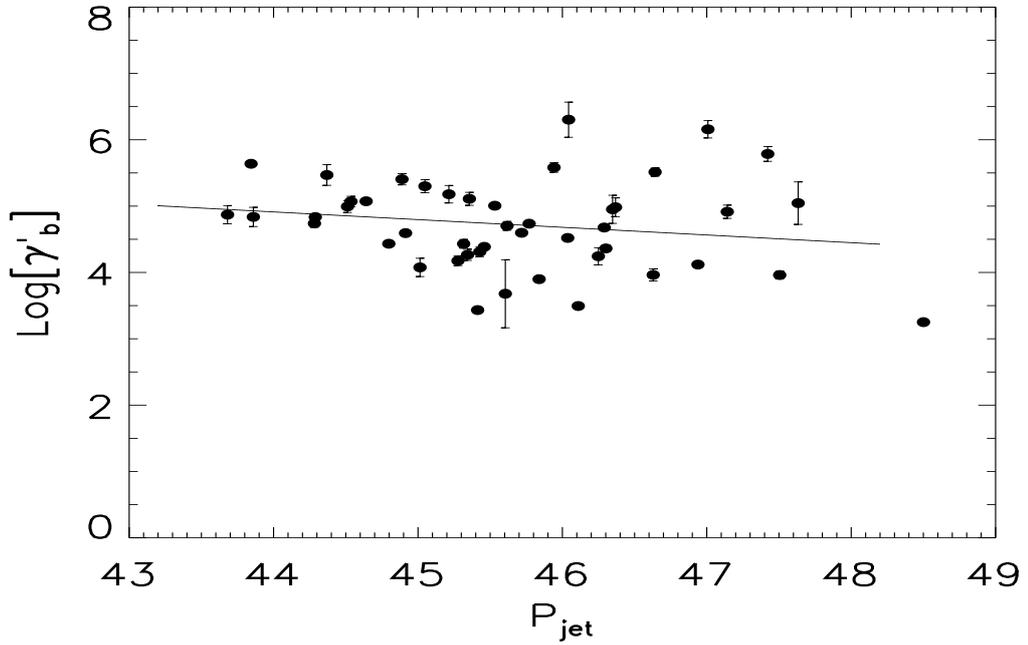}
\end{center}
\caption{ The top panel: the peak luminosity $L_{\rm {pk}}$ as a
function of the peak frequency $\nu_{\rm {pk}}$. The correlation
coefficients of the best-fit line for the sample containing low
(line A) and high states (line B), are -0.47 and -0.42 respectively.
The solid black squares and circles are BL Lacs in low stage and
high stage, respectively. The open triangle represents the FSRQs.
The bottom panel: the power $P'_{\rm {jet}}$ of the jet as a
function of the break Lorentz factor $\gamma'_{\rm b}$, where the
correlation coefficient of the best-fit line is - 0.116. The solid
black circles are BL Lacs.}

\end{figure}

Based on the correlations of $\nu_{\rm {pk}}-L_{\rm {pk}}$ and
$\gamma'_{\rm b}-P'_{\rm {jet}}$, we support the blazar sequence, in
which the radiation energy density causes a particle energy
distribution to be a break at low energies \citep{fos1998,
ghi08,fan2016}. Despite of the anti-correlation presented in Fig.5,
the flare could shift $\nu_{\rm b}$ to the high bands. In addition,
for several BL Lacs, their types do not match what given in the
literature, and their SEDs may fail to be reproduced by the simply
one-zone SSC model. Furthermore, we also find that for some BL Lacs,
such as 1ES 1101+496, their peak luminosity in the low hump decrease
even they are in the high state, the main reason is that the
flare does not happen in the optical to the X-ray bands.

\section{SUMMARY}
We have reproduced the SEDs of 46 GeV-TeV BL Lacs upon one-zone SSC
model using the MCMC technology, then we use the best-fitting model
parameters to analyze the jet powers, the accretion rates, and their
correlations. Based on our sample, we also test the blazar sequence
and the proper structure of the jet.

Firstly, the MCMC technology enable us to reproduce GeV-TeV BL Lacs
SEDs based on the simultaneous or quasi-simultaneous observation
data in a large range of the parameter space, avoiding visual
identification.  One zone SSC model can well produce the observed
SEDs, however, for some objects such as LBLs, other radiation
mechanism should be considered. In addition, some distributions show
bimodal phenomenon, which reflect the limited number of objects or
the emission mechanism should be revisited. Secondly, GeV-TeV BL
Lacs are old blue quasars, and they have weak magnetic field and
large Doppler factor, which cause ineffective cooling and shift the
SEDs to higher bands. Their jet powers are around $4.0\times 10^{45}
\rm{ erg\cdot s}^{-1}$, comparing with the radiation power,
$5.0\times 10^{42} \rm{ erg\cdot s}^{-1}$,  indicating that only a
small fraction of the jet power is transformed into the emission
power. Thirdly, we argue that for some BL Lacs with large Dopplers,
their jet components could have two substructures, e.g., the fast
core and the slow sheath. For most GeV-TeV BL Lacs, the
Kelvin-Helmholtz instability is suppressed by the higher magnetic
fields, leading few micro-variability or intro-day variability in
the optical bands.  Finally, $L_{\rm {pk}}-\nu_{\rm {pk}}$ and
$\gamma_{\rm b} - P_{\rm {jet}}$ have the anti-correlations,
favoring the blazar sequence.

\section*{Acknowledgments}
We thank anonymous referee for useful comments and suggestions. This
research has made use of the NASA/IPAC Extragalactic Database (NED)
which is operated by the Jet Propulsion Laboratory, California
Institute of Technology, under contract with the National
Aeronautics and Space Administration. The authors gratefully
acknowledge the financial supports from the National Natural Science
Foundation of China 11673060, 11661161010,11763005, and the Natural
Science Foundation of Yunnan Province under grant 2016FB003. The
authors gratefully acknowledge the computing time granted by the
Yunnan Observatories, and provided on the facilities at the Yunnan
Observatories Supercomputing Platform.


\section{appendix: Parameters Distributions And Spectral Energy Distributions}
\begin{figure*}
  \begin{center}
   \begin{tabular}{cc}
\includegraphics[width=5.8cm,height=4.9cm]{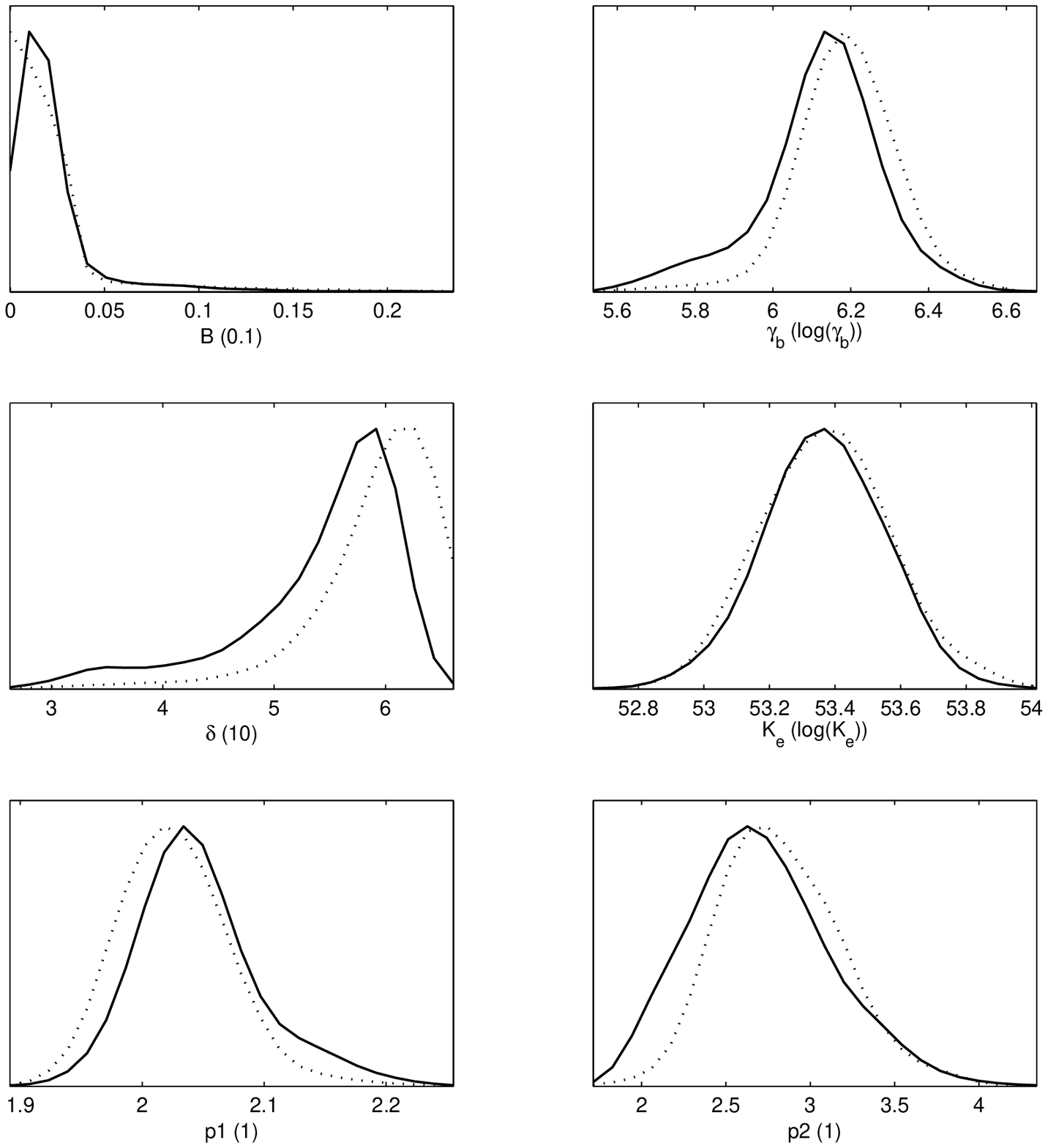}
\includegraphics[width=8.0cm,height=5.2cm]{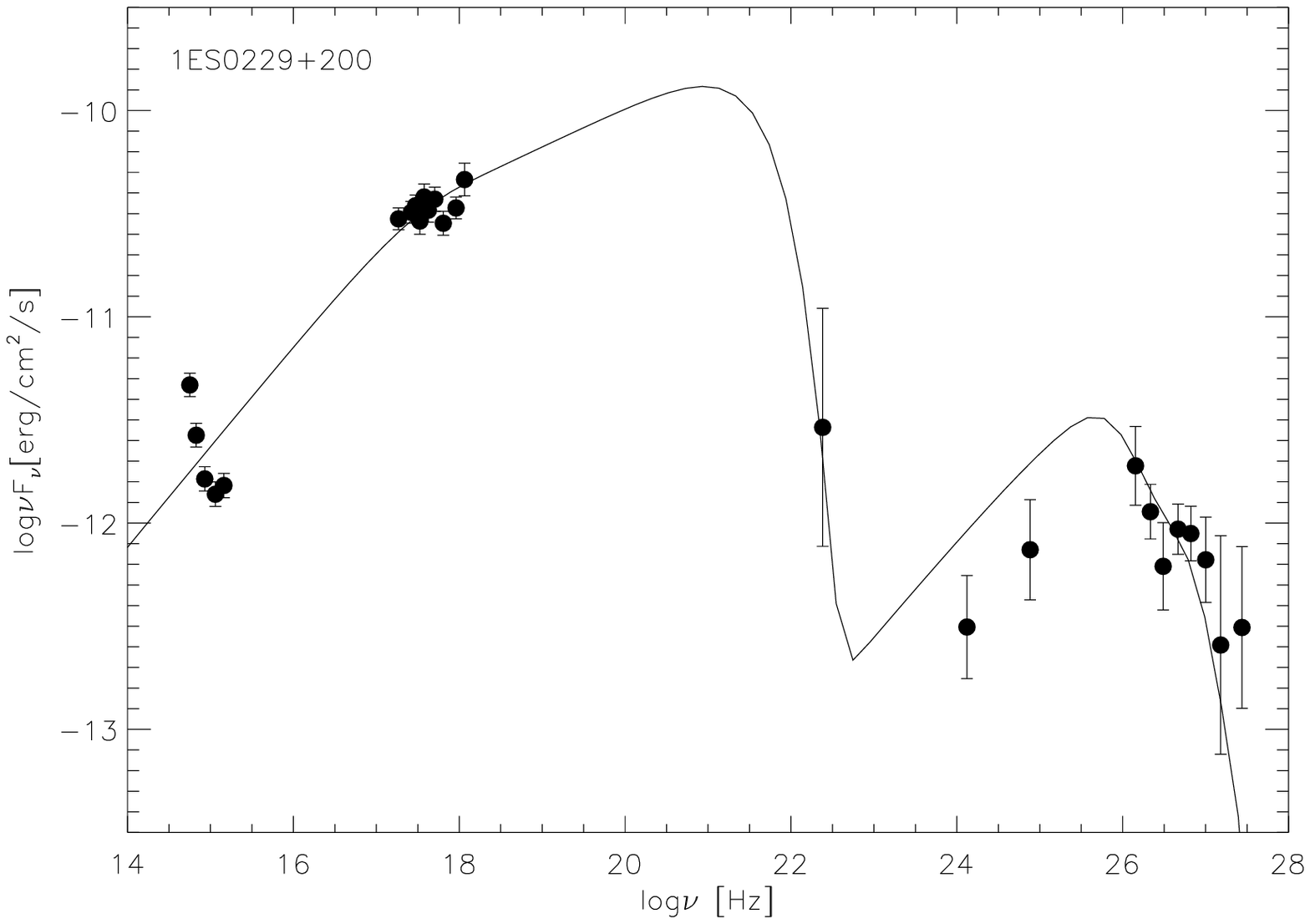}\\
\includegraphics[width=5.8cm,height=4.9cm]{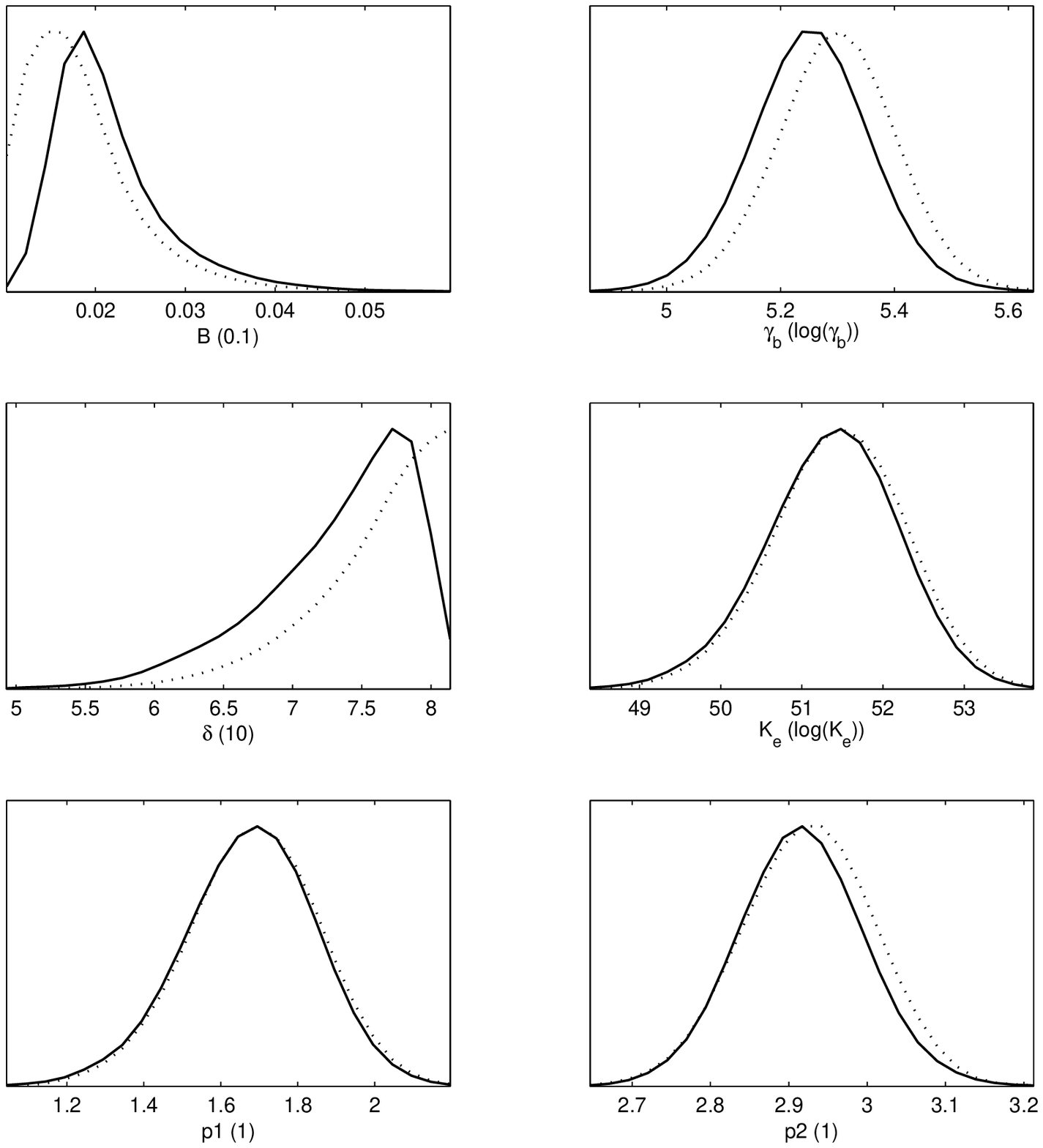}
\includegraphics[width=8.0cm,height=5.2cm]{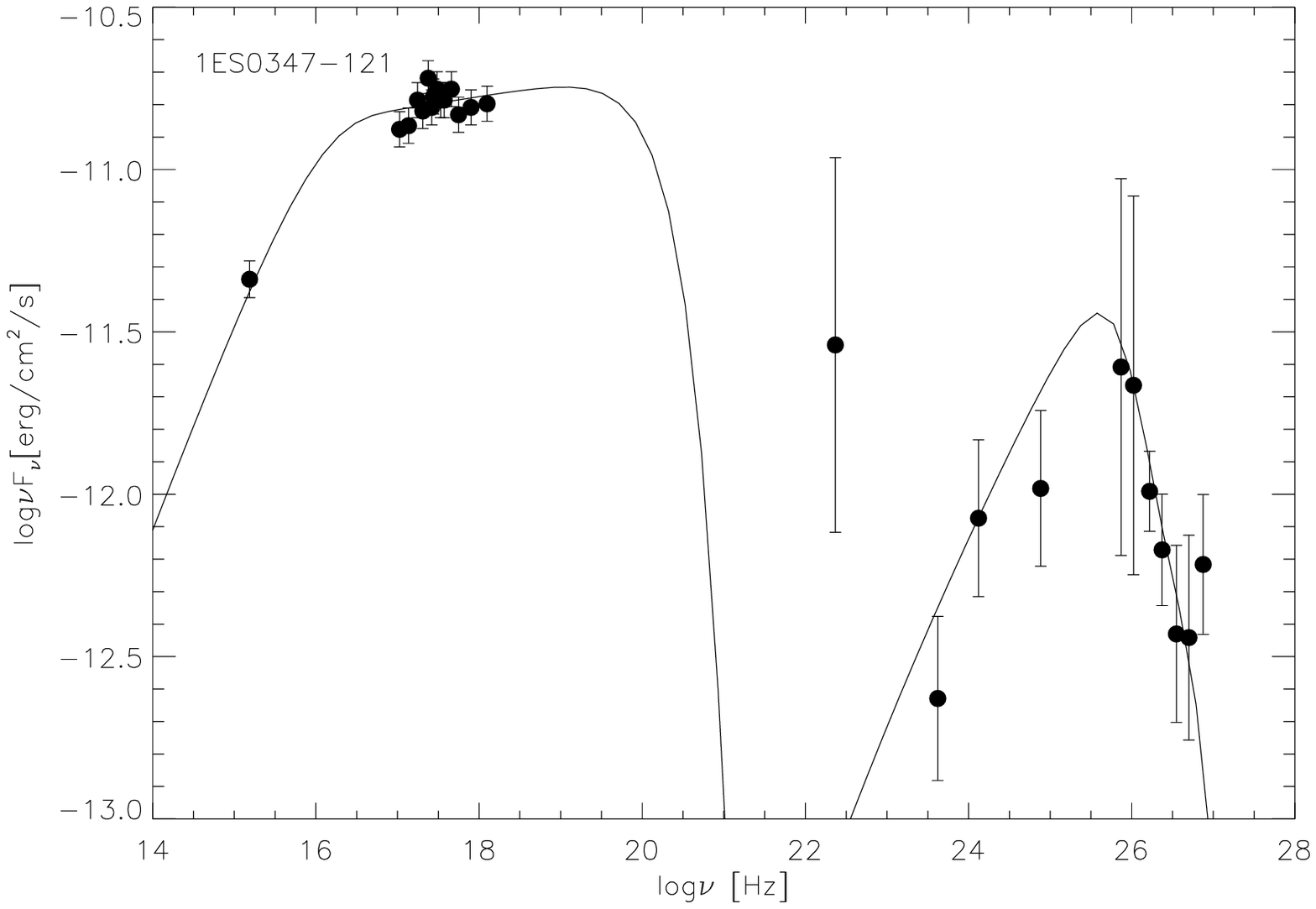}\\
\includegraphics[width=5.8cm,height=4.9cm]{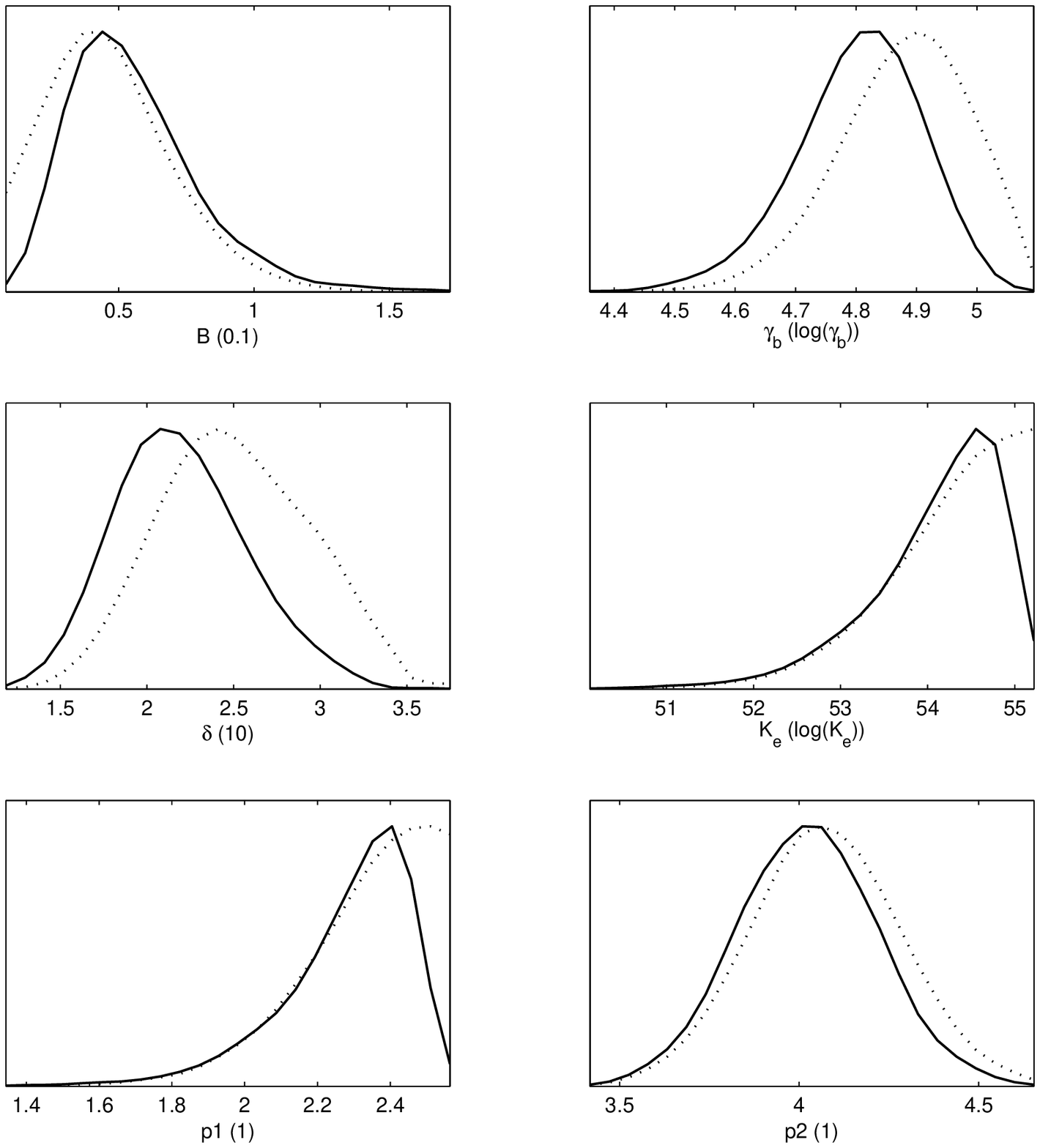}
\includegraphics[width=8.0cm,height=5.2cm]{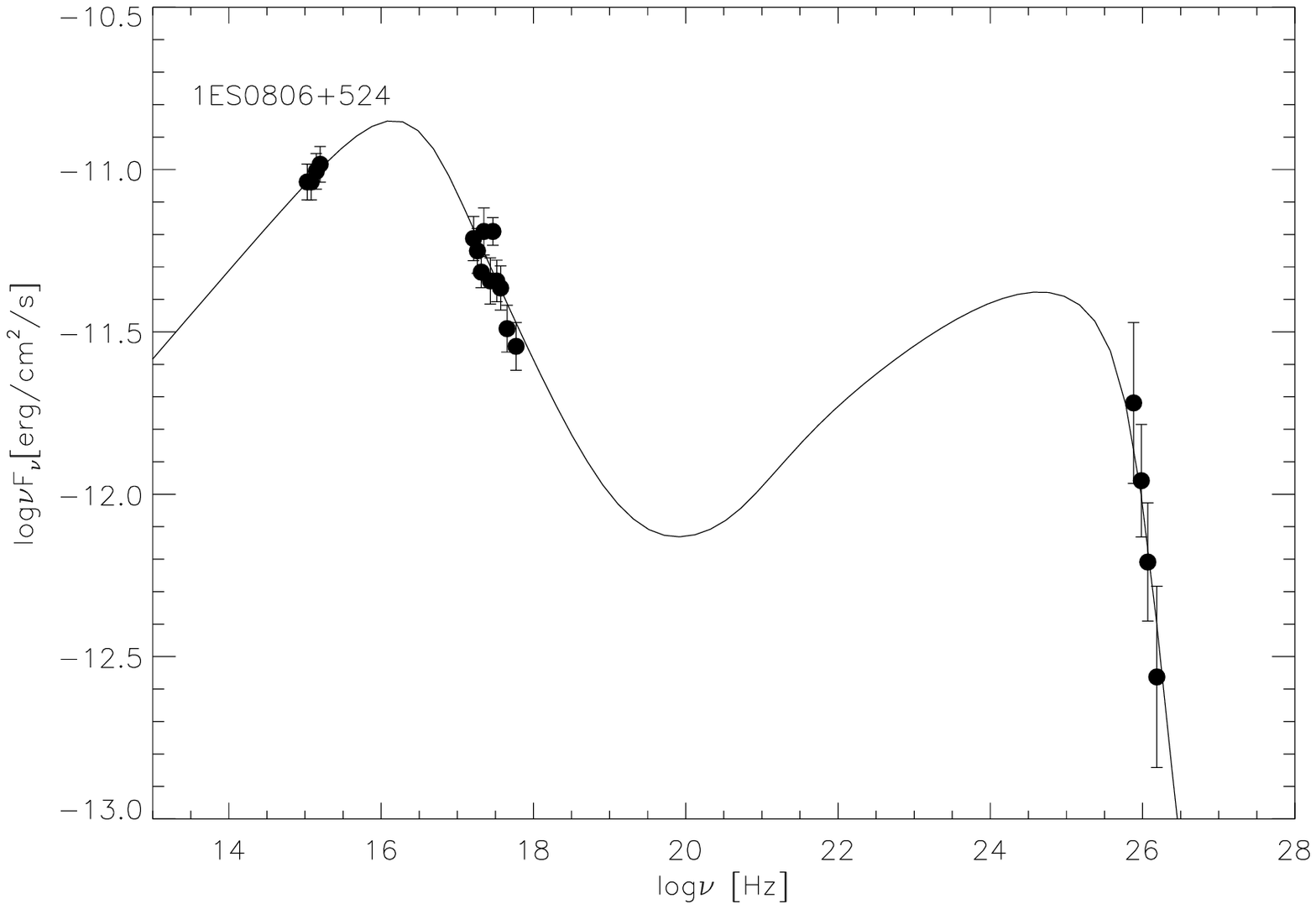}\\
\includegraphics[width=5.8cm,height=4.9cm]{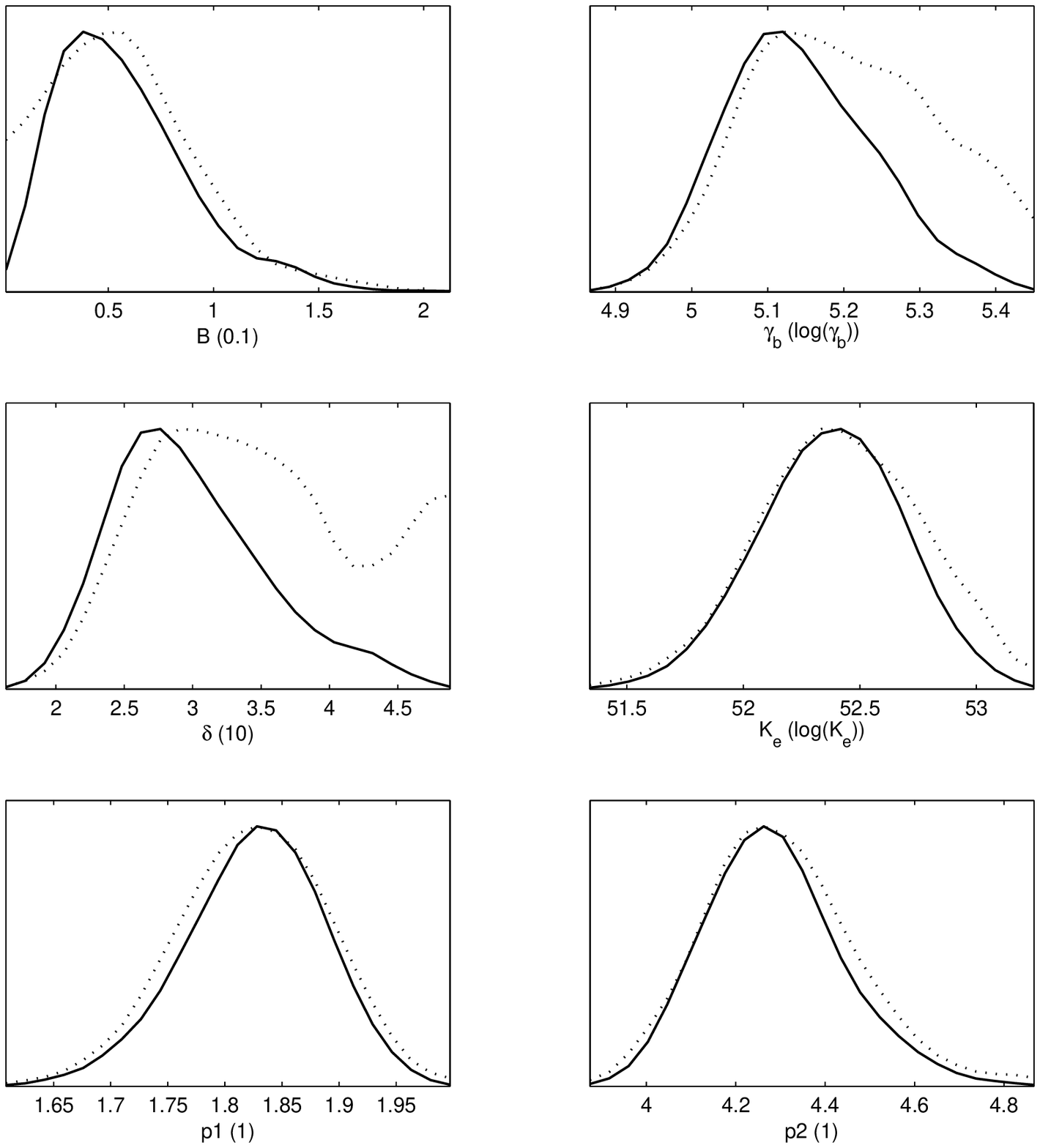}
\hfill
\includegraphics[width=8.0cm,height=5.2cm]{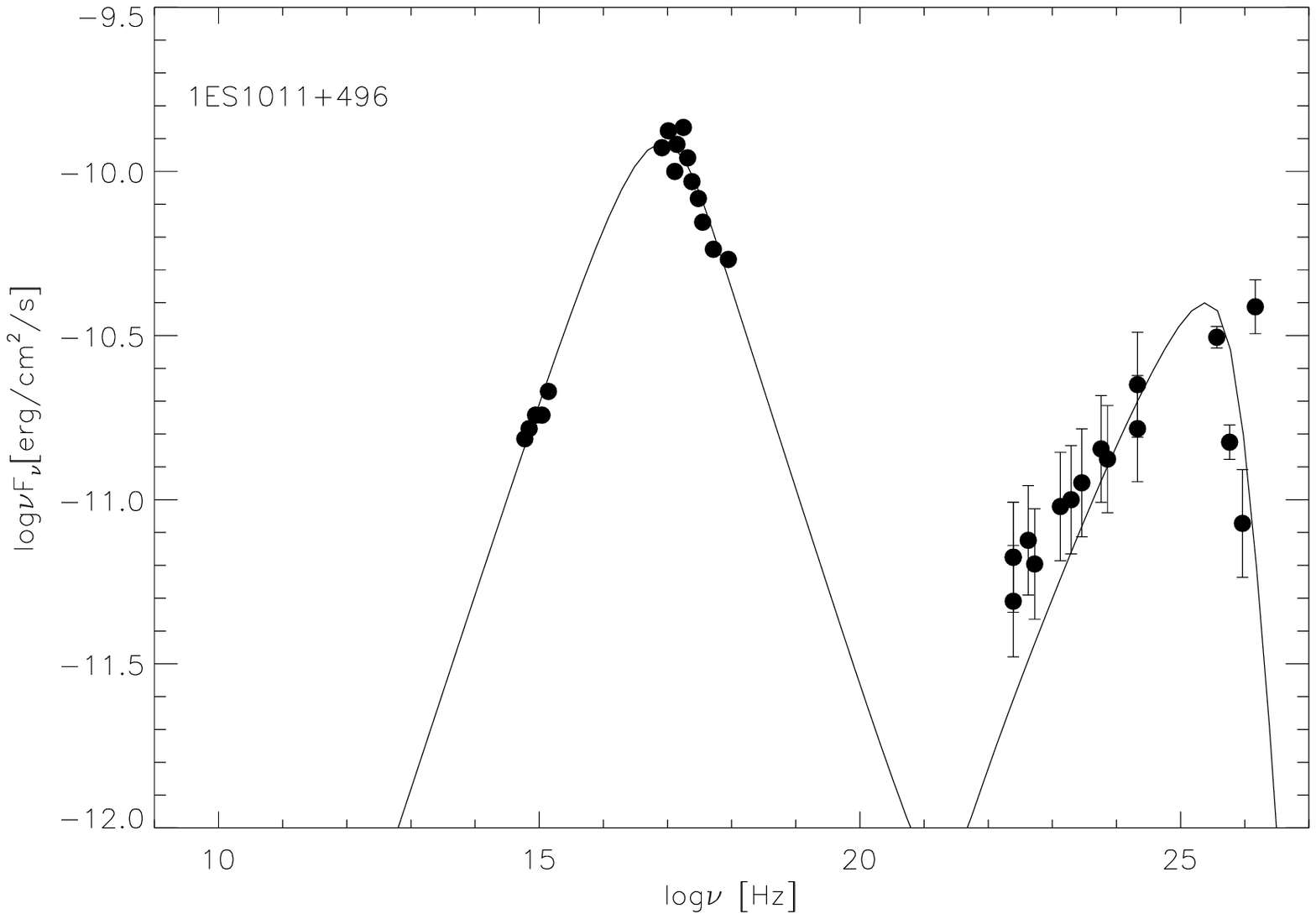}
    \end{tabular}
  \end{center}
    \center{\textbf{Fig. 6.} Left panels: the distributions of the model parameters, where the dotted lines show the maximum likelihood distributions, the solid lines show the
marginalized probability distributions. Right panels: the SEDs of
GeV-TeV objects.}
\end{figure*}

\begin{figure*}
  \begin{center}
   \begin{tabular}{cc}
\includegraphics[width=5.8cm,height=5.0cm]{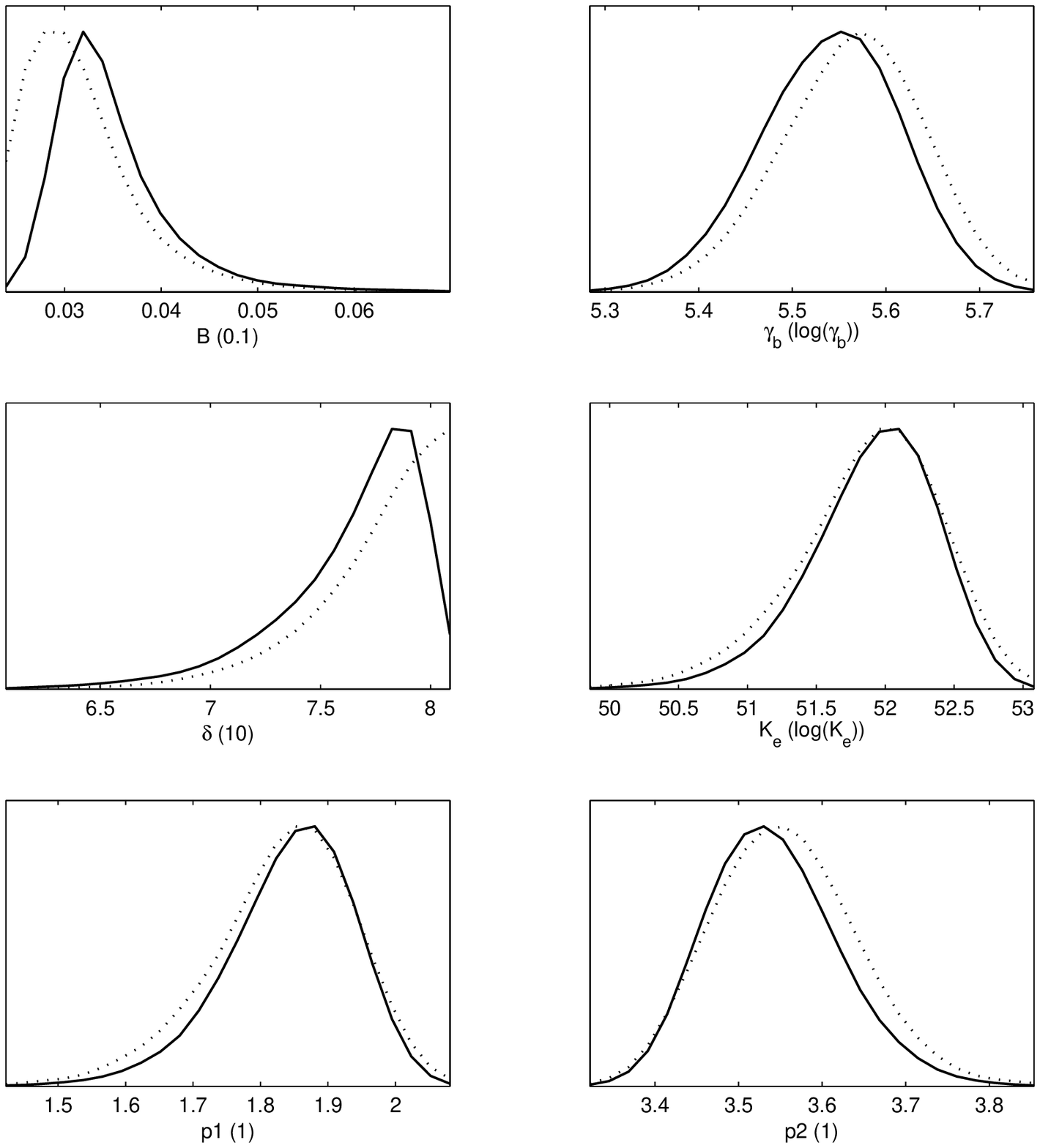}
\includegraphics[width=8.0cm,height=5.3cm]{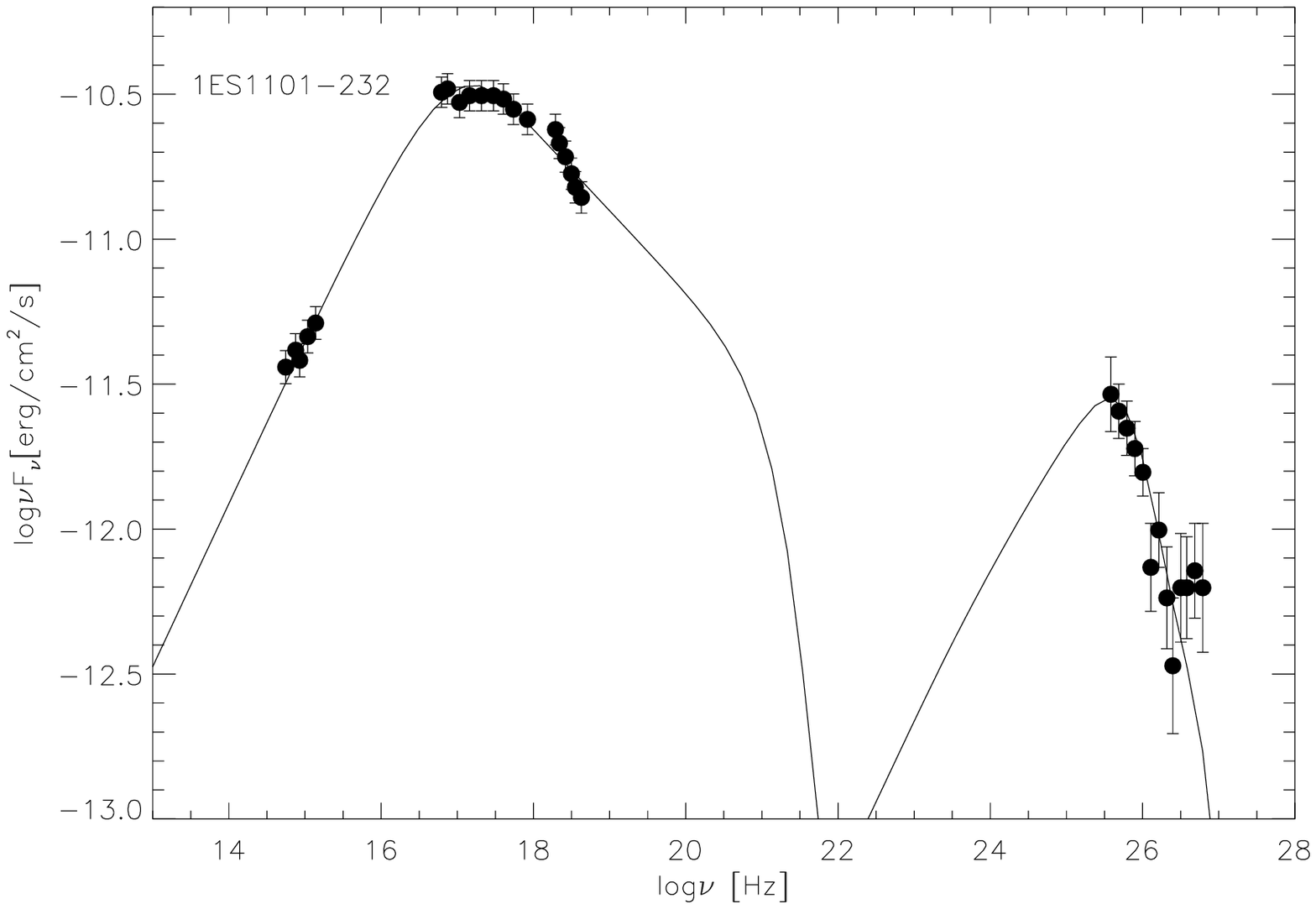}\\
\includegraphics[width=5.8cm,height=5.0cm]{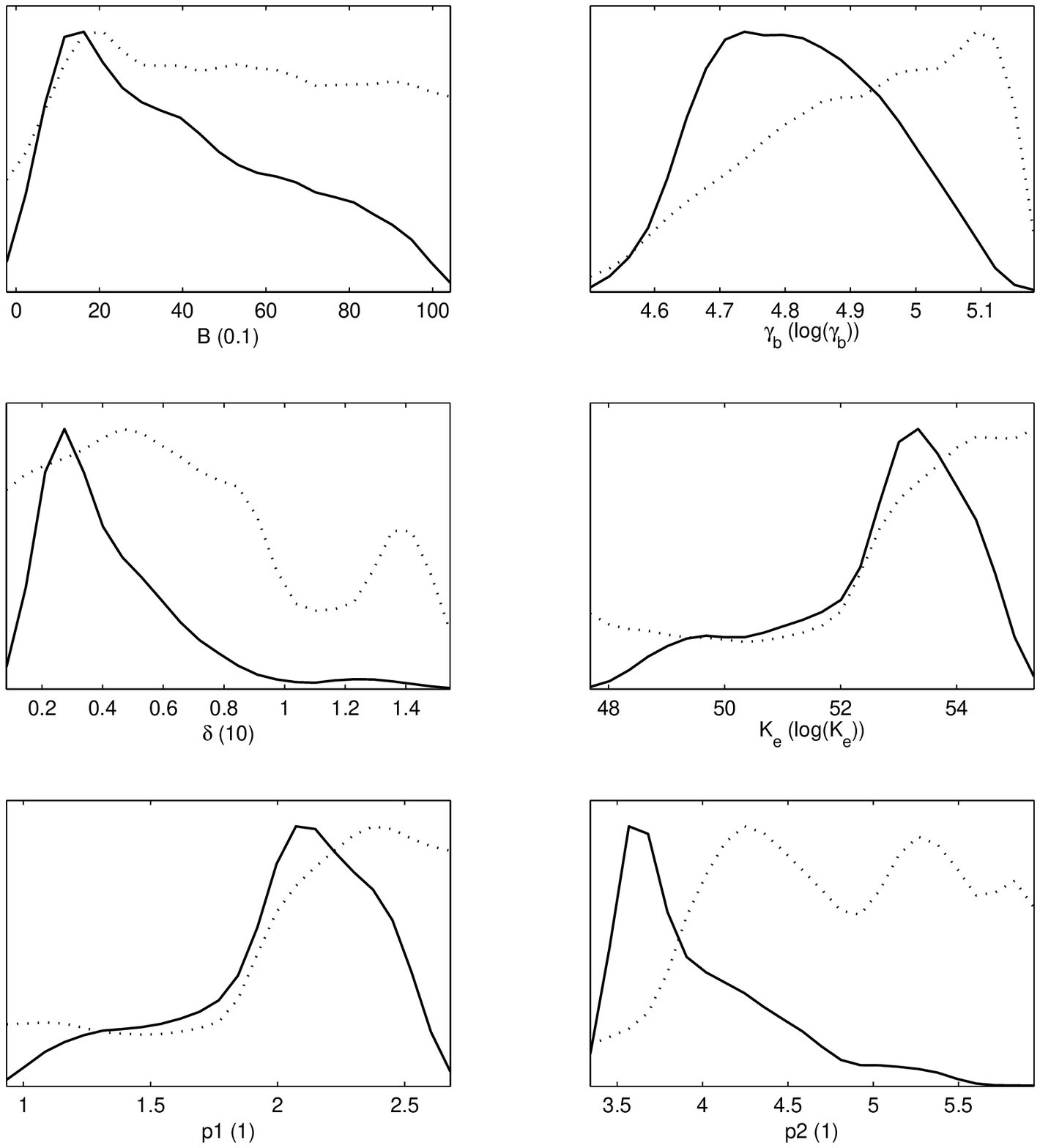}
\includegraphics[width=8.0cm,height=5.3cm]{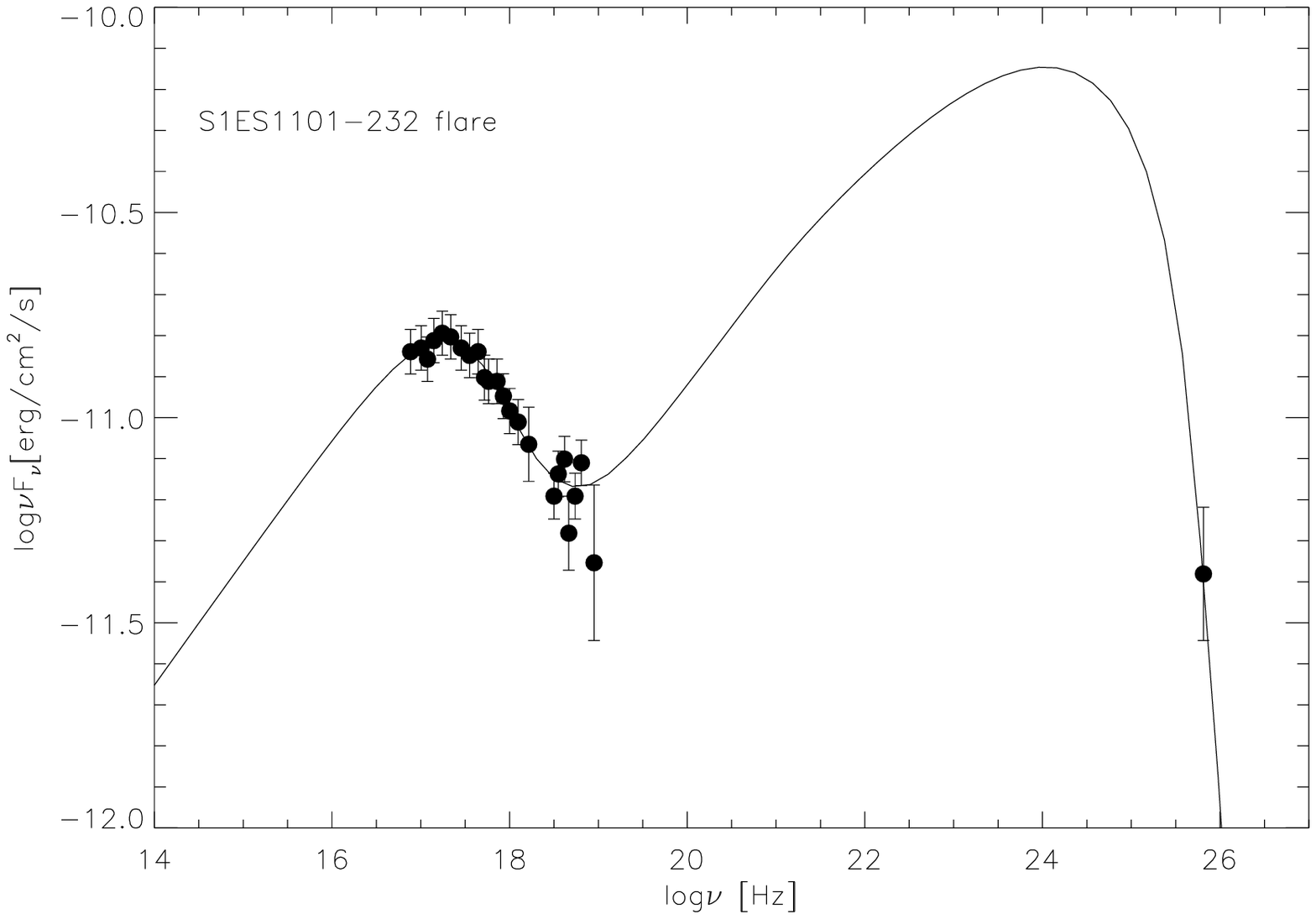}\\
\includegraphics[width=5.8cm,height=5.0cm]{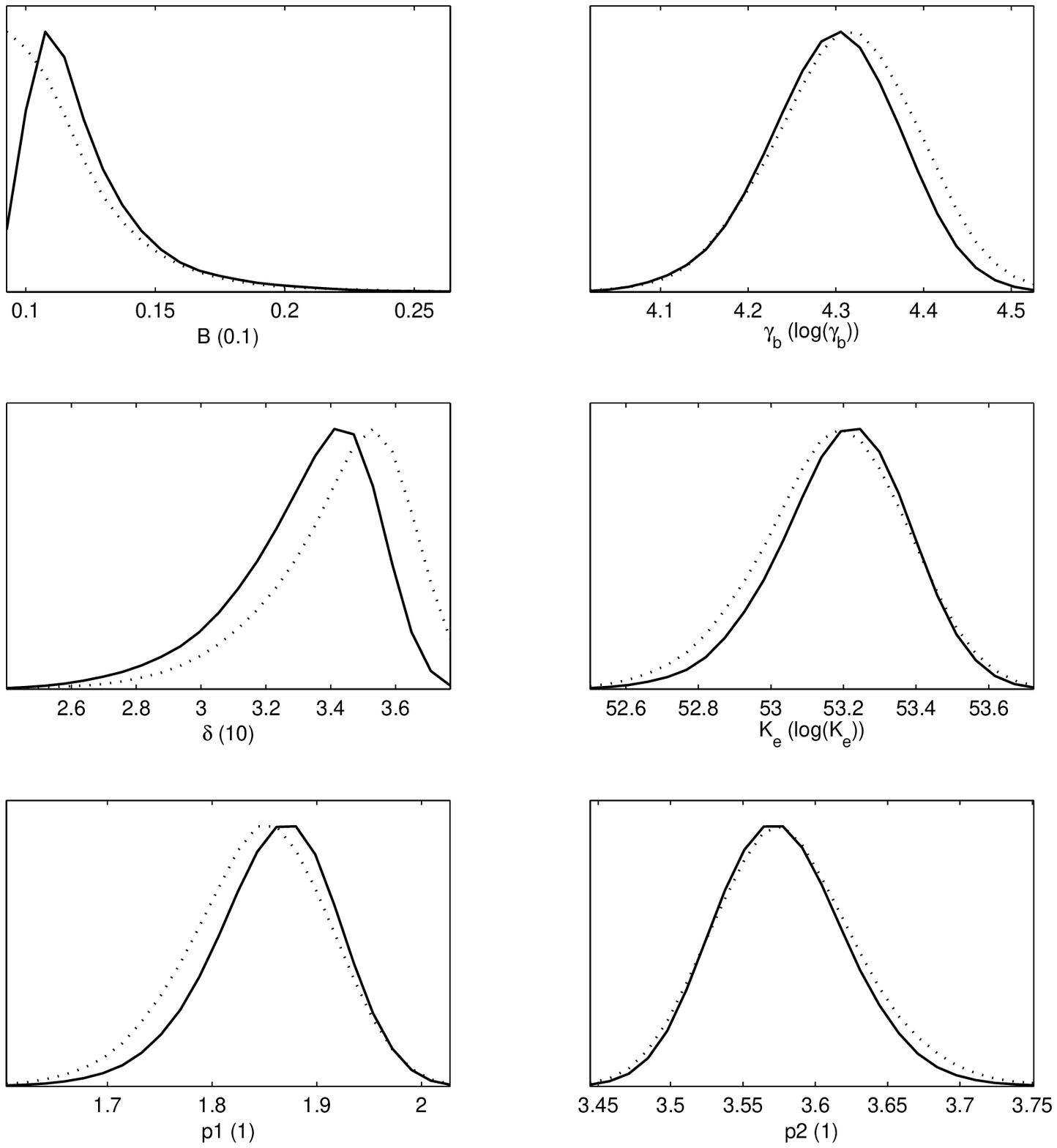}
\includegraphics[width=8.0cm,height=5.3cm]{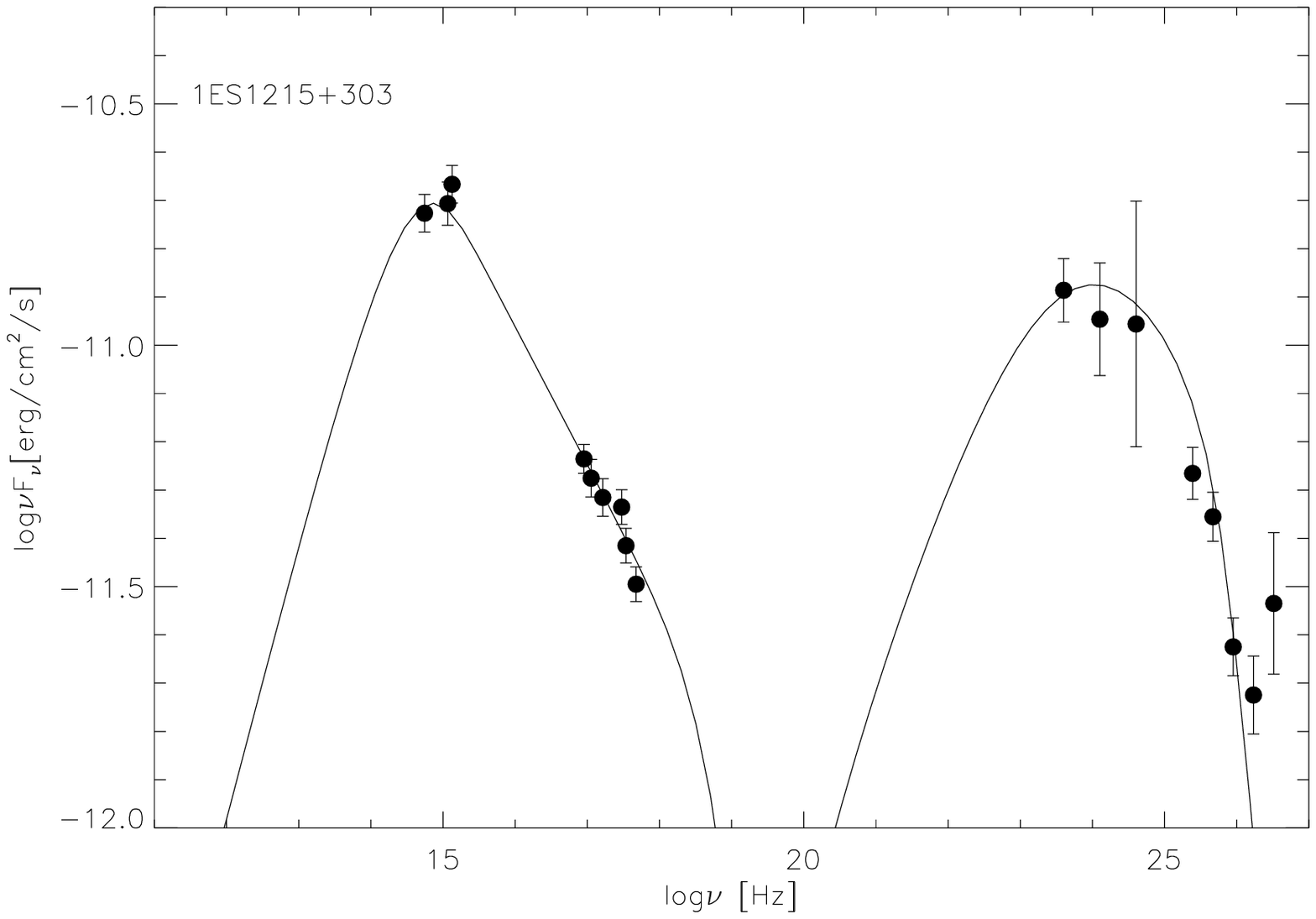}\\
\includegraphics[width=5.8cm,height=5.0cm]{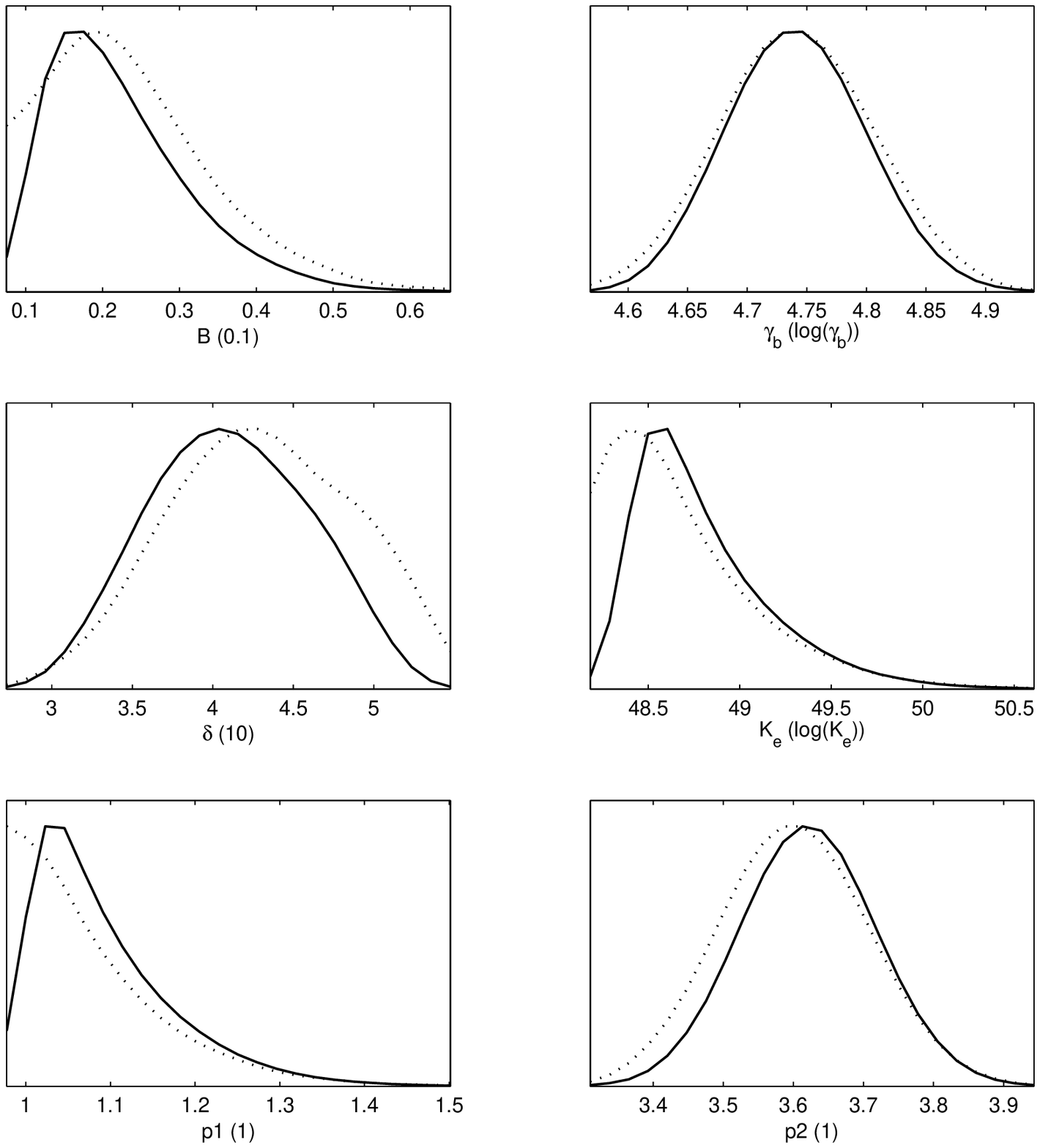}
\hfill
\includegraphics[width=8.0cm,height=5.3cm]{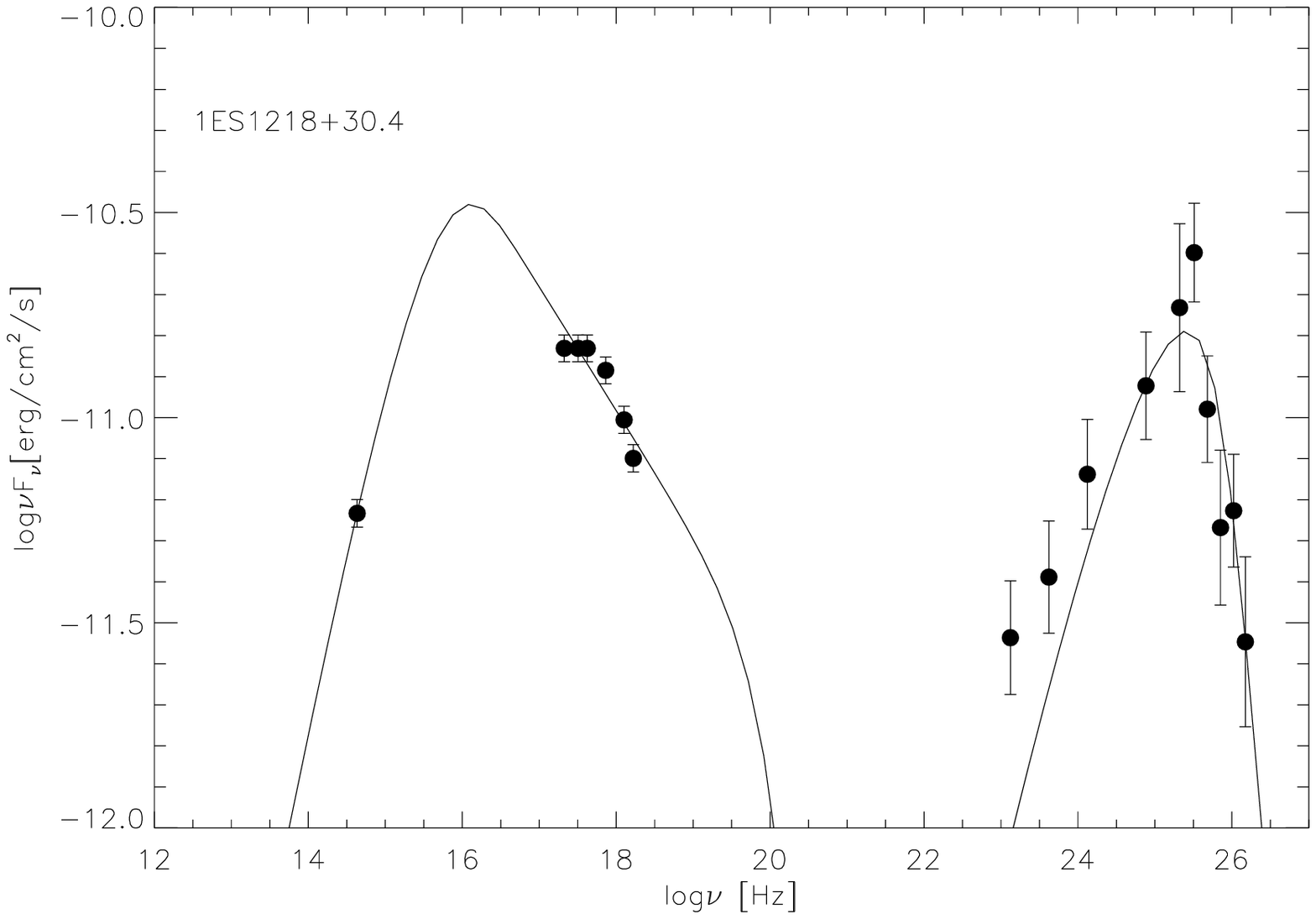}

\hfill

    \end{tabular}
  \end{center}
    \center{\textbf{Fig. 6.}---  continued}

\end{figure*}

\begin{figure*}
  \begin{center}
   \begin{tabular}{cc}
\includegraphics[width=5.8cm,height=5.0cm]{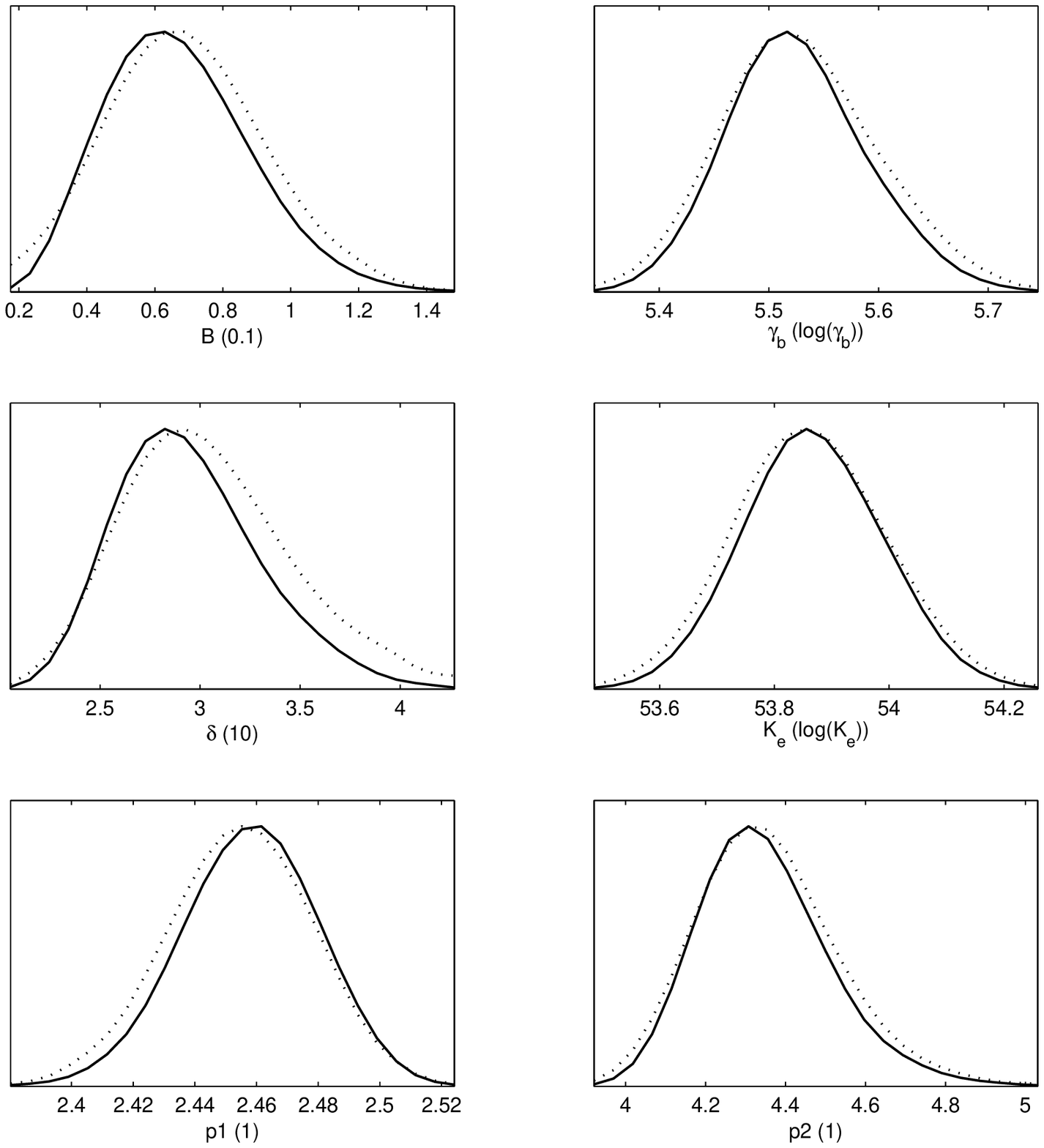}
\includegraphics[width=8.0cm,height=5.3cm]{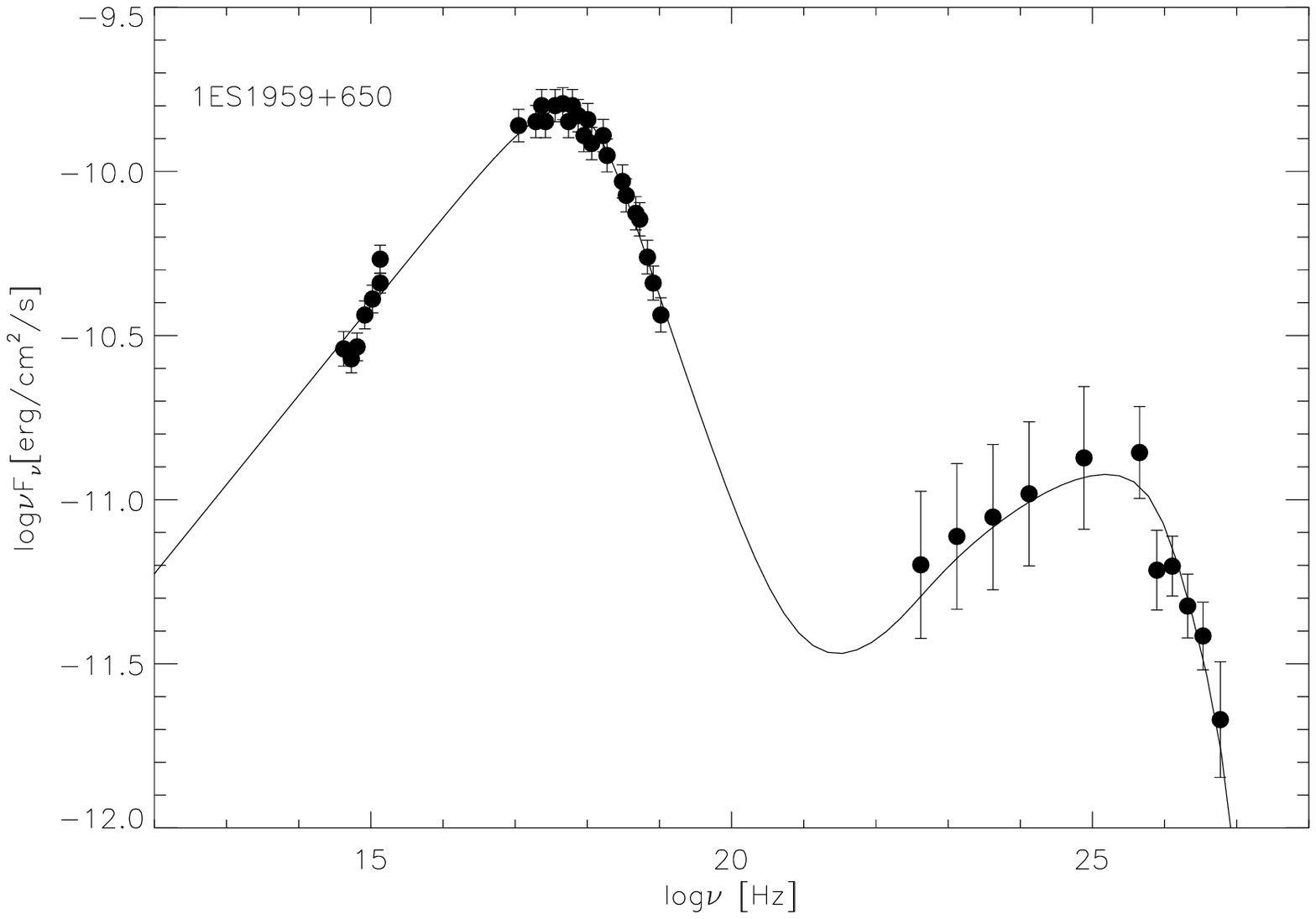}\\
\includegraphics[width=5.8cm,height=5.0cm]{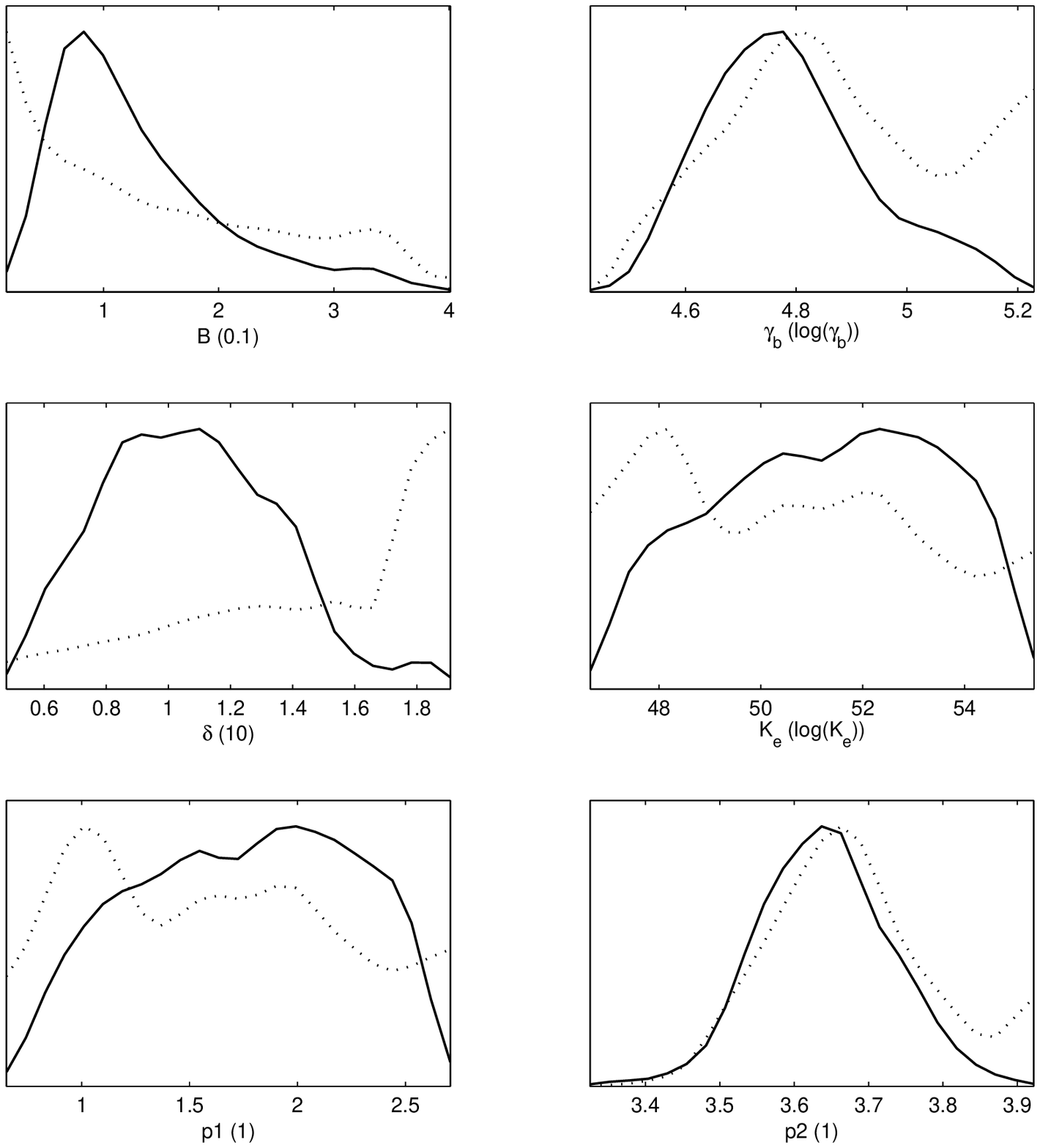}
\includegraphics[width=8.0cm,height=5.3cm]{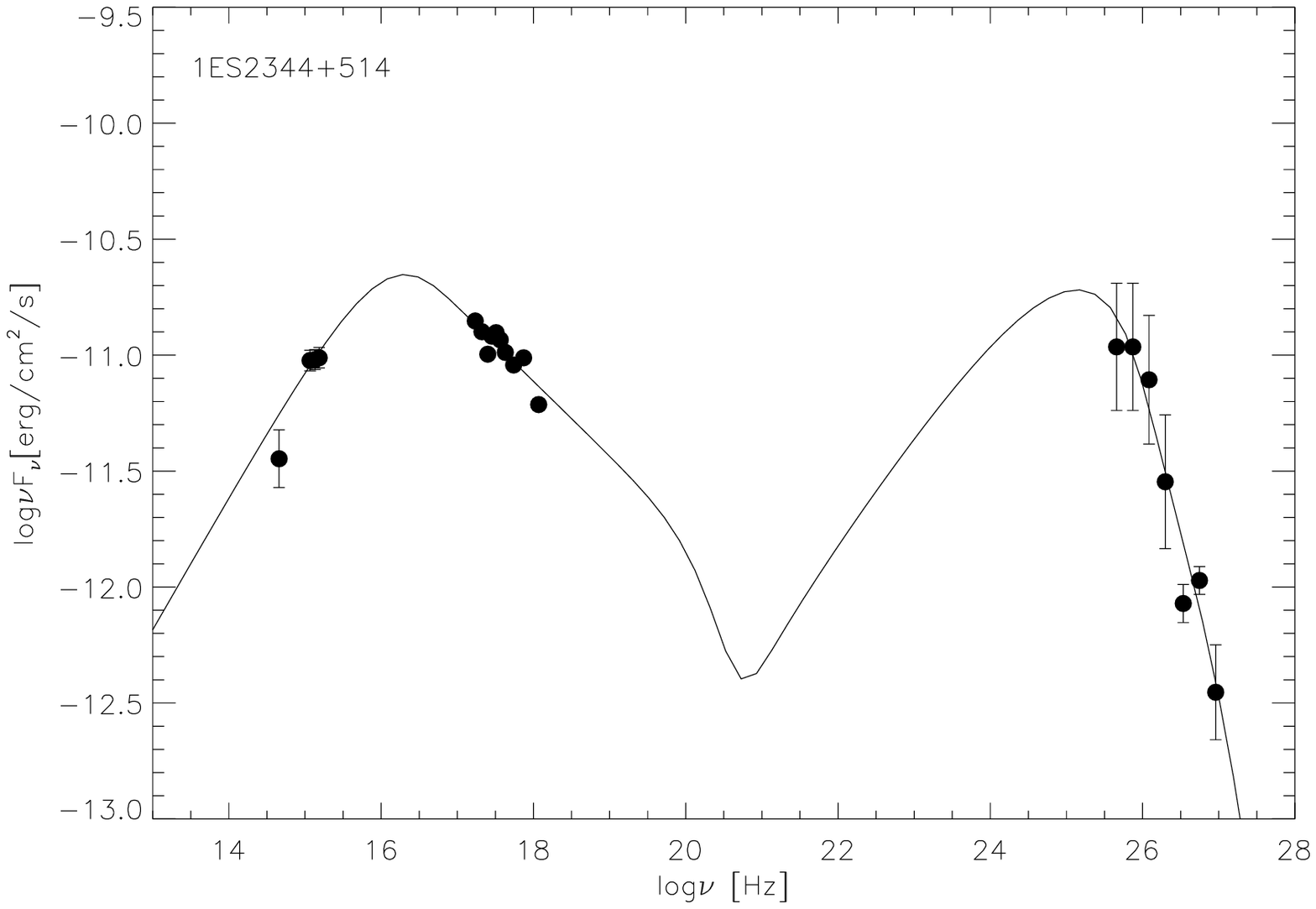}\\
\includegraphics[width=5.8cm,height=5.0cm]{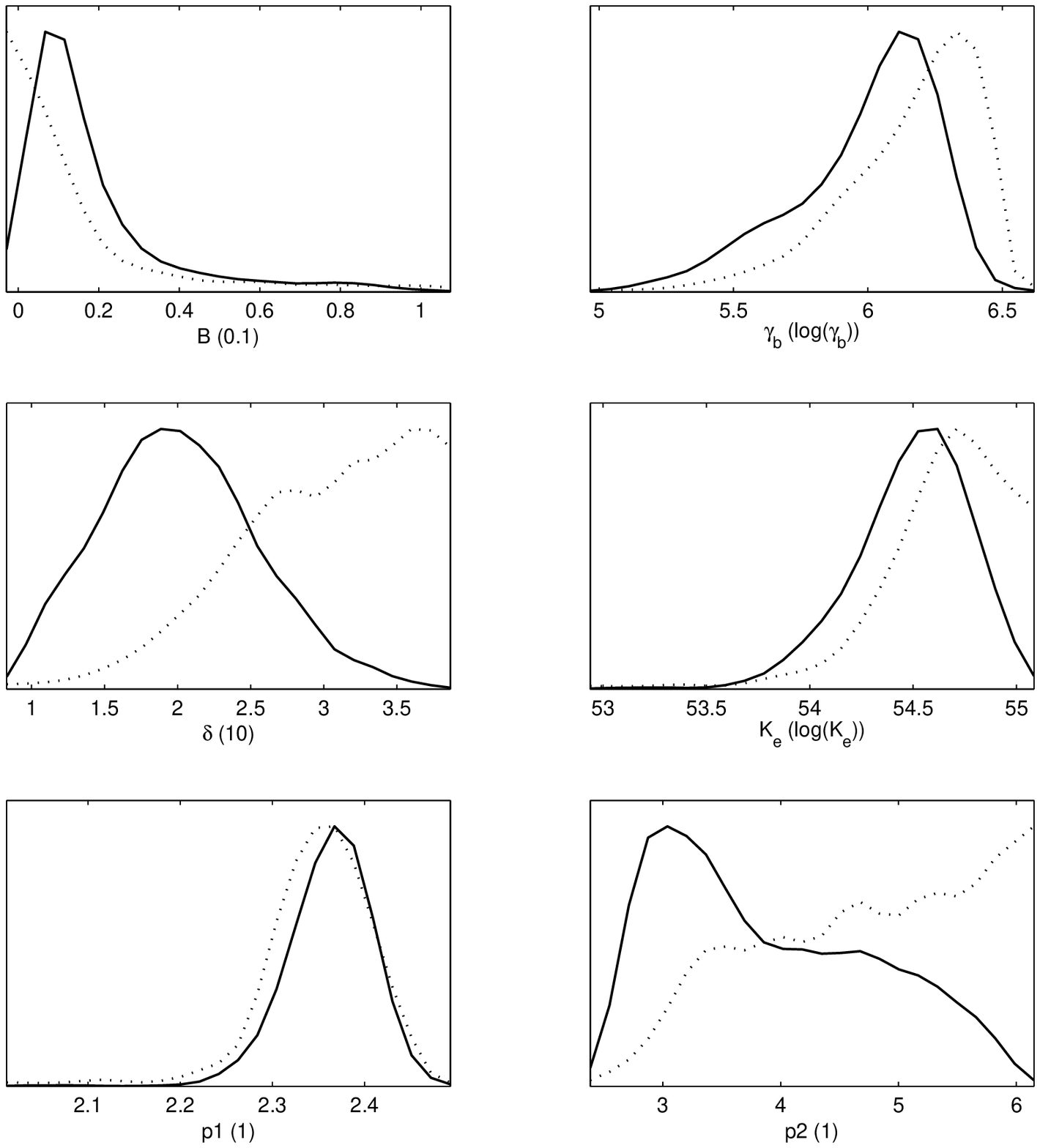}
\includegraphics[width=8.0cm,height=5.3cm]{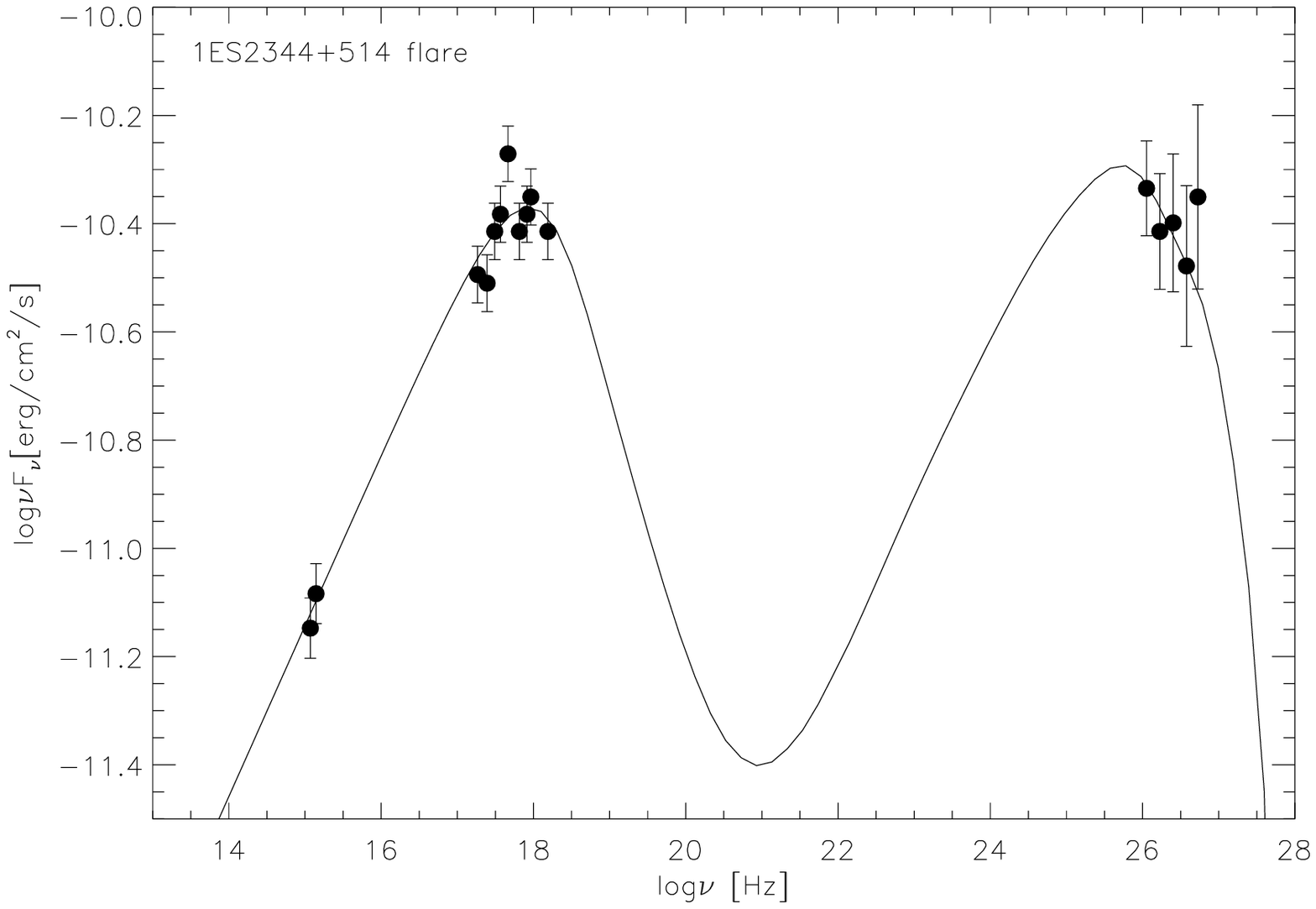}\\
\includegraphics[width=5.8cm,height=5.0cm]{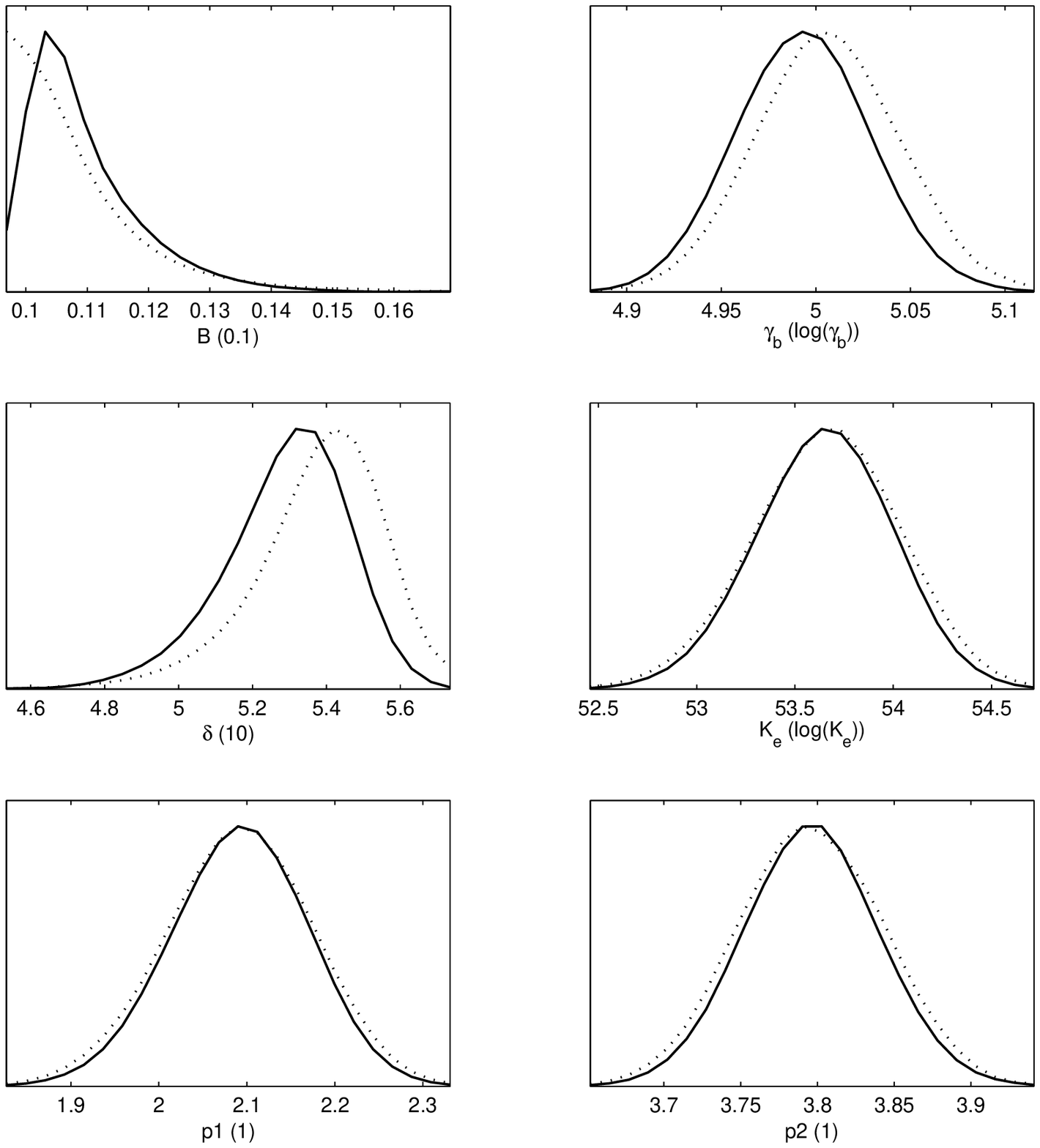}
\hfill
\includegraphics[width=8.0cm,height=5.3cm]{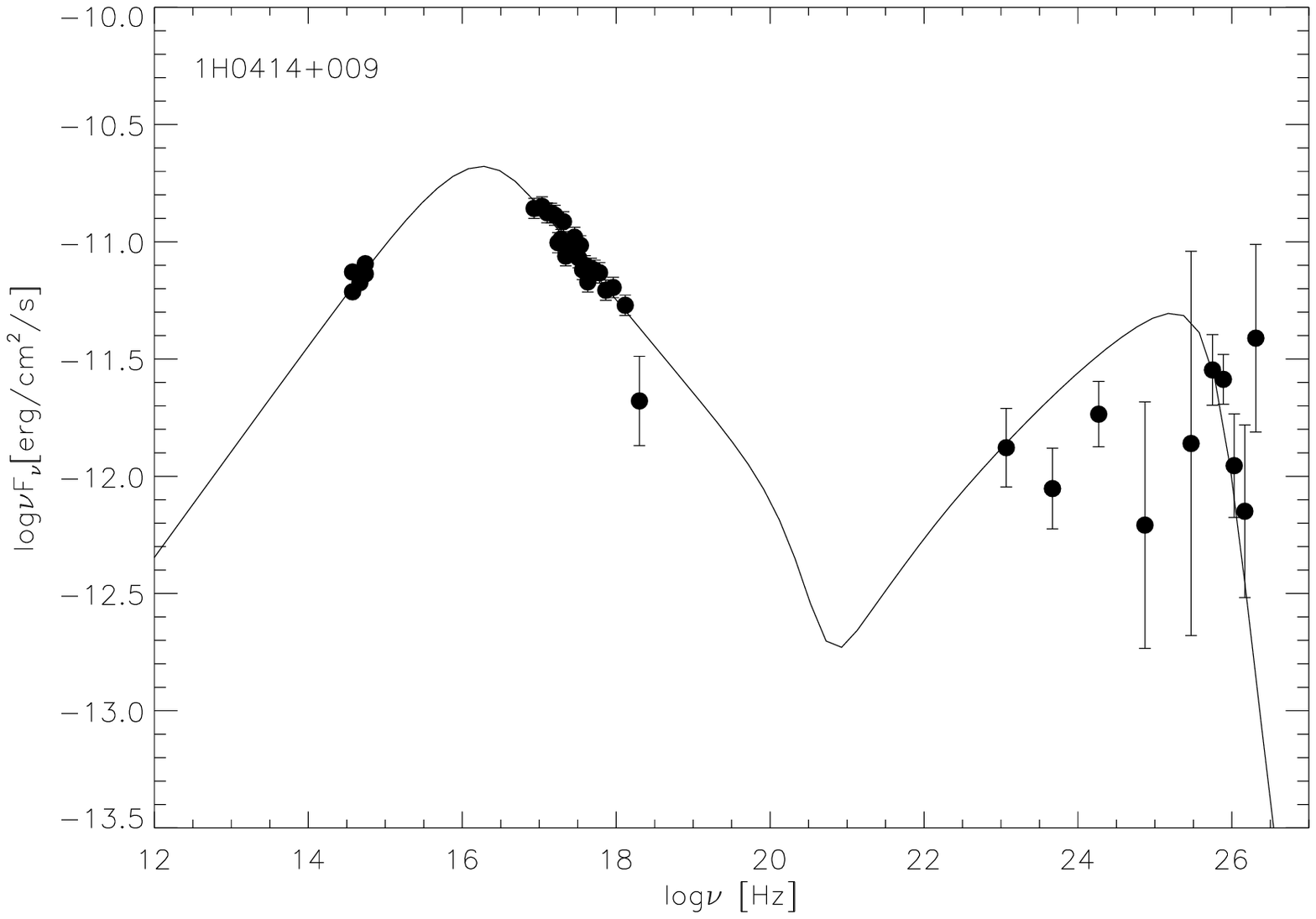}
\hfill

    \end{tabular}
  \end{center}
    \center{\textbf{Fig. 6.}---  continued}

\end{figure*}

\begin{figure*}
  \begin{center}
   \begin{tabular}{cc}
\includegraphics[width=5.8cm,height=5.0cm]{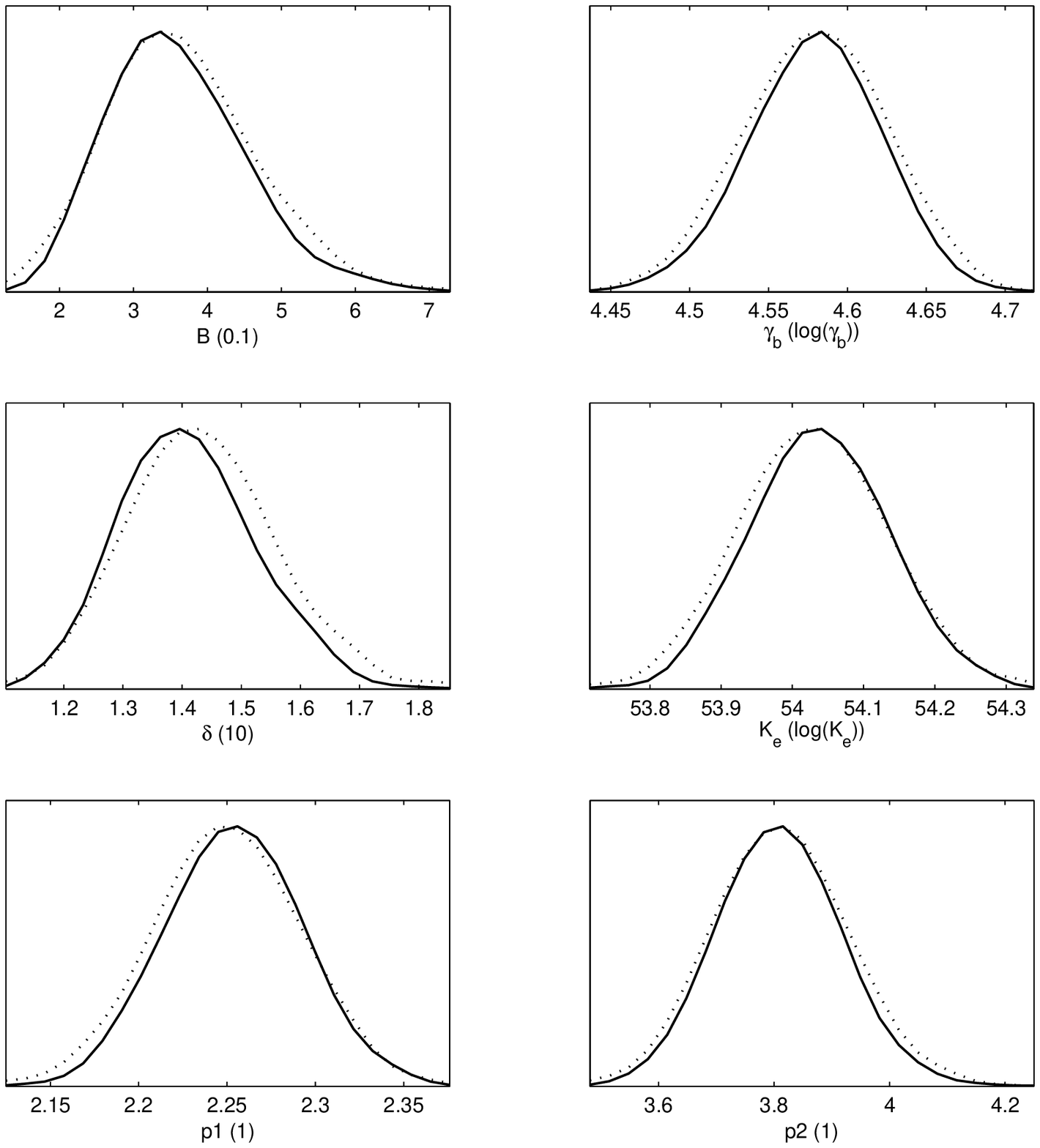}
\includegraphics[width=8.0cm,height=5.3cm]{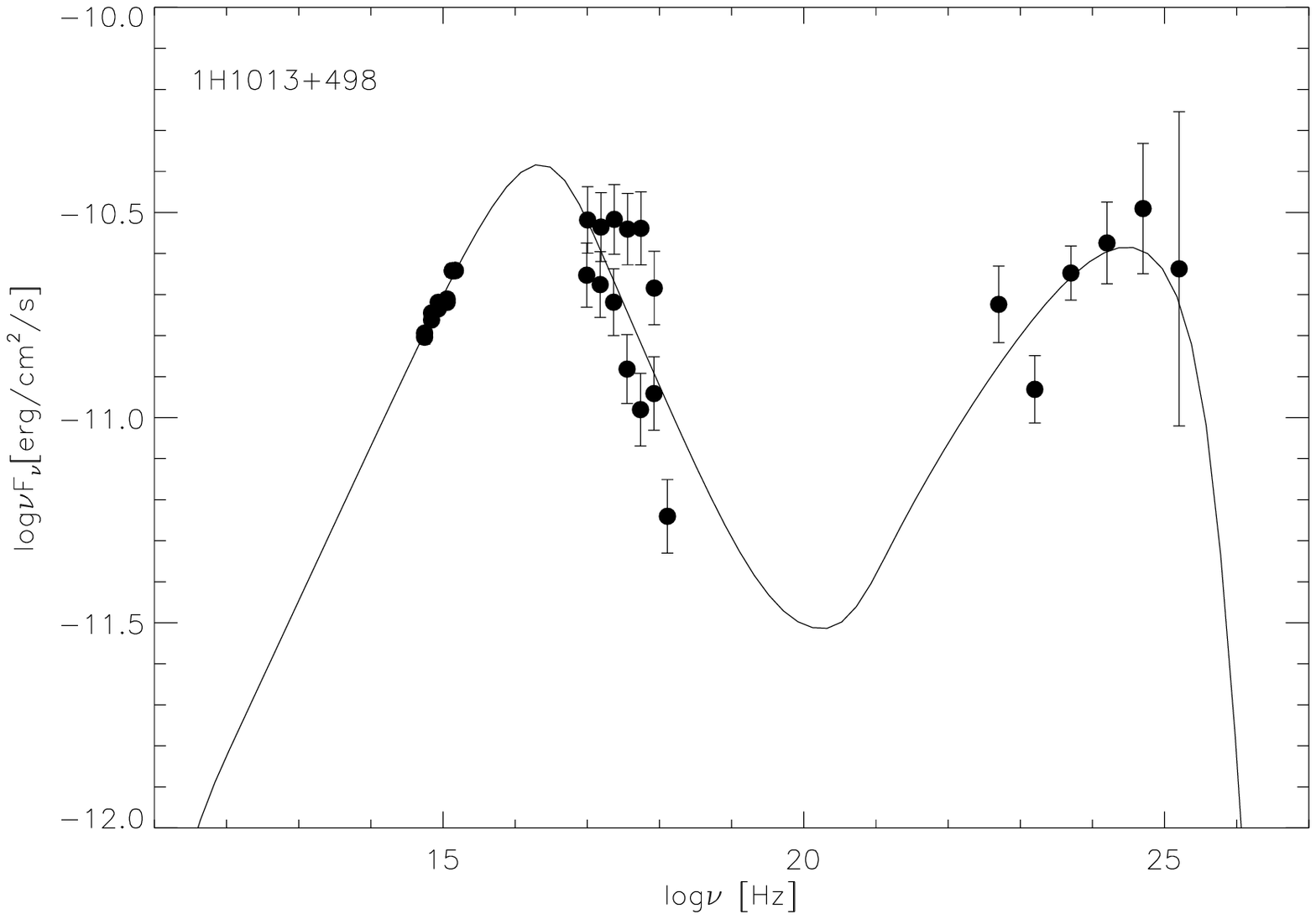}\\
\includegraphics[width=5.8cm,height=5.0cm]{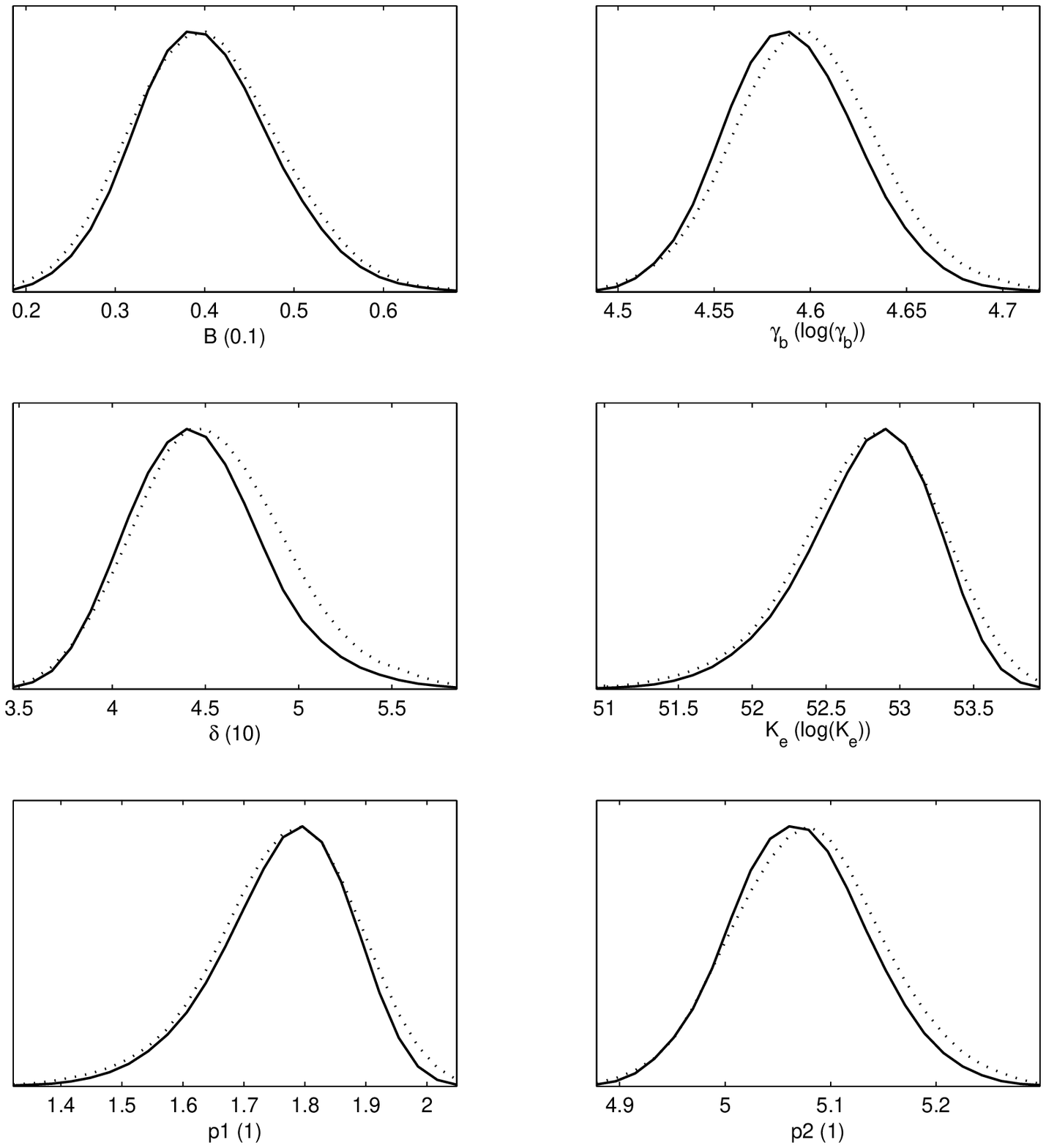}
\includegraphics[width=8.0cm,height=5.3cm]{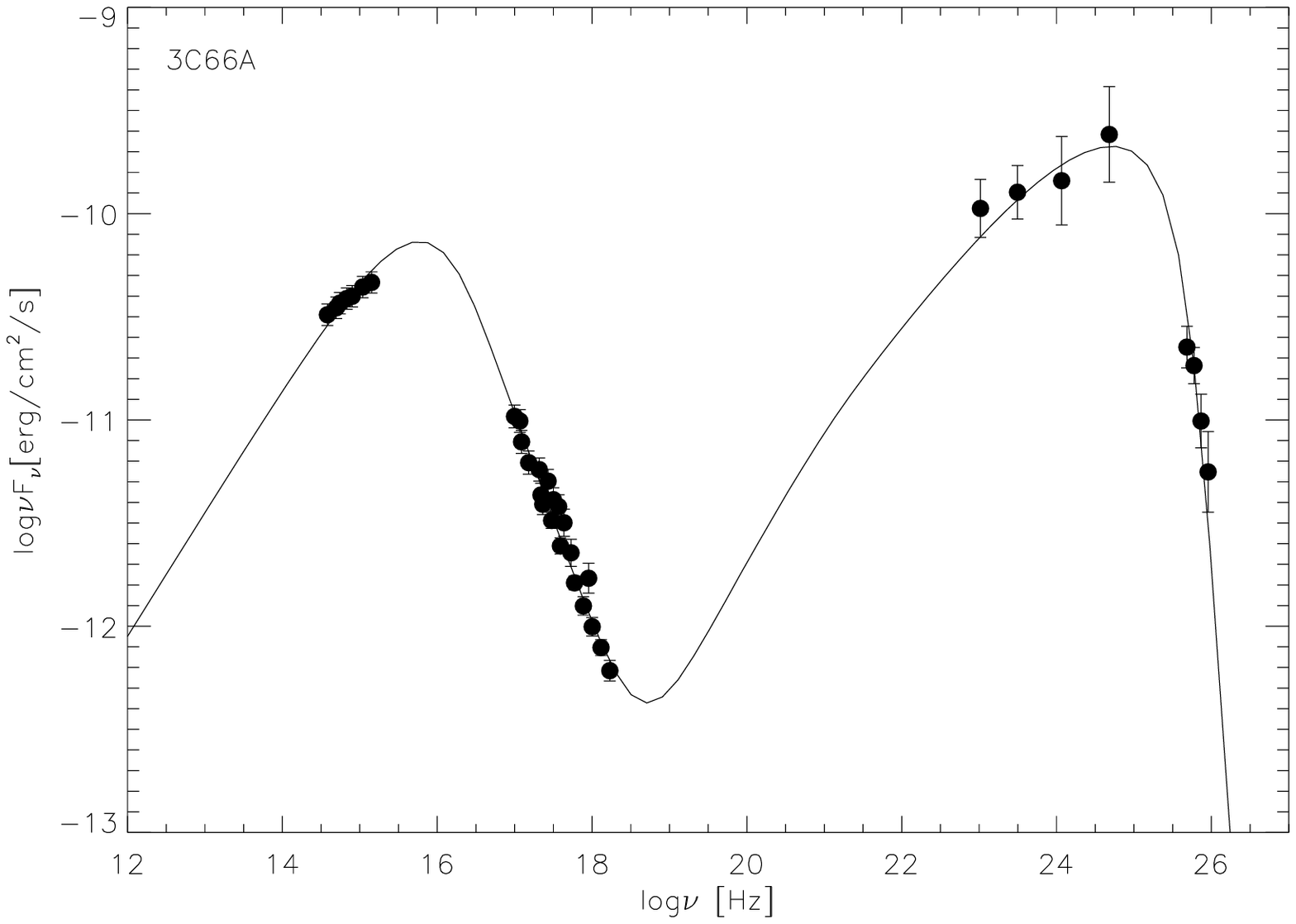}\\
\includegraphics[width=5.8cm,height=5.0cm]{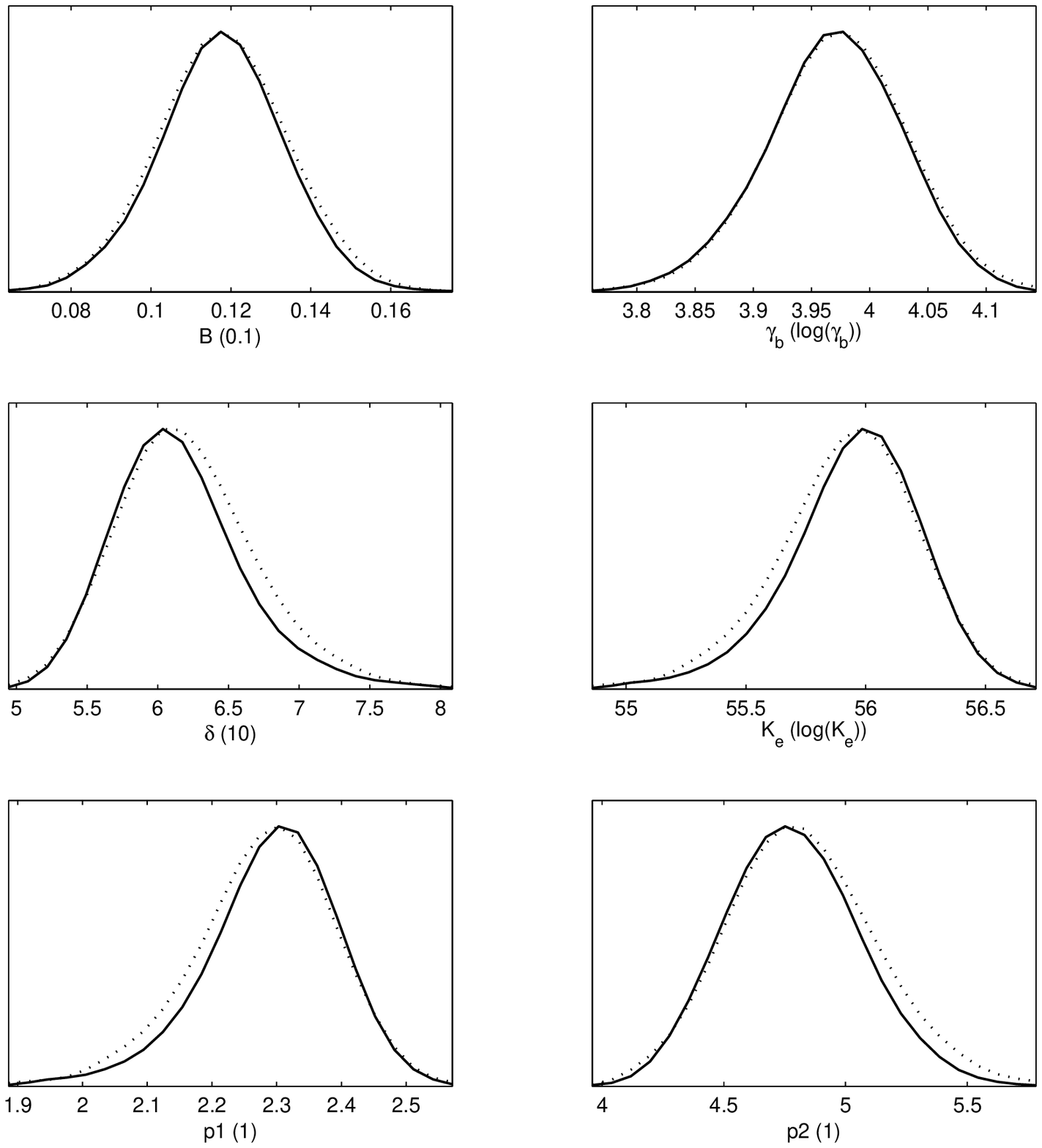}
\includegraphics[width=8.0cm,height=5.3cm]{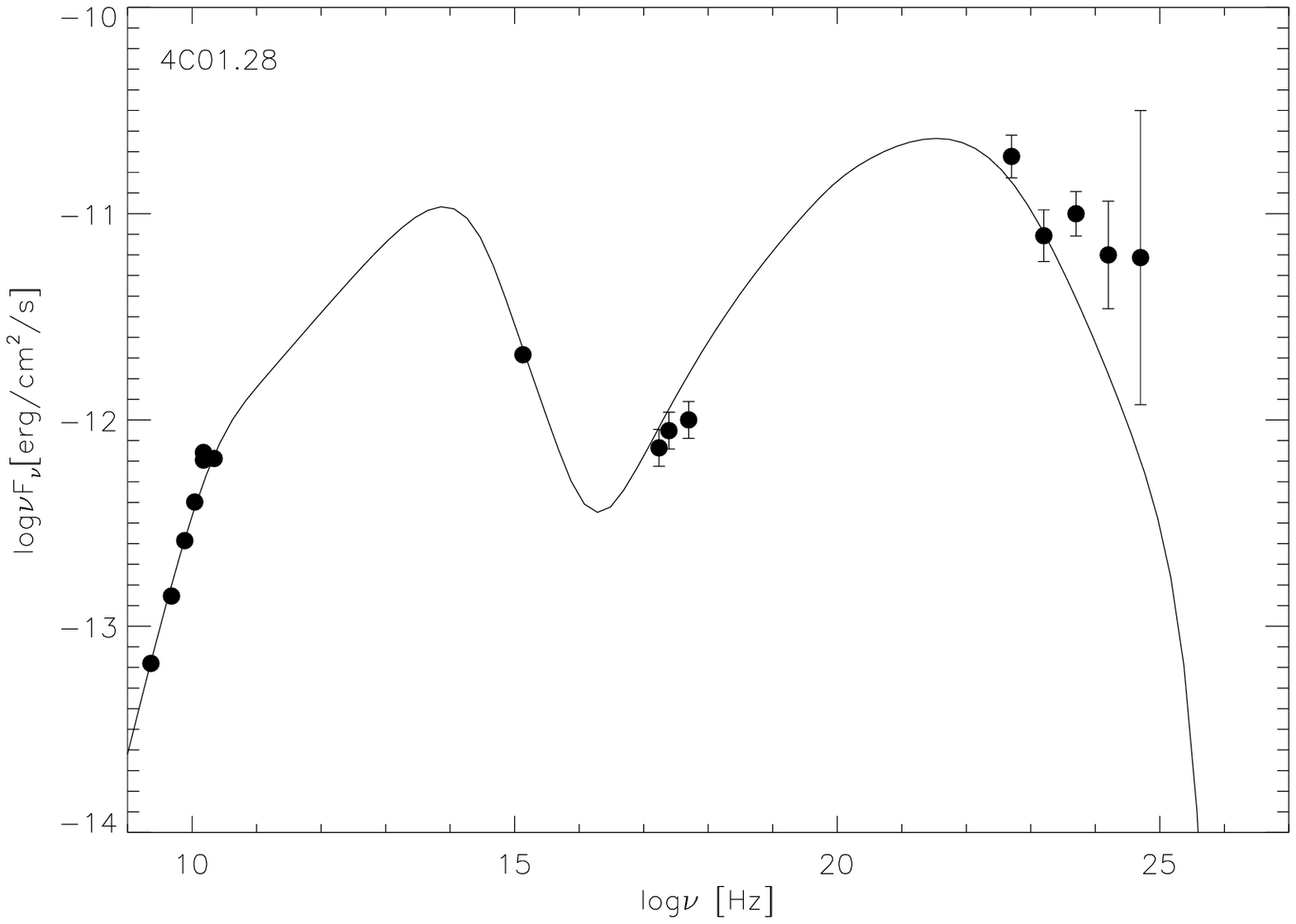}\\
\includegraphics[width=5.8cm,height=5.0cm]{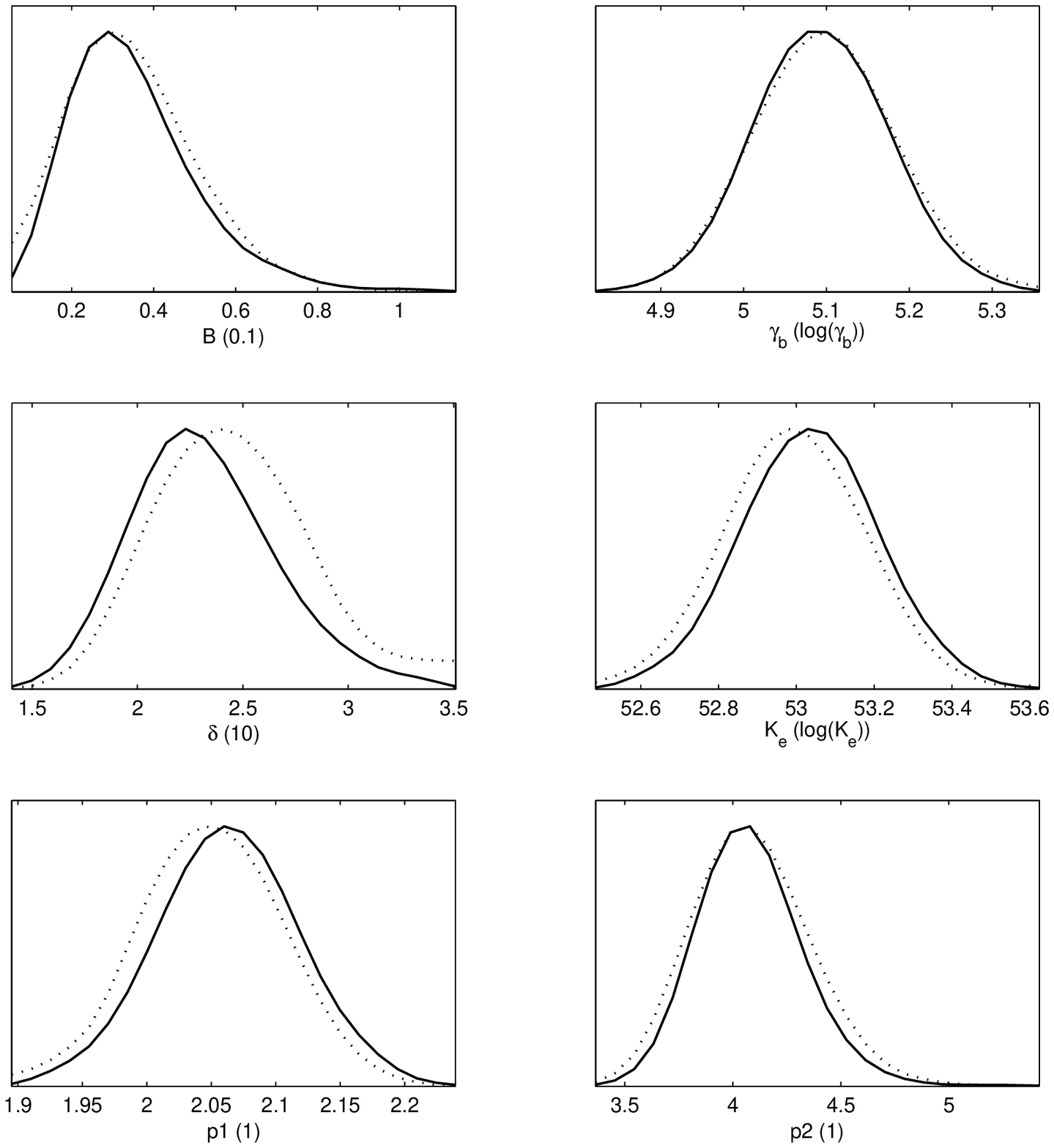}
\hfill
\includegraphics[width=8.0cm,height=5.3cm]{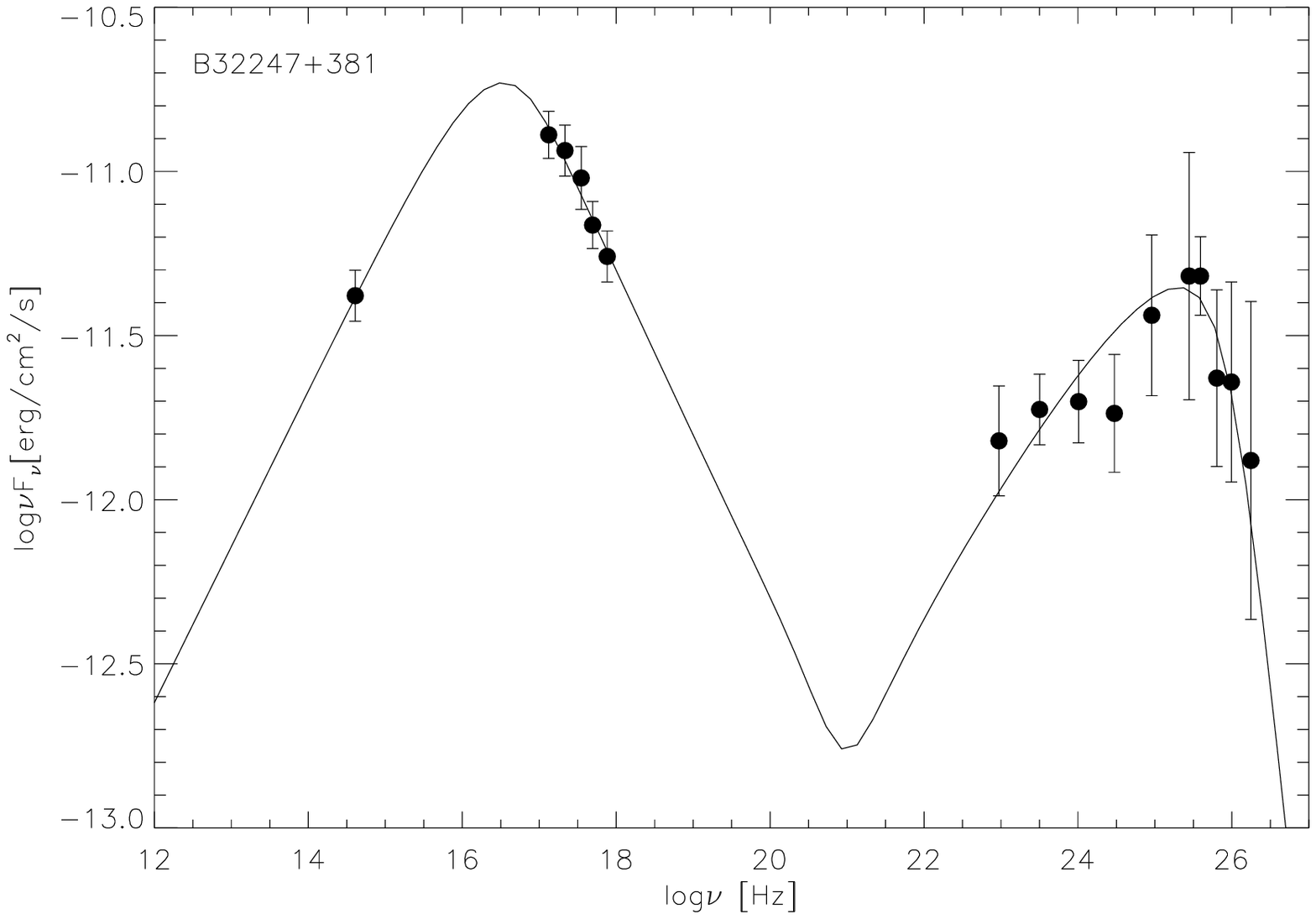}
\hfill

    \end{tabular}
  \end{center}
    \center{\textbf{Fig. 6.}---  continued}

\end{figure*}

\begin{figure*}
  \begin{center}
   \begin{tabular}{cc}
\includegraphics[width=5.8cm,height=5.0cm]{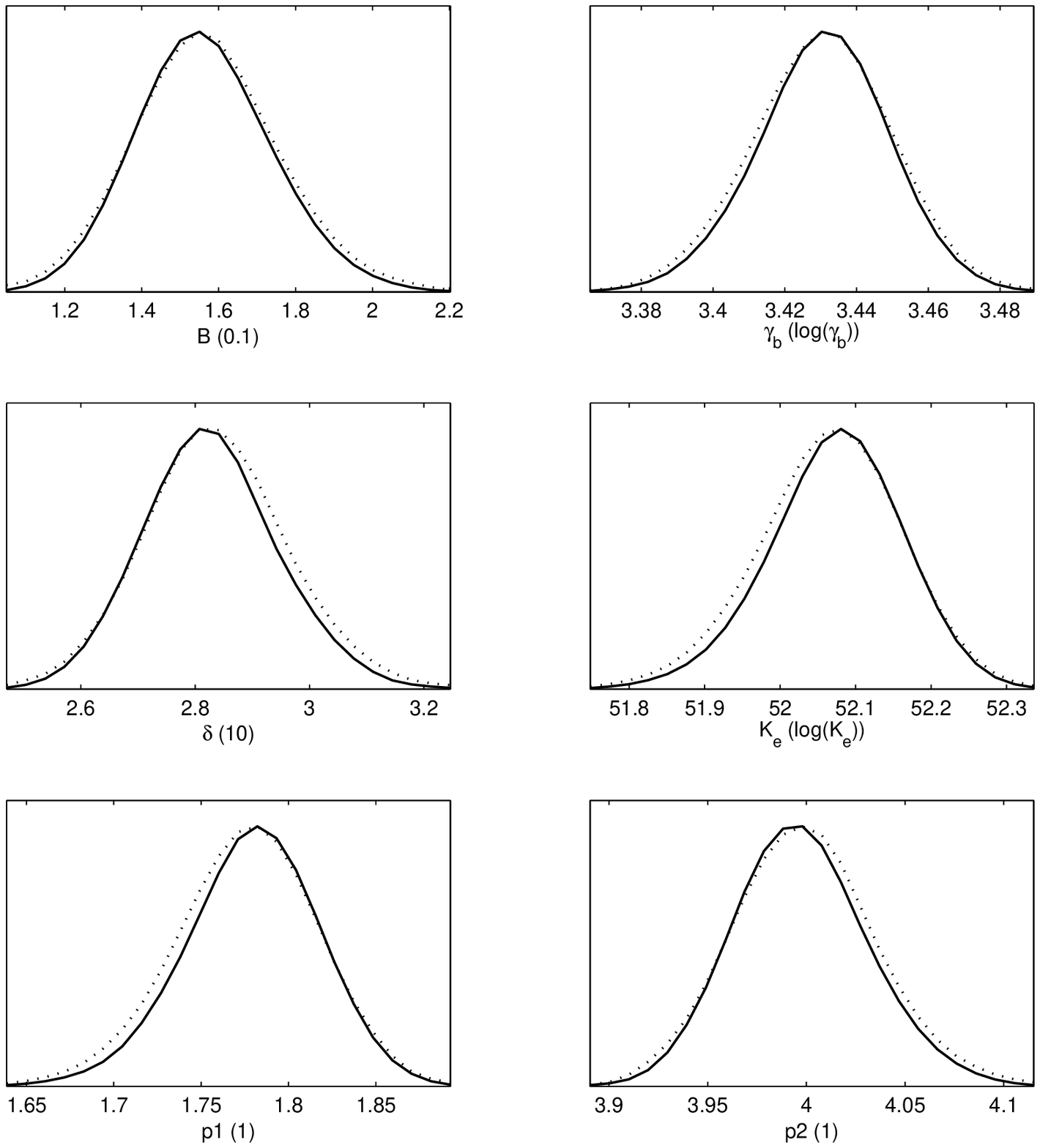}
\includegraphics[width=8.0cm,height=5.3cm]{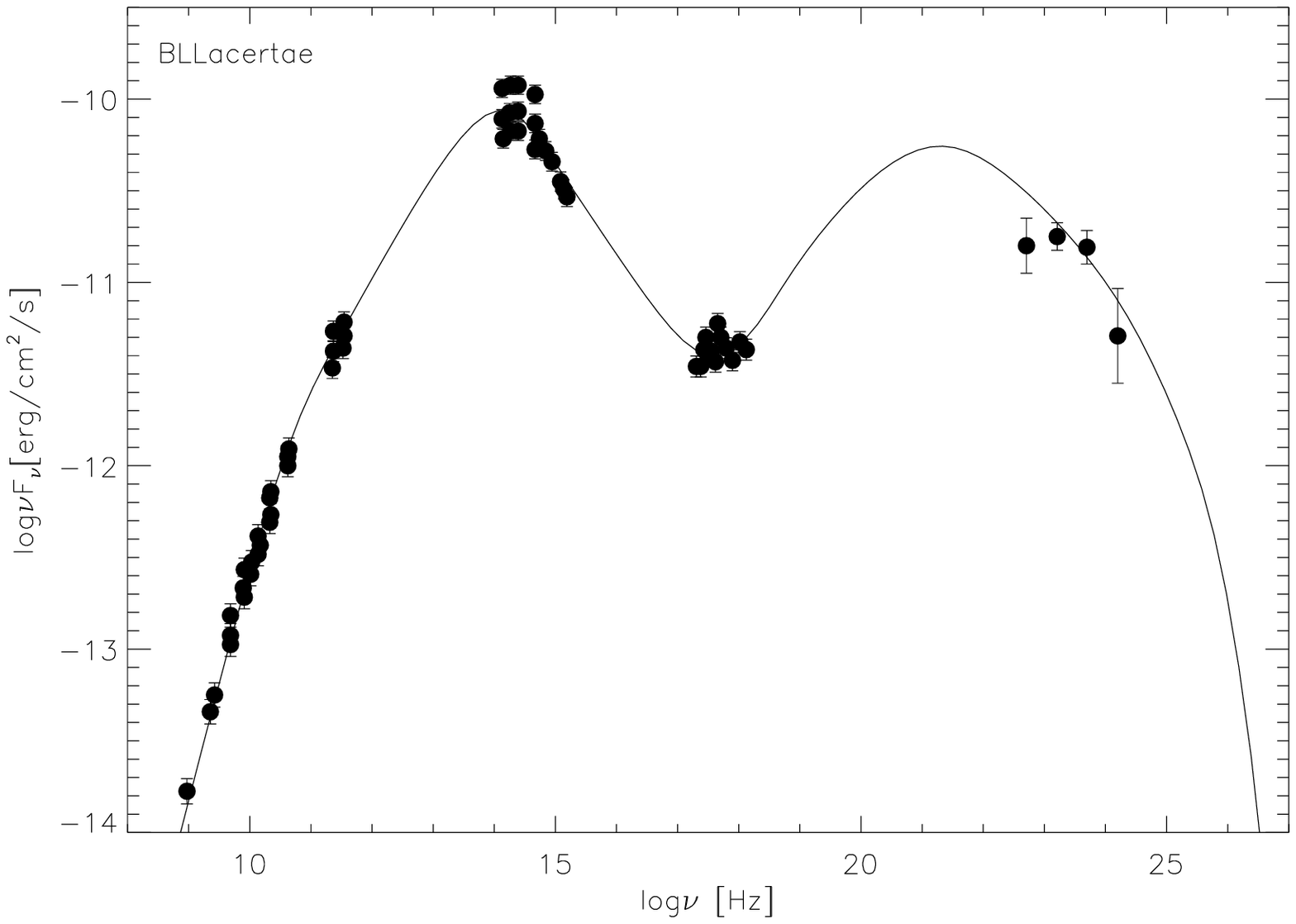}\\
\includegraphics[width=5.8cm,height=5.0cm]{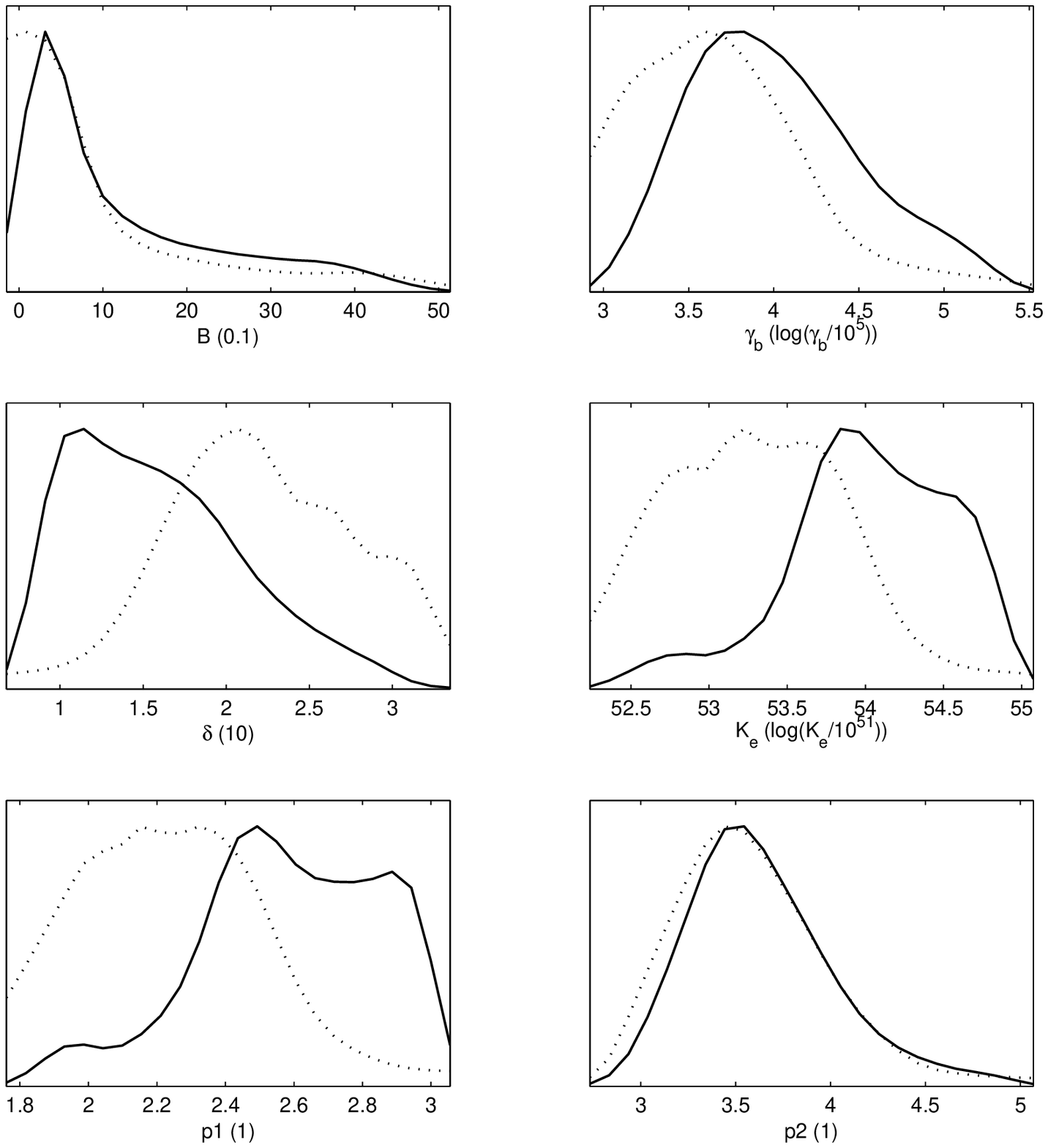}
\includegraphics[width=8.0cm,height=5.3cm]{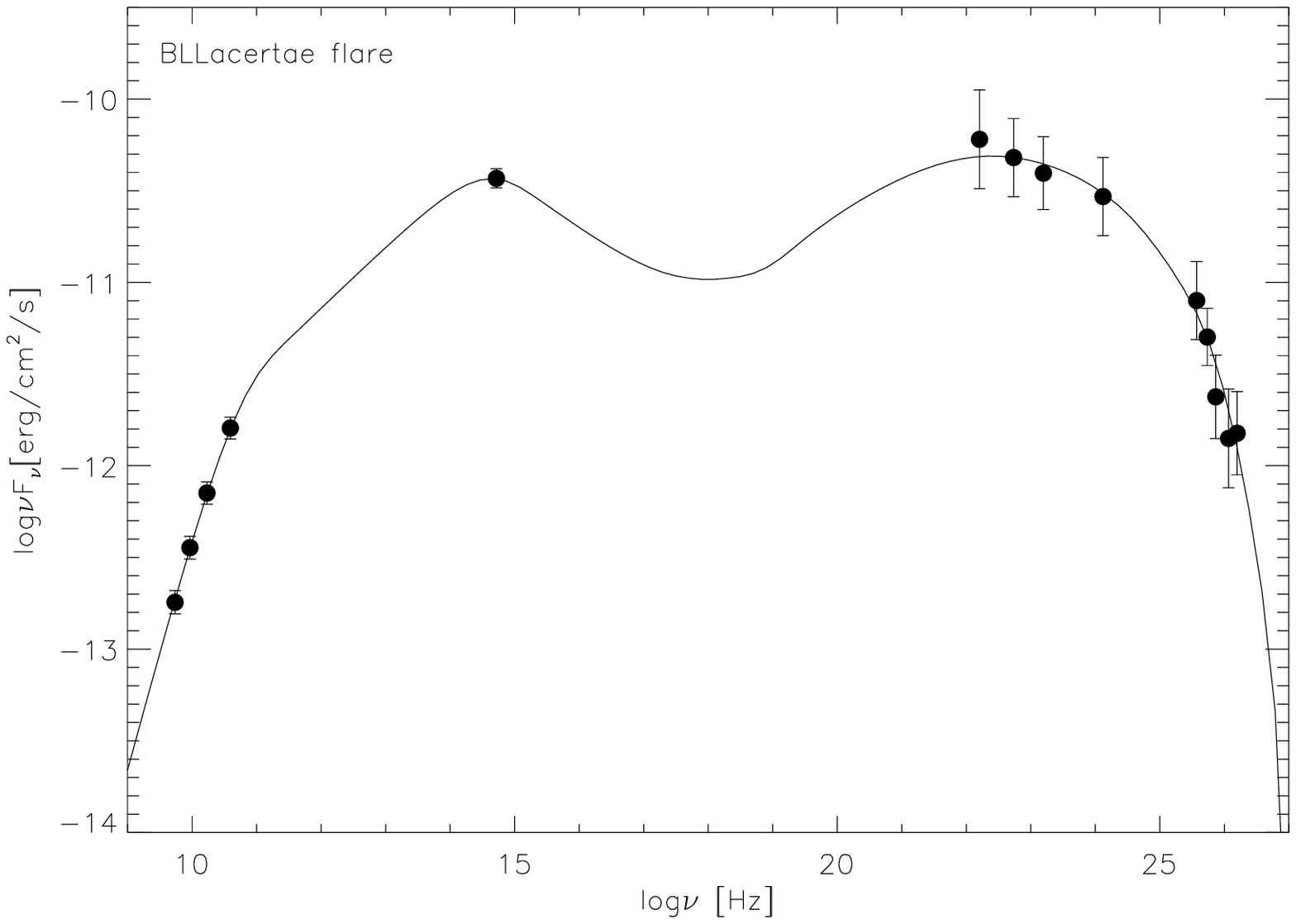}\\
\includegraphics[width=5.8cm,height=5.0cm]{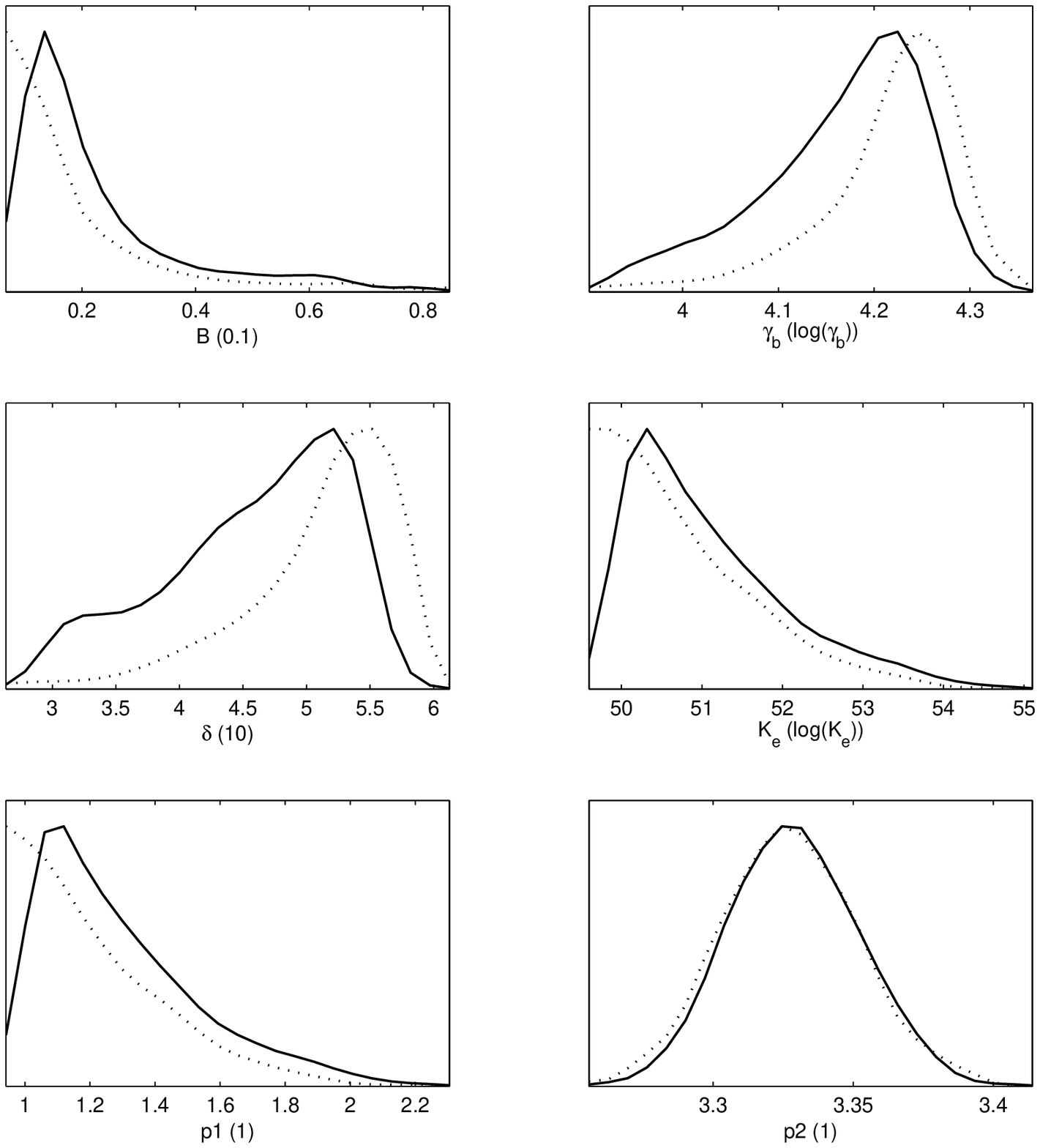}
\includegraphics[width=8.0cm,height=5.3cm]{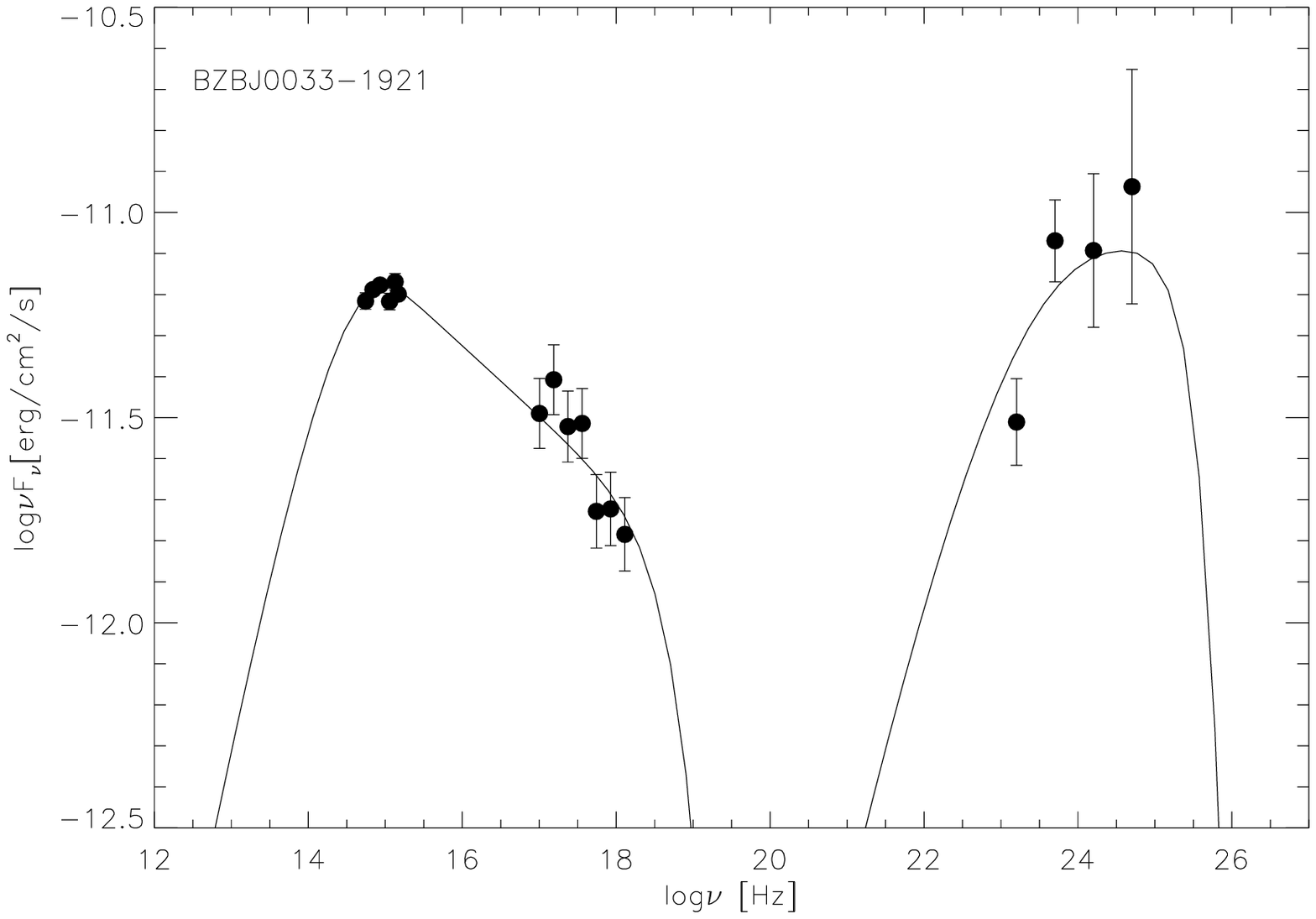}\\
\includegraphics[width=5.8cm,height=5.0cm]{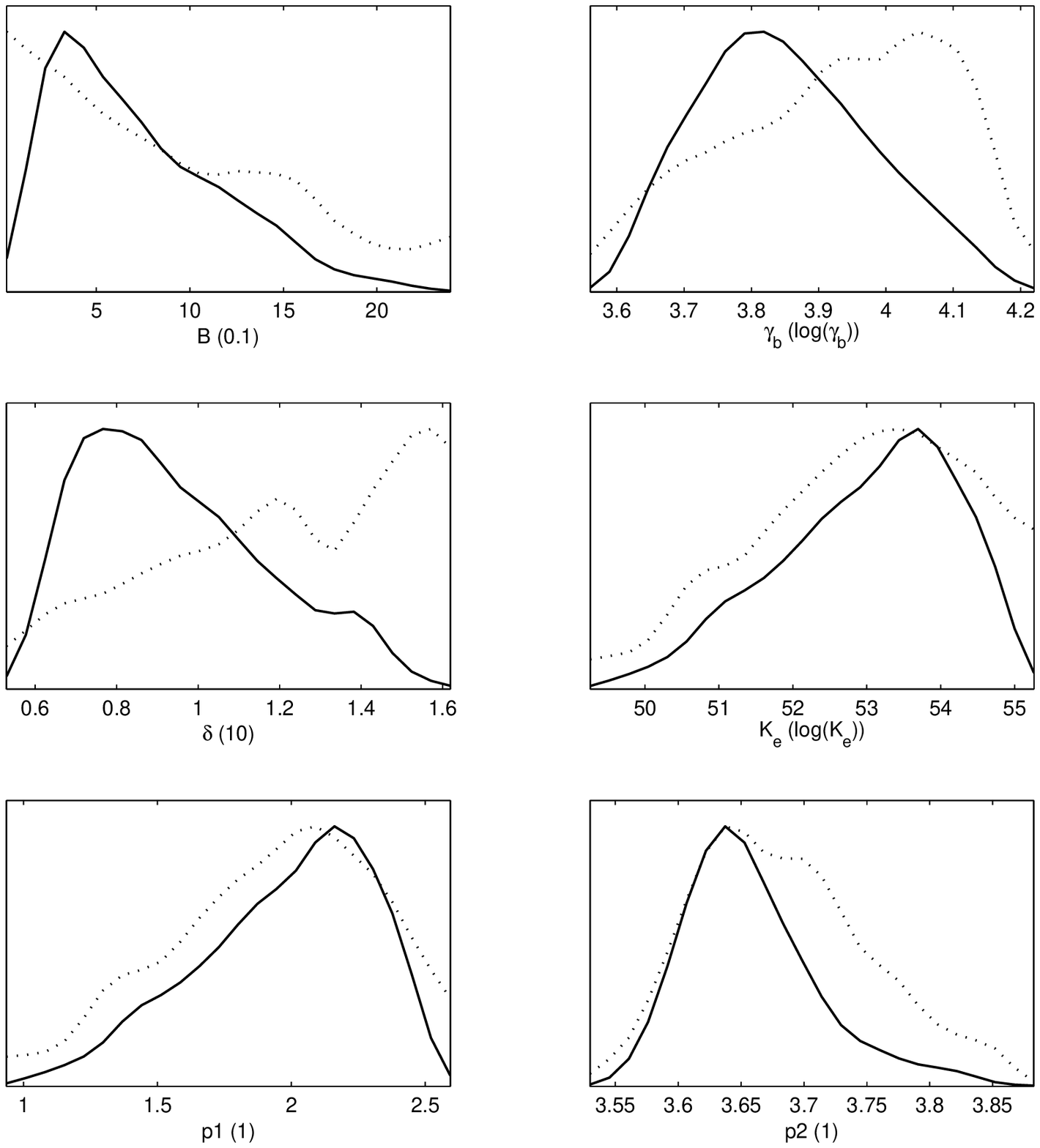}
\hfill
\includegraphics[width=8.0cm,height=5.3cm]{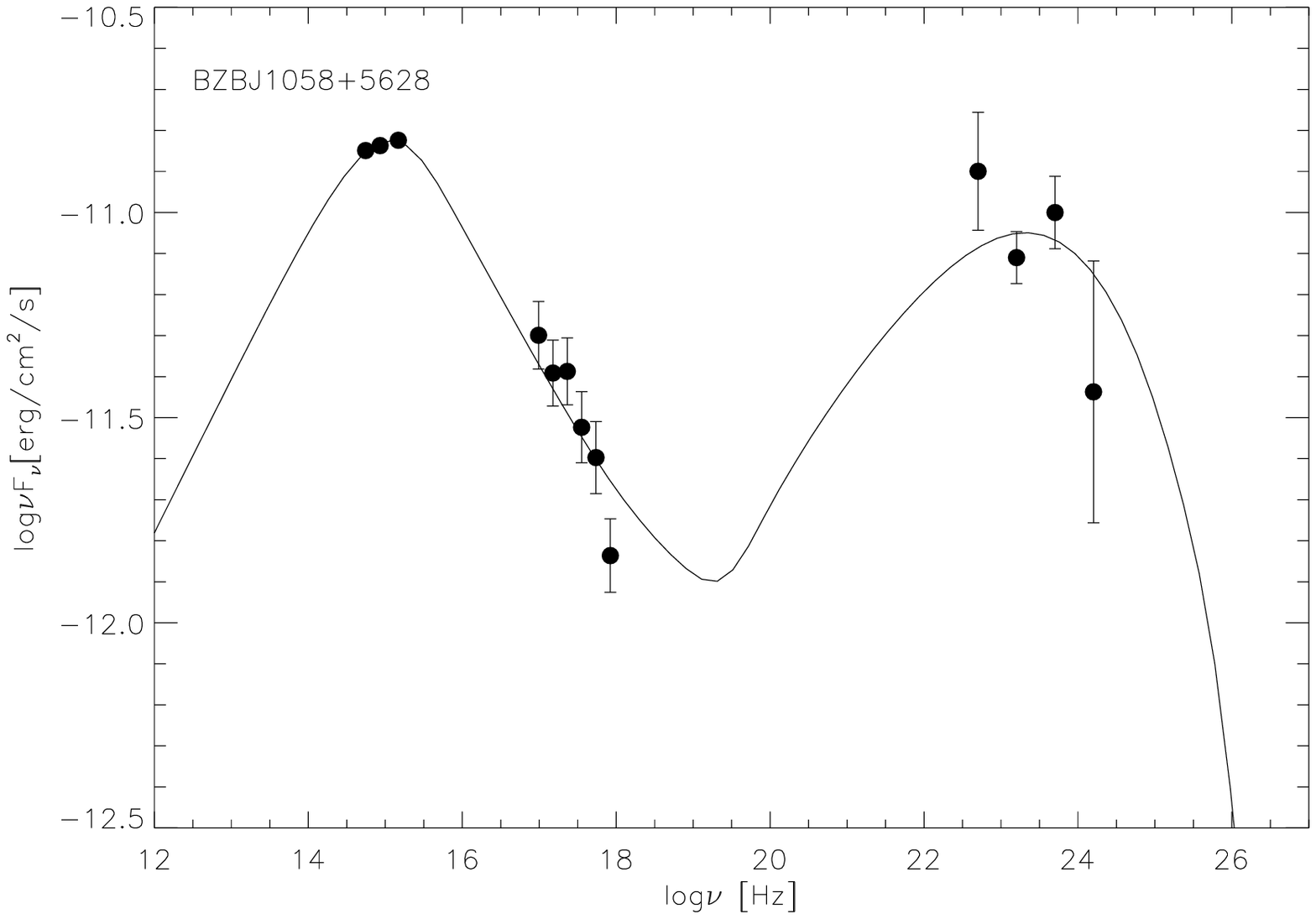}
\hfill

    \end{tabular}
  \end{center}
    \center{\textbf{Fig. 6.}---  continued}

\end{figure*}

\begin{figure*}
  \begin{center}
   \begin{tabular}{cc}
\includegraphics[width=5.8cm,height=5.0cm]{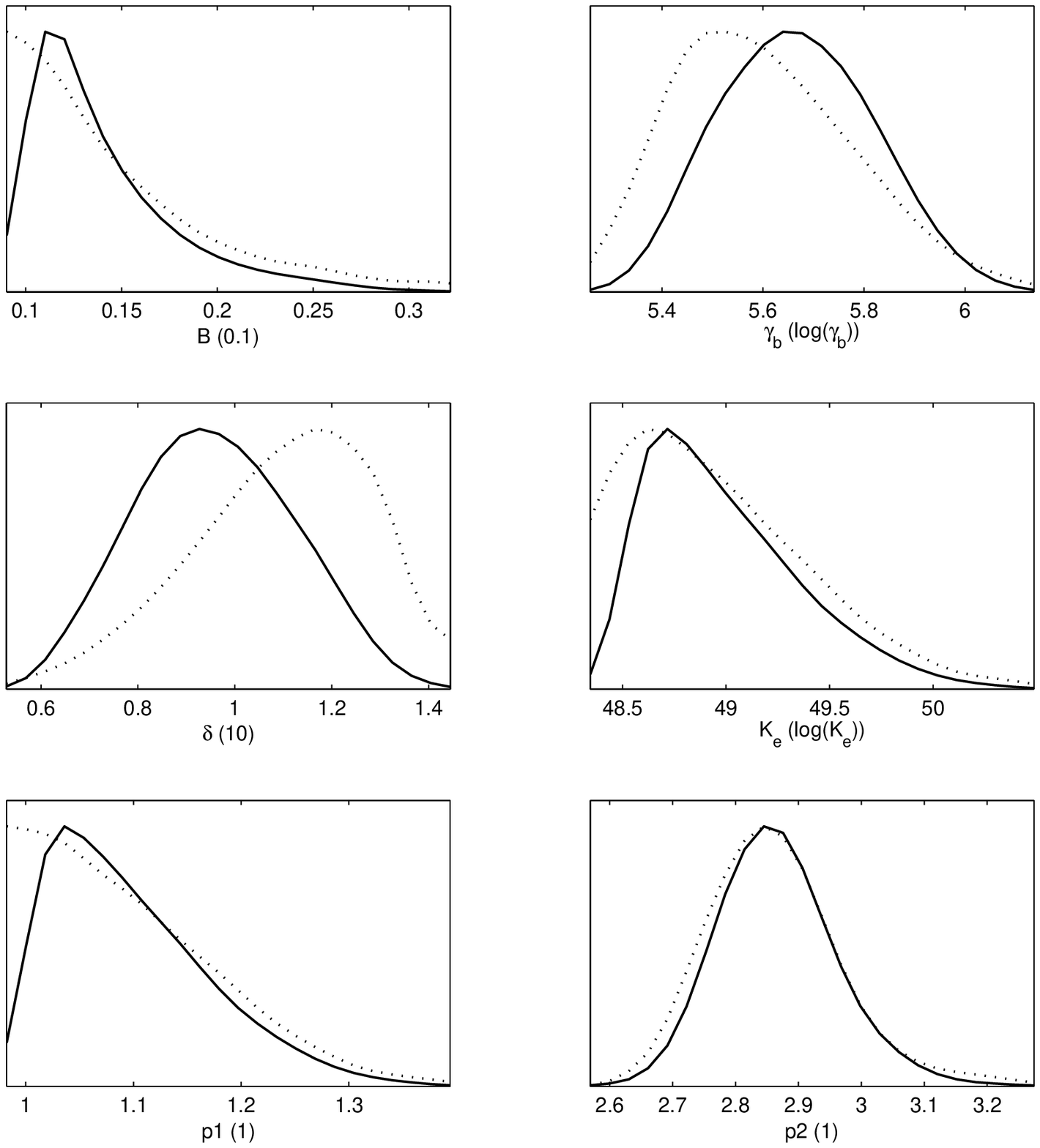}
\includegraphics[width=8.0cm,height=5.3cm]{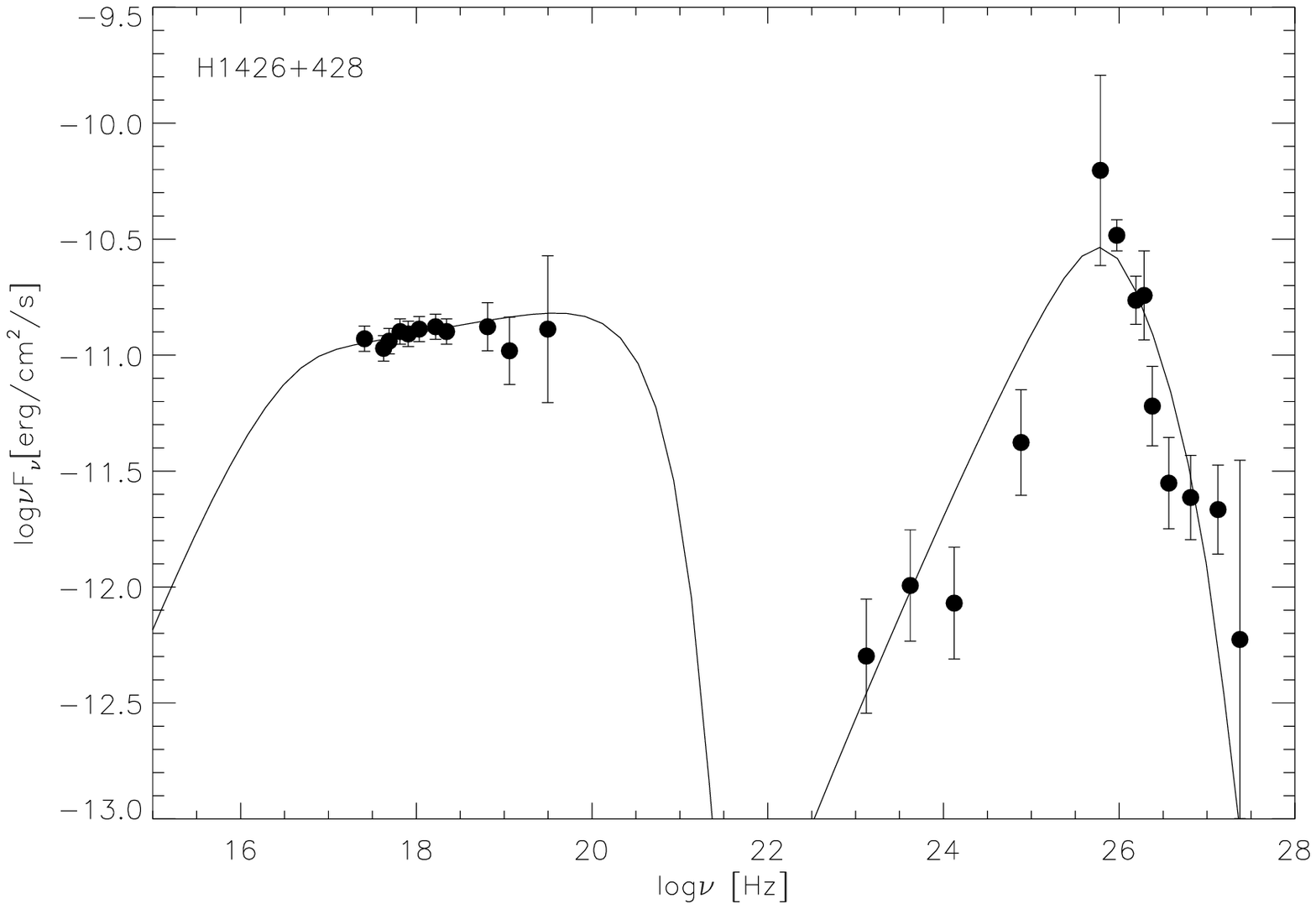}\\
\includegraphics[width=5.8cm,height=5.0cm]{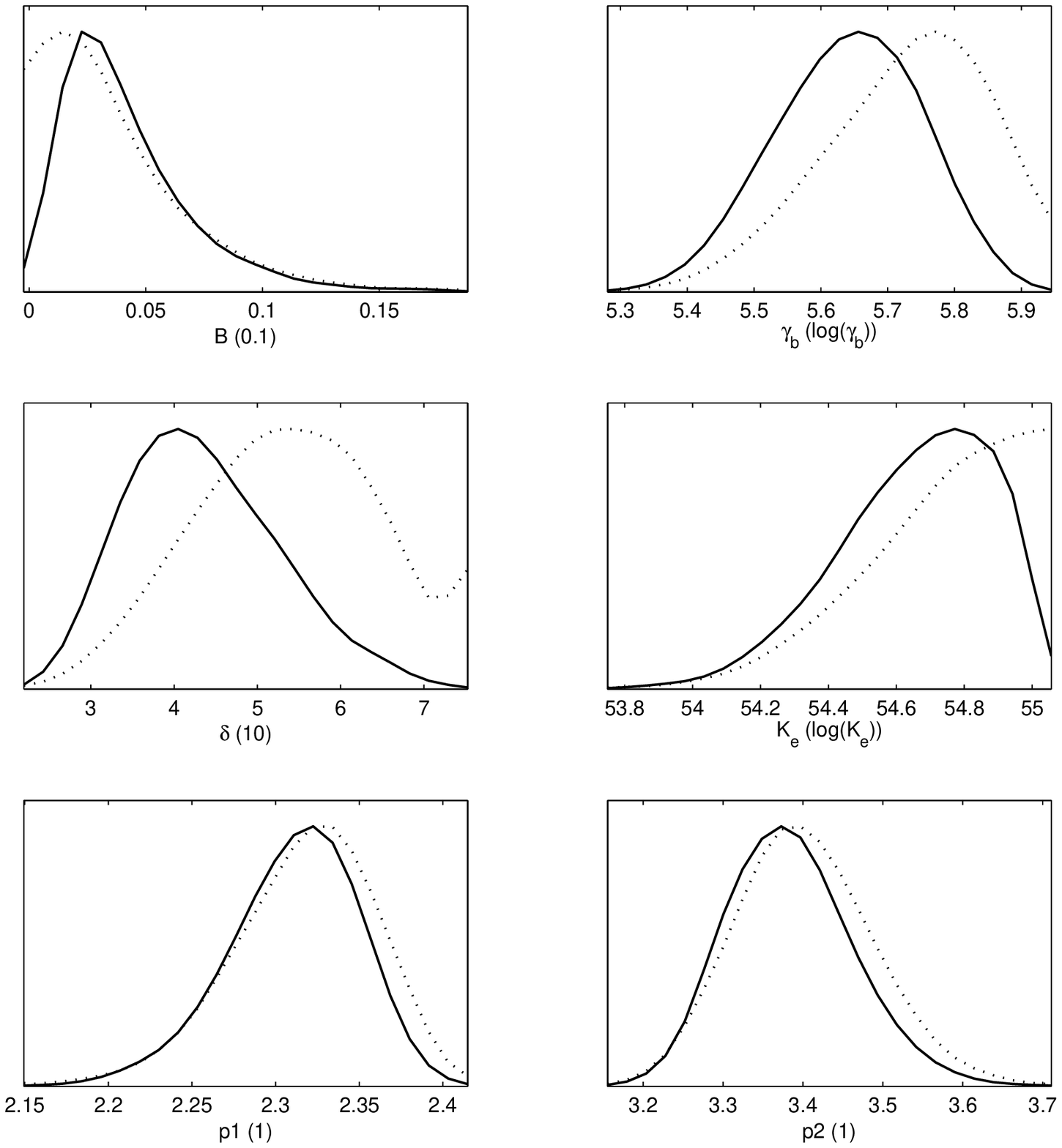}
\includegraphics[width=8.0cm,height=5.3cm]{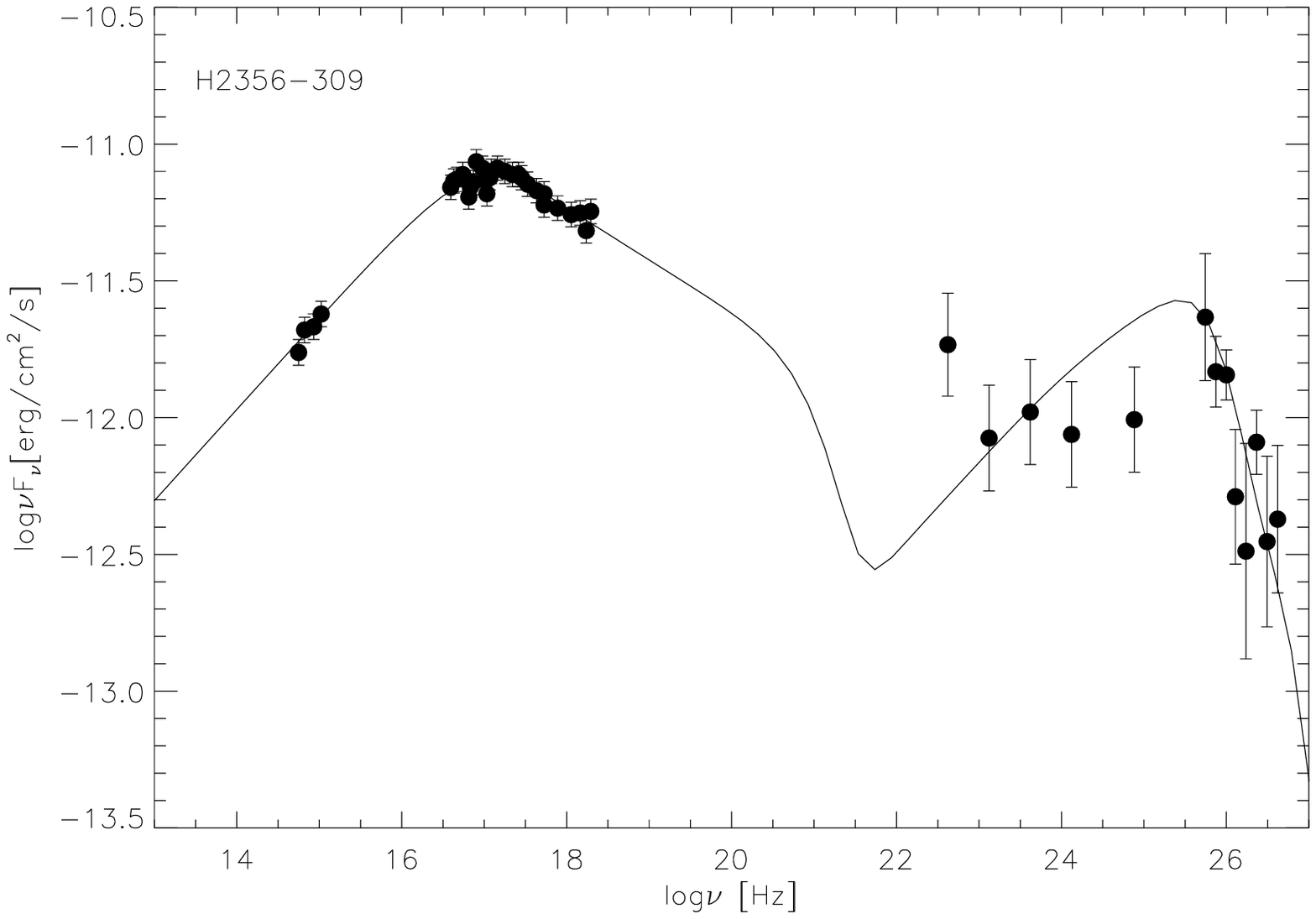}\\
\includegraphics[width=5.8cm,height=5.0cm]{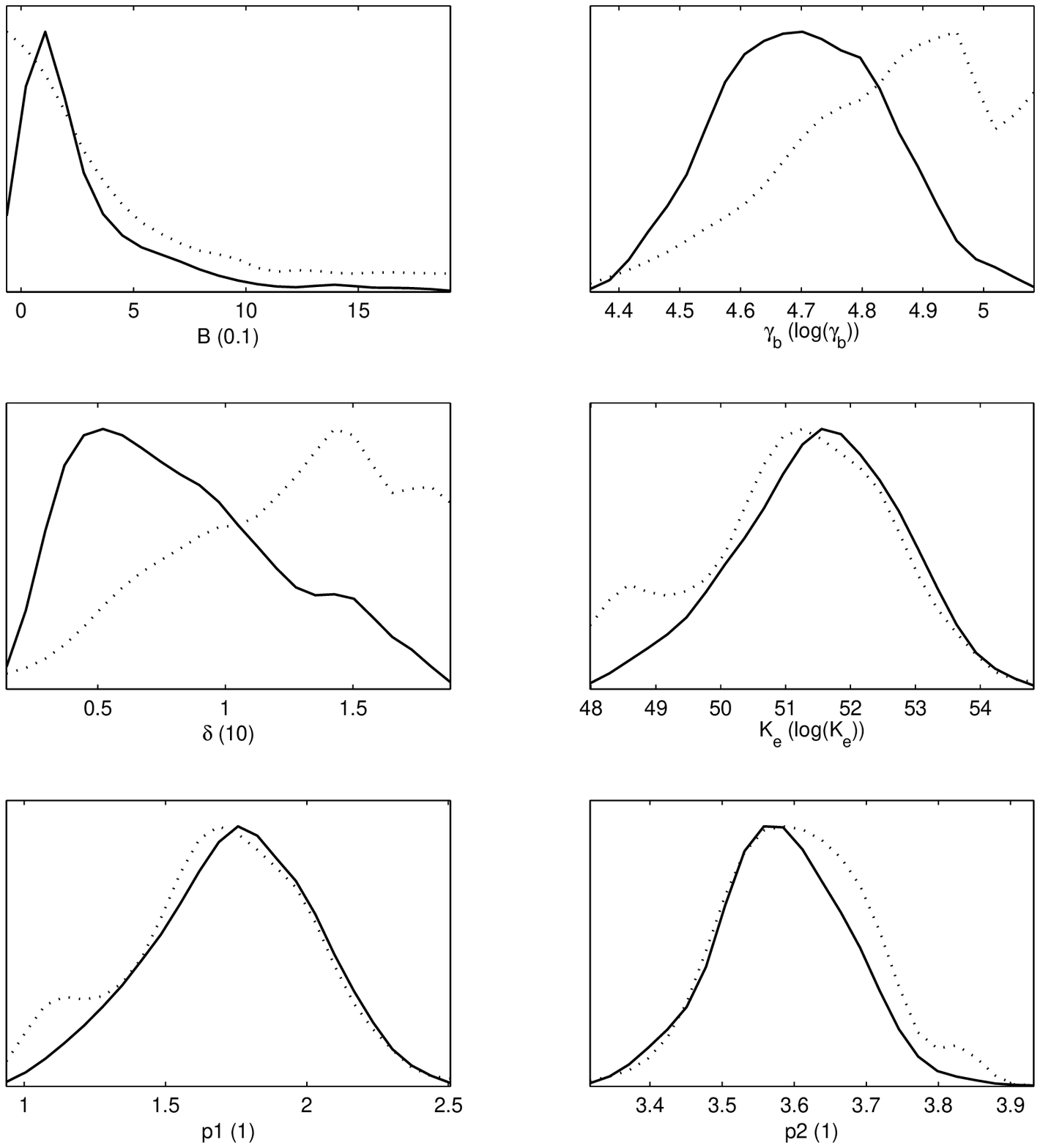}
\includegraphics[width=8.0cm,height=5.3cm]{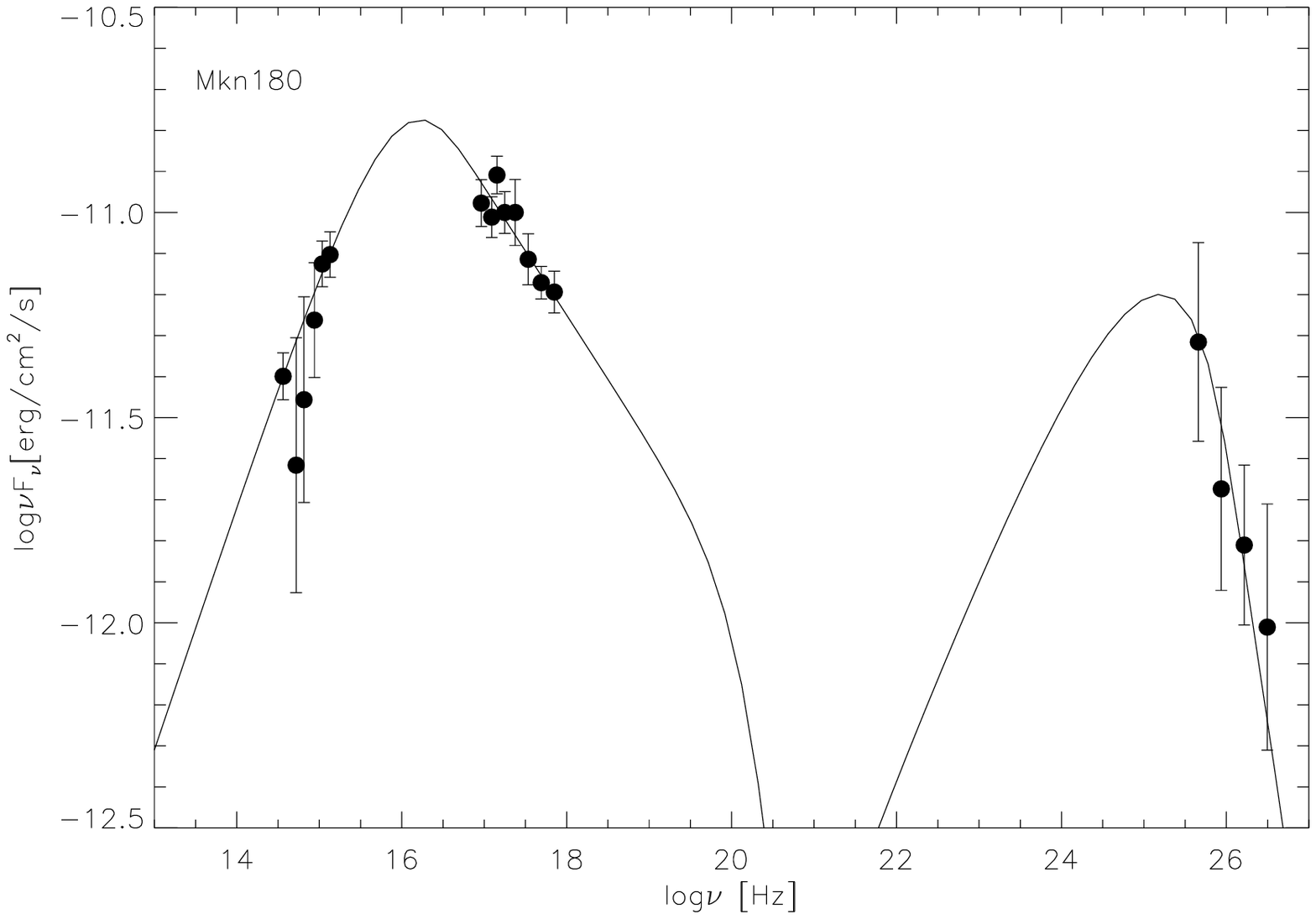}\\
\includegraphics[width=5.8cm,height=5.0cm]{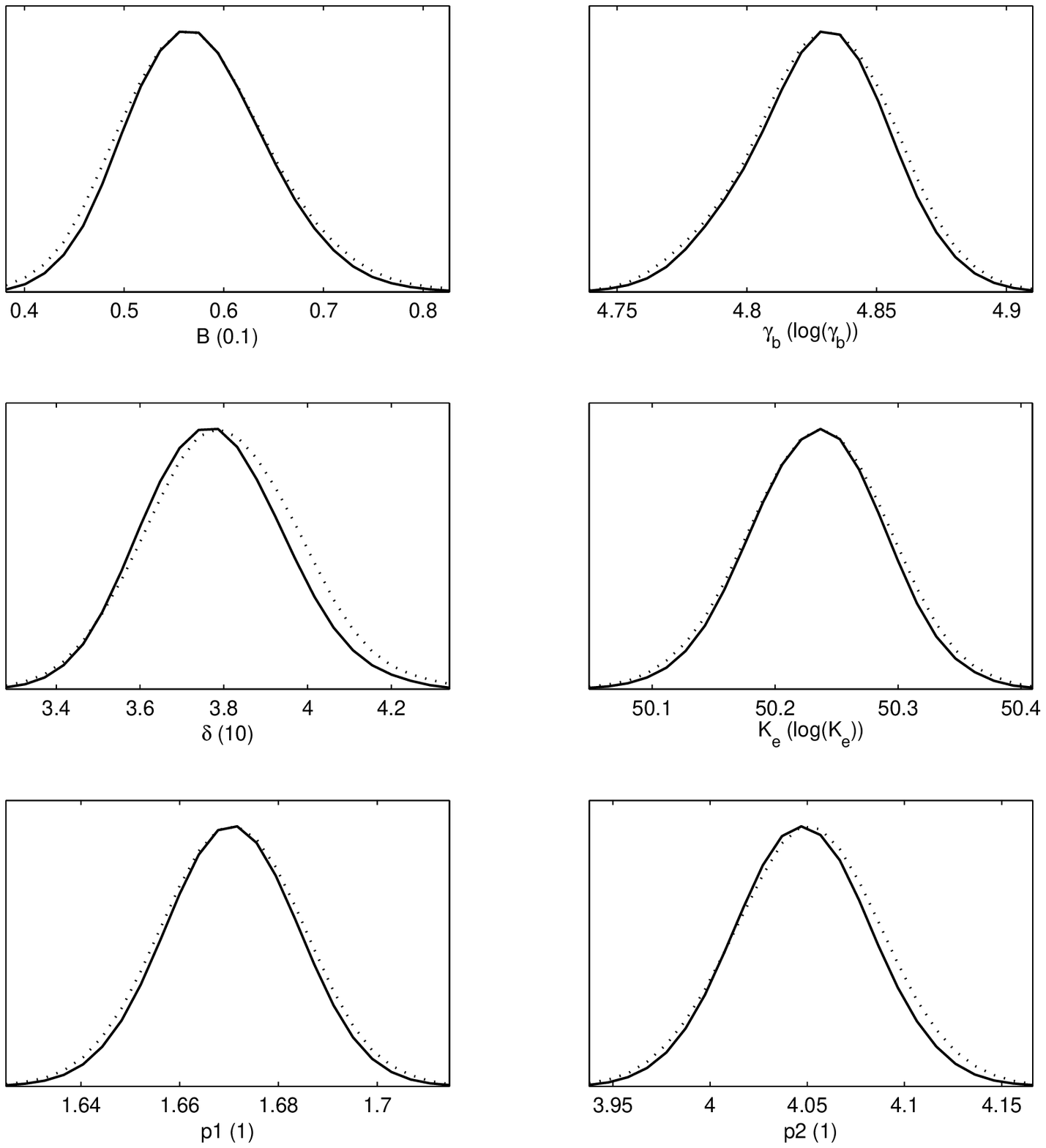}
\hfill
\includegraphics[width=8.0cm,height=5.3cm]{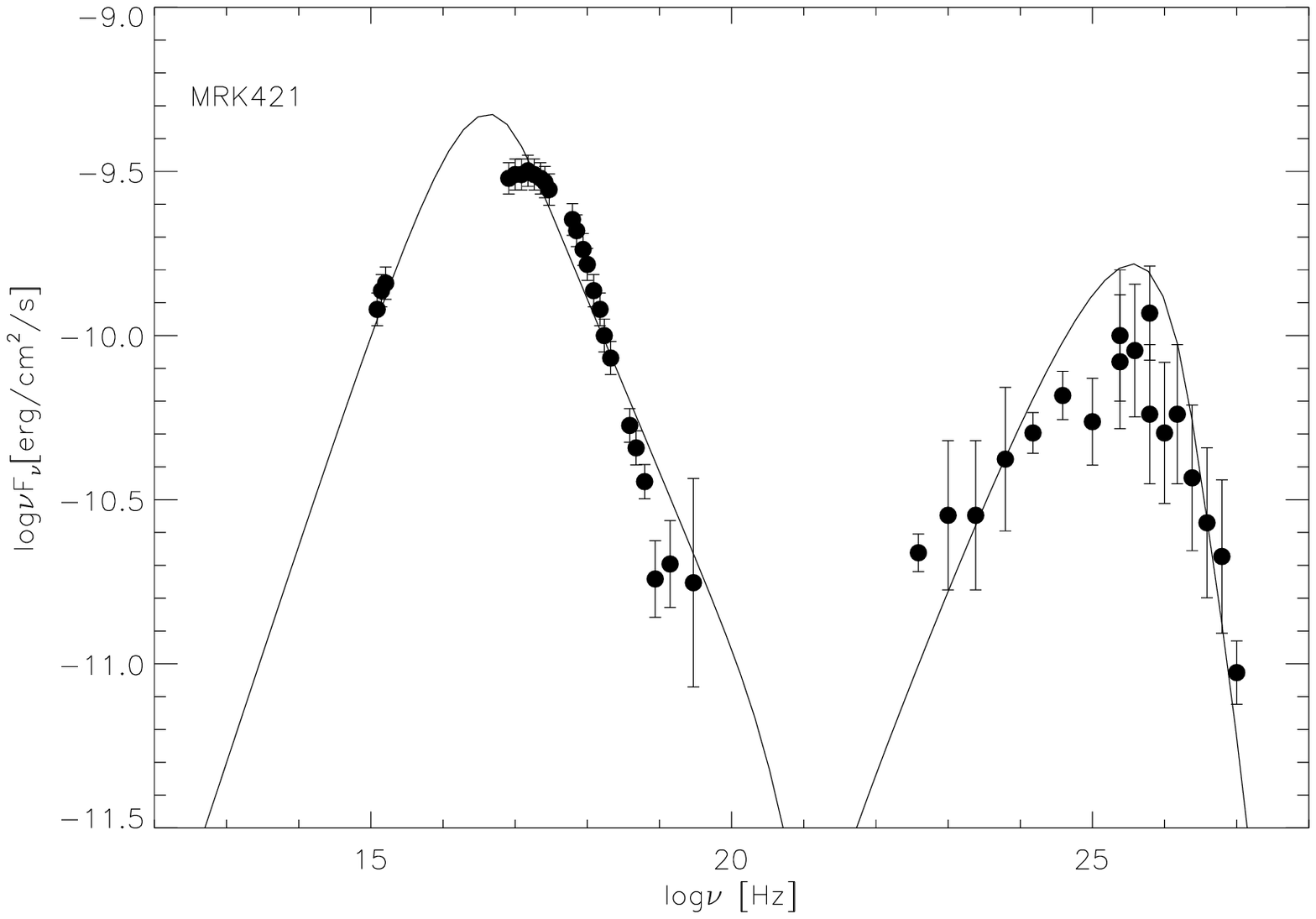}
\hfill

    \end{tabular}
  \end{center}
    \center{\textbf{Fig. 6.}---  continued}

\end{figure*}

\begin{figure*}
  \begin{center}
   \begin{tabular}{cc}
\includegraphics[width=5.8cm,height=5.0cm]{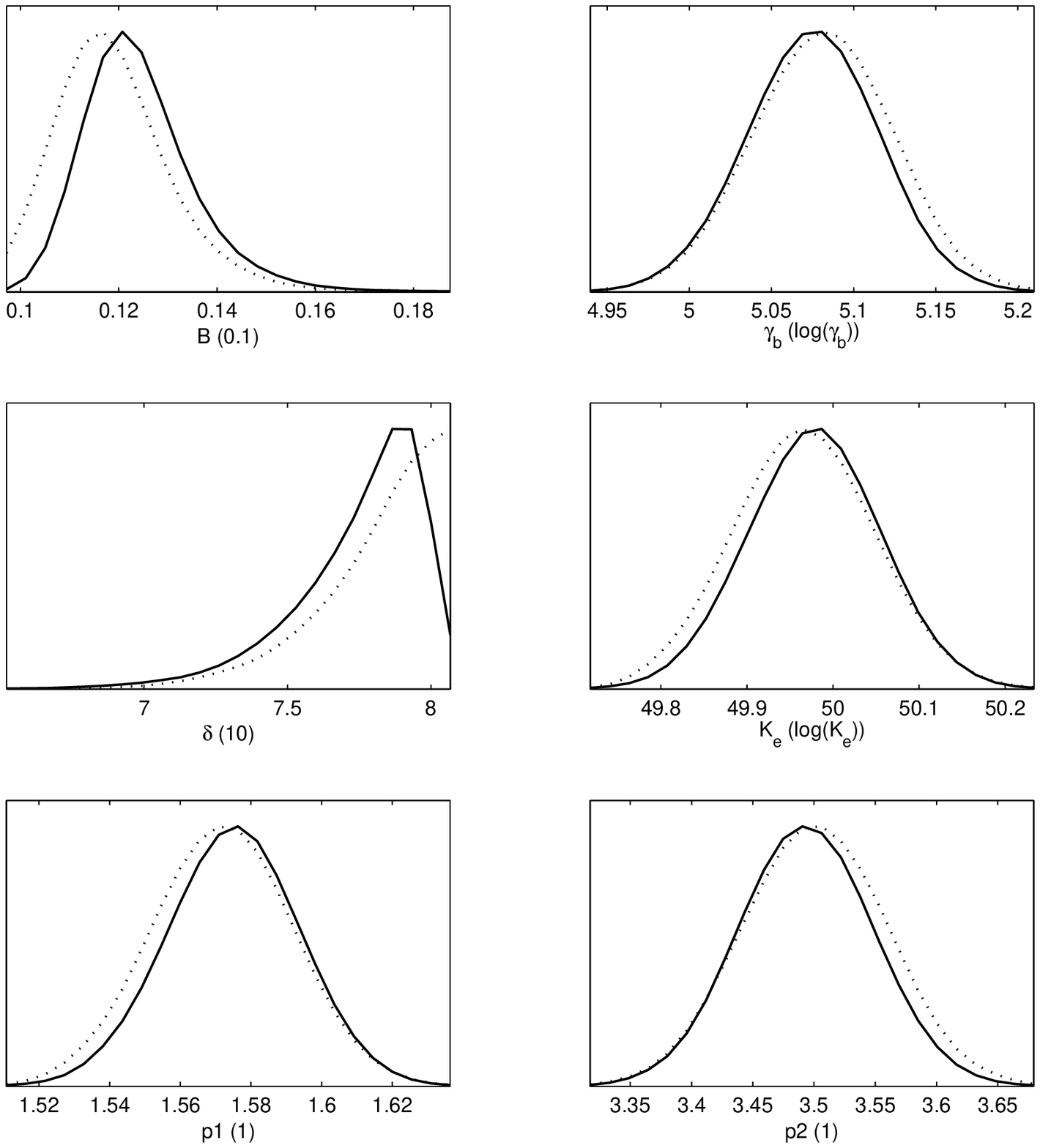}
\includegraphics[width=8.0cm,height=5.3cm]{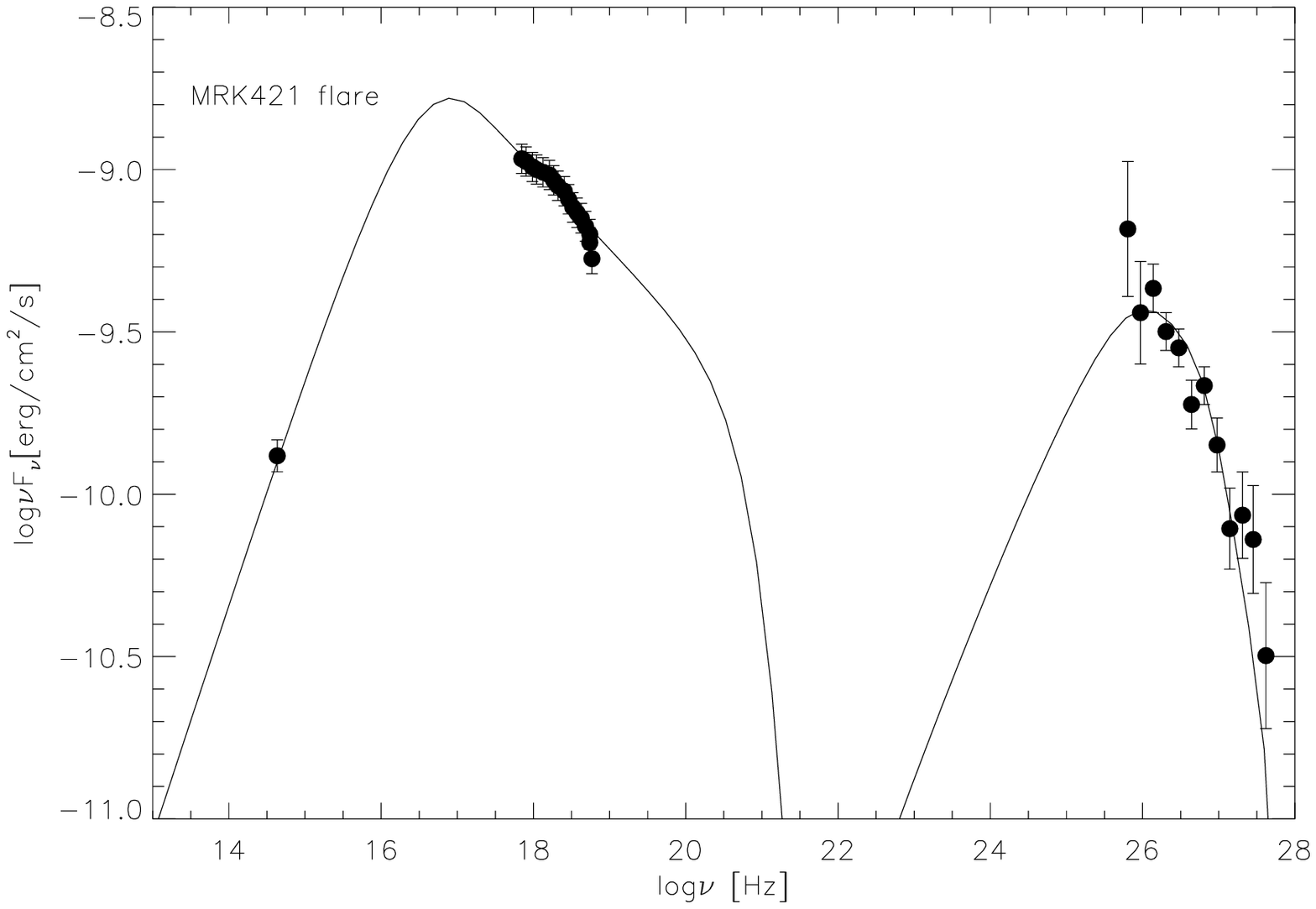}\\
\includegraphics[width=5.8cm,height=5.0cm]{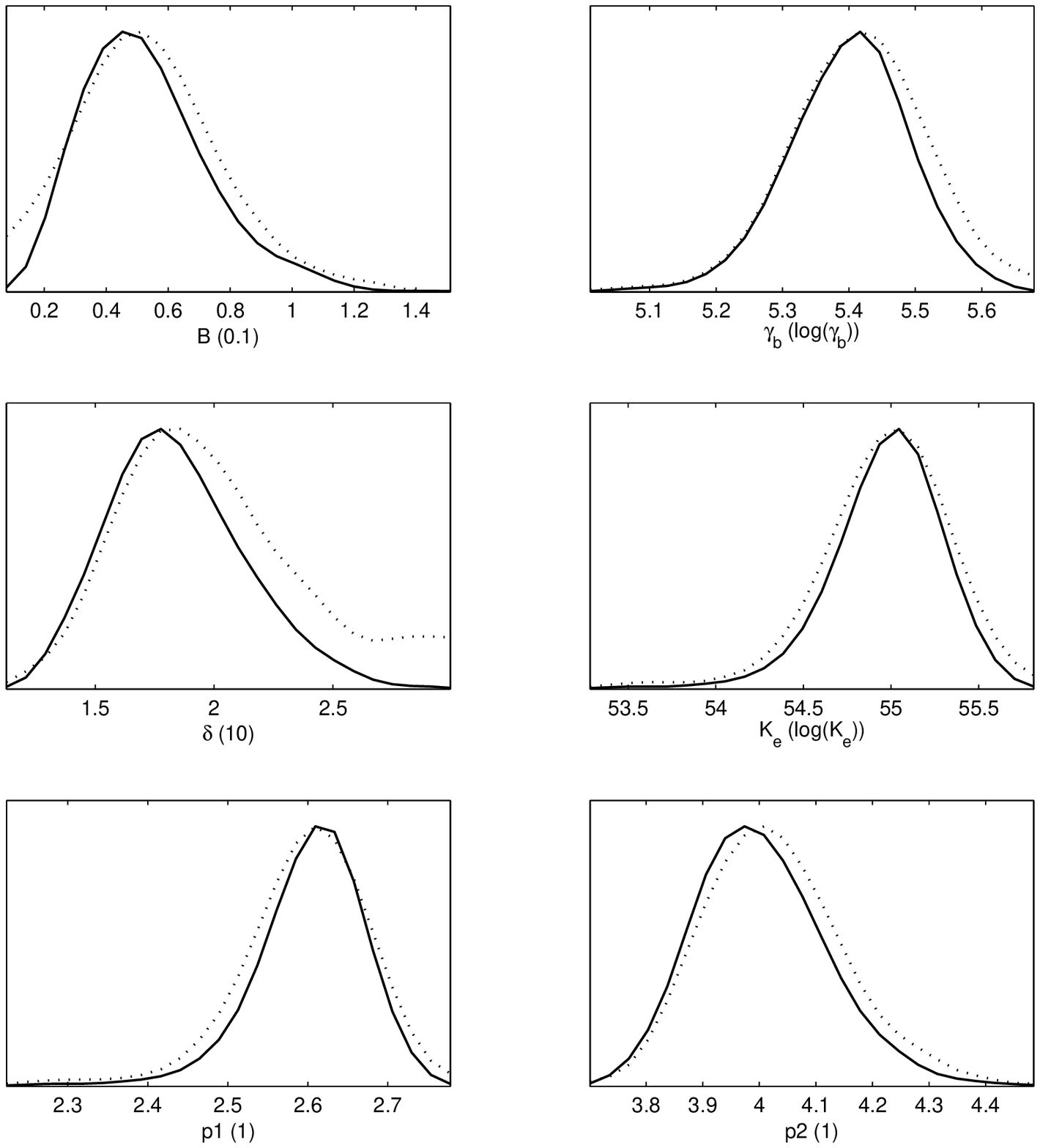}
\includegraphics[width=8.0cm,height=5.3cm]{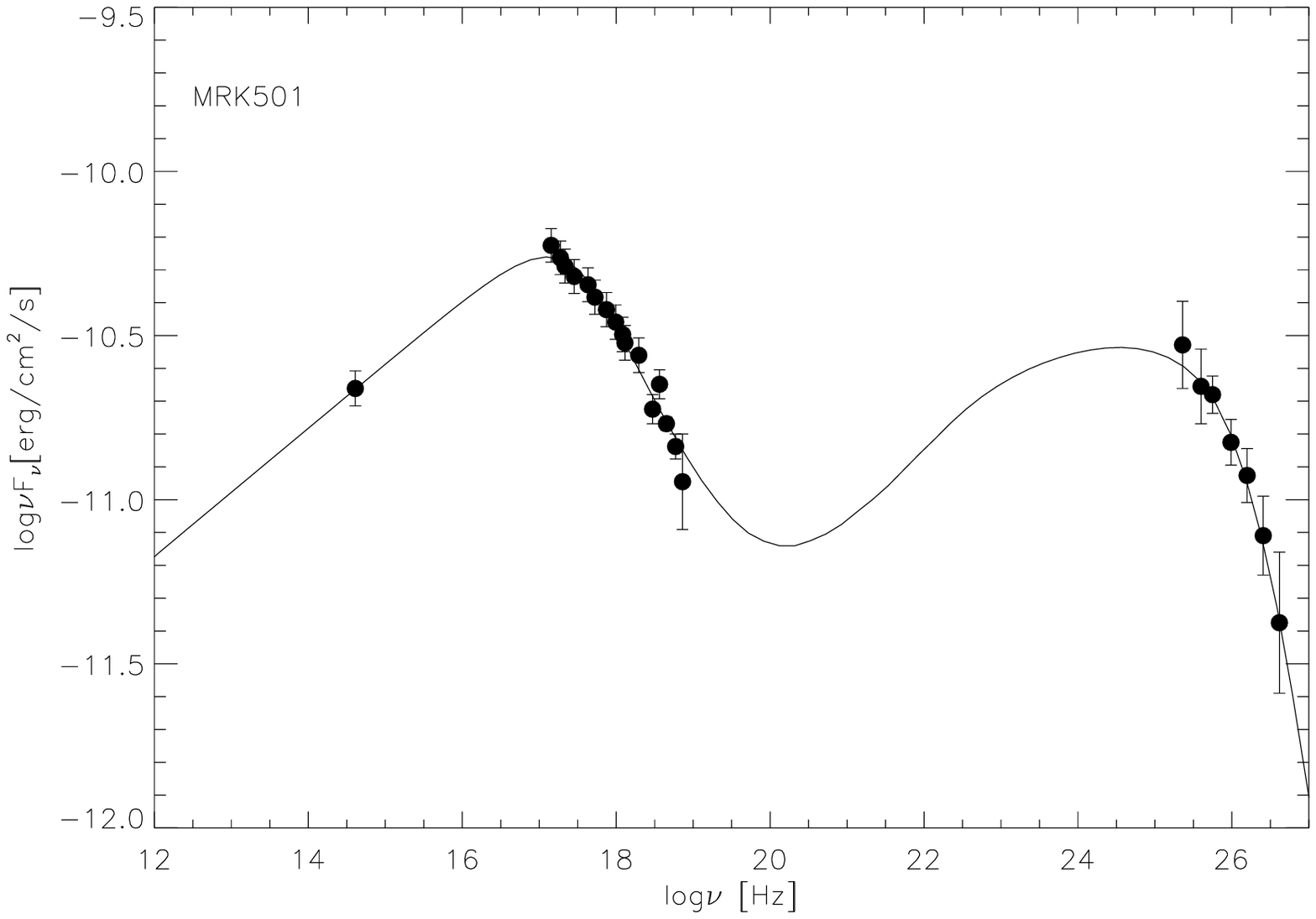}\\
\includegraphics[width=5.8cm,height=5.0cm]{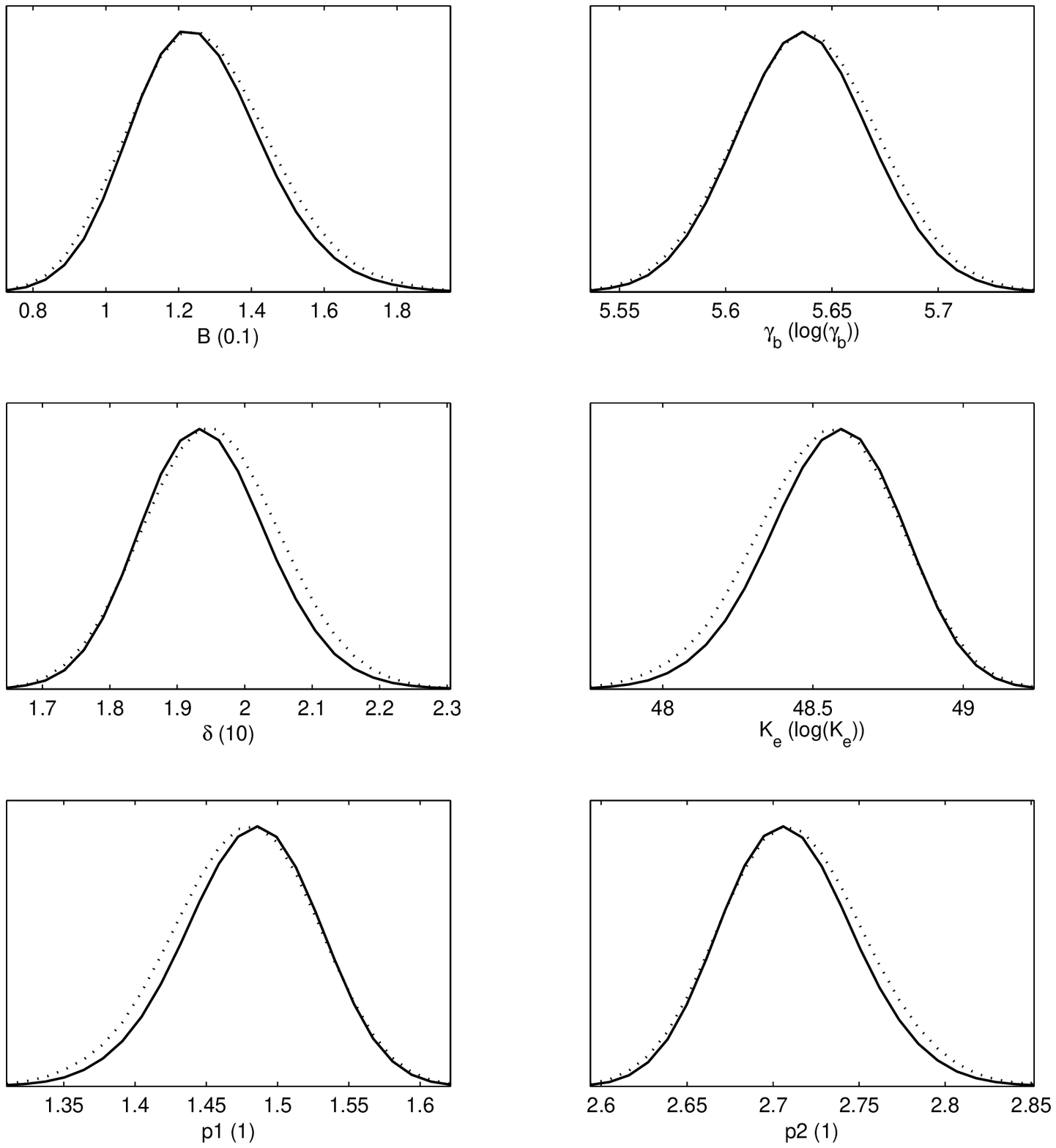}
\includegraphics[width=8.0cm,height=5.3cm]{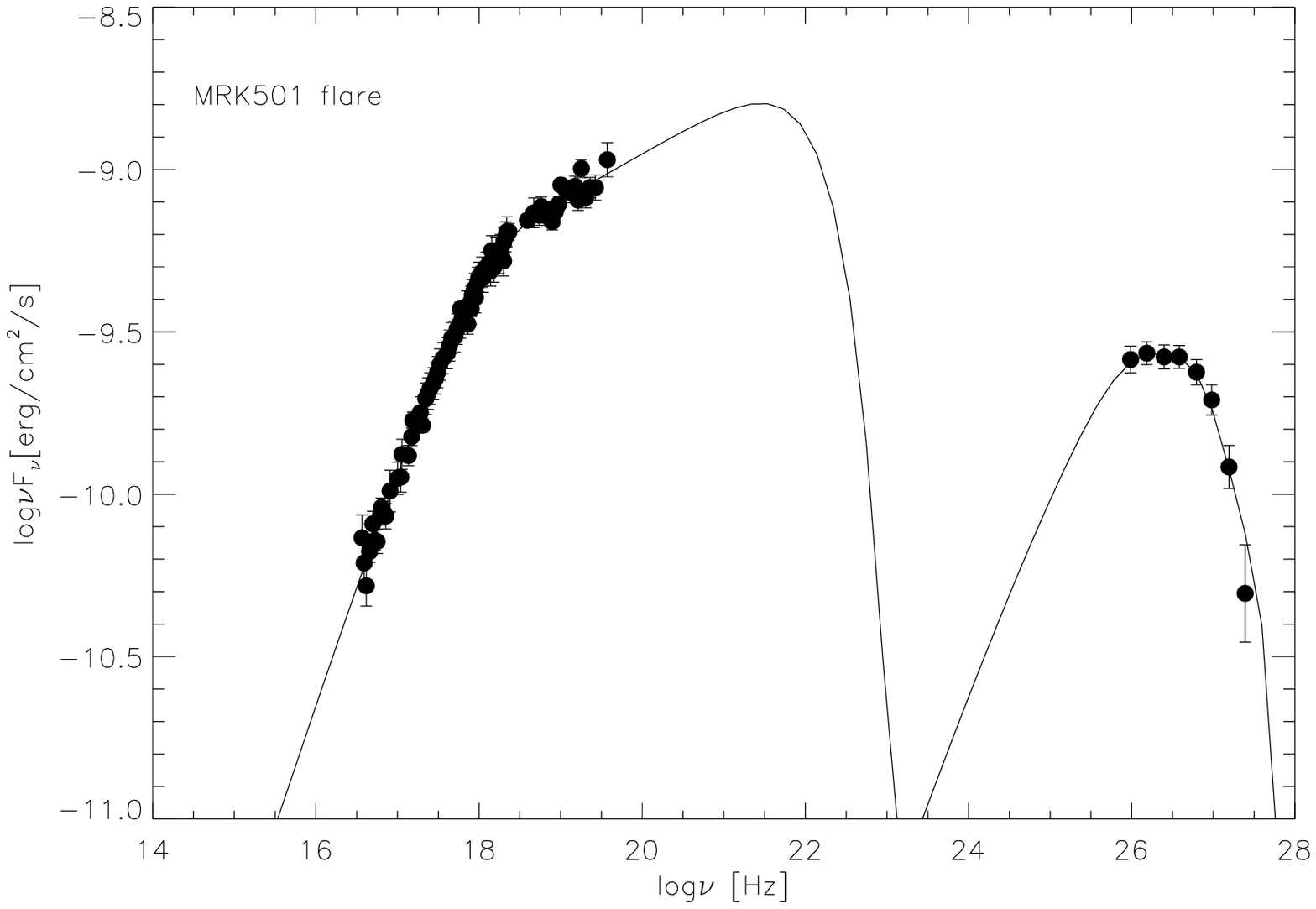}\\
\includegraphics[width=5.8cm,height=5.0cm]{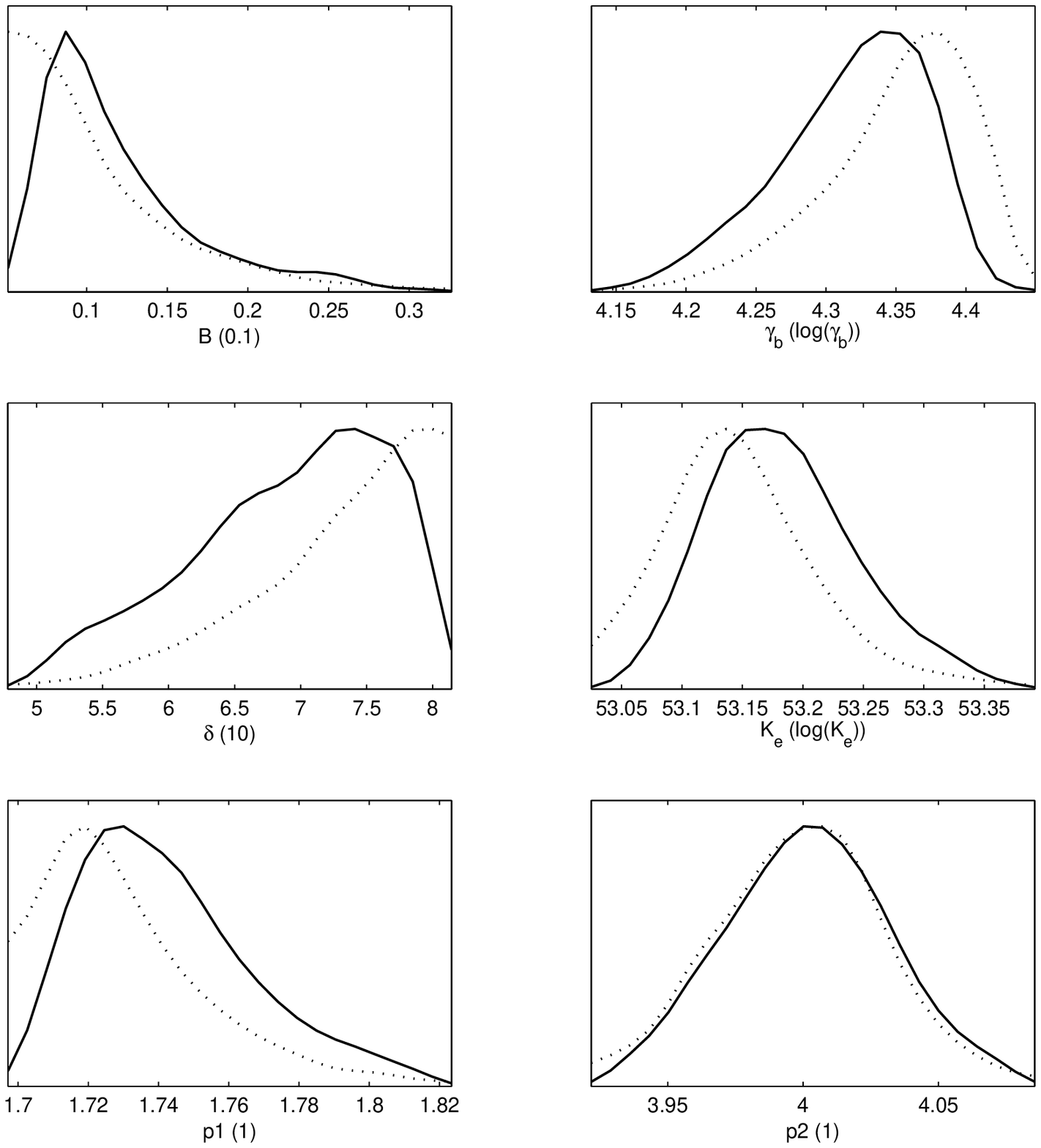}
\hfill
\includegraphics[width=8.0cm,height=5.3cm]{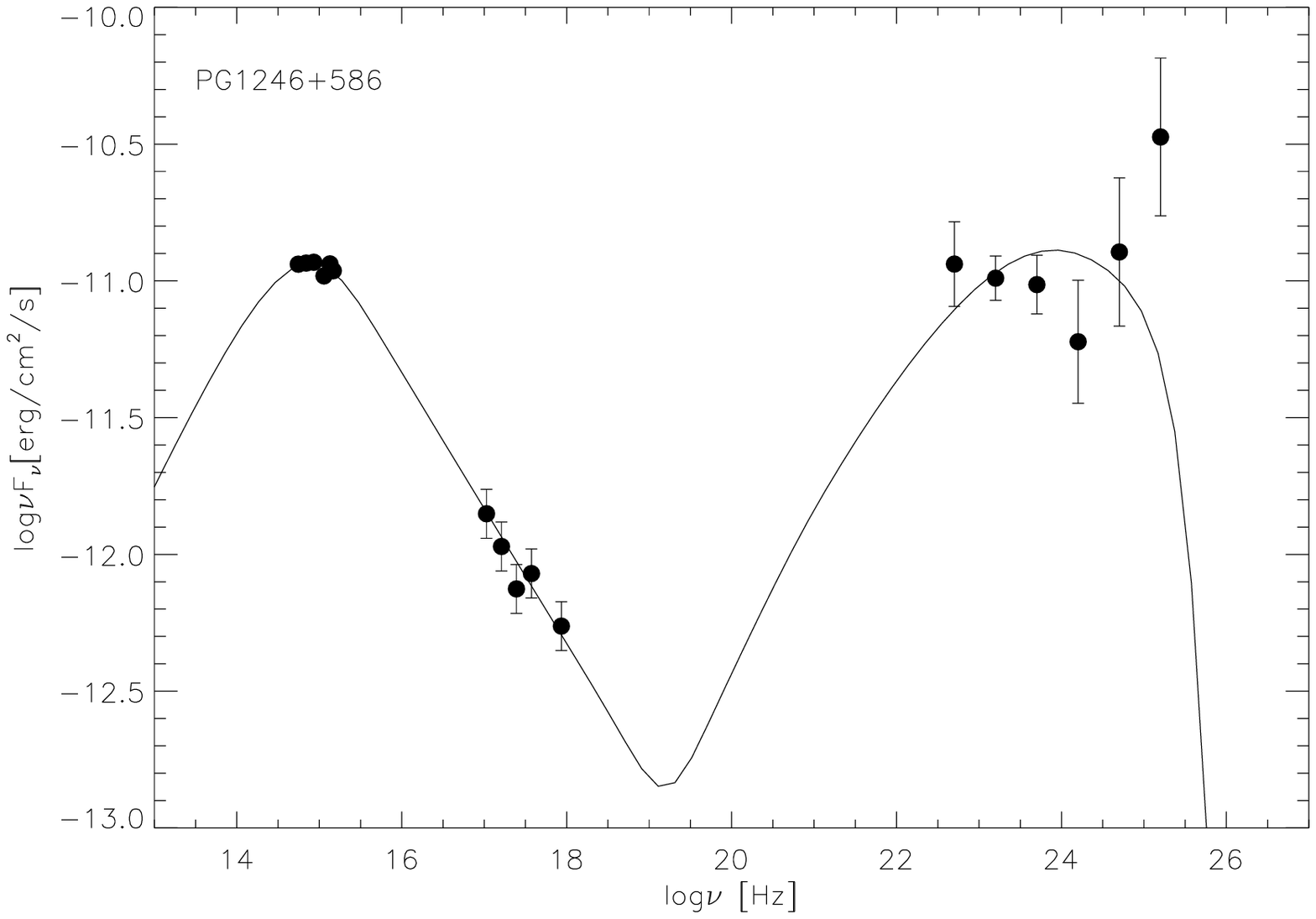}
\hfill

    \end{tabular}
  \end{center}
    \center{\textbf{Fig. 6.}---  continued}

\end{figure*}

\begin{figure*}
  \begin{center}
   \begin{tabular}{cc}
\includegraphics[width=5.8cm,height=5.0cm]{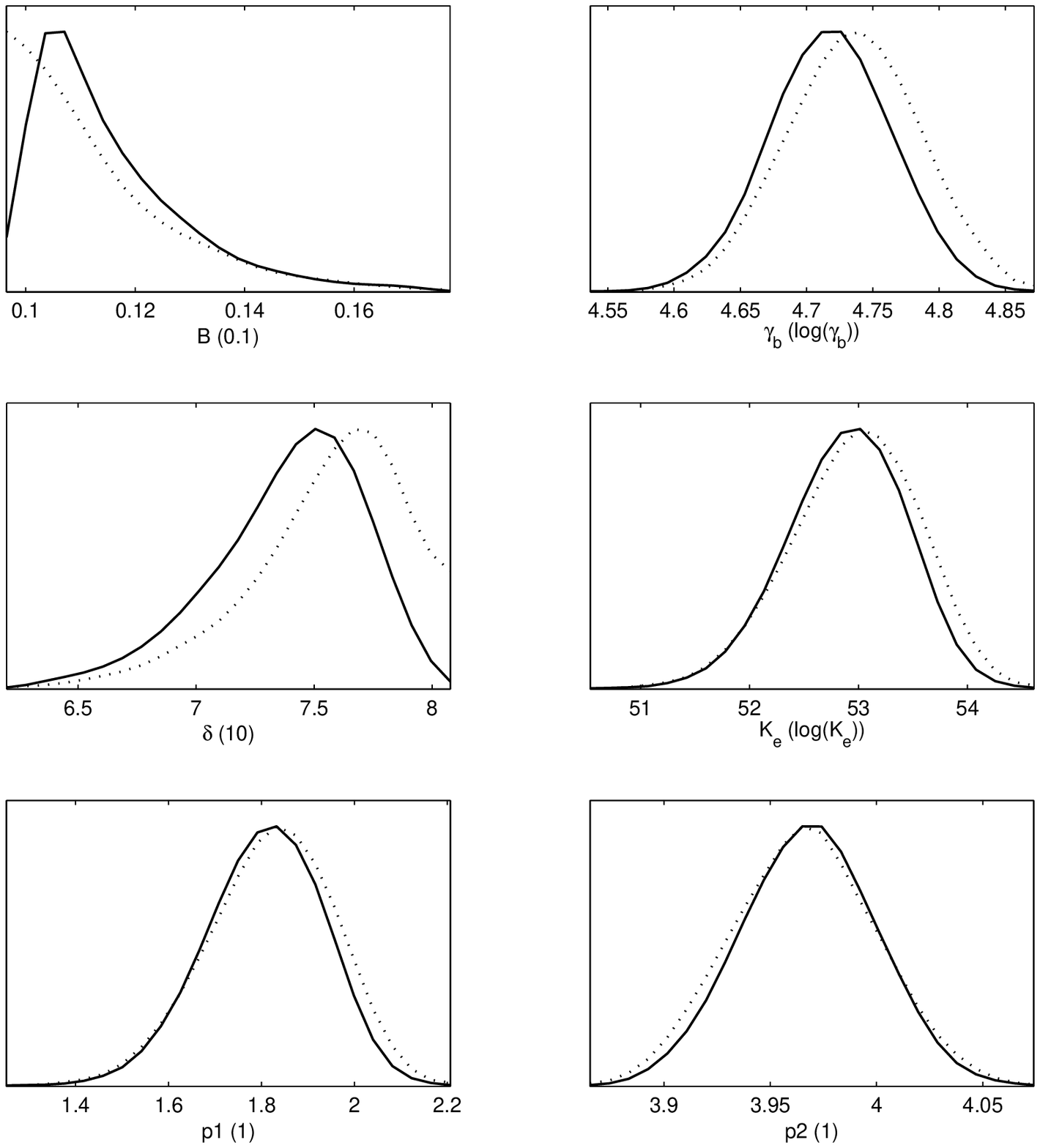}
\includegraphics[width=8.0cm,height=5.3cm]{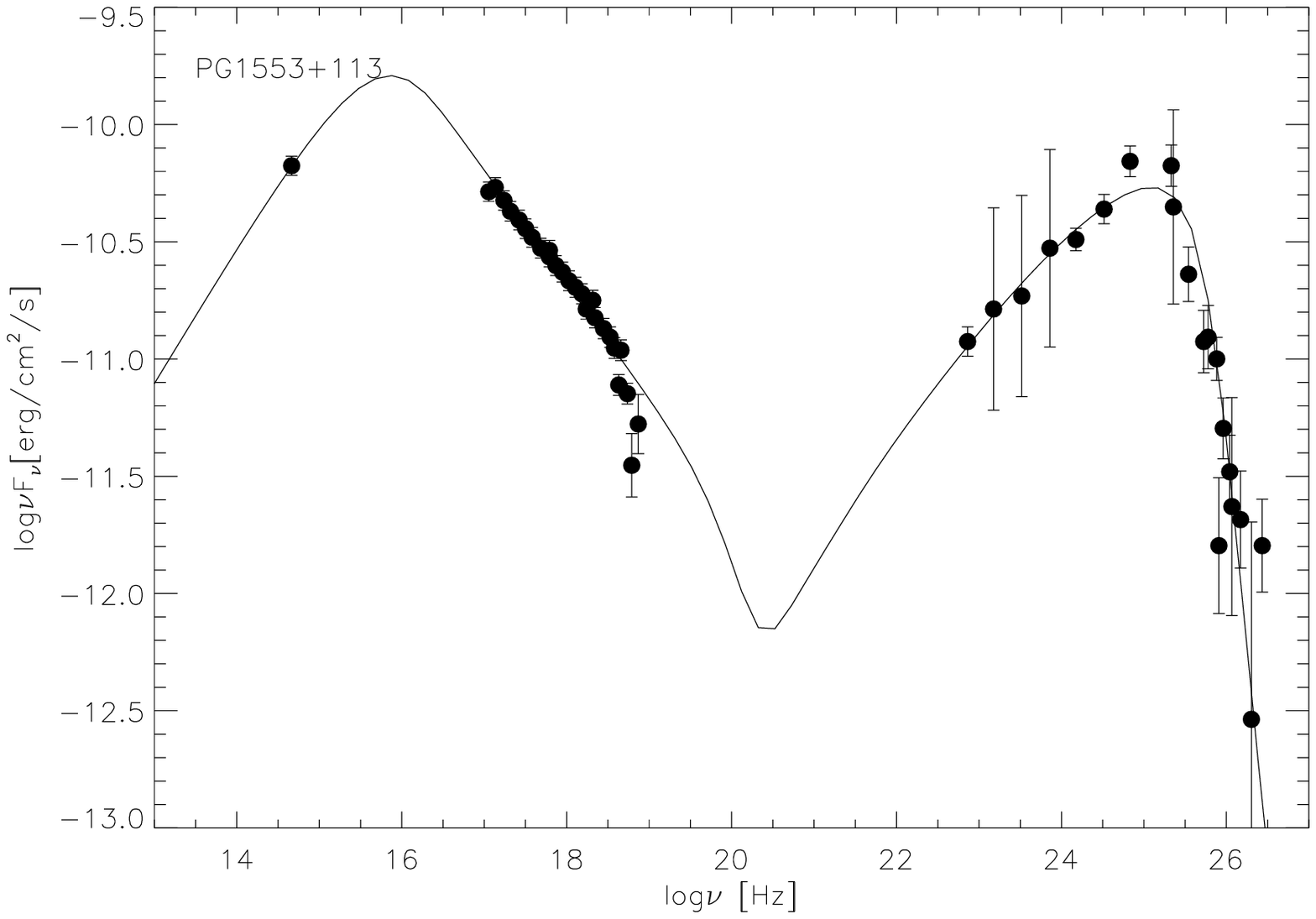}\\
\includegraphics[width=5.8cm,height=5.0cm]{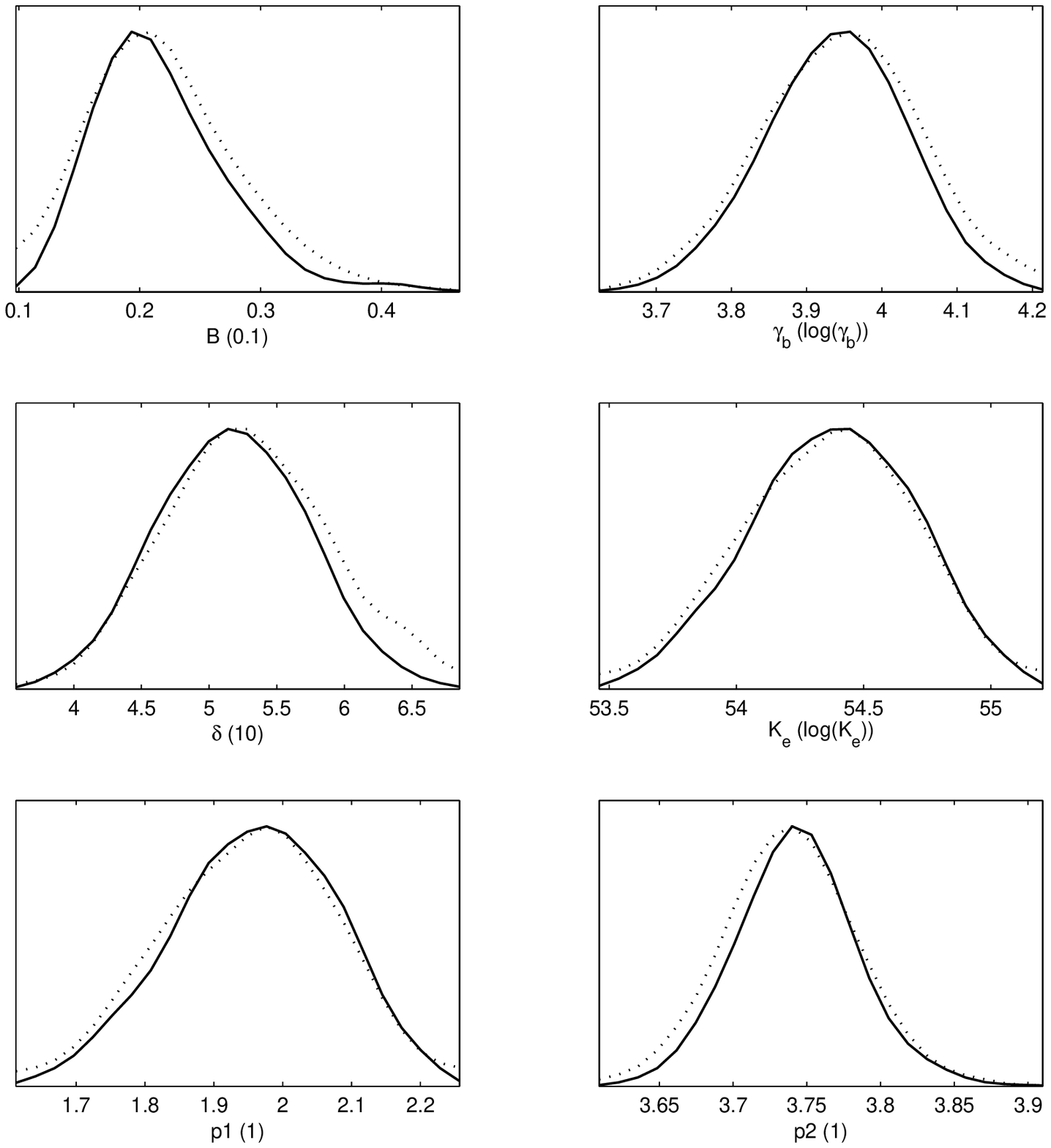}
\includegraphics[width=8.0cm,height=5.3cm]{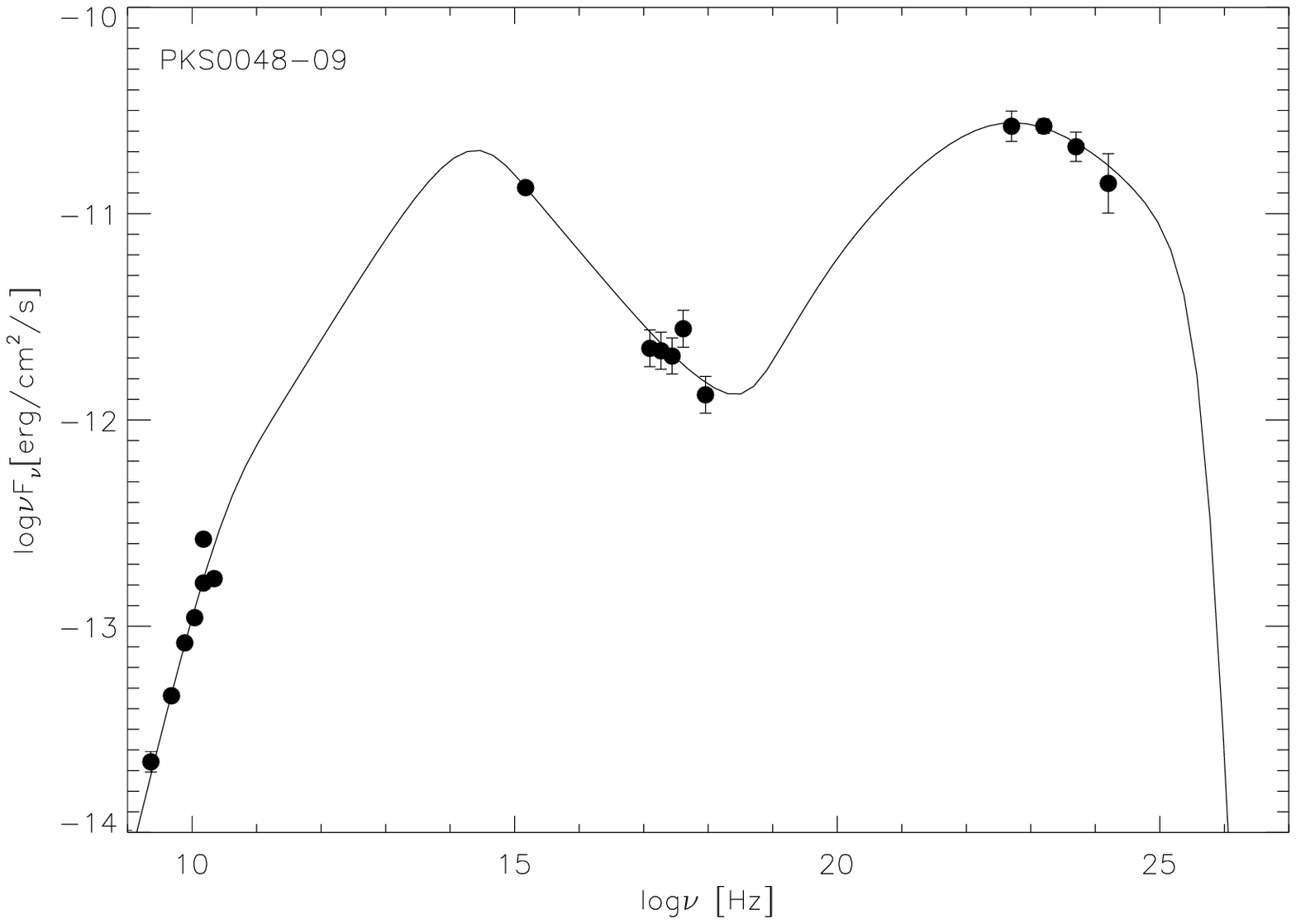}\\
\includegraphics[width=5.8cm,height=5.0cm]{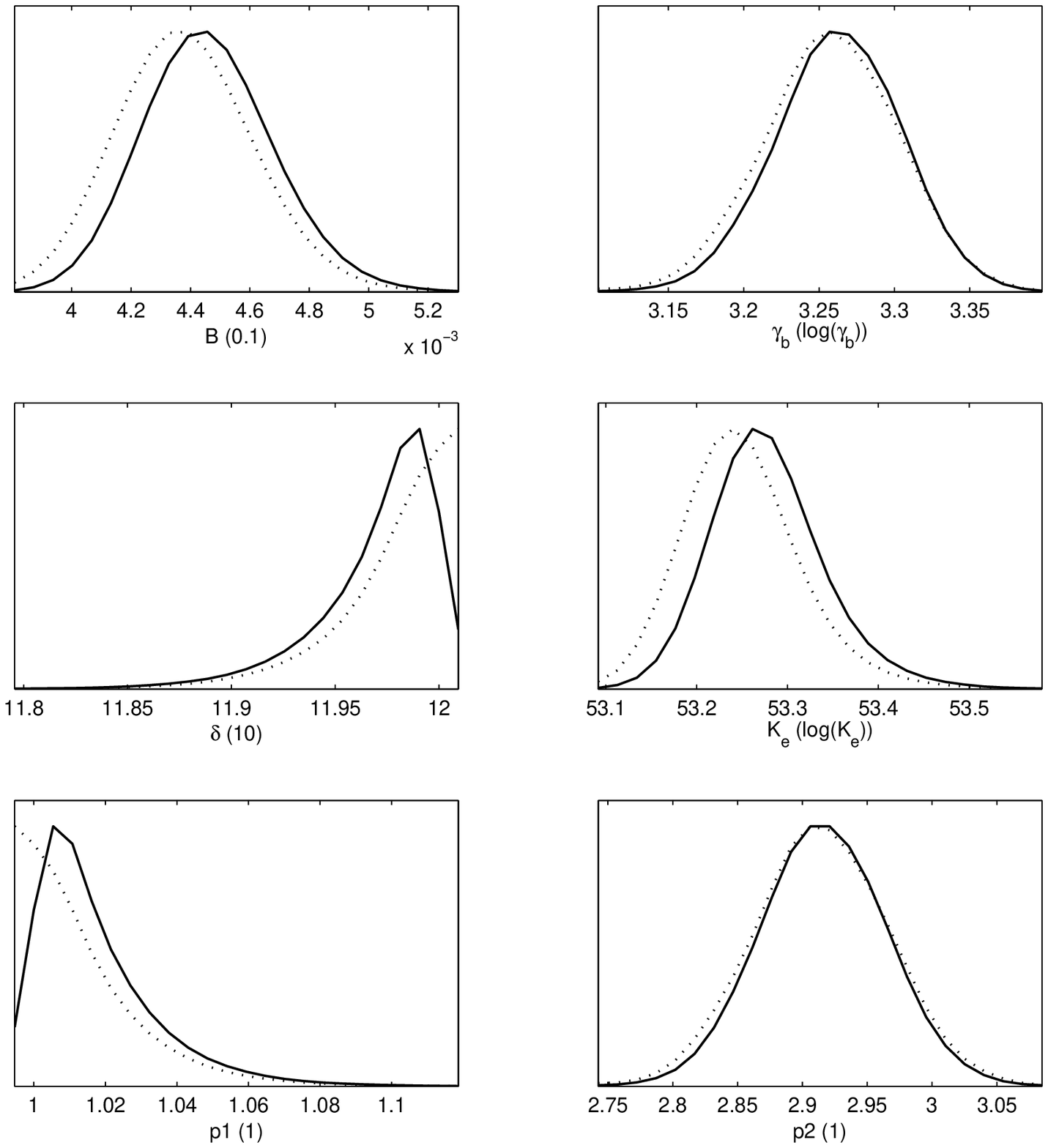}
\includegraphics[width=8.0cm,height=5.3cm]{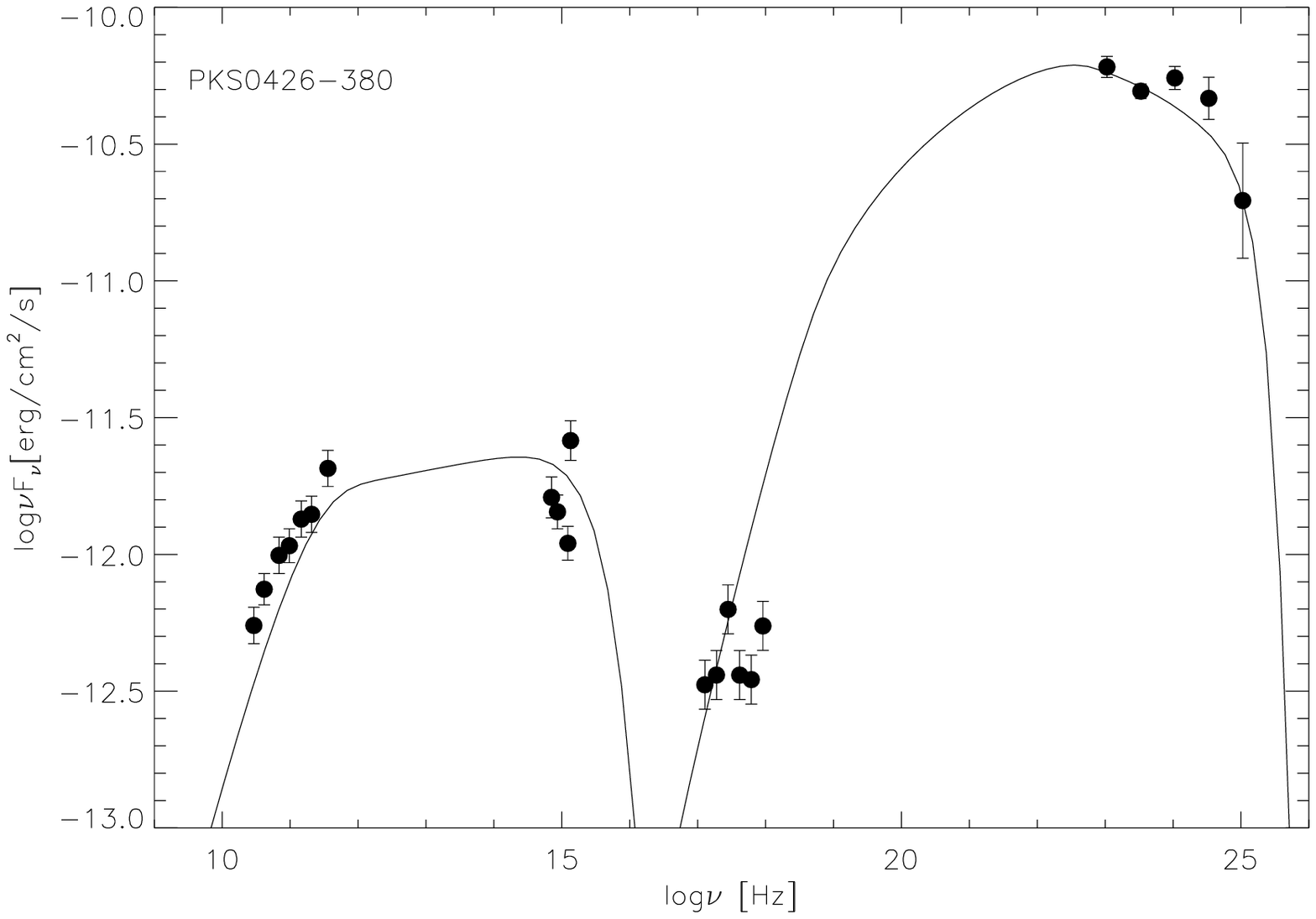}\\
\includegraphics[width=5.8cm,height=5.0cm]{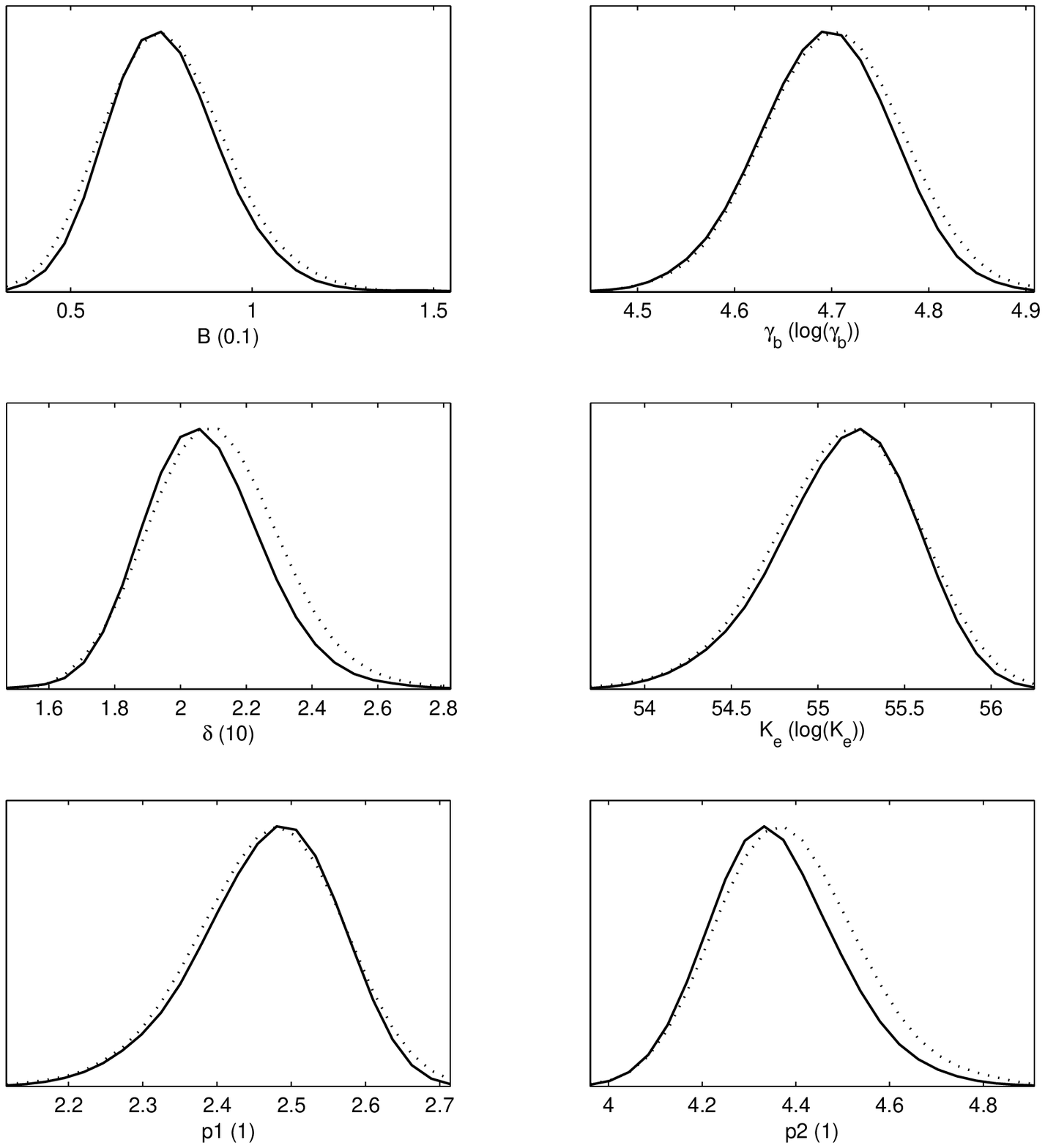}
\hfill
\includegraphics[width=8.0cm,height=5.3cm]{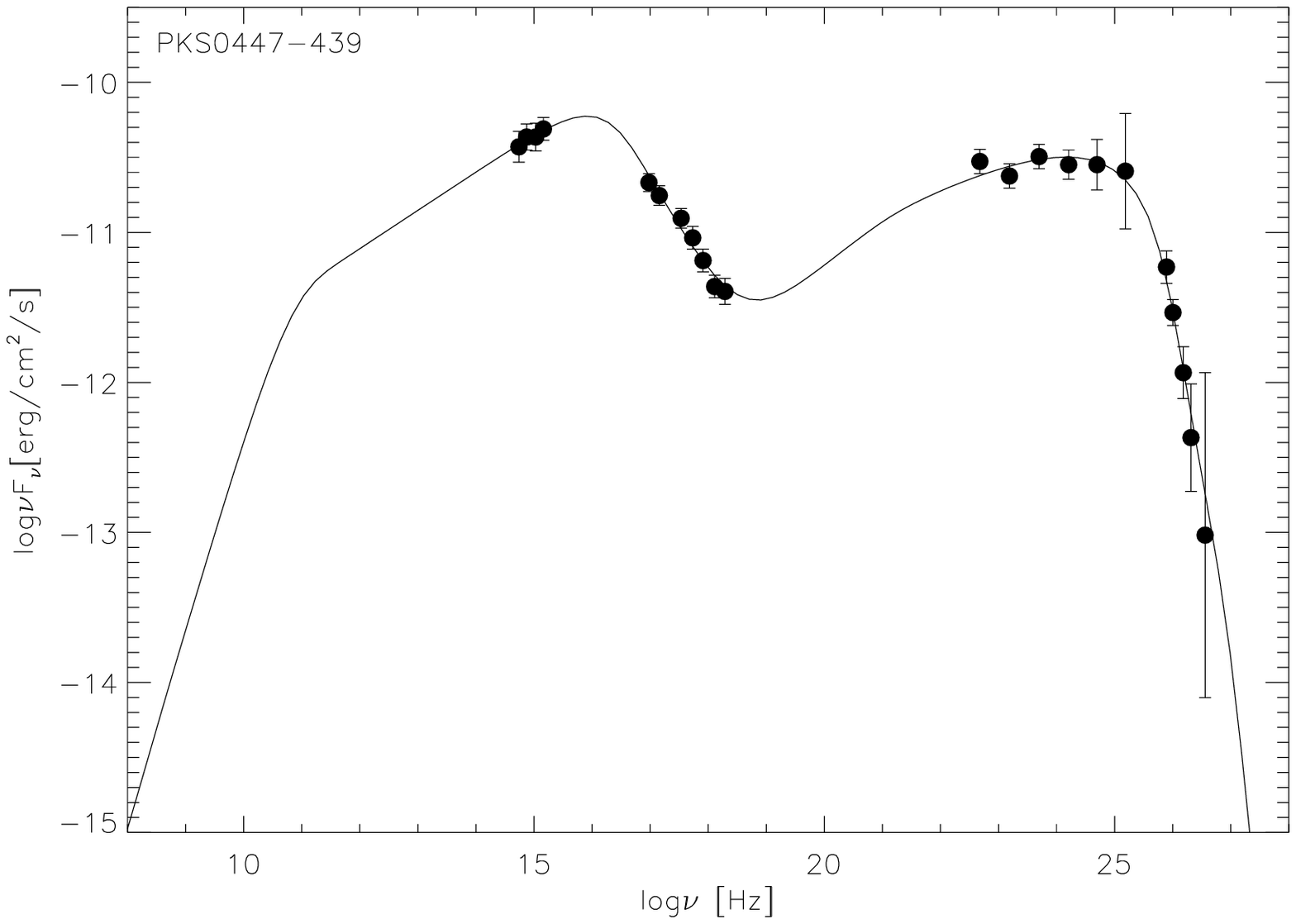}
\hfill

    \end{tabular}
  \end{center}
    \center{\textbf{Fig. 6.}---  continued}

\end{figure*}

\begin{figure*}
  \begin{center}
   \begin{tabular}{cc}

\includegraphics[width=5.8cm,height=5.0cm]{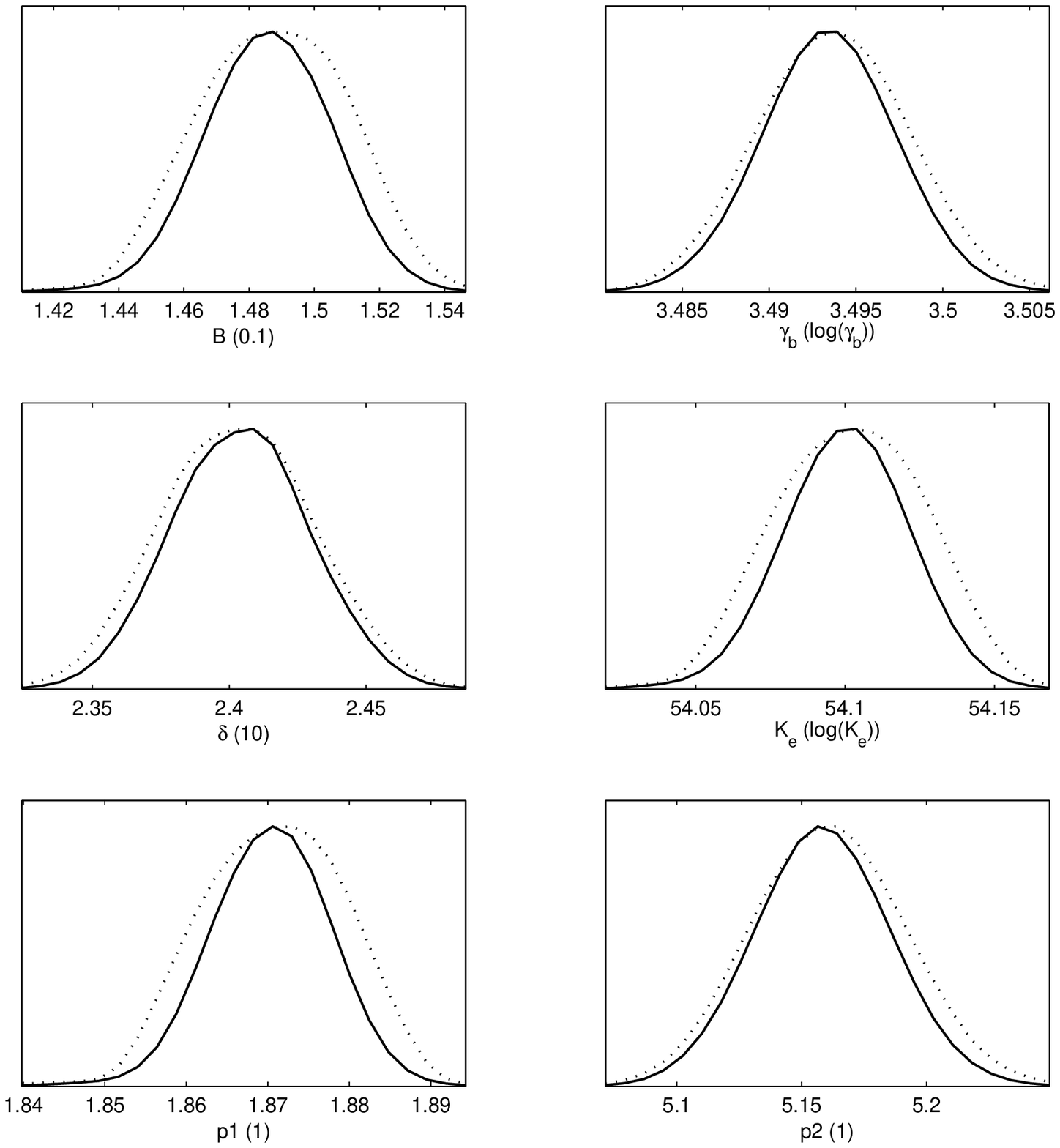}
\includegraphics[width=8.0cm,height=5.3cm]{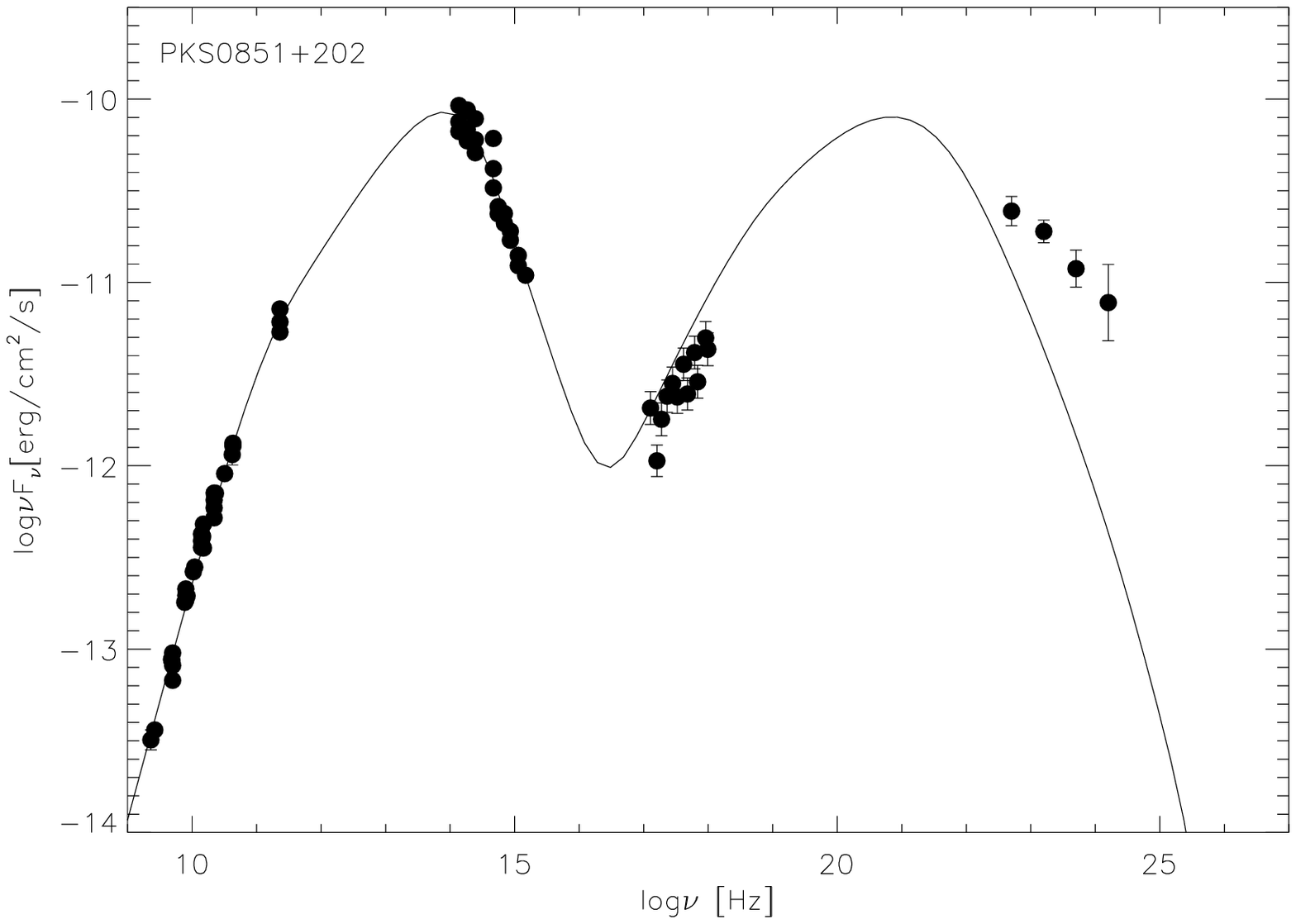}\\
\includegraphics[width=5.8cm,height=5.0cm]{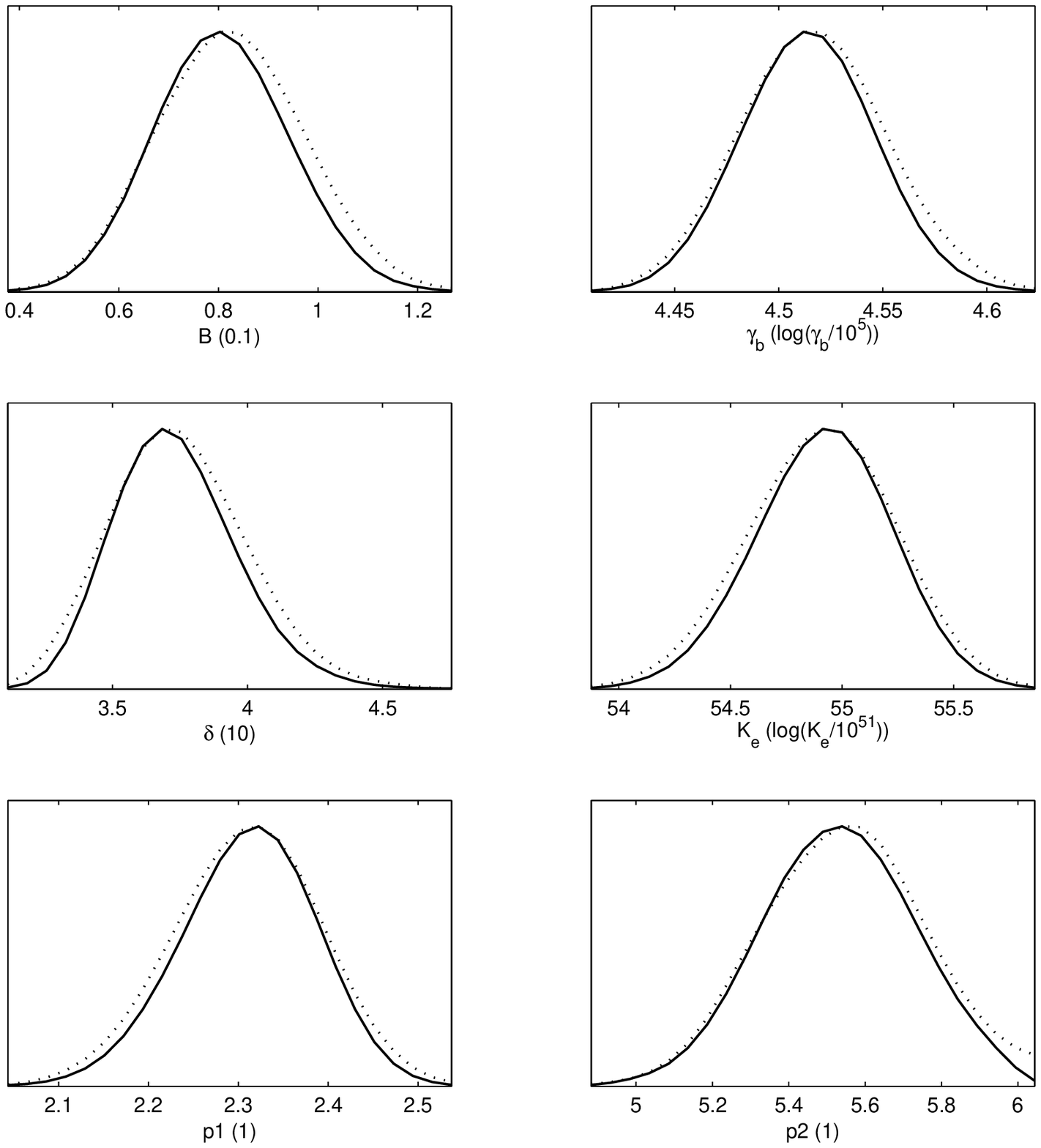}
\includegraphics[width=8.0cm,height=5.3cm]{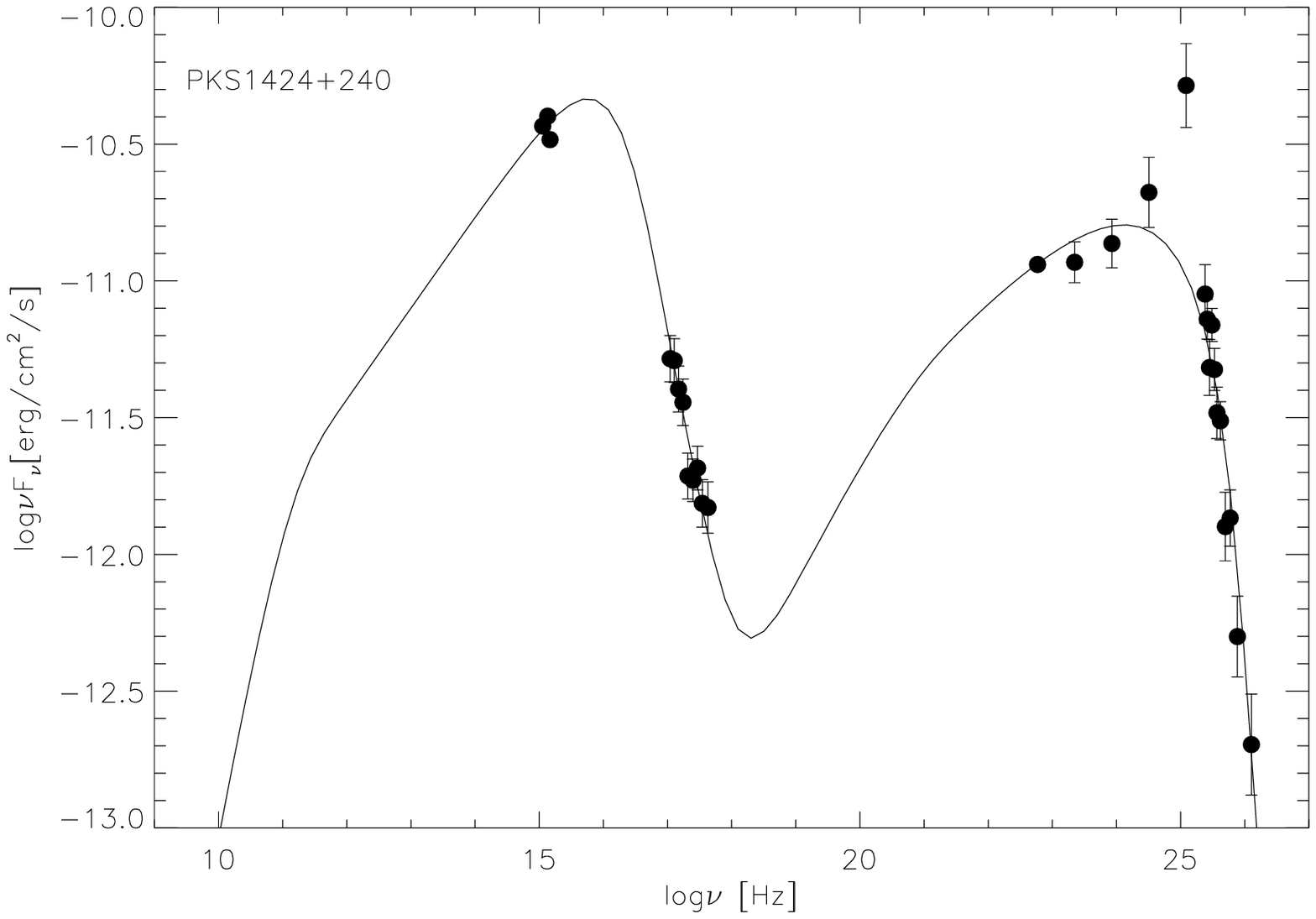}\\
\includegraphics[width=5.8cm,height=5.0cm]{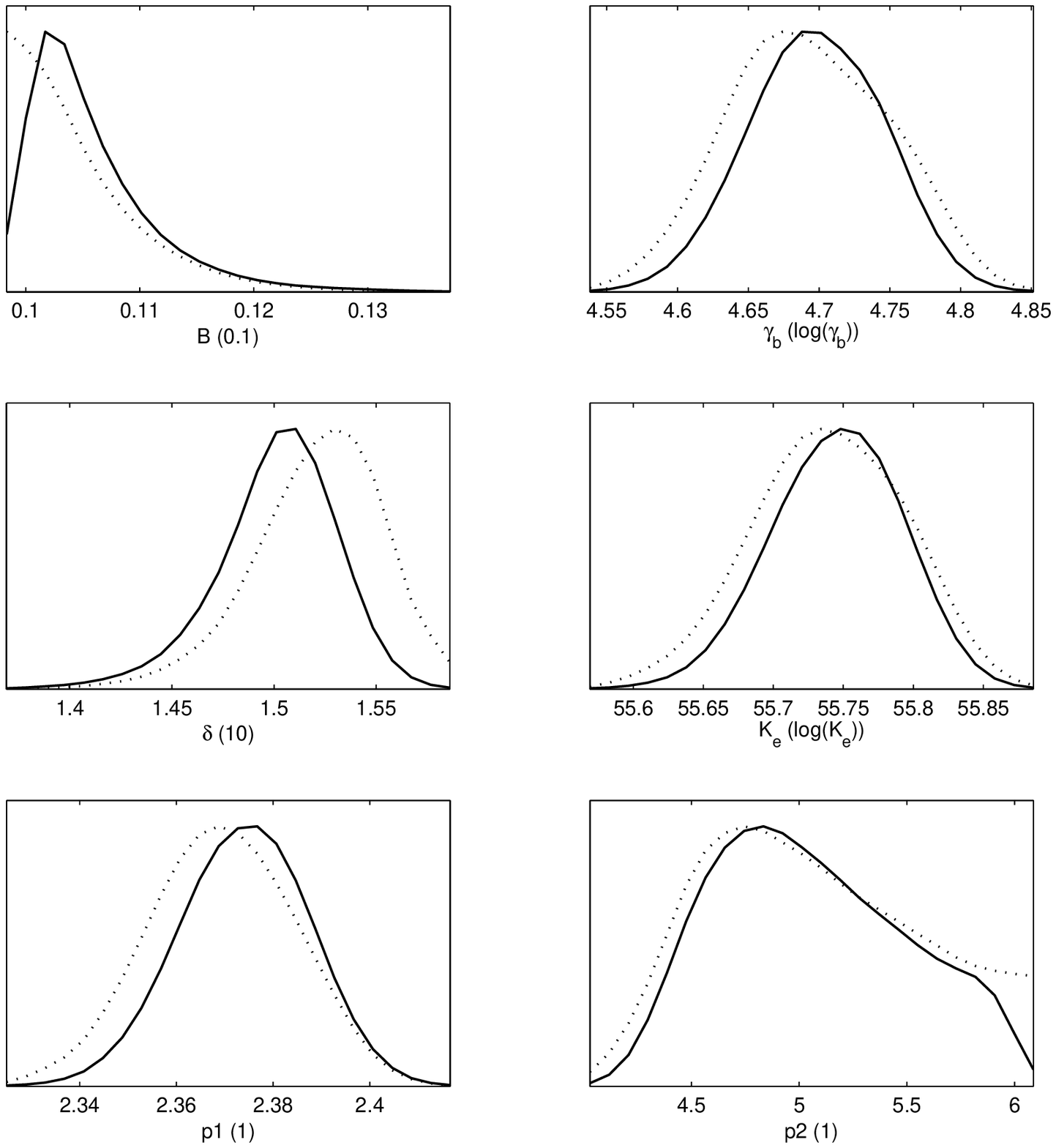}
\includegraphics[width=8.0cm,height=5.3cm]{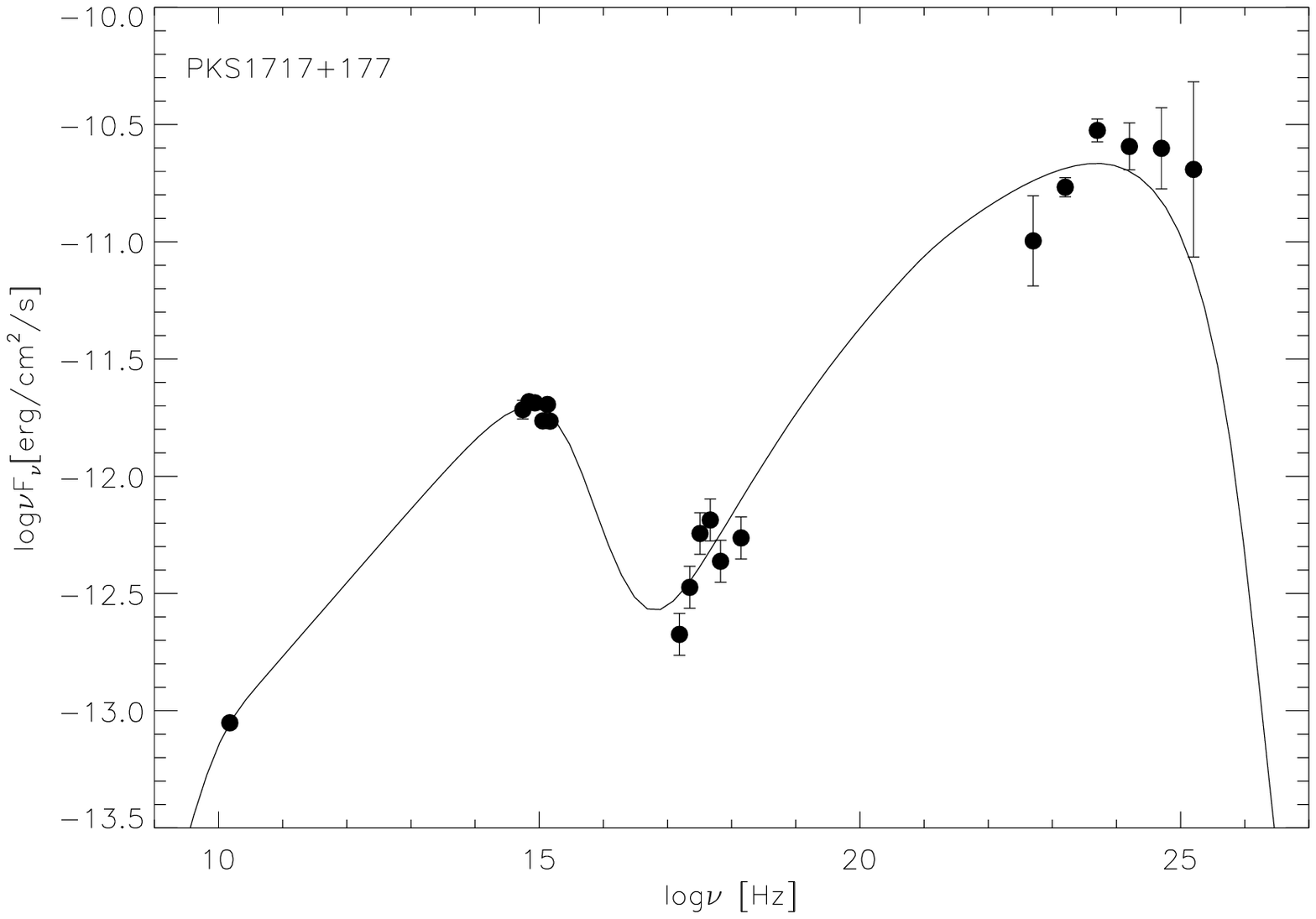}\\
\includegraphics[width=5.8cm,height=5.0cm]{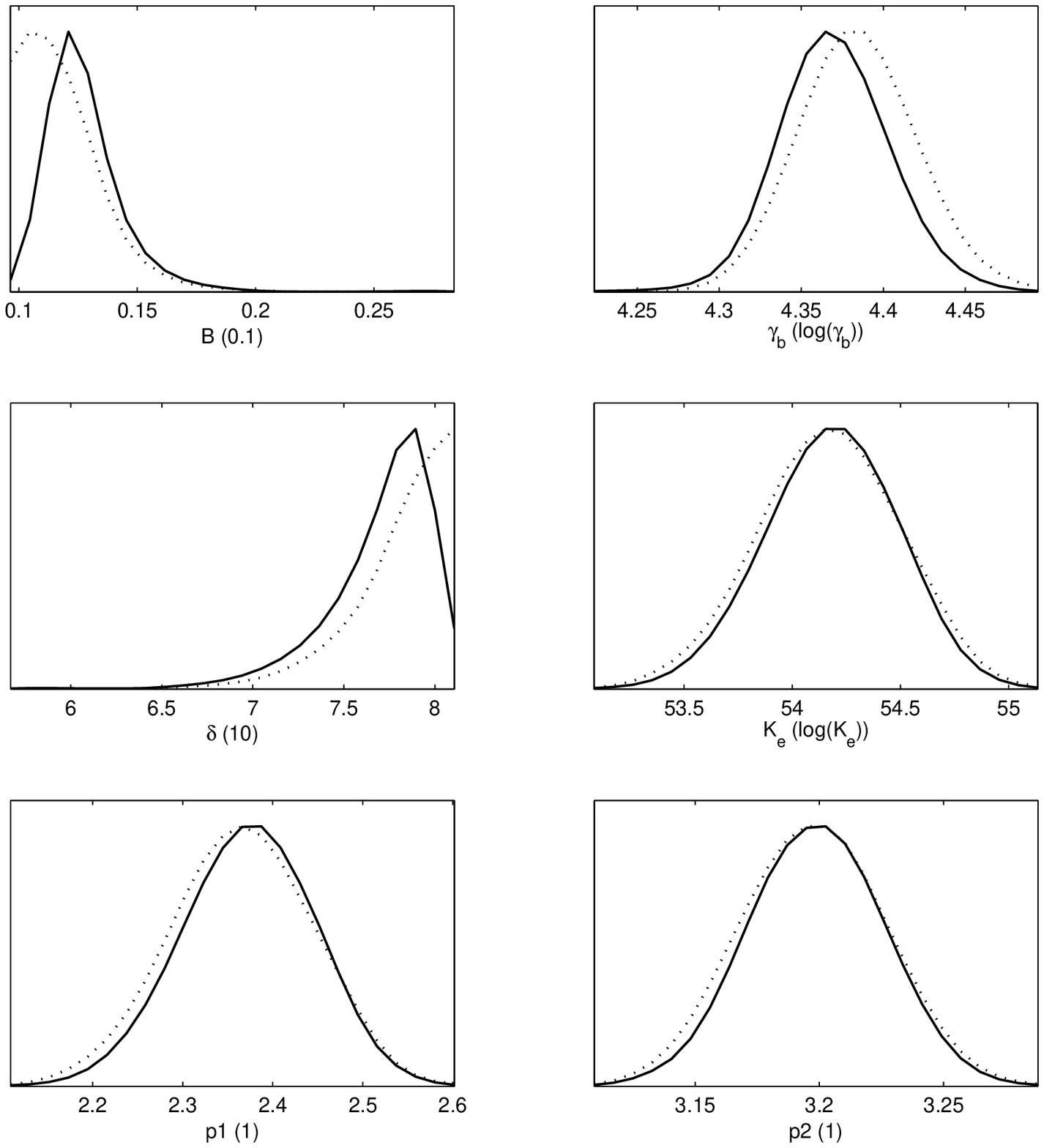}
\hfill
\includegraphics[width=8.0cm,height=5.3cm]{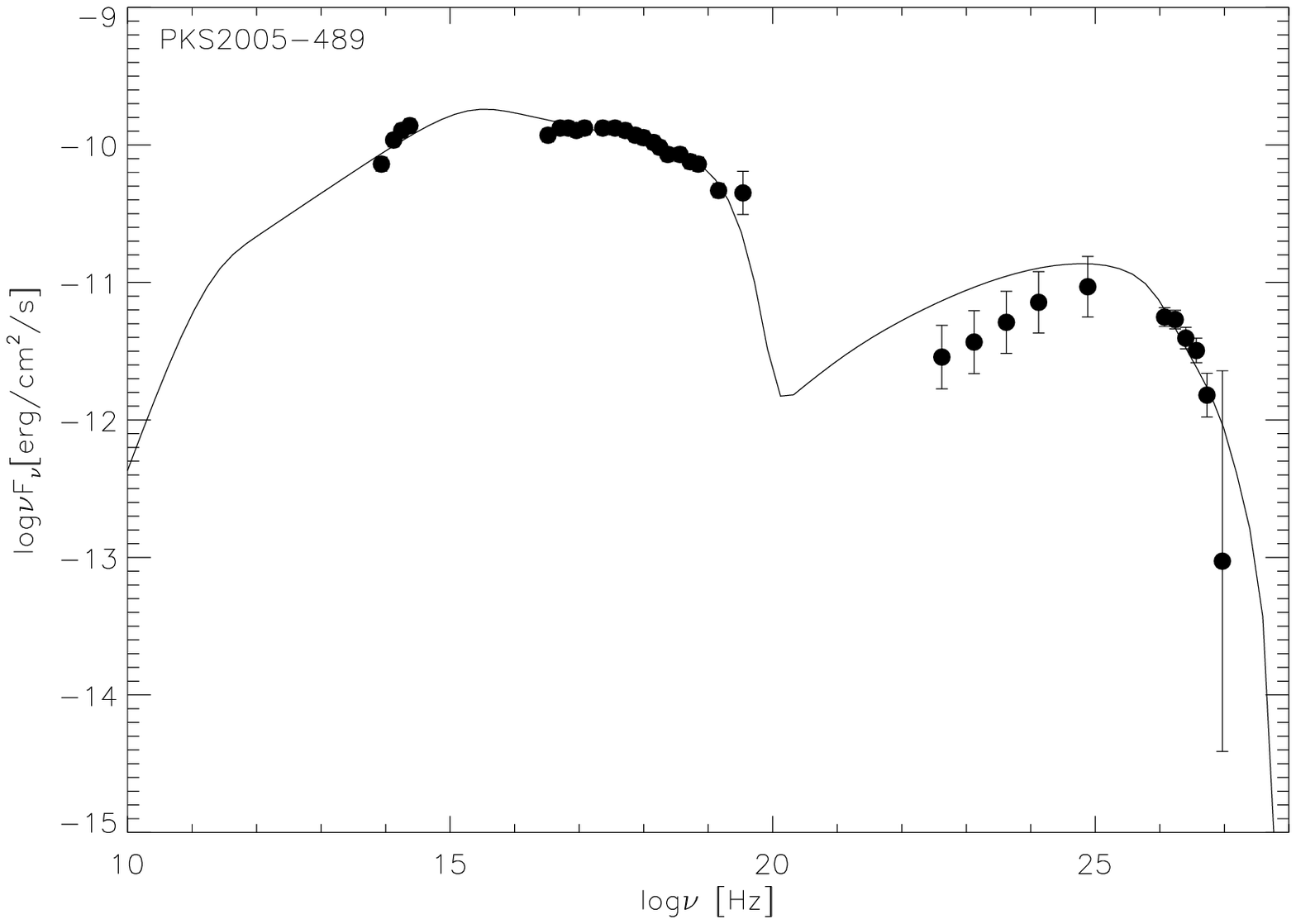}
\hfill

    \end{tabular}
  \end{center}
    \center{\textbf{Fig. 6.}---  continued}

\end{figure*}

\begin{figure*}
  \begin{center}
   \begin{tabular}{cc}

\includegraphics[width=5.8cm,height=5.0cm]{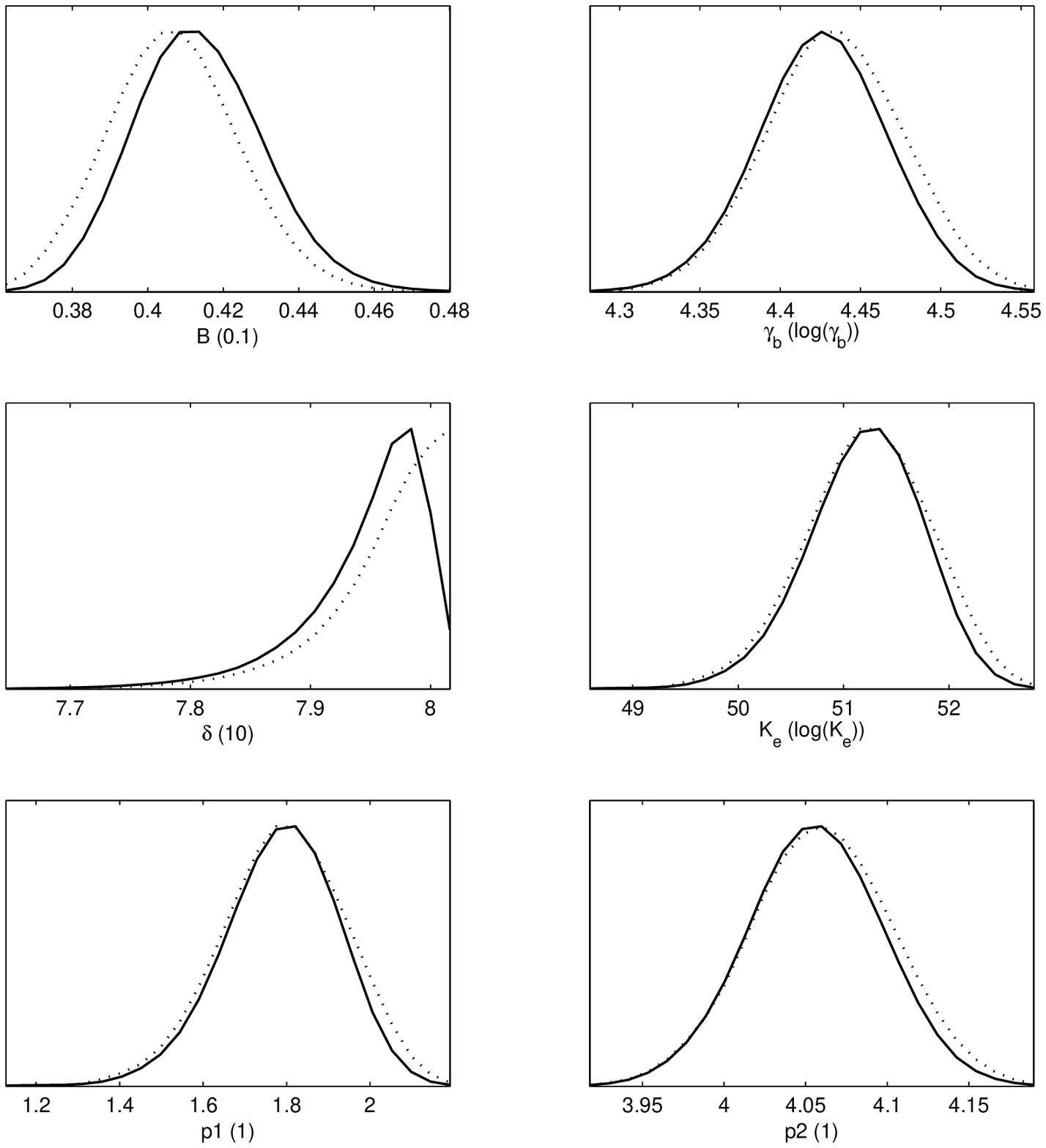}
\includegraphics[width=8.0cm,height=5.3cm]{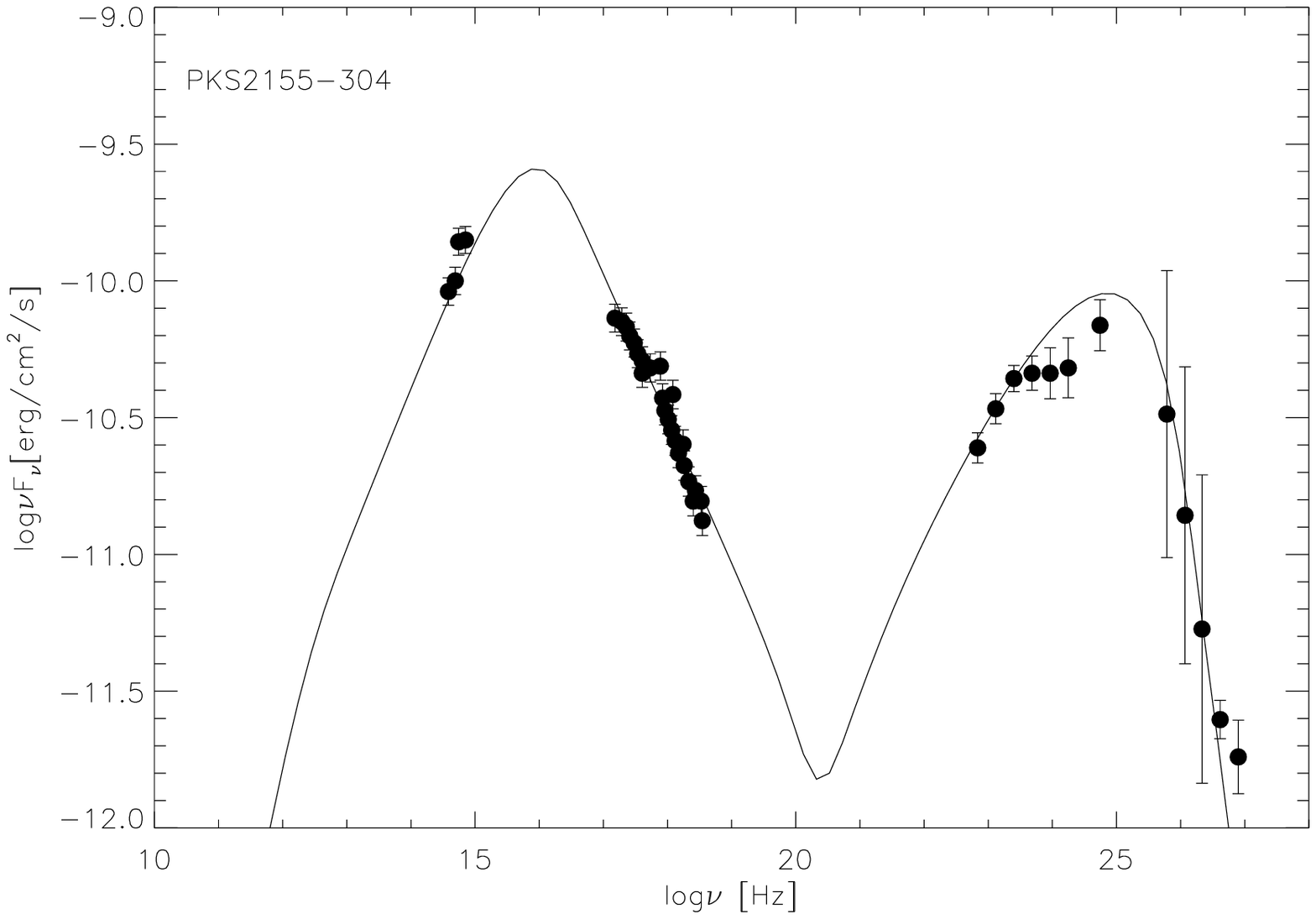}\\
\includegraphics[width=5.8cm,height=5.0cm]{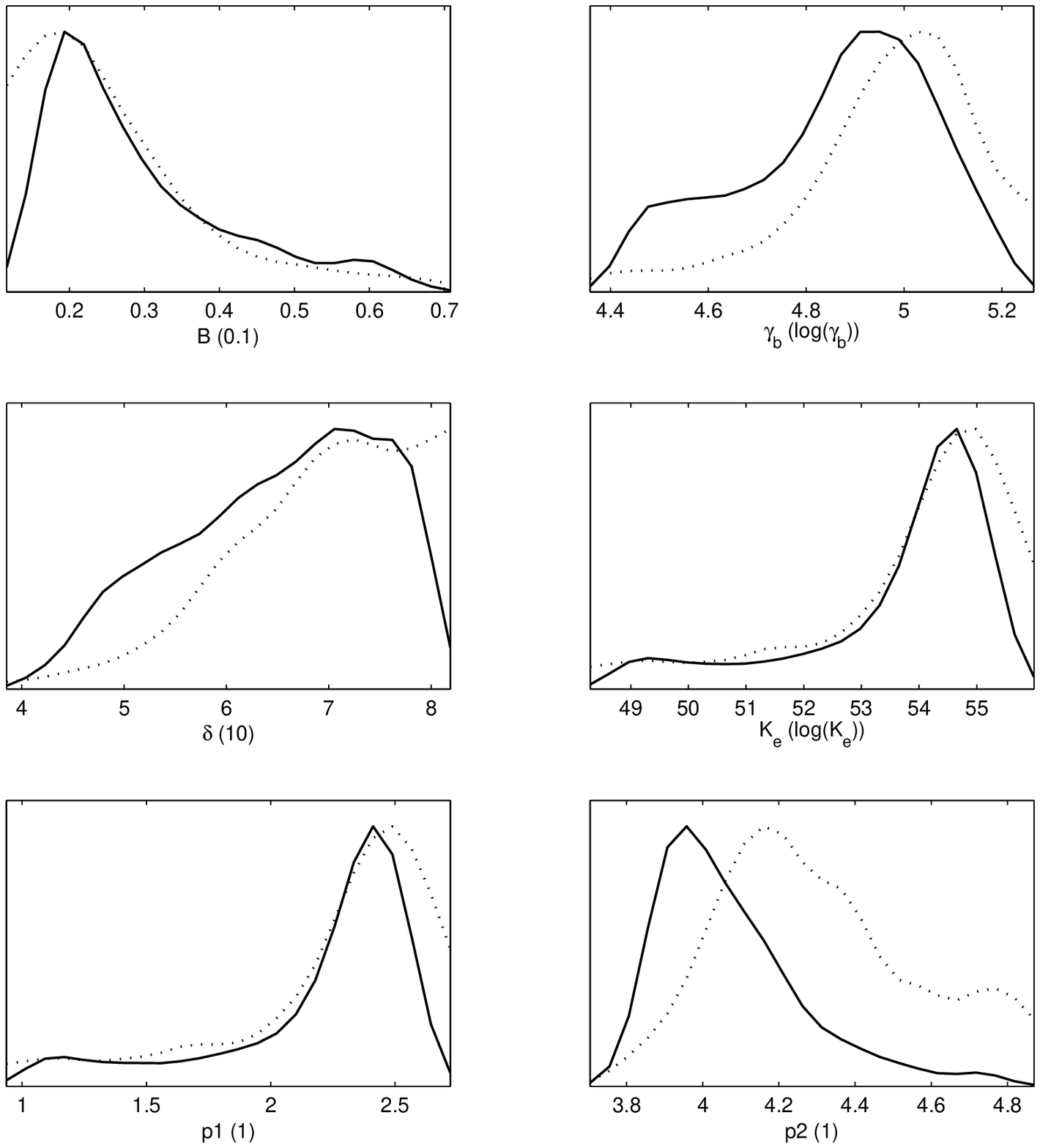}
\includegraphics[width=8.0cm,height=5.3cm]{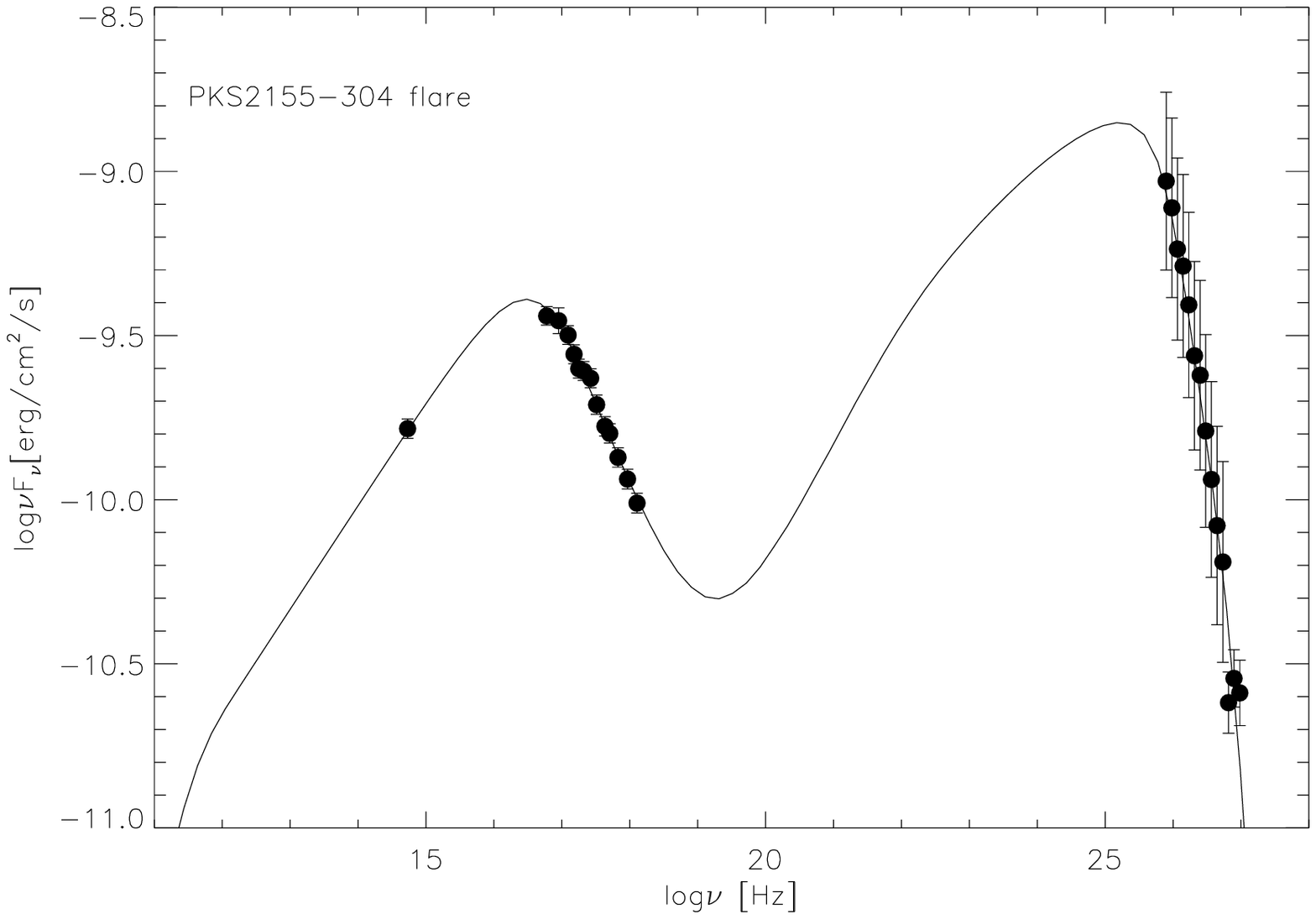}\\
\includegraphics[width=5.8cm,height=5.0cm]{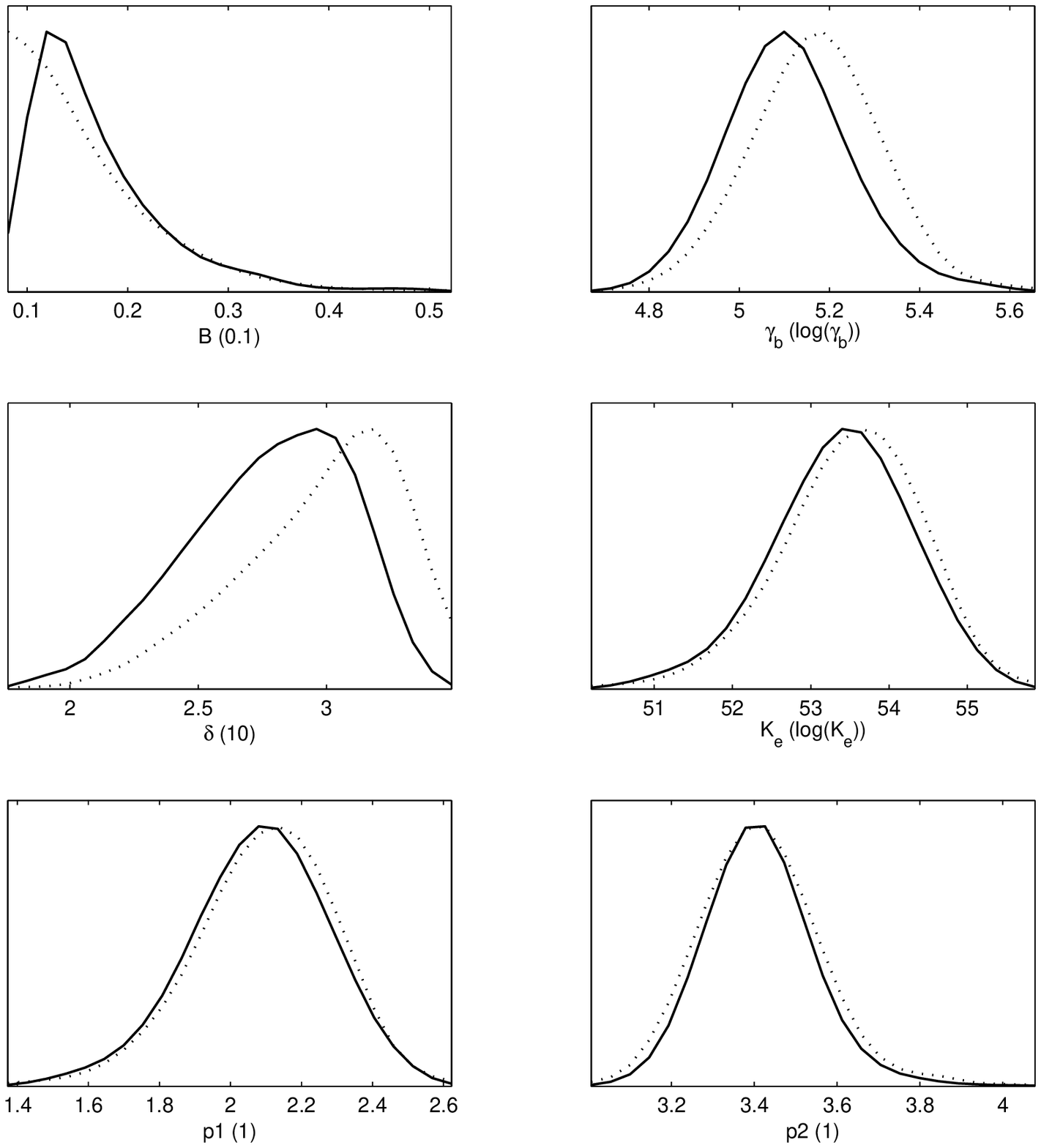}
\includegraphics[width=8.0cm,height=5.3cm]{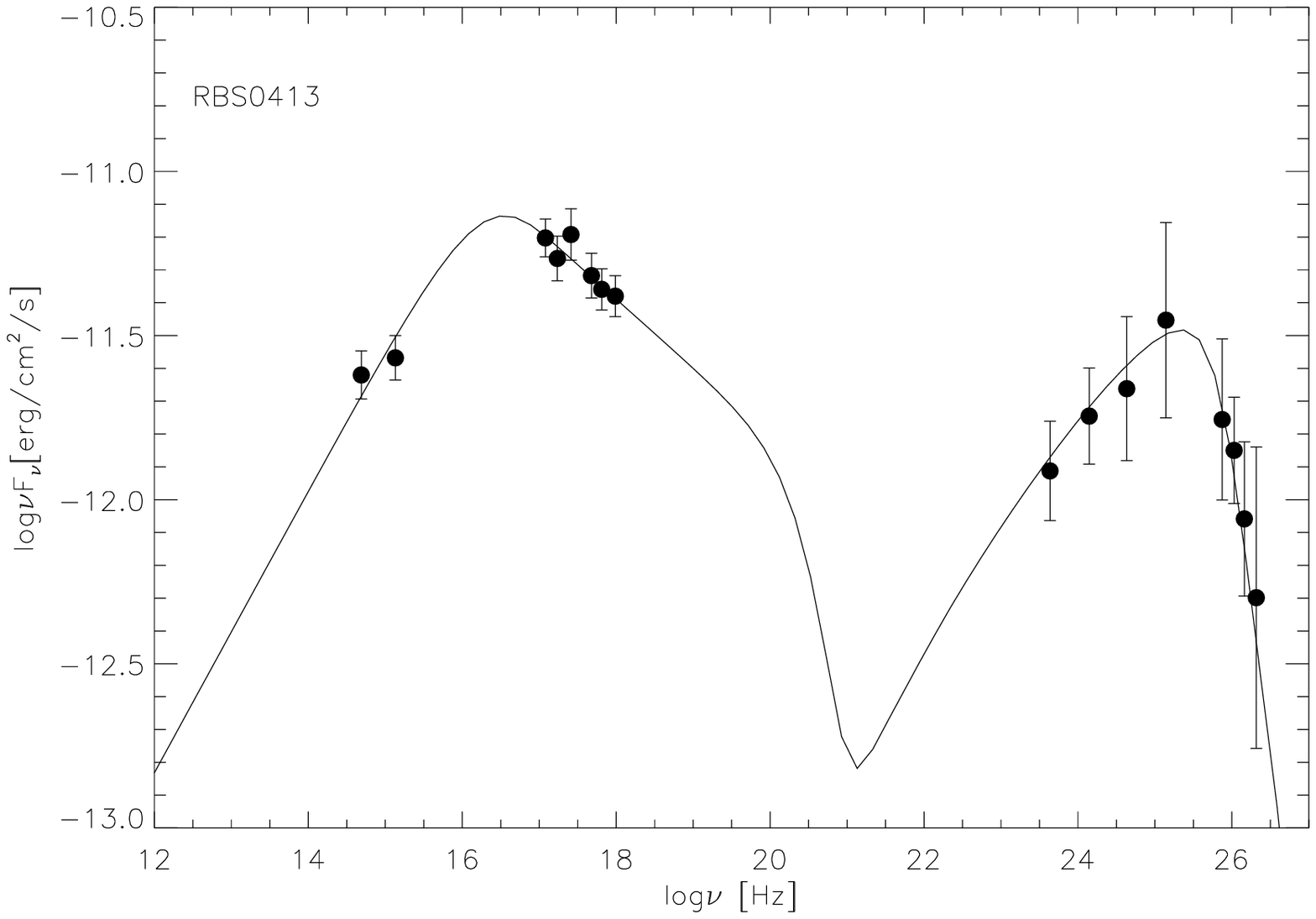}\\
\includegraphics[width=5.8cm,height=5.0cm]{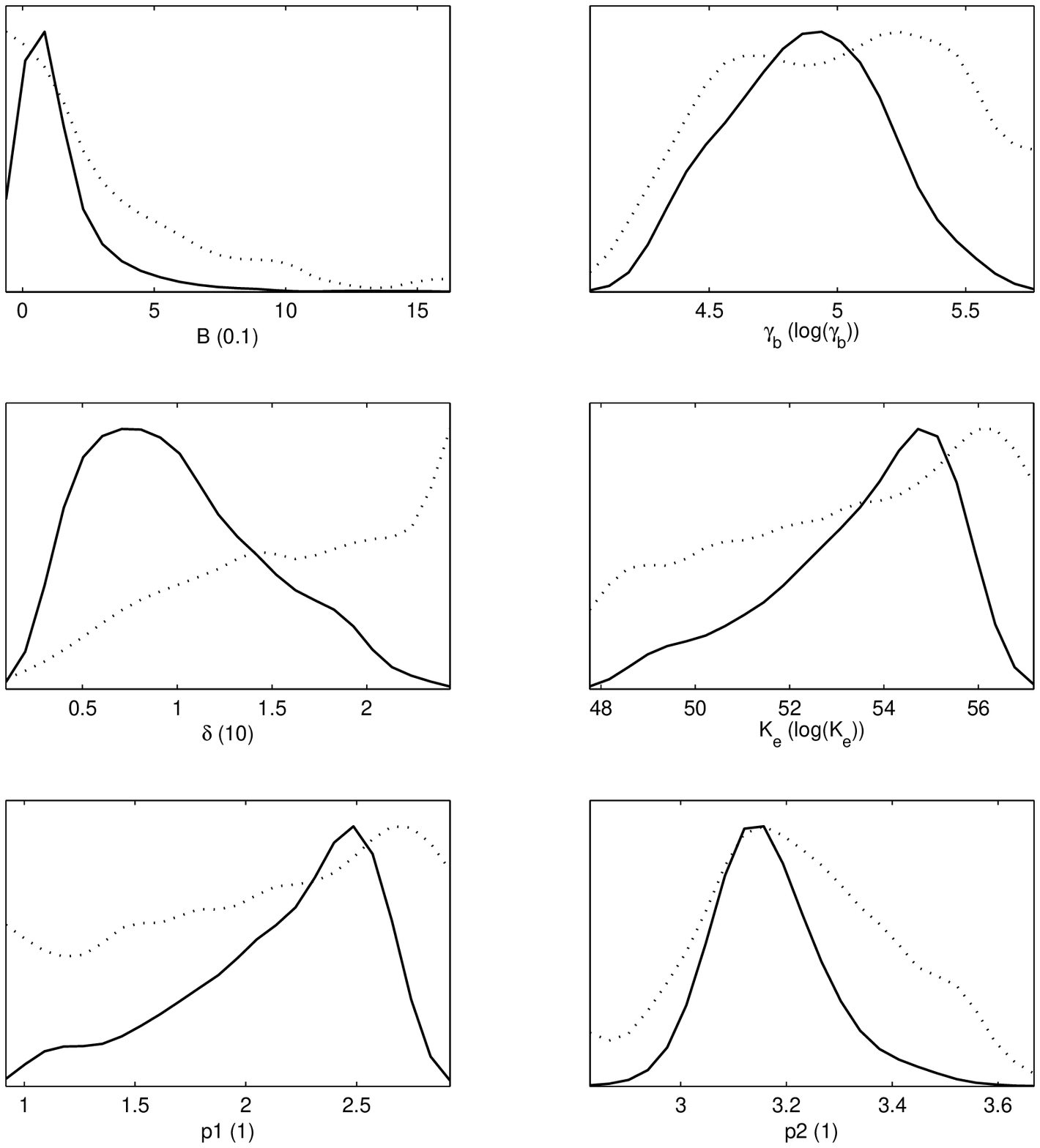}
\hfill
\includegraphics[width=8.0cm,height=5.3cm]{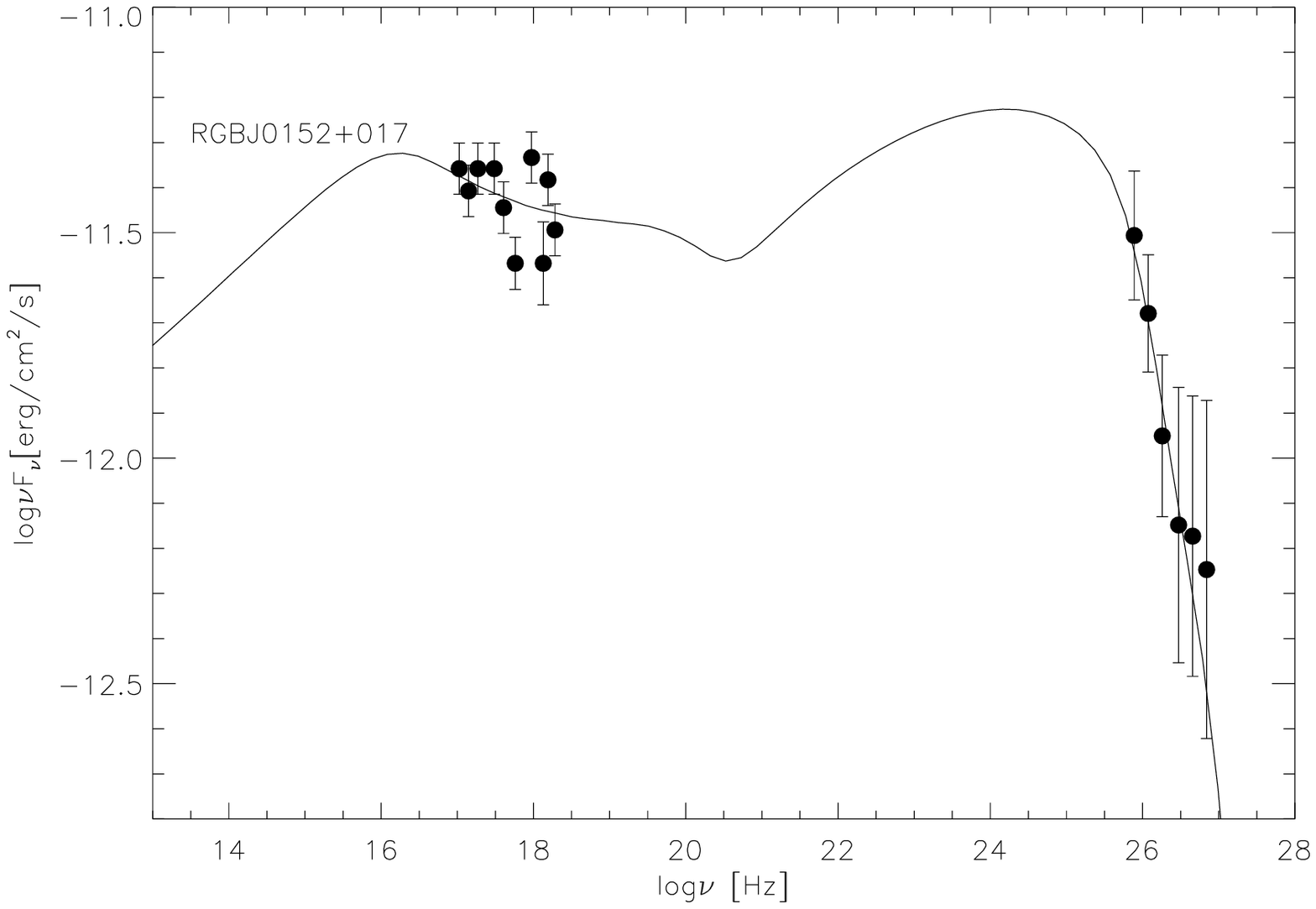}
\hfill

    \end{tabular}
  \end{center}
    \center{\textbf{Fig. 6.}---  continued}

\end{figure*}

\begin{figure*}
  \begin{center}
   \begin{tabular}{cc}

\includegraphics[width=5.8cm,height=5.0cm]{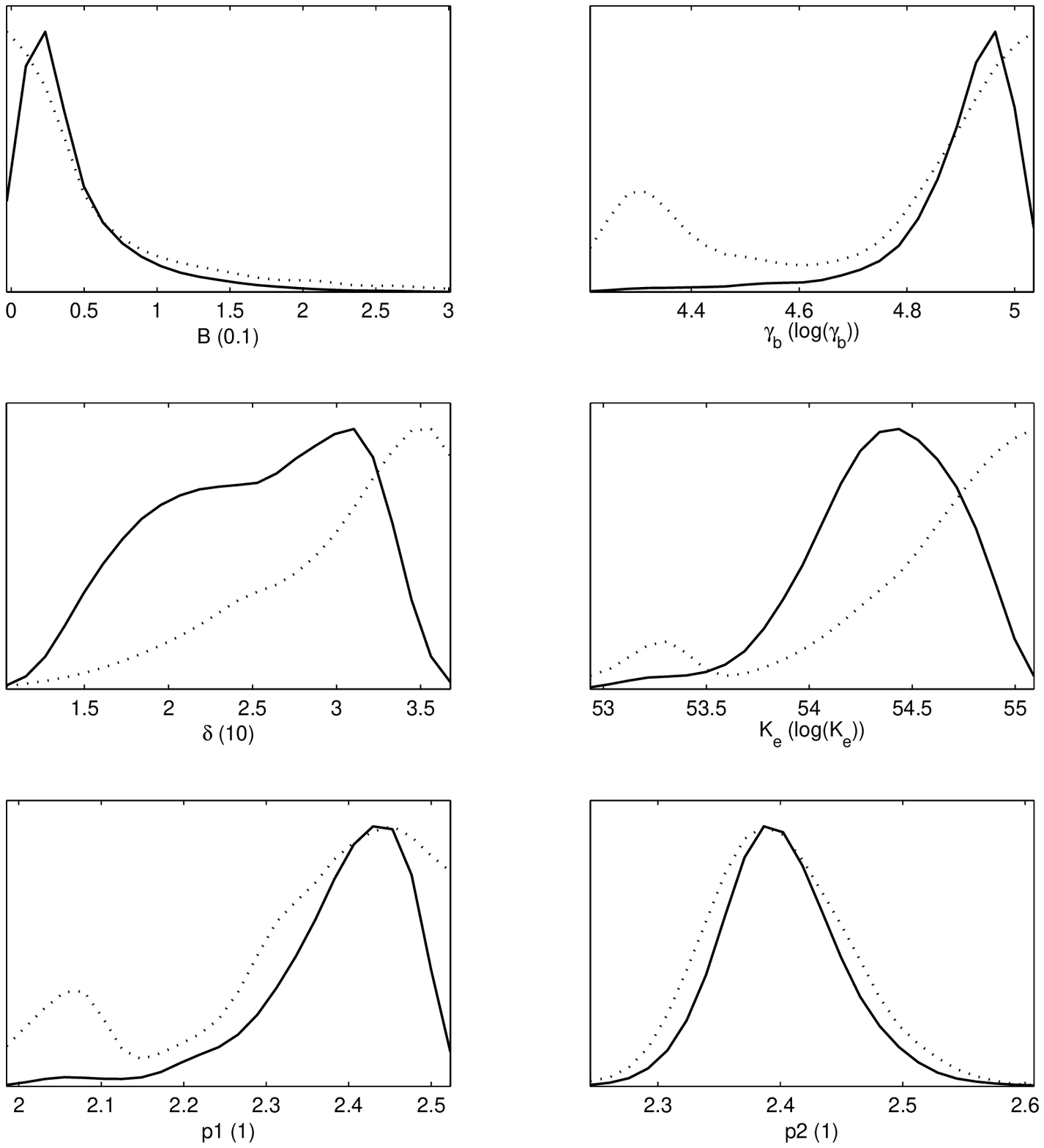}
\includegraphics[width=8.0cm,height=5.3cm]{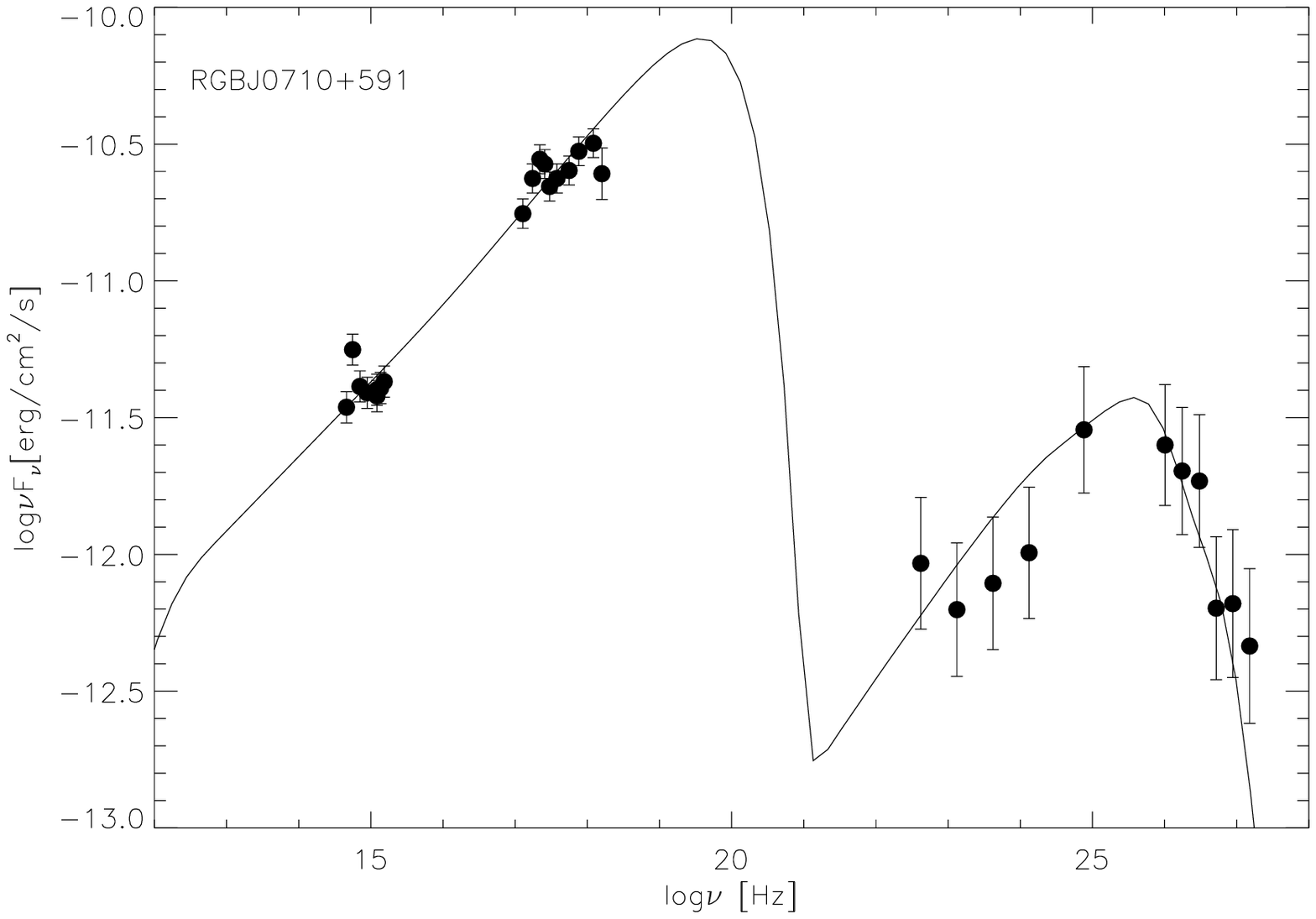}\\
\includegraphics[width=5.8cm,height=5.0cm]{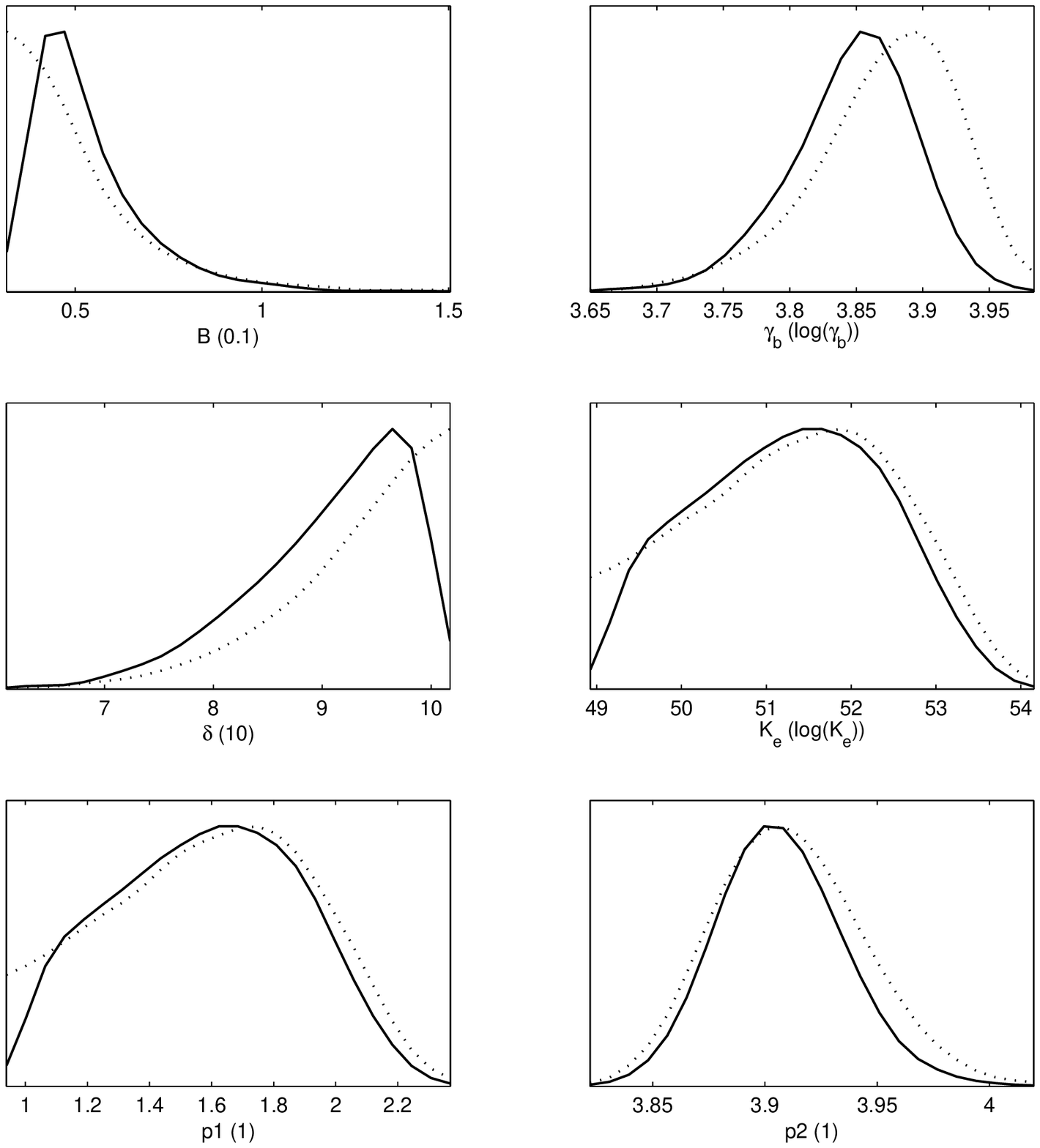}
\includegraphics[width=8.0cm,height=5.3cm]{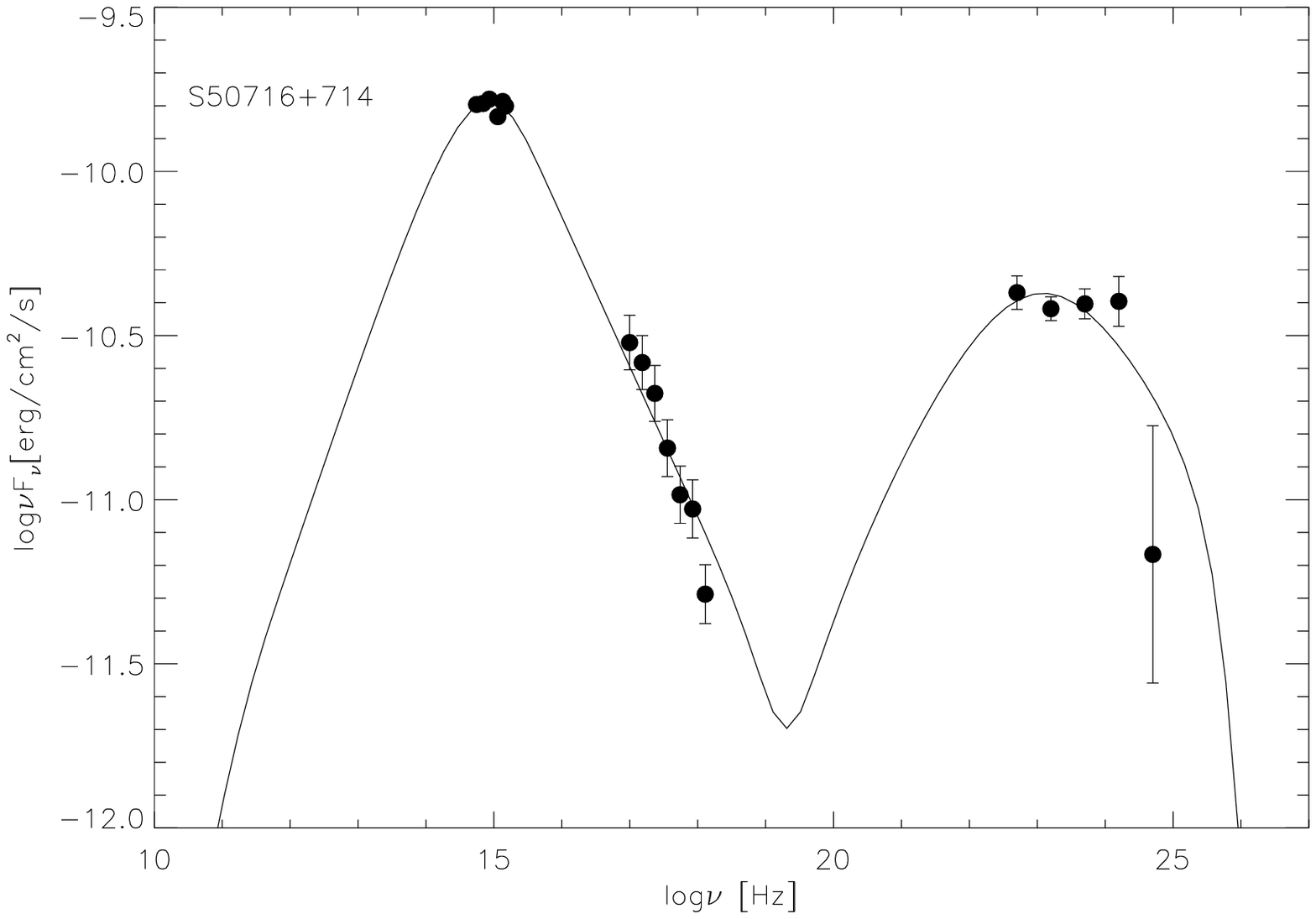}\\
\includegraphics[width=5.8cm,height=5.0cm]{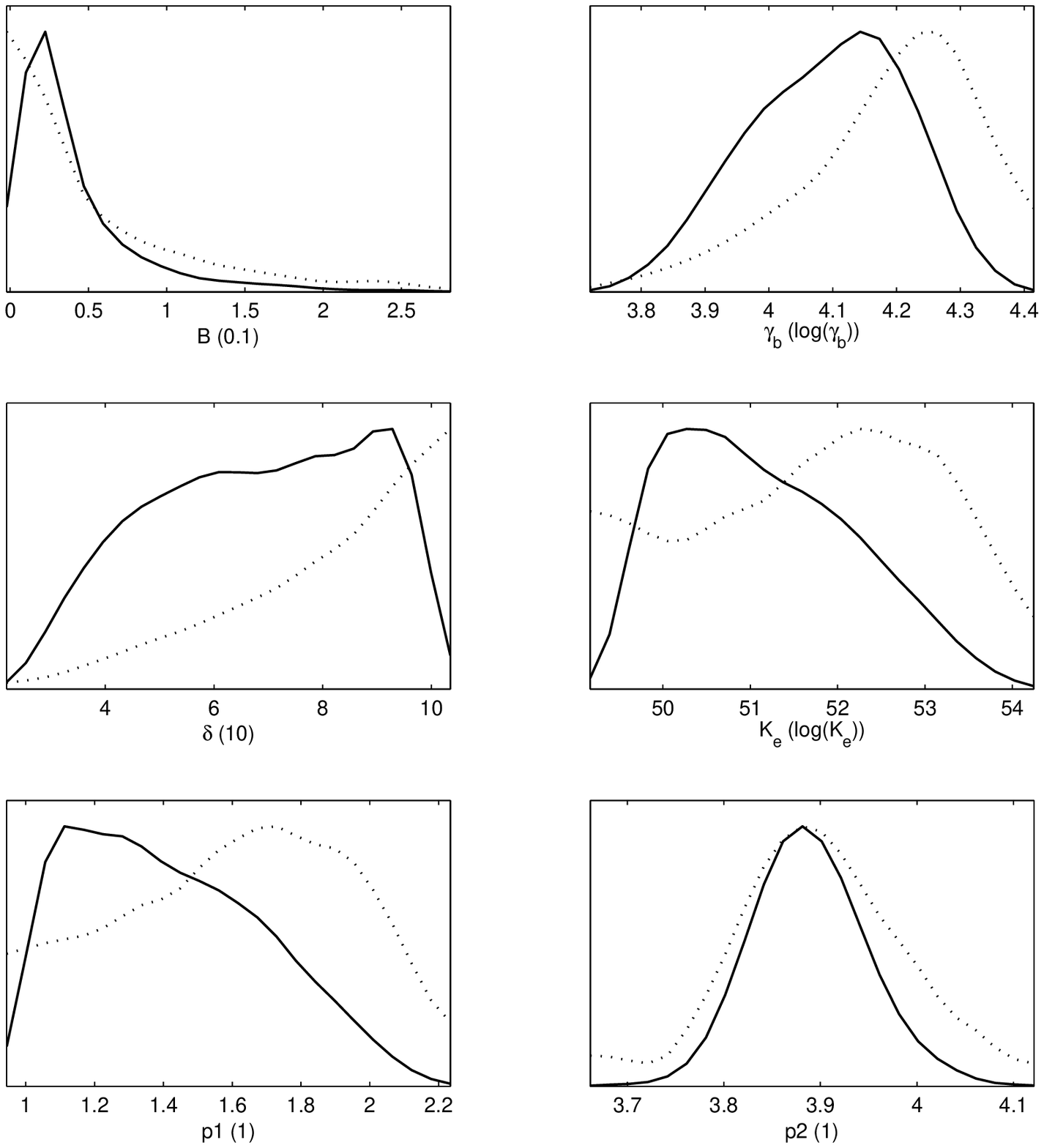}
\includegraphics[width=8.0cm,height=5.3cm]{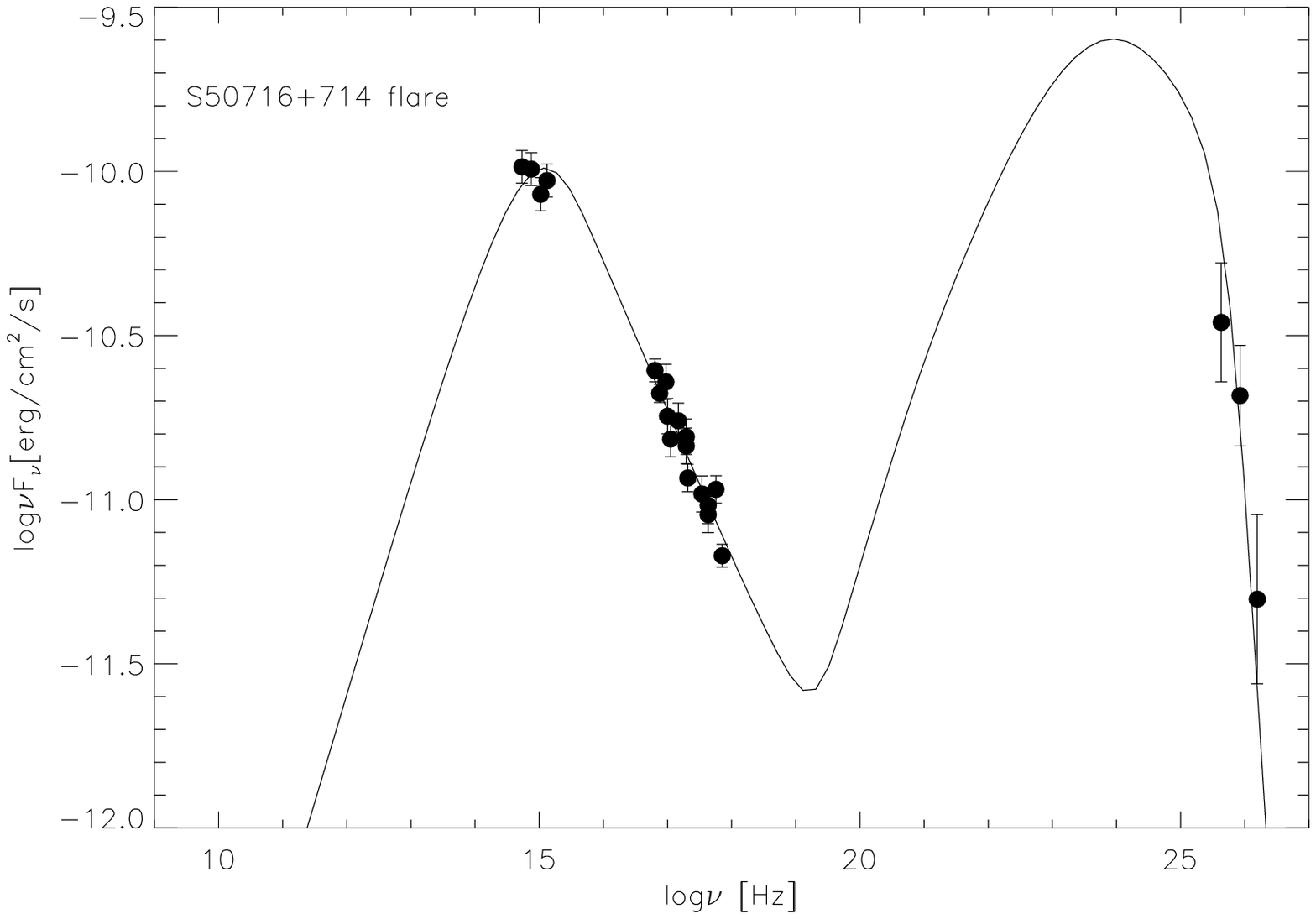}\\
\includegraphics[width=5.8cm,height=5.0cm]{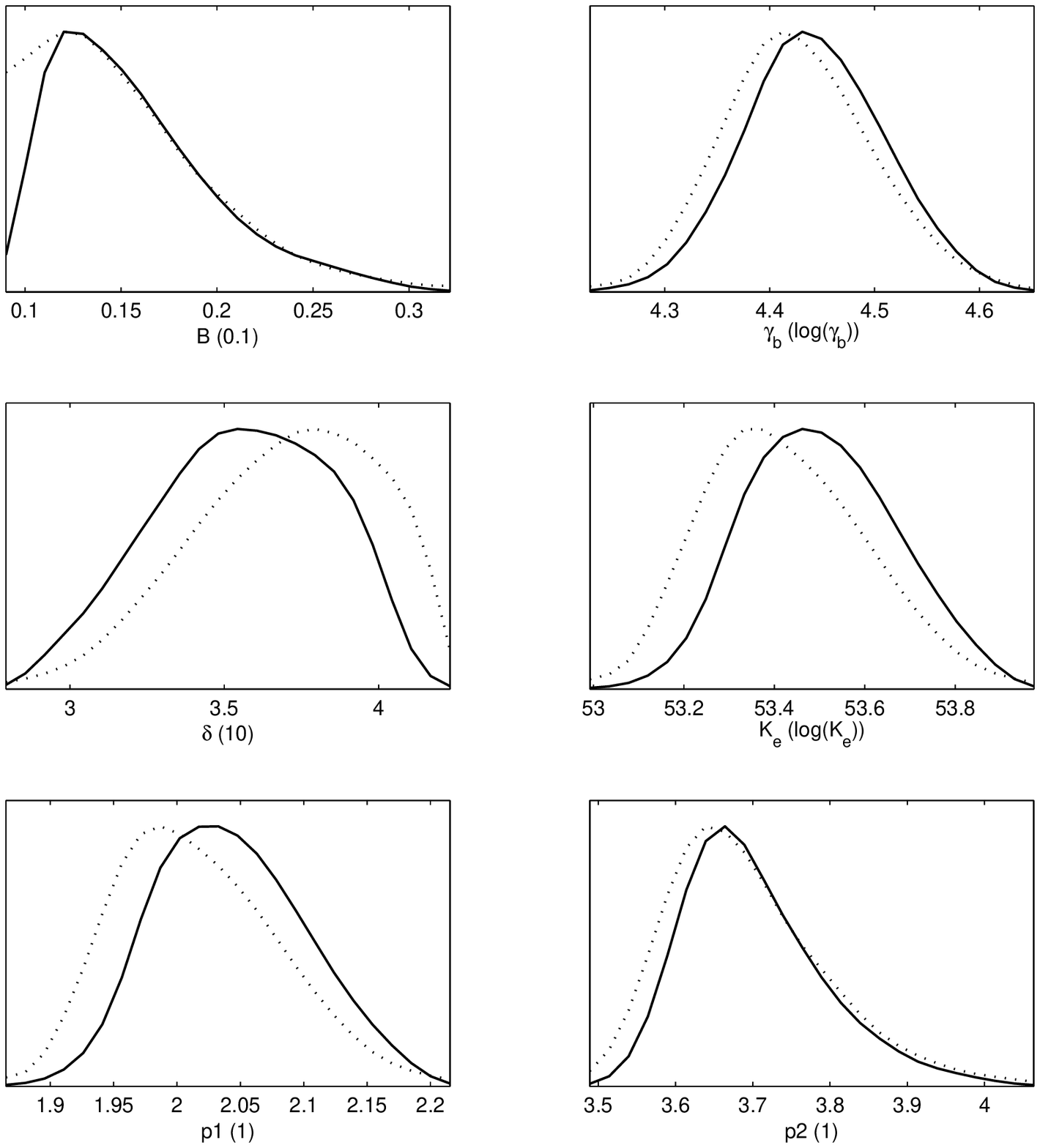}
\hfill
\includegraphics[width=8.0cm,height=5.3cm]{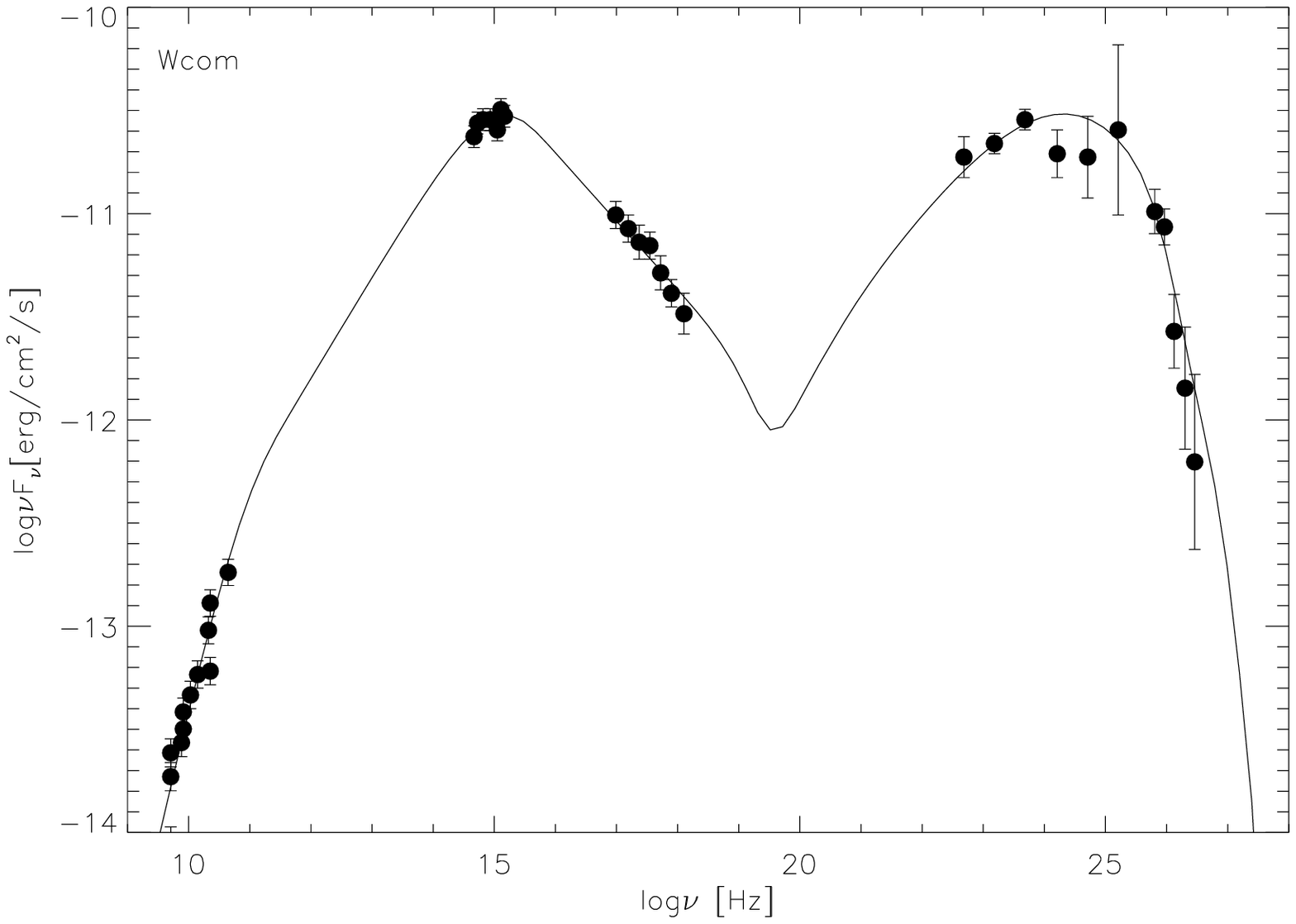}
\hfill

    \end{tabular}
  \end{center}
    \center{\textbf{Fig. 6.}---  continued}

\end{figure*}

\begin{figure*}
  \begin{center}
   \begin{tabular}{cc}

\includegraphics[width=5.8cm,height=5.0cm]{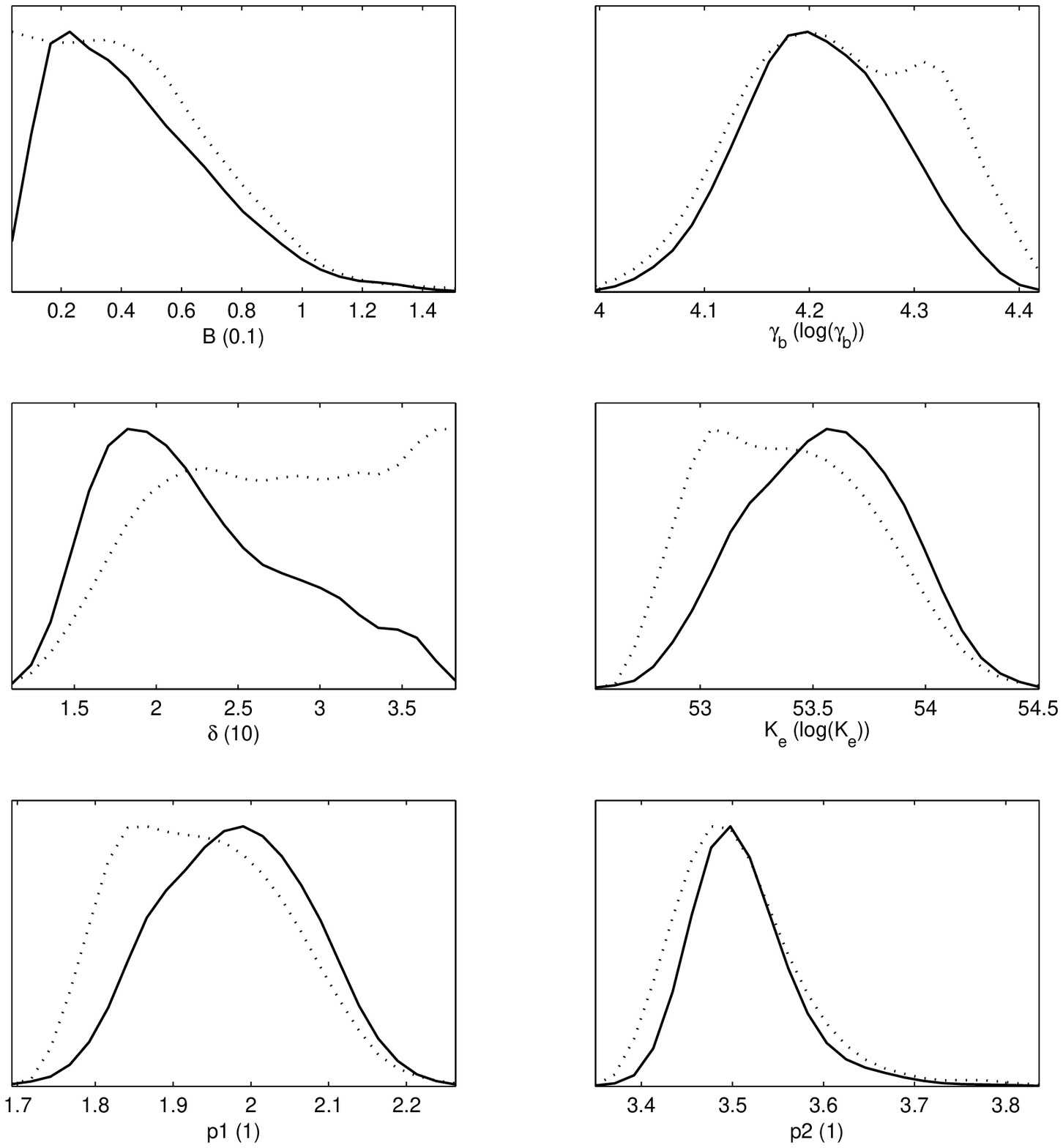}
\includegraphics[width=8.0cm,height=5.3cm]{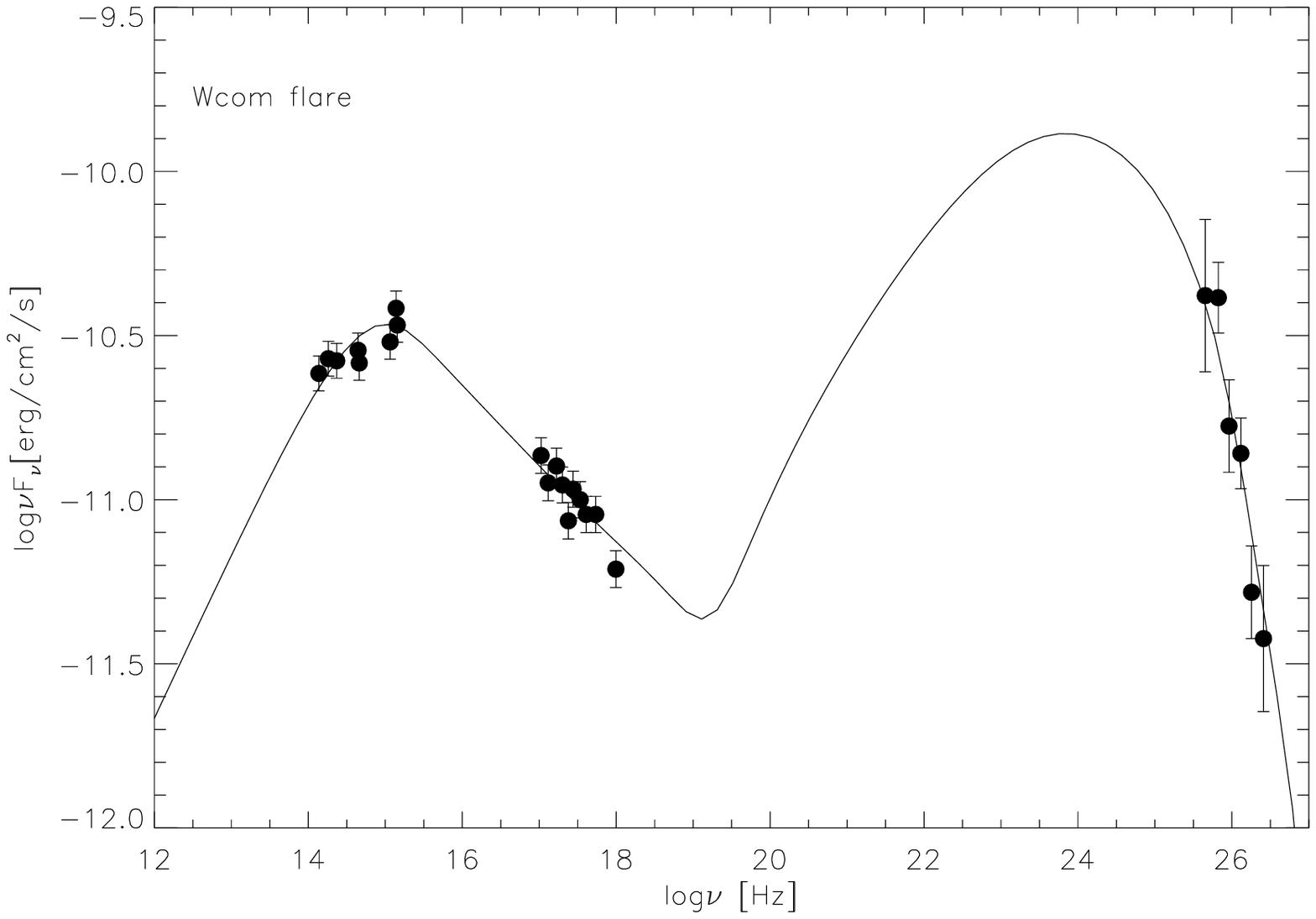}\\
\includegraphics[width=5.8cm,height=5.0cm]{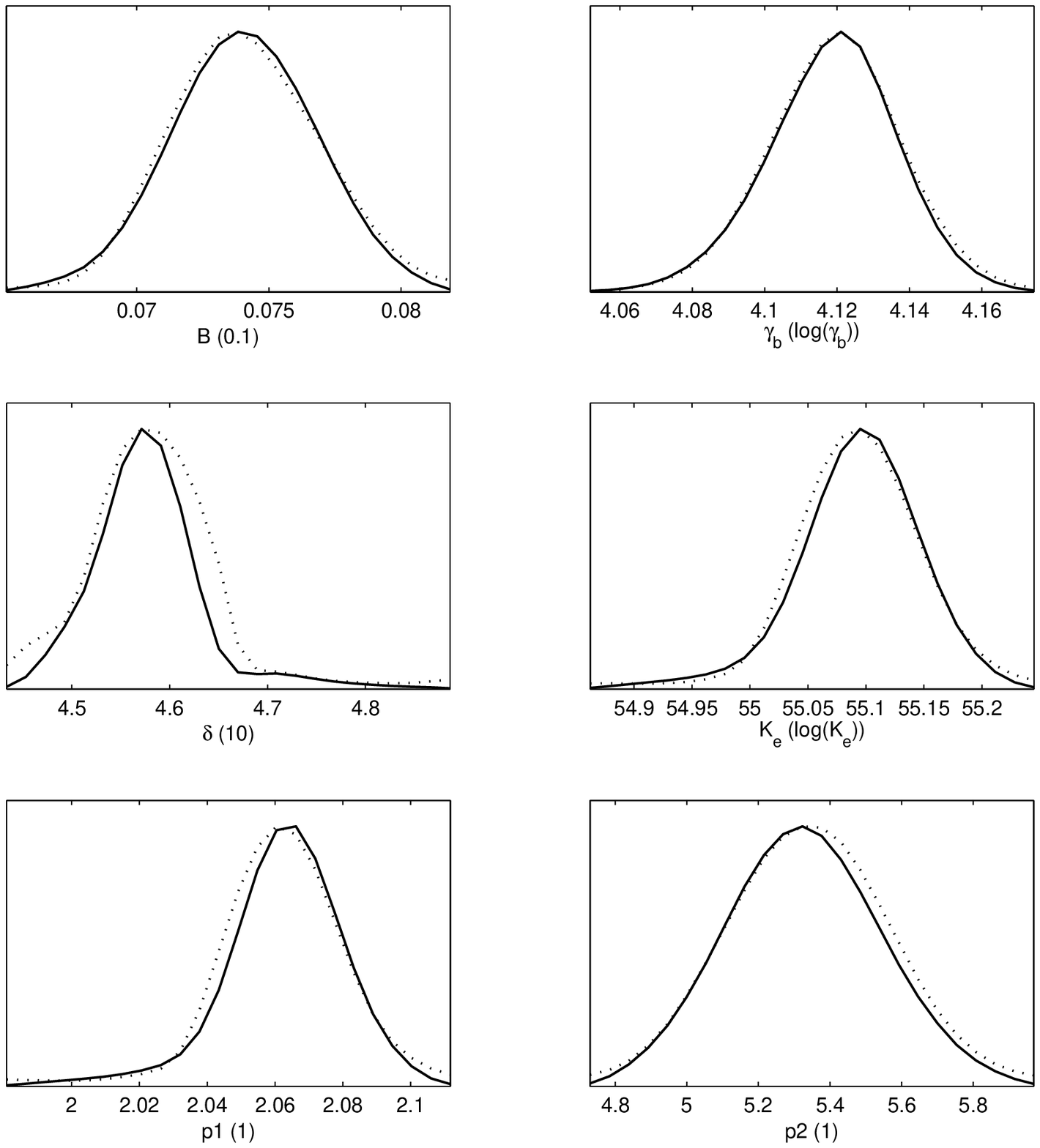}
\hfill
\includegraphics[width=8.0cm,height=5.3cm]{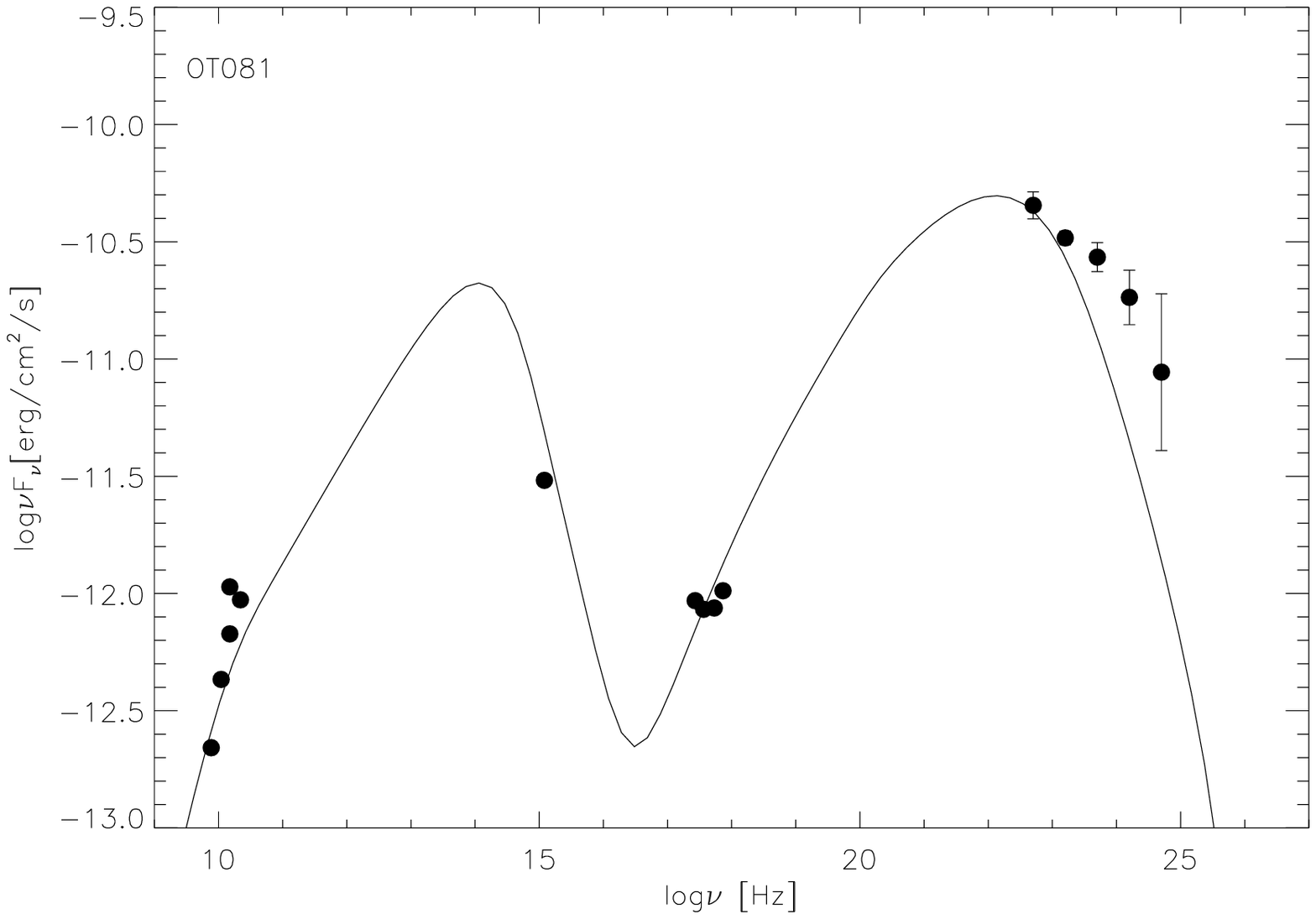}
\hfill

    \end{tabular}
  \end{center}
    \center{\textbf{Fig. 6.}---  continued}

\end{figure*}

\clearpage

\end{document}